\documentclass[aps,prc,amsmath,amssymb,aps,showpacs,superscriptaddress,nofootinbib]{revtex4-1}

\usepackage{graphicx,amssymb,amsmath,float}
\usepackage[dvips]{xcolor}
\usepackage{gensymb}
\usepackage{hyperref}
\usepackage{slashed}
\usepackage[utf8]{inputenc}

\usepackage{titlesec}

\hypersetup{
    colorlinks=false,
    linktocpage,
    }

\newcommand{\be}{\begin{equation}}
\newcommand{\ee}{\end{equation}}
\newcommand{\kbar}{\not{\!k}}
\newcommand{\pbar}{\not{\!p}}

\newcommand{\Qbar}{\not{\!Q}}

\newcommand{\Pbar}{\not{\!P}}

\newcommand{\ba}{\begin{eqnarray}}
\newcommand{\ea}{\end{eqnarray}}
\newcommand{\uu}{u({\bf p},s)}

\newcommand{\tauvec}{\mbox{\boldmath $\tau$}}

\newcommand{\nsigma}{\mbox{\boldmath $\sigma$}}

\newcommand{\nl}{{\bf      l}}

\newcommand{\nh}{{\bf      h}}

\newcommand{\nk}{{\bf      k}}
\newcommand{\np}{{\bf      p}}
\newcommand{\nq}{{\bf      q}}
\newcommand{\nr}{{\bf      r}}

\newcommand{\nx}{{\bf      x}}

\newcommand{\hy}{{\bf \hat{y}}}

\newcommand{\Kslash}{\slashed{K}}

\newcommand{\Qslash}{\slashed{Q}}

\newcommand{\non}{\nonumber}

\newcommand{\Ivec}{\mbox{\boldmath $I$}}

\newcommand{\munit}{\mbox{\boldmath $1\!\!1$}}

\newcommand{\psib}{\overline{\psi}}

\newcommand{\phib}{\phi^*}
\newcommand{\ubarN}{\overline{u}({\bf p}_N,s_N)}

\newcommand{\bma}{\begin{pmatrix}}
\newcommand{\ema}{\end{pmatrix}}

\newcommand{\half}{\frac{1}{2}}

\begin{document}
\title{Electron- versus neutrino-nucleus scattering}
\author{J.E.~Amaro} \affiliation{Departamento de F\'isica At\'omica,
  Molecular y Nuclear, and Instituto de F\'isica Te\'orica y
  Computacional Carlos I, Universidad de Granada, Granada 18071,
  Spain}
 \author{M.B.~Barbaro} \affiliation{Dipartimento di Fisica,
   Universit\`{a} di Torino and INFN, Sezione di Torino, Via P. Giuria
   1, 10125 Torino, Italy}
 \author{J.A.~Caballero} \affiliation{Departamento de F\'{i}sica
   At\'omica, Molecular y Nuclear, Universidad de Sevilla, 41080
   Sevilla, Spain}
\author{R.~Gonz\'alez-Jim\'enez} \affiliation{Grupo de F\'isica
  Nuclear, Departamento de Estructura de la Materia, F\'isica
  T\'ermica y Electr\'onica, Facultad de Ciencias F\'isicas,
  Universidad Complutense de Madrid and IPARCOS, Madrid 28040, Spain}
\author{G.D.~Megias} \affiliation{University of Tokyo, Institute for
  Cosmic Ray Research, Research Center for Cosmic Neutrinos, Kashiwa,
  Japan} \affiliation{Departamento de F\'{i}sica At\'omica, Molecular
  y Nuclear, Universidad de Sevilla, 41080 Sevilla, Spain}
\author{I. Ruiz Simo} \affiliation{Departamento de F\'isica At\'omica,
  Molecular y Nuclear, and Instituto de F\'isica Te\'orica y
  Computacional Carlos I, Universidad de Granada, Granada 18071,
  Spain}

\date{\today}

\begin{abstract}
We illustrate the connection between electron and neutrino scattering
off nuclei and show how the former process can be used to constrain
the description of the latter. After reviewing some of the nuclear
models commonly used to study lepton-nucleus reactions, we describe in
detail the SuSAv2 model and show how its predictions compare with the
available electron- and neutrino-scattering data over the kinematical
range going from the quasi-elastic peak to pion-production and highly
inelastic scattering.
\end{abstract}


\maketitle

\tableofcontents

\section{Introduction}

\label{sec:intro}

Electron scattering is a powerful probe to study the
internal structure of atomic nuclei, due to the weakness of the
electromagnetic interaction as compared to the strong forces
\cite{Boffi:1996ikg,Waleckacvc}.  In the
last years the field has seen renewed interest due to its connection
to neutrino-nucleus scattering and the importance of nuclear physics
for the interpretation of long baseline neutrino oscillation
experiments~\cite{Alvarez-Ruso17,Katori17}.  These experiments (MiniBooNE, T2K,
NOvA, DUNE, T2HK, \ldots), will improve our knowledge
of the Pontecorvo-Maki-Nakagawa-Sakata matrix, in particular the
CP-violating phase which is related to the matter-antimatter asymmetry
in the Universe.
The achievement of the
general goal requires a good control of the systematic
uncertainties coming from the modelling of
neutrino-nucleus interactions. Hence the accurate description of these
reactions has become one of the top challenges for theoretical nuclear
physics \cite{Alvarez-Ruso17}.

In the energy regime of the above mentioned experiments, going from
about 100 MeV up to a few GeV, several mechanisms contribute to the
nuclear response: from the excitation of nuclear collective states in
the lowest energy part of the spectrum, up to the deep inelastic
scattering at the highest energy transfers, encompassing the
quasi-elastic region, corresponding to one-nucleon knockout, and the
resonance region, corresponding to the excitation of nucleon
resonances followed by their decay and subsequent production of pions
and other mesons. Nuclear models should be able to describe, as
consistently as possible, all these processes with a few percent
accuracy, a very challenging task.  Moreover, these models should be
relativistic, as requested by the high energies involved in the
experimental setup.

Due to the rareness of previous neutrino-nucleus experiments in this
energy regime, nuclear models must be validated against
electron scattering experiments, closely
related to neutrino scattering.
The huge amount of
electron scattering data allows to discriminate between
theoretical models much better than comparing with neutrino
scattering data.
 Therefore, it is mandatory that any model used in the
analysis of neutrino oscillation experiment is first validated against
electron scattering data. In this paper we focus on the
connection between these two processes, illustrating and
comparing some of the models used to describe them
simultaneously.

The paper is organized as follows: in Sect.~\ref{sec:genform} we
introduce the general formalism for lepton-nucleus scattering and we
comment on the similarities and differences between the electron and
neutrino cases.  In Sect.~\ref{sec:QE} we present different models for
the quasielastic (QE) reaction, going from the simplest Relativistic
Fermi Gas to more realistic descriptions of the nucleus. Then we
discuss relativistic models for the reactions occurring at higher
energy transfer: two-particle-two-hole (2p2h) excitations
(Sect.~\ref{sec:2p2h}), pion production (Sect.~\ref{sec:pion}), higher
inelastic and deep inelastic scattering (Sect.~\ref{sec:DIS}). In
Sect.~\ref{sec:susav2-results} we concentrate on the SuSAv2 model,
developed by our group for both electron and neutrino reactions,
collecting in a coherent way the main results obtained in our past
work and illustrating the possible implementation of the model in
Monte Carlo generators. Finally, in Sect.~\ref{sec:concl}, we draw our
conclusions and outline the future developments of our approach.

\section{Connecting electron and neutrino scattering}
\label{sec:genform}

In order to emphasize the similarities and differences between
electron- and (anti)neutrino-nucleus scattering, here we present in
parallel the general formalism for the two processes.  Let us consider
the scattering problem 
$$ l+A \rightarrow l^\prime+X+B $$ where an incident lepton $l$ of
mass $m$ and 4-momentum $K^\mu=(\epsilon,\bf k)$~\footnote{In this
  paper we shall use the units $\hbar=c$=1.} scatters off a nuclear
target of mass number $A$ (a nucleus or a single nucleon) and a lepton
$l^\prime$ of mass $m^\prime$ and 4-momentum
$K^{\prime\mu}=(\epsilon^\prime,\bf k^\prime)$ emerges, with
$\epsilon=\sqrt{{\bf k}^2+m^2}$ and $\epsilon^\prime=\sqrt{{\bf
    k^\prime}^2+m^{\prime 2}}$, together with a system $X$ and
residual nucleus $B$.

For electron scattering $m=m'=m_e$, whereas for neutrino scattering
$m=m_{\nu_l}$ and $m'=m_l$ for charged-current (CC) processes, where a
$W$ boson is exchanged, or $m'=m_{\nu_l}$ for neutral-current (NC)
processes, associated to the exchange of a $Z^0$ boson.

Three kinds of cross sections can be considered in these reactions:
(i) The {\it inclusive} cross
section $(l,l^\prime)$, where only the outgoing lepton is detected. It
corresponds to the integration over all the final particles, except
the outgoing lepton, and a sum over all open channels compatible with
the kinematics. (ii) The {\it semi-inclusive} cross
section $(l,l^\prime X)$, which corresponds to detecting, in coincidence, the
lepton $l^\prime$ and the particle (or a system of particles) $X$.
(iii) Finally, the {\it exclusive} cross section, corresponding to the
      simultaneous detection of {\it all} the final scattering
      products.
      
    Exclusive processes are more challenging from both experimental
    and theoretical points of view: they are more difficult to measure
    and more sensitive to the details of the nuclear model
    description.
    Different nuclear models can
    describe equally well inclusive data, while exclusive or
    semi-inclusive data constitute a more stringent test for the
    validity of a model and can better discriminate between different
    theories.
 
     In this article we mainly focus on the case of inclusive
     reactions, where more data are available and nuclear models are
     more solid.  It should be kept in mind, however, that the
     validity of nuclear models is better tested against
     semi-inclusive and exclusive data, and that a good description of
     the latter is desirable for models to be implemented in Monte
     Carlo generators used to analyze neutrino oscillation
     experiments.
 
     In Born approximation
   \cite{Donnelly:1985ry,Moreno:2014kia}
     the lepton-nucleus differential cross section
   $d\sigma$ is proportional to the contraction of the leptonic
   $(\eta_{\mu\nu})$ and hadronic $(W^{\mu\nu})$ tensors, defined as
    \begin{eqnarray}
    \eta_{\mu\nu}(K,K') &=& 2mm^\prime \overline\sum_{ss'} j_\mu^*(K's',Ks) j_\nu(K's',Ks) 
    \label{eq:eta}
    \\
    W^{\mu\nu} &=& \overline\sum_{SS'} J^{\mu\, *}_{S'S} J^\nu_{S'S} \,,
    \label{eq:W}
    \end{eqnarray}
    where $j^\mu_{s's}$ and $J^\mu_{S'S}$ are the leptonic
    and hadronic 4-current matrix elements, respectively.
    The symbol $\overline\sum$ represents average and sum over
    the initial and final leptonic ($s,s'$) or hadronic
    ($S,S'$) spin quantum numbers if the
     initial lepton and hadron
    are not polarized and the final
    polarizations are not measured.

The leptonic current entering the tensor $\eta^{\mu\nu}$ can be
    written as
    \begin{equation}
        j^\mu(K's',Ks) = \overline u(K',s')(a_V\gamma^\mu+a_A\gamma^\mu\gamma_5)u(K,s)\,,
        \label{eq:jmu}
    \end{equation}
    where $u(K,s)$ is the lepton Dirac spinor and, according to the
    Standard Model of electroweak interactions, $a_V=1$ while $a_A=0$
    for unpolarized electron scattering (assuming only pure
    electromagnetic interaction~\footnote{Electrons can also interact
      weakly: in this case $a_V=4\sin^2\theta_W - 1$ and $a_A=-1$.})
    and $a_A=\mp 1$ in the neutrino $(-)$ or antineutrino $(+)$
    case. By inserting Eq.~(\ref{eq:jmu}) into (\ref{eq:eta}) and
    performing the traces one gets
    \begin{equation}
    \label{lepton_tensor}
        \eta_{\mu\nu}(K,K') = \eta_{\mu\nu}^{VV}+\eta_{\mu\nu}^{AA}+\eta_{\mu\nu}^{VA}
    \end{equation}
    with
    \begin{eqnarray}
        \eta_{\mu\nu}^{VV} &=& a_V^2\left[K_\mu K'_\nu+K'_\mu K_\nu-g_{\mu\nu} \left(K\cdot K'-mm'\right)\right]
        \label{eq:eta_VV}
        \\
        \eta_{\mu\nu}^{AA} &=& a_A^2\left[K_\mu K'_\nu+K'_\mu K_\nu-g_{\mu\nu} \left(K\cdot K'+mm'\right)\right]
        \\
        \eta_{\mu\nu}^{VA} &=&
        -2 i a_V a_A\varepsilon_{\mu\nu\alpha\beta} K^\alpha K'^\beta \,, \label{eq:eta_VA}
    \end{eqnarray}
    where the upper labels $VV$, $AA$ and $VA$ correspond
    to the vector-vector, axial-axial, and vector-axial
    components of the  leptonic tensor.
  
      The number of variables upon which the hadronic tensor
      $W^{\mu\nu}$, and hence the differential cross section, depend
      varies with the type of reaction considered: the inclusive
      $(l,l^\prime)$ cross section is a function of only 2 independent
      kinematic variables, {\it e.g.} the energy and scattering angle
      of the outgoing lepton; the semi-inclusive $(l,l^\prime N)$
      cross section, where the final lepton and a nucleon $N$ are
      detected in coincidence, depends on 5 variables, and so on. For
      semi-inclusive processes in which $n$ final particles are
      detected beyond the lepton, the kinematical variables are
      2+3$n$. Therefore the degree of complexity of nuclear response
      increases with the "exclusiveness" of the process.
    
    The number of components of the complex, hermitian $(W^{\mu\nu
      *}=W^{\nu\mu})$, hadronic matrix contributing to the cross
    section also depends on the specific process. In the most general
    case, the contraction $\eta_{\mu\nu}W^{\mu\nu}$ gives rise to 16
    independent terms. These are usually organized into real linear
    combinations of the hadronic tensor components, the {\it nuclear
      response functions}
    \begin{equation}
        R_K, \ \ \ \ \ K=CC,CL,LL,T,TT,TC,TL,T',TC',TL',\underline{TT},\underline{TC},\underline{TL},\underline{CL'},\underline{TC'},\underline{TL'} \,,
    \end{equation}
    where the labels $C$, $L$, $T$ and $T'$ correspond to a reference
    frame where the $z$-axis ($\mu=3$) is parallel to the momentum
    transfer ${\bf q}={\bf k}-{\bf k'}$ and refer to the charge
    ($\mu=0$), longitudinal ($\mu=3$) and transverse ($\mu=1,2$)
    projections of the two currents building up the response.  A
    similar decomposition can be performed for the leptonic tensor, in
    such a way that the tensor contraction can be written as
    \begin{equation}
        \eta_{\mu\nu} W^{\mu\nu} = v_0 \sum_K V_K R_K \,,
    \end{equation}
    where the coefficients $v_0 V_K$ are linear combinations of the
    leptonic tensor components.  Their explicit expressions are given
    in Sect.~\ref{sec:QE}.
    
    In the case of semi-inclusive unpolarized electron scattering
    (pure electromagnetic interaction), due to absence of axial
    current and to the conservation of the vector current (CVC), the
    response functions reduce to 4:
    \begin{equation}
        d\sigma(e,e'N) \propto V_{L} R_{L}+V_{T} R_{T}+V_{TT} R_{TT}+V_{TL} R_{TL}  \,.
    \label{eq:dsem}
    \end{equation}
  
    In the case of inclusive lepton scattering, as a consequence of
    the integration over the full phase space of the outgoing hadron,
    some of the responses do not contribute and one has:
   \begin{eqnarray}
        d\sigma(\nu_l,l') &\propto& V_{CC} R_{CC}+ 2 V_{CL} R_{CL}+V_{LL} R_{LL}+V_{T} R_{T}+ V_{T'} R_{T'} \\
          d\sigma(e,e') &\propto& V_{L} R_{L}+V_{T} R_{T} \,.
    \end{eqnarray}
 In the neutrino case, each response can be decomposed as
 \begin{equation}
     R_K = R_K^{VV}+R_K^{AA}+R_K^{VA}
 \end{equation} 
 where the upper labels denote the vector or axial nature of the two
 hadronic currents entering the response in Eq.~(\ref{eq:W}).
The above decomposition into response functions is valid in all
kinematical regions (elastic, quasielastic, 2p2h excitations,
inelastic), characterized by different nuclear currents.  Further
details for the response functions will be illustrated in Section
\ref{sec:QE}, where calculations of the quasielastic electromagnetic
and weak cross sections will be presented in some specific nuclear
models.

\subsection*{What can we learn from electron scattering?}
\label{sec:electron}

    The formalism presented above shows that
    electron and neutrino scattering are closely related and
    validation against electron scattering data constrains models to
    be applied to neutrino scattering.
    Comparison with inclusive electron scattering data is a necessary,
    but not sufficient, test for nuclear models, because of two main
    differences:
    \begin{itemize}
        \item The weak current carried by neutrinos has a vector and
          an axial component, while the electromagnetic current is
          purely vector. As a consequence, neutrinos can probe the
          axial nuclear response, not accessible via unpolarized
          electron scattering.  This introduces uncertainties related
          to the knowledge of the axial and pseudoscalar form
          factors. In principle valuable information on the axial
          response could also be extracted from parity-violating (PV)
          electron scattering off
          nuclei~\cite{Musolf:1992xm,Donnelly:1991qy,PRRaul}, which
          could also provide important complementary information on
          nuclear
          correlations~\cite{Alberico:1993ur,Barbaro:1993jg,Barbaro:1995ez,Barbaro:1995gp}
          and on the radiative corrections entering in the isovector
          axial-vector
          sector~\cite{Gonzalez-Jimenez:2015tla}. However, few PV data
          on medium-heavy nuclei exist and are mostly limited to the
          elastic part of the spectrum.
        \item In typical electron scattering experiments the incident
          beam energy is known with good accuracy, hence the
          transferred energy, $\omega$, and momentum, $q$, can be
          precisely determined by measuring the outgoing lepton
          kinematics. In long baseline neutrino experiments the beam
          is not monochromatic: neutrinos are produced from meson
          decay with a more or less - depending on the experiment -
          broad distribution around an average value. This implies
          that each kinematic (momentum and scattering angle) of the
          outgoing lepton corresponds in general to a finite range of
          $\omega$ and $q$, which can in turn correspond to various
          overlapping processes. An important example of this
          difference is represented by the mixing of 1p1h and 2p2h
          excitations: in the $(e,e')$ the former correspond to the
          quasielastic peak, while the latter are peaked in the "dip"
          region between the QE and the $\Delta$ resonance peaks. In
          $(\nu_l,l)$ data there is no way to disentangle the two
          channels and one has to rely on nuclear models to
          reconstruct the neutrino energy from lepton kinematics.
    \end{itemize}

A large set of high quality inclusive electron-nucleus scattering
data has been collected in the past \cite{Benhar:2006wy},
covering an energy range from $~$100
MeV to several GeV, and various nuclei (from $^3He$ up to
$^{238}U$). Recently \cite{Dai:2018xhi,Dai:2018gch} the $(e,e')$ cross
section on argon and titanium has been measured at JLab, with the
specific purpose of constraining models used in the analysis of
neutrino-nucleus scattering. Some data are also available for the
separated longitudinal and transverse responses. These response functions
allow for a more stringent test of the models because nuclear effects
are different in the two channels.

\subsection*{Inclusive electron-nucleus scattering and Superscaling}
\label{subsec:scaling}

A powerful mean of extracting information on nuclear effects from
$(e,e')$ data at different beam energies and on different nuclei is
the superscaling analysis proposed in
\cite{Donnelly99a,Donnelly99b,Alberico:1988bv}. This approach
factorizes the cross section into a single-nucleon term
times a function - the {\it scaling function} - which embodies the
nuclear dynamics. Assuming that the latter is independent of the
specific probe, scaling allows one to predict inclusive
neutrino-nucleus cross sections using electron-nucleus
data~\cite{Amaro:2004bs}.

Scaling phenomena occur in various different fields, including atomic,
nuclear and hadronic physics, whenever a new scale is probed in a
certain process.  For example, $x$-scaling occurs in lepton-nucleon
scattering when the lepton, at high values of $Q^2$, is resolving the
inner structure of the proton (or neutron), interacting with its
constituents, the quarks. Scaling manifests when
the nucleon's structure
function, $F_{1,2}(Q^2,\nu)$, becomes a function of only one variable
(in this case the Bjorken variable $x$). In
lepton-nucleus scattering scaling occurs when the nuclear response,
depending in general on the two variables $q,\omega$, becomes function
of only one scaling variable, indicating that the probe interacts with
the nucleus' constituents, the nucleons. Since these, unlike the
quarks, are not point-like, the cross section has to be "reduced",
{\it i.e.}, divided by a single-nucleon function which takes into
account the internal structure of the nucleon.  In this context the
{\it scaling function} is defined as
the ratio between the double differential cross section and a
single-nucleon cross section $\overline{\sigma_{eN}}$:
\begin{equation}
  f(q,\omega) = \frac{k_F}{\overline{\sigma_{eN}}}
  \frac{d^2\sigma}{d\Omega_e d\omega}.
  \label{fcross}
\end{equation}
The single-nucleon cross section is not unique and takes into account
the added contribution from all the separate nucleons in average.  In
this work we use a single nucleon cross section averaged over the
relativistic Fermi gas to take into account that the initial nucleon is not at
rest:
\begin{equation}
  \overline{\sigma_{eN}}=
  \sigma_{Mott}(v_LG_L^{ee'}+v_TG_T^{ee'}).
\end{equation}
The single-nucleon factors
$G_K^{ee'}$ 
are taken from the relativistic Fermi gas model (RFG), and
are defined in Eqs. (\ref{eq:Rem},\ref{ul},\ref{ut})
of  Sect.\ref{section-rfg}.

The factor $k_F$ (Fermi momentum) in Eq. (\ref{fcross})
is introduced to remove the
dependence on the particular nucleus, also motivated by
the RFG model. 

The {\it scaling variable} is a particular combination of $q$ and
$\omega$ (see Appendix A).  In this work we use the scaling variable
$\psi=\psi(q,\omega)$, defined in detail in Appendix
\ref{scaling-appendix}.  Another common choice in the quasielastic
domain is the $y$-scaling variable, the difference being that $y$ is
defined in terms of nucleon momentum (or missing momentum) and $\psi$
is defined in terms of nucleon kinetic energy.  Scaling is satisfied
when $f(q,\omega) = f(\psi(q,\omega))$.  This occurs for high enough
values of the momentum transfer $q$, able to resolve the internal
structure of the nucleus. Qualitatively this corresponds to the region
$q\gtrsim$ 300-400 MeV/c, where low energy collective effects (like
giant resonances) become negligible.

The analysis of $(e,e')$ experimental data performed in
Refs.~\cite{Day:1990mf,Donnelly99a,Donnelly99b} shows that scaling is
satisfied with good accuracy in the "scaling region"
$\omega<\omega_{QEP}$, where $\omega_{QEP}$ corresponds to the maximum
of the quasielastic peak. It is broken at higher energies loss,
indicating that processes other than QE scattering, like pion
production and excitation of two-particle two-hole (2p2h) states
contribute to the cross section.

The above described $q$-independence of the scaling function $f$ when
expressed in terms of $\psi$ is also called {\it scaling of first kind},
in order to distinguish it from {\it scaling of second kind},
concerning the dependence of the function $f$ upon the specific
nucleus.
Scaling of the second kind means that $f(q,\omega)$ is the same for
different nuclei for conveniently chosen values of the Fermi momentum
$k_F$.
The analysis of
data on different nuclei (with mass number A ranging from 4 to 197)
proves that scaling of the second kind is respected with very good
accuracy in the scaling region.

{\it Superscaling} is the simultaneous occurrence of scaling of first and
second kind. The function \ref{fcross} is referred to as {\it superscaling function}.

The superscaling analysis of  cross sections
can be extended
to the separate longitudinal
and transverse 
responses. In this case two superscaling functions can be introduced
by dividing 
the longitudinal and  transverse response functions
by the corresponding  single-nucleon factor $G_L$, $G_T$
\begin{equation}
  f_L(\psi) = k_F \frac{R_L(q,\omega)}{G^{e,e'}_L(q,\omega)} \,,
  \kern 1cm
f_T(\psi) = k_F \frac{R_T(q,\omega)}{G_T^{e,e'}(q,\omega)} \,.
\label{eq:fT}
\end{equation}
The analysis of the separated $L$ and $T$ responses shows that scaling
violations mainly reside in the transverse response, whereas in the
longitudinal response scaling works quite well in the full quasielastic
region, {\it i.e.} also at $\omega>\omega_{QEP}$. The reason for this
difference is that the main contributions which violate scaling,
namely the 2p2h and $\Delta$ resonance excitations, are essentially
transverse, with small longitudinal contamination of relativistic
origin.  From the analysis of the L/T separated $(e,e')$ data a
phenomenological longitudinal superscaling function has been
extracted~\cite{Jourdan:1996aa}. The agreement with this function
represents a strong constrain for nuclear models used in
neutrino-nucleus scattering simulations. This topic is examined in
detail in subsequent sections.

\subsection*{Semi-inclusive electron-nucleus scattering}

The study of $(e,e'N)$ reactions on nuclei provides essential
information about nuclear structure and dynamics, not only on
single-particle properties such as the spectral function and the
nucleon's momentum distribution, but also on more complex mechanisms
as nucleon-nucleon correlations and meson-exchange currents
~\cite{Frullani:1984nn,Boffi:1996ikg,Kelly:1996hd}.  These are
essential inputs for the simulation of $\nu$-$A$ scattering and
therefore neutrino energy reconstruction.

The theoretical prediction of semi-inclusive reactions is a much
harder task than modelling the inclusive process since the related
observables are far more sensitive to the details of the dynamics.
This allows one to better discriminate
among different models.

In Plane Wave Impulse Approximation (PWIA) the 6-th differential
$(e,e'N)$ cross section can be written as~\cite{Donnelly:2017aaa}
\begin{equation}
    \frac{d^6\sigma}{dp_N d\Omega_N dk'_e d\Omega_e} = 
    \frac {p_N m_N M_{A-1}} {\sqrt{\left(M_{A-1}\right)^2+p^2}} \,\sigma_{eN}(q,\omega;p,{\cal E}) S(p_m,E_m) \,,
    \label{eq:dseepn}
\end{equation}
where $p_N$, $\Omega_N$, $k'_e$, $\Omega_e$ are the momenta and solid
angles of the outgoing nucleon and electron, respectively, $M_{A-1}$
is the ground state mass of the recoiling system, ${\bf p_m}={\bf
  p}_{A-1}\equiv -{\bf p}={\bf q}-{\bf p}_N$ is the missing momentum,
$E_m=W_{A-1}+m_N-M_A$ is the missing energy, ${\cal E}=
\sqrt{W_{A-1}^2+p^2}-\sqrt{M_{A-1}^2+p^2}$ is the excitation energy of
the residual nucleus having invariant mass $W_{A-1}$, related to the
missing energy $E_m$ and to the separation energy
$E_s=M_{A-1}+m_N-M_A$ by ${\cal E}\simeq E_m-E_s$. Finally,
$\sigma_{eN}$ is the half-off-shell single-nucleon cross
section~\cite{DeForest:1983ahx} and $S(p_m,E_m)$ the nuclear {\it
  spectral function}, which yields the probability of removing a
nucleon of momentum $p=p_m$ from the nuclear ground state leaving the
residual system with excitation energy ${\cal E}$.  The spectral
function is related to the nucleon's momentum distribution
\begin{equation}
    n(p) = \int_0^\infty S(p,{\cal E})\,d{\cal E} \,.
\end{equation}
The factorization \eqref{eq:dseepn} breaks if effects beyond the PWIA,
as Final State Interactions (FSI) of the knocked-out nucleon with the
residual nucleus and effects of two- or many-body currents, are taken
into account. Moreover, even in the plane wave limit, the factorized
expression \eqref{eq:dseepn} no longer holds if dynamical relativistic
effects in the bound nucleons, arising from the lower components in
the relativistic wave functions, are incorporated.  As long as all
these effects are properly taken into account in a Relativistic
Distorted Wave Impulse Approximation (RDWIA) framework, which
describes the distortion of the ejected proton wave function, the
experimental data for $(e,e'p)$ can provide reliable information on
the nuclear spectral function.

Several $(e,e'p)$ experiments were performed in the past in various
laboratories (Saclay, NIKHEF, MIT/Bates, Mainz, JLab) at different
kinematic conditions and on a variety of nuclei, from $^2 H$ to
$^{208}Pb$. 
New data on $^{40}Ar$ have recently been taken at
JLab~\cite{Dai:2018gch} and will provide a valuable input for the
analyses of neutrino experiments.  
The comparison of existing data
with theoretical models points to a systematic overestimation of the
data from shell model based calculations. This discrepancy is
interpreted as a probe of the limitations of the mean field approach
and of the importance of NN correlations in the nuclear wave function.
It is often quantified in terms of {\it spectroscopic factors},
$Z_\alpha$, which measure the actual occupancy (different from 1) of
each shell (being $\alpha$ the set of quantum numbers characterizing a
given orbital).  The spectroscopic factors extracted from $(e,e'p)$
data are typically of the order of 0.5-0.7, depending on the orbital
and on the nucleus. The knowledge of these factors and the
understanding of their microscopic origin are important ingredients
for the modelling of semi-inclusive reactions.

In the next sections we provide a systematic study of the models
developed by our group comparing our predictions with inclusive as
well as more exclusive reactions. In the latter, a much more
challenging task, we will develop reliable models for the description
of semi-inclusive $(\nu_l,lN)$ reactions, which require a more
detailed knowledge of the nuclear structure and dynamics.

\section{Models for QE in the Impulse Approximation}
\label{sec:QE}

\subsection{Relativistic Fermi Gas (RFG)}
\label{section-rfg}

The quasielastic electroweak cross section is proportional to the
hadronic tensor or response function for single-nucleon excitations
transferring momentum $\nq$ and energy $\omega$
\cite{Alberico:1988bv}.  The relativistic Fermi gas gives the
simplest approach to a fully relativistic nuclear system whose
response functions can be analytically computed, because the
uncorrelated single-nucleon wave functions are free plane waves
multiplied by Dirac spinors
\begin{equation}
\psi_{\np,s}(\nr) = \frac{1}{\sqrt{V}}{\rm e}^{i\np\cdot\nr} u_s(\np)
\end{equation}
with single nucleon energy $E=(m_N^2+\np^2)^{1/2}$. The positive
energy spinor, $u_s(\np)$, with mass $m_N$, is normalized to
$\overline{u}u = 1$. In the ground state of the RFG all the momenta
$\np$ with $p<k_F$ (the Fermi momentum) are filled in the Lab system.
In the impulse approximation, the electroweak current is approximated
by a one-body operator, which can produce only one-particle one-hole
(1p1h) excitations. Therefore the hadronic tensor is given in this
model by
\begin{eqnarray}
W^{\mu\nu}(q,\omega) 
&=& \sum_{\np} \sum_{s,s'} \delta(E'-E-\omega)
\frac{m_N^2}{EE'} 
J^{\mu *}_{s's}(\np',\np)J^{\nu}_{s's}(\np',\np)
\theta(k_F-p)\theta(p'-k_F)
\end{eqnarray}
where $J^{\mu}$ is the electroweak current matrix element while
$E=\sqrt{\np^2+m_N^2}$ is the initial nucleon energy in the Fermi gas.
The final momentum of the nucleon is $\np'=\np+\nq$ and its energy is
$E'=\sqrt{\np'{}^2+m_N^2}$. Initial and final nucleons have
spin component $s$ and $s'$, respectively.

In the thermodynamic limit we substitute the
sums by momentum integrations. Then the volume $V=3\pi^2 {\cal N}/k_F^3$ of
the system is related to the Fermi momentum $k_F$ and proportional to
the number ${\cal N}$ of protons or neutrons participating in the
process:
\begin{eqnarray}
W^{\mu\nu}(q,\omega) 
&=& \frac{V}{(2\pi)^3} \int d^3p \,\delta(E'-E-\omega)
\frac{m_N^2}{EE'} \,
2w^{\mu\nu}_{s.n.}(\np',\np)
\theta(k_F-p)\theta(p'-k_F)\,,
\label{hadronicorfg}
\end{eqnarray}
where we have defined the single-nucleon tensor for the 1p1h excitation
\begin{equation}
w^{\mu\nu}_{s.n.}(\np',\np)=\frac12\sum_{ss'}
J^{\mu *}_{s's}(\np',\np)J^{\nu}_{s's}(\np',\np) \,.
\label{single-nucleon-tensor}
\end{equation}

In the case of electron scattering
 the electromagnetic current matrix element is given by
\begin{equation}
J^\mu_{s's}(\np',\np)=
\overline{u}_{s'}(\np')
\left[ 
F_1(Q^2)\gamma^\mu 
+F_2(Q^2)i\sigma^{\mu\nu}\frac{Q_\nu}{2m_N}
\right]u_{s}(\np),
\label{electromagnetic-current}
\end{equation}
where $F_1$ and $F_2$ are, respectively, the Dirac and Pauli
electromagnetic form factors of proton or neutron.

In the case of neutrino or antineutrino CC scattering the weak current
matrix element is the sum of vector and axial-vector terms $J^\mu=
V^\mu- A^\mu $, where the vector current is
\begin{equation}
V^\mu_{s's}(\np',\np)=
\overline{u}_{s'}(\np')
\left[ 
2F_1^V\gamma^\mu 
+2F_2^Vi\sigma^{\mu\nu}\frac{Q_\nu}{2m_N}
\right]u_{s}(\np)
\label{vector-current}
\end{equation}
being $F_i^V=(F_i^P-F_i^N)/2$ the isovector form factors of the
nucleon.  The axial current is
\begin{equation}\label{axial-current}
A^\mu_{s's}(\np',\np)=
\overline{u}_{s'}(\np')
\left[ 
G_A\gamma^\mu\gamma_5 
+G_P\frac{Q^\mu}{2m_N}\gamma_5
\right]u_{s}(\np),
\end{equation}
where $G_A$ is the nucleon axial-vector form factor and $G_P$ is the
pseudo-scalar axial form factor.  It is usual to assume the dipole
parametrization of the axial form factor, with the axial mass $M_A =
1.032$ GeV.  From partial conservation of the axial current (PCAC) and
pion-pole dominance, $G_P$ and $G_A$ are related by
\begin{equation}
G_P =  \frac{4m_N^2}{m_\pi^2+|Q^2|}G_A ,
\end{equation}
where $Q^2=\omega^2-q^2 < 0$.

The single nucleon tensor for the electroweak current is computed by
performing the traces in Appendix \ref{sn-appendix}.
To obtain the quasielastic cross section for $(e,e')$ reaction we use
the standard expansion in terms of response functions
\begin{equation}
\frac{d\sigma}{d\Omega'd\epsilon'}
= \sigma_{\rm Mott}\left( v_L R^{e.m.}_L + v_T R^{e.m.}_T \right)
\end{equation}
where $\sigma_{\rm Mott}$ is the Mott cross section, $v_L= Q^4/q^4$
and $v_T=\tan^2(\theta/2)-Q^2/2q^2$, with $\theta$ the scattering
angle.  The electromagnetic longitudinal and transverse response
functions are the following components of the e.m. hadronic tensor in
the scattering coordinate system with the $z$-axis in the $\nq$
direction (longitudinal)
\begin{eqnarray}
R^{e.m.}_L(q,\omega) &=& W_{e.m.}^{00} \\
R^{e.m.}_T(q,\omega) &=& W_{e.m.}^{11}+ W_{e.m.}^{22} .
\end{eqnarray}

Analogously the $(\nu_l,l^-)$ charged-current quasielastic (CCQE) cross section for neutrino energy
$E_\nu=\epsilon$ and final lepton energy $\epsilon'=E_l$, has been
expanded in terms of five response functions.  If the lepton
scattering angle is $\theta_l$, the double-differential cross section
can be written as \cite{Amaro:2004bs,Amaro:2005dn}
\begin{equation}
\frac{d^2\sigma}{dE_l d\cos\theta_l}
=
\sigma_0
\left\{
V_{CC} R_{CC}+
2{V}_{CL} R_{CL}
+{V}_{LL} R_{LL}+
{V}_{T} R_{T}
\pm
2{V}_{T'} R_{T'}
\right\} \, , 
\end{equation}
where we have defined the cross section 
\begin{equation}
\sigma_0= 
\frac{G^2\cos^2\theta_c}{4\pi}
\frac{k'}{\epsilon}v_0. 
\label{eq:sig0}
\end{equation}
In \eqref{eq:sig0} the Fermi weak constant is $G=1.166\times
10^{-11}\quad\rm MeV^{-2} 
$, the Cabibbo angle is
$\cos\theta_c=0.975$, $k'$ is the final lepton momentum, and we have
defined the factor $v_0= (\epsilon+\epsilon')^2-q^2$.  Note that the
fifth response function $R_{T'}$ is added (+) for neutrinos and
subtracted ($-$) for antineutrinos.  The lepton $V_K$ coefficients
depend only on the lepton kinematics and they are defined by
\begin{eqnarray}
{V}_{CC}
&=&
1-\delta^2\frac{|Q^2|}{v_0}
\label{vcc}\\
{V}_{CL}
&=&
\frac{\omega}{q}+\frac{\delta^2}{\rho'}
\frac{|Q^2|}{v_0}
\\
{V}_{LL}
&=&
\frac{\omega^2}{q^2}+
\left(1+\frac{2\omega}{q\rho'}+\rho\delta^2\right)\delta^2
\frac{|Q^2|}{v_0}
\\
{V}_{T}
&=&
\frac{|Q^2|}{v_0}
+\frac{\rho}{2}-
\frac{\delta^2}{\rho'}
\left(\frac{\omega}{q}+\frac12\rho\rho'\delta^2\right)
\frac{|Q^2|}{v_0}
\\
{V}_{T'}
&=&
\frac{1}{\rho'}
\left(1-\frac{\omega\rho'}{q}\delta^2\right)
\frac{|Q^2|}{v_0}.
\label{vtp}
\end{eqnarray}
Here we have defined the dimensionless factors $\delta =
m_l/\sqrt{|Q^2|}$, proportional to the charged lepton mass $m_l$,
$\rho = |Q^2|/q^2$, and $\rho' = q/(\epsilon+\epsilon')$.

The five nuclear response functions $R_K$, $K=CC, CL, LL, T, T'$
($C$=Coulomb, $L$=longitudinal, $T$=transverse) in the same scattering
coordinate system as for electrons are then given by the following
components of the hadronic tensor:
\begin{eqnarray}
R_{CC } & = & W^{00 } \label{rcc}  \\
R_{CL } & = & -\frac12 (W^{03 } + W^{30 } ) \\
R_{LL } & = & W^{33 }  \\
R_{T } & = & W^{11 } + W^{22 } \\
R_{T' } & = & -\frac{i}{2}(W^{12 } - W^{21 }). \label{rtprima}
\end{eqnarray}

It is convenient to introduce dimensionless variables measuring the
energy and momentum in units of $m_N$, namely $\lambda= \omega/2m_N$,
$\kappa=q/2m_N$, $\tau=\kappa^2-\lambda^2$, $\eta_F= k_F/m_N$, and
$\xi_F=\sqrt{1+\eta_F^2}-1$.

In Appendix \ref{RFG-appendix} we calculate analytically the response
functions of the RFG by integrating Eq. (\ref{hadronicorfg}) for the
different components of the single nucleon tensor computed in Appendix
\ref{sn-appendix}. The RFG responses are given as an universal
function, called superscaling function, times a response that depends
mainly on the nucleon-boson vertex, although it also
incorporates some "minor" contributions linked to the Fermi
momentum. To make simpler the discussion that follows we simply denote
this as "single-nucleon" response (see Appendix \ref{RFG-appendix} and
equations below for details). The superscaling function in the RFG
model is given by
\begin{equation} \label{scalingfun}
f(\psi)= \frac34 ( 1 - \psi^2 ) \theta ( 1 - \psi^2 ) \,.
\end{equation}
As shown in Appendix \ref{RFG-appendix}, the scaling variable,
$\psi\equiv\psi(q,\omega)$, is a specific combination of the two
variables $q$ and $\omega$ (or, equivalently, $\kappa$, $\lambda$ and
$\tau$) related to the minimum energy $\varepsilon_0$ allowed for the initial
nucleon to absorb the momentum and energy transfer $(q,\omega)$. This is
given by
\begin{equation}  \label{epsilon0}
\varepsilon_0={\rm Max}
\left\{ 
       \kappa\sqrt{1+\frac{1}{\tau}}-\lambda, \varepsilon_F-2\lambda
\right\} , 
\end{equation}
where $\varepsilon_F=\sqrt{1+\eta_F^2}$ is the Fermi energy in units
of $m_N$.  The scaling variable is defined by
\begin{equation} \label{psi}
  \psi\equiv \psi(\lambda,\tau) =
  \sqrt{\frac{\varepsilon_0-1}{\varepsilon_F-1}}
       {\rm sgn} (\lambda-\tau)\,.
\end{equation}
Note that $\psi < 0 $ for $\lambda < \tau$ (on the left side of the
quasielastic peak).  The meaning of $\psi{}^2$ is the following: it is
the minimum kinetic energy of the initial nucleon divided by the
kinetic Fermi energy.

Using this definition, the electromagnetic responses can be written as
\begin{eqnarray}
  R^{e.m.}_K(\lambda,\tau) & = &
  \frac{1}{k_F}
  G_K^{ee'}(\lambda,\tau) f(\psi(\lambda,\tau)), \label{rfge} \\
  G_K^{ee'}(\lambda,\tau) &=&
  \frac{\xi_F}{\eta_F^2 \kappa}
  (Z U^p_K(\lambda,\tau)+NU^n_K(\lambda,\tau)) \,,
\label{eq:Rem}
\end{eqnarray}
where $Z$ ($N$) is the proton (neutron) number. 
Analogously, for the weak responses 
\begin{eqnarray}
  R_K(\lambda,\tau) &=&
\frac{1}{k_F}
G_K(\lambda,\tau) f(\psi)
\\
G_K(\lambda,\tau)  &=&
\frac{\xi_F}{\eta_F^2 \kappa}
{\cal N}   U_K(\lambda,\tau) , 
\label{eq:RK}
\end{eqnarray}
where ${\cal N}= N$ ($Z$) for neutrino (antineutrino) scattering. The explicit expressions for the functions $G_K$ (likewise $U_K$) are given below.
This factorization of the scaling function inspires the superscaling
models discussed in Sect. \ref{sec:susav2} using a phenomenological
scaling function instead of the universal superscaling function of the
RFG.

The remaining quantities are the integrated single nucleon response
functions. The electromagnetic ones, for protons or neutrons, are
given by
\begin{eqnarray}
U^{p,n}_L(\lambda,\tau) &=& \frac{\kappa^2}{\tau}
\left[ (G^{p,n}_E(\tau))^2 + \frac{(G_E^{p,n}(\tau))^2
+ \tau (G_M^{p,n}(\tau))^2}{1+\tau}\Delta(\lambda,\tau) \right]
\label{ul}\\
U^{p,n}_T(\lambda,\tau) &=&
2\tau  (G_M^{p,n}(\tau))^2 
+ \frac{(G_E^{p,n}(\tau))^2 +
  \tau (G_M^{p,n}(\tau)) ^2}{1+\tau}\Delta(\lambda,\tau) ,
\label{ut}
\end{eqnarray}
where the quantity $\Delta$ has been introduced
\begin{equation}\label{amaro-delta}
\Delta(\lambda,\tau)= \frac{\tau}{\kappa^2}
\left[ -\frac{(\lambda-\tau)^2}{\tau}
     +\xi_F[  (1+\lambda)(1+\psi^2)+\frac{\xi_F}{3}(1+\psi^2+\psi^4)]
\right].
\end{equation}

In what follows we present the explicit expressions for the weak
integrated single-nucleon responses.

The $U_{CC}$ is the sum of vector and axial contributions. The vector
part implements the conservation of the vector current (CVC).  The
axial part can be written as the sum of conserved (c.)  plus non
conserved (n.c.) parts. Then (the arguments of each function are omitted for brevity)
\begin{eqnarray} \label{ucc}
U_{CC} &=& U_{CC}^{VV}+
\left(U_{CC}^{AA}\right)_{\rm c.}
+\left(U_{CC}^{AA}\right)_{\rm n.c.} \, .
\end{eqnarray}
For the vector CC response we have
\begin{eqnarray}
U_{CC}^{VV} &=&
\frac{\kappa^{2}}{\tau}
\left[ (2G_E^{V})^2+\frac{(2G_E^{V})^2+\tau (2G_M^{V})^2}{1+\tau}\Delta
\right]\ ,
\label{uccv}
\end{eqnarray}
where $G_E^{V}$ and $G_M^{V}$ are the isovector electric and magnetic
nucleon form factors
\begin{eqnarray}
G_E^{V}  &=&  F_1^V-\tau F_2^V \\
G_M^{V}  &=& F_1^V+ F_2^V.  \label{GM}
\end{eqnarray}

The axial-vector CC responses are 
\begin{eqnarray}
\left(U_{CC}^{AA}\right)_{\rm c.}
&=&
\frac{\kappa^{2}}{\tau}G_A^2\Delta
\\
\left(U_{CC}^{AA}\right)_{\rm n.c.}
&=&
\frac{\lambda^{2}}{\tau}(G_A - \tau G_P )^2.
\label{amaro-uccnc}
\end{eqnarray}

Using current conservation for the conserved part of the integrated
single-nucleon responses $K=CL,LL$ we can express them as
\begin{eqnarray}
U_{CL} &=& -\frac{\lambda}{\kappa}
\left[U_{CC}^{VV}+\left(U_{CC}^{AA}\right)_{\rm c.}\right]
+\left(U_{CL}^{AA}\right)_{\rm n.c.}
\label{ucl} \\
U_{LL} &=& 
\frac{\lambda^{2}}{\kappa^{2}}
\left[ U_{CC}^{VV}+\left(U_{CC}^{AA}\right)_{\rm c.} \right]
+\left(U_{LL}^{AA}\right)_{\rm n.c.}\ ,
\label{ull}
\end{eqnarray}
where the n.c. parts are
\begin{eqnarray}
\left(U_{CL}^{AA}\right)_{\rm n.c.}
&=& -\frac{\lambda\kappa}{\tau}(G_A - \tau G_P )^2
\\
\left(U_{LL}^{AA}\right)_{\rm n.c.}
&=& \frac{\kappa^{2}}{\tau}(G_A - \tau G_P )^2\ .
\end{eqnarray}
Finally, the transverse responses are given by
\begin{eqnarray}
U_T &=& U_T^{VV}+U_T^{AA} 
\\
U_T^{VV} &=&  2\tau(2G_M^{V})^2+\frac{(2G_E^{V})^2+\tau 
(2G_M^{V})^2}{1+\tau}\Delta \label{utv}
\\
U_T^{AA} &=& G_A^2 \left[2(1+\tau)+ \Delta\right] \label{uta}
\\
U_{T'} &=& 2G_A(2G_M^{V}) \sqrt{\tau(1+\tau)}[1+\widetilde{\Delta}]
\label{utp}
\end{eqnarray}
with
\begin{equation} \label{amaro-deltatilde}
\widetilde{\Delta}=
\frac{1}{ \sqrt{ \tau(1+\tau) }} 
\left[ 
\frac{\tau}{\kappa}(1+\lambda) 
- \sqrt{\tau(\tau+1)}
+\frac{\tau}{\kappa} \frac12 \xi_F(1+\psi^{2})
\right]\,.
\end{equation}

\begin{figure}[H]
      \begin{center}
\includegraphics[scale=0.23,angle=270]{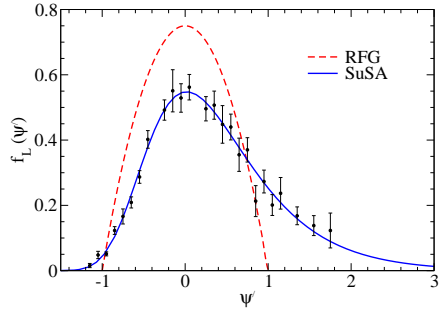}
\end{center}
\vspace*{-0.5cm}
\caption{\label{fig:scaling} The experimental longitudinal scaling
  function displayed versus the scaling
  variable $\psi^\prime$ and compared to the RFG prediction (red
  dashed curve). The solid blue curve is a fit of the
  data~\cite{Jourdan:1996aa}.} \label{fig:SuSA}
\end{figure}

For completeness we show in Fig.~\ref{fig:SuSA} the comparison of the
experimental superscaling function extracted from the analysis of the
separate longitudinal response data (denoted as SuSA) with the
relativistic Fermi gas prediction. The variable $\psi^\prime$ differs
from $\psi$ by an energy shift related to the nucleon separation
energy (see Appendix A for details). Note the striking difference
between the RFG result and the experimental data: the RFG predicts a
symmetric function around $\psi^\prime=0$ , with a maximum value of
0.75 and restricted to the region $[-1,1]$ (see
Eq.~\eqref{scalingfun}), while the data display a pronounced
asymmetric tail at large $\psi^\prime$ (large $\omega$) and a maximum
of $\sim$0.6.

\subsection{Semi-relativistic shell models}

 Nuclear models based on the Fermi gas describe the single nucleon
 wave functions as plane waves, thus neglecting the finite size of the
 nucleus, assuming that the response per particle is the same as for
 infinite nuclear matter. The validity of this assumption has been
 investigated theoretically in the nuclear shell model (SM) for the
 kinematics of interest
 \cite{Amaro94,Amaro:1995kk,Amaro:2005dn,Amaro:2006if,Amaro:2006tf}.
 The Fermi gas cannot describe the very low $\omega$
 region of the nuclear response, dominated by excitation of
 discrete states and giant resonances. Besides, Pauli blocking, for
 small momentum transfer $q< 2k_F$, forbids emission from a subset of
 occupied momentum states in the Fermi gas, even if there is enough
 energy transfer, by momentum conservation. In a real nucleus,
 however, the nucleons are not plane waves. This
 makes the RFG inappropriate to precise modelling of the nuclear
 response at low momentum transfer $q < 500$ MeV/c \cite{Amaro94}.
 
 In Ref.~\cite{Amaro94} the finite size effects on the electromagnetic
 QE response functions were studied using a continuum shell model
 (CSM) in the non relativistic regime for $q \leq 550$ MeV/c. In the
 CSM the single nucleon wave functions were obtained by solving the
 Schr\"odinger equation with a mean field potential $V(\nr)$ of
 Woods-Saxon type, including central and spin orbit interactions, plus
 a Coulomb part for protons
\begin{equation}
    V(\nr)= 
    \frac{-V_0}{1+e^{(r-R)/a}}+
    \frac{1}{m_\pi^2} \frac{1}{r} \frac{d}{dr}
\left(\frac{-V_{ls}}{1+e^{(r-R)/a}}\right) \nl \cdot \nsigma
+ V_{\rm Coul}(r).
\end{equation} 
This potential is solved numerically for bound and continuum nucleon
states, and the parameters $V_0, V_{ls}, R, a$ are fitted to the
experimental energies of the valence shells for protons and neutrons,
obtained from the masses of the neighboring nuclei. The SM is the
simplest model where the response of a confined quantum system of
nucleons can be evaluated, and thus, it allows us to explore the
importance of finite size against the infinite Fermi Gas (FG)
model. This study was made theoretically by comparing the separate
longitudinal and transverse response functions.  While the FG model
reproduces the position and width of the separate SM responses, it
cannot fully account for the detailed energy dependence. Specifically,
the SM responses present tails in the low and high ends of the energy
transfer, reminiscent of the momentum distribution of a finite system.

\begin{figure}[t]
\begin{center}
\includegraphics[scale=0.5,  bb= 50 420 540 805]{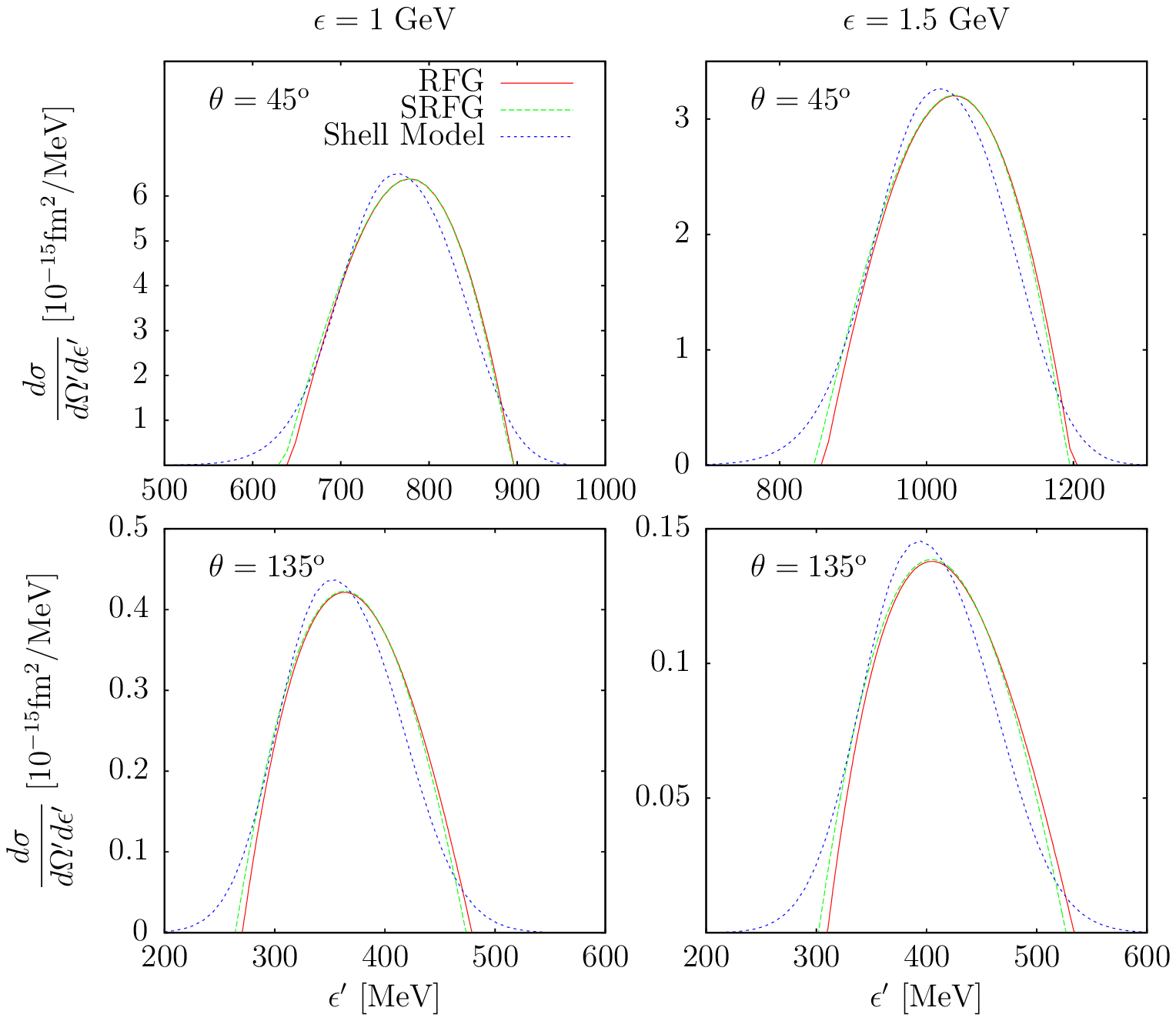}
\vspace{-0.3cm}
\caption{Differential cross section of the reaction
  $^{12}{\rm C}(\nu_\mu,\mu^-)$ for incident neutrino energies $\epsilon=1$
  and 1.5 GeV and for two scattering angles: RFG with $k_F=215$
  MeV/c. SRFG and shell model from \cite{Amaro:2005dn}.
\label{csr5} }
\end{center}
\end{figure}

In Ref.~\cite{Amaro:1995kk} the shell model was extended to a
semi-relativistic (SR) shell model by implementing relativistic
kinematics and using a semi-relativistic expansion of the
electromagnetic current. The relativistic electroweak current was
expanded in powers of the initial nucleon momentum, while treating
exactly the relativistic kinematics of the final ejected particle.
The SR expansion of the current and relativistic kinematics was tested
in the Fermi gas model by comparison of the RFG and semi relativistic
Fermi gas (SRFG) response functions. The two models are in great
accord for all values of the momentum transfer. This is seen in
Fig.~\ref{csr5}, where results for the semi relativistic shell model
(SRSM) are compared to the RFG and the SRFG models for the reaction
$^{12}{\rm C}(\nu_\mu,\mu^-)$ \cite{Amaro:2005dn}.
The RFG and SRFG
results are essentially equal, while the SRSM also gives similar
results, with the exception of the characteristic finite size tails
and a small shift of the maximum.

The semi-relativistic shell model was extended in \cite{Amaro:2006if}
to improve the description of the final-state interaction of
relativistic nucleons by performing a Schr\"odinger reduction of the Dirac
equation with scalar and vector potentials, taken from the
relativistic mean field (RMF) model \cite{Caballero05}.  The
longitudinal scaling function computed with the SR shell model with Dirac Equation Based
(DEB) potentials was shown to be very similar to the experimental data,
with the appropriate asymmetry and long tail for large $\omega$. This
suggests that relativistic dynamics take into account the effects
necessary to describe QE lepton scattering. In this respect the RMF of
the next section is a clear candidate to achieve this goal.

\subsection{Relativistic Mean Field (RMF)}
\label{sec:RMF}

The high energy and momentum transfer involved in electroweak reactions
require a fully relativistic formalism, not only for
the kinematics but also for the nuclear dynamics. This
was the basic motivation to introduce the Relativistic
Mean Field (RMF) approach to describe 
electron and neutrino scattering with nuclei. In this
section we summarize the formalism of RMF approach
\cite{Maieron03,Caballero05,Caballero:2006wi,Udias93,Udias01,Alberico97plus,Udias:1996iy,Udias:1993zs,Udias93},
and show its capability to describe
inclusive and semi-inclusive reactions.
The RMF model
is described here within the framework of the Impulse
Approximation (IA), where the scattering
is described as an incoherent sum of single-nucleon
scattering reactions. Although this is an oversimplified description
of the real process, it accounts for the main effects
for quasielastic (QE) kinematics. Ingredients
beyond the IA, such as nucleon correlations, meson exchange currents
(MEC), etc., should be taken into account for a proper description of
the data. This topic is discussed in subsequent sections.

The analysis of the QE data for $(e,e')$
shows that scaling is fulfilled at high level. Thus
any theoretical model aiming at explaining electron scattering is
constrained to satisfy scaling, and to describe the particular
shape of the experimental 
scaling function
with a
long tail extended to high values of the energy transfer (high
positive values of the scaling variable). Most of the
models based on the IA satisfy scaling; this is the case of the RFG,
the semirelativistic approach presented in the previous section, and
non relativistic spectral
function approaches.
However, most of them lead to symmetrical scaling functions
that depart significantly from the data analysis, without the high energy
tail.

On the contrary, the RMF model leads to an asymmetrical superscaling
function in accordance with data. This proves the capability of the
RMF model to describe inclusive electron scattering reactions and its
extension to neutrino processes. Not only that, the model has been
also applied to semi-inclusive $(e,e'p)$ reactions providing a good
description of the reduced cross section (distorted momentum
distribution), as well as the spectroscopic factors. Within the
general framework of $(e,e'p)$ processes, the reduced cross section is
defined as the semi-inclusive differential cross section divided by
the single-nucleon cross section evaluated using different options for
the nucleon current operators and gauges
\cite{Udias:1993zs,Udias:1996iy,Udias:1999tm,Caballero98a,Frullani:1984nn,Kelly:1996hd}.

The single-nucleon current matrix element entering in the electromagnetic and weak
tensors is
\be
\langle
J^\mu(Q)\rangle = \int d\nr
e^{i\nq\cdot\nr}\overline{\psi}_F(\np_F,\nr)\hat{\Gamma}^\mu\psi_B^{jm}(\nr)
\ee
with $\hat{\Gamma}^\mu$ the corresponding single-nucleon current
operator for electron or weak neutrino scattering, and $\Psi_B^{jm}$
($\Psi_F(\np_F,\nr)$) the wave function for the initial bound
(emitted) nucleon. The RMF model incorporates a fully relativistic
description of the scattering reaction based on the impulse
approximation. We use the relativistic free nucleon expressions for
the current operators corresponding to the usual options, denoted as
CC1 and CC2,
\begin{eqnarray}
  \hat{\Gamma}^\mu_{CC1} &=& (F_1+F_2)\gamma^\mu
-\frac{F_2}{2m_N}(\overline{P}+P_F)^\mu \\ \hat{\Gamma}^\mu_{CC2} &=&
F_1\gamma^\mu +\frac{iF_2}{2m_N}\sigma^{\mu\nu}Q_\nu \, ,
\end{eqnarray}
where
$F_1$ and $F_2$ are the Pauli and Dirac nucleon form factors,
respectively, that depend only on $Q^2$. We have introduced the
on-shell four-momentum $\overline{P}^\mu=(\overline{E},\np)$ with
$\overline{E}=\sqrt{p^2+m_N^2}$ and $\np$ the bound nucleon
momentum. These two operators are equivalent for
on-shell nucleons, as occurs in the RFG model. However,
 bound and ejected nucleons are off-shell, and hence the $CC1$ and
$CC2$ operators lead to different results. Moreover, the current is
not strictly conserved and uncertainties linked to the election of the
gauge also occur~\cite{Caballero:2006wi,Caballero98a}. In the case of
neutrino scattering processes the current operator also includes the axial
term,
\be
\hat{\Gamma}^\mu_A=\left[G_A\gamma^\mu+\frac{G_P}{2m_N}Q^\mu\right]\gamma^5
\ee
with $G_A$ and $G_P$ the axial-vector and pseudoscalar nucleon
form factors, respectively.

\begin{figure}[t]
\centering
\includegraphics[scale=0.3]{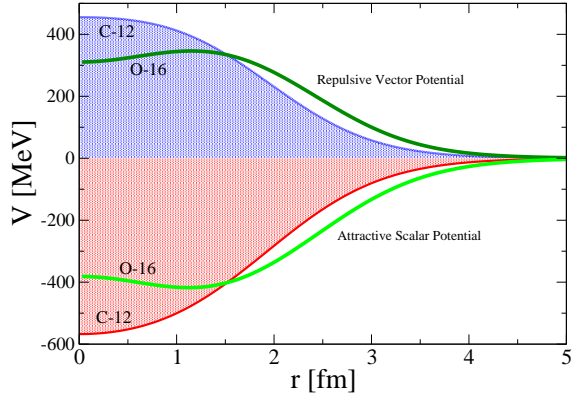}
\caption{Relativistic attractive scalar and repulsive vector
  potentials corresponding to $^{12}$C and $^{16}$O.}
\label{RMF_POTENTIALS}
\end{figure}

Concerning the nucleon wave functions, they are given as solutions of
the Dirac equation in presence of phenomenological relativistic
potentials with scalar and vector terms. We describe the bound nucleon
states, $\psi_B^{jm}$, as self-consistent Dirac-Hartree solutions,
using a Lagrangian with local potentials fitted to saturation properties of
nuclear matter, radii and nuclear masses. In Fig.~\ref{RMF_POTENTIALS}
we show the potentials  of $^{12}$C and
$^{16}$O. As shown, large scalar (attractive) and vector (repulsive)
potentials, that do not depend on the energy, are present. It is
important to point out that the RMF does get saturation, even with no
Fock terms neither explicit nucleon correlations included. This comes
from the combination of the strong scalar and vector potentials that
incorporate repulsive and attractive interactions, and it makes a big
difference with most non-relativistic approaches where correlations
are needed in order to get saturation.

\begin{figure}
\begin{center}
\includegraphics[scale=0.35]{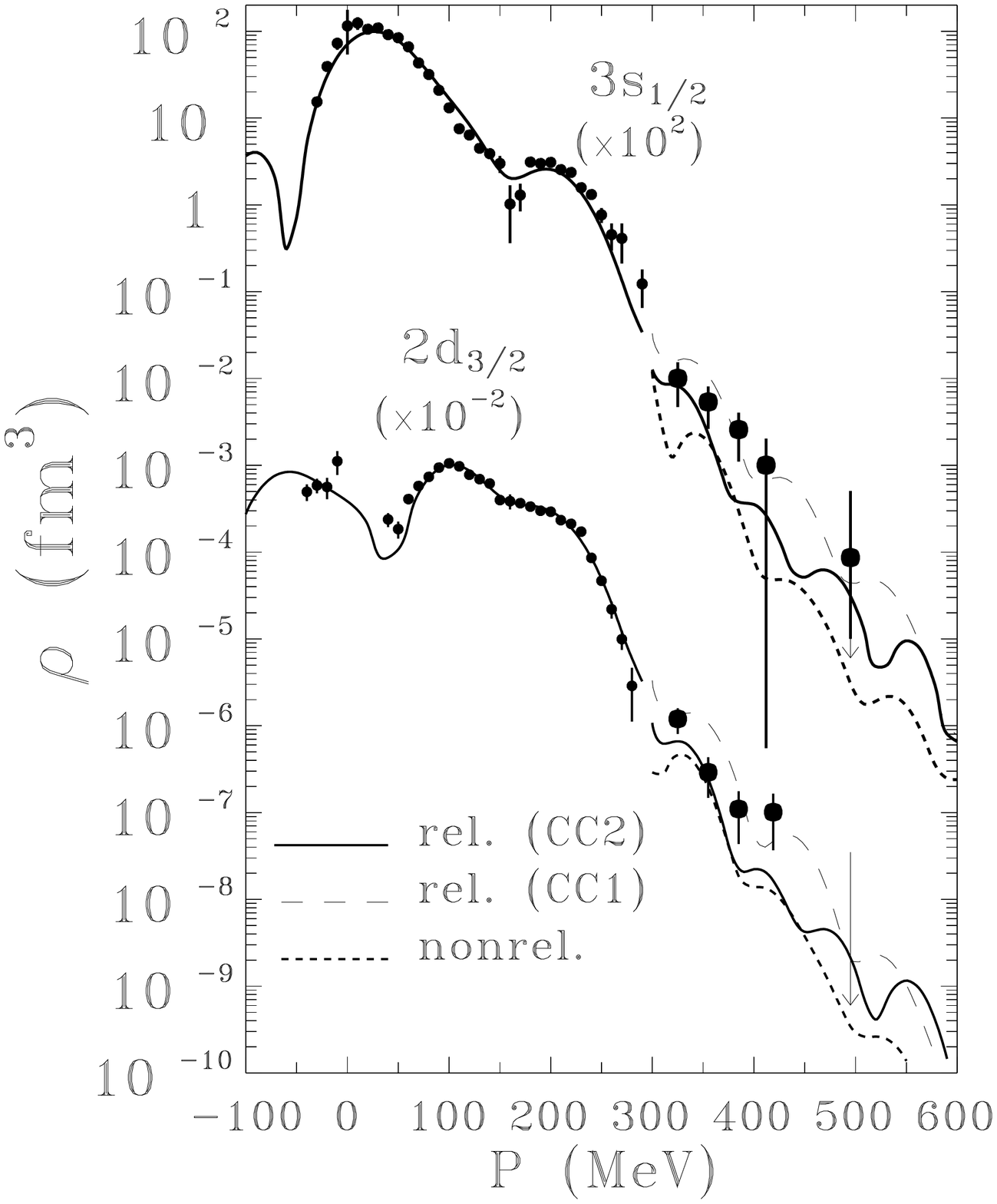}
\hspace{0.3cm}
\includegraphics[scale=0.4]{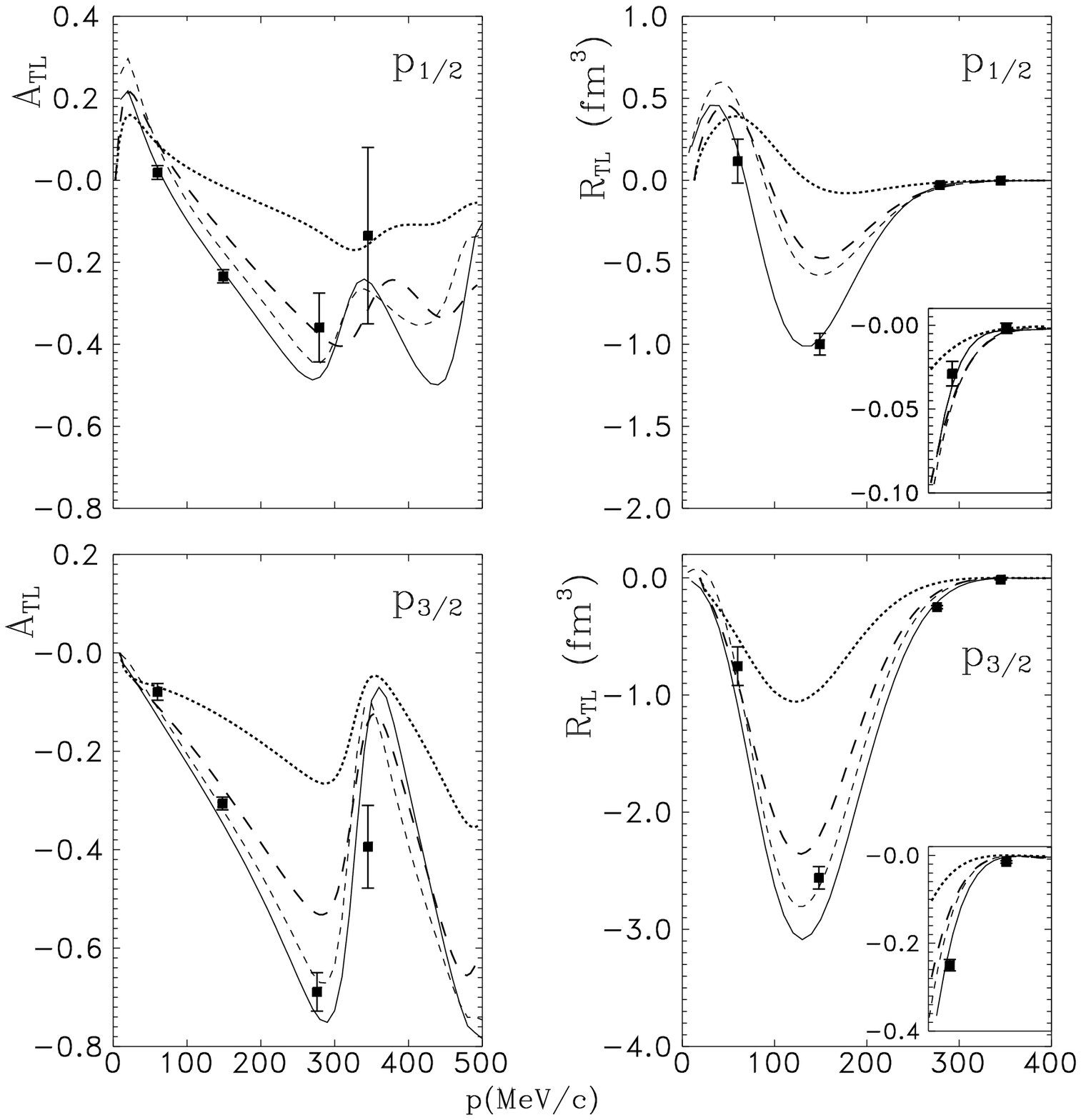}
\caption{(Left panel) Reduced cross sections versus missing momentum
  for the shells 3s$_{1/2}$ and 2d$_{5/2}$ of $^{208}$Pb. Small
  circles with error bars are data from Ref.~\cite{Quint88}. In the
  high missing momentum region the relativistic results obtained with
  the currents CC2 (solid line) and CC1 (long-dashed lines), as well
  as the nonrelativistic results (short-dashed lines) are compared
  with data from Ref.~\cite{Bobeldijk:1994zz}. Figure taken from
  Ref.~\cite{Udias:1996iy}. (Middle and right panels) Interference
  longitudinal-transverse response $R_{TL}$ and $TL$ asymmetry
  $A_{TL}$ for proton knockout from $^{16}$O for the 1p$_{1/2}$ (top
  panels) and 1p$_{3/2}$ (bottom panels) orbits. Results correspond to
  a fully relativistic calculation using the Coulomb gauge and the
  current operator CC2 (solid line), a calculation performed by
  projecting the bound and scattered proton wave functions over
  positive-energy states (short-dashed line) and two nonrelativistic
  calculations with (long-dashed) and without (dotted) the spin-orbit
  correction term in the charge density operator.
   }
\label{fig1}
\end{center}
\end{figure}

In the exclusive $(e,e'p)$ reactions, the final-state interaction (FSI)
of the ejected nucleon is described using phenomenological
energy-dependent complex optical potentials fitted to elastic
proton-nucleus scattering.  The imaginary part of the potential 
produces absorption, namely, flux lost into the unobserved
non-exclusive channels.
The RMF model with complex optical potentials yields
very good agreement with $(e,e'p)$ data. Not only reasonable values
for the spectroscopic factors are
given~\cite{Udias:1993zs,Udias93} but also the 
cross section describes the data even at high values of
the missing momentum~\cite{Udias:1996iy}. It should be emphasized that
the high $p$ region is very sensitive to theoretical models,
and in particular, to the strong enhancement of the lower
components in the Dirac wave functions produced by the relativistic
potentials mainly in the final scattering state. The remarkable
agreement between the RMF predictions and data for the reduced cross
sections (divided by the single nucleon cross section)
and, particularly, for the
interference longitudinal-transverse response and $TL$ asymmetry,  is a clear signal
of dynamical relativistic effects~\cite{Udias:1999tm} (see Fig.~\ref{fig1}).
\begin{figure}
\begin{center}
\includegraphics[scale=0.4]{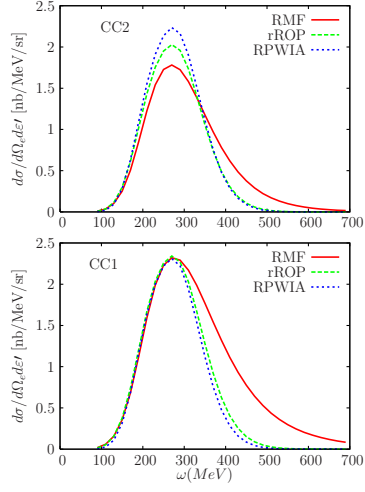}
\caption{Double differential cross section for $(e,e')$ scattering on
  $^{12}$C. Results correspond to the electron beam energy fixed to 1
  GeV and scattering angle $\theta=45^o$. Top (bottom) panel refers to
  results obtained with the CC2 (CC1) current operators. Coulomb gauge
  has been considered.}
\label{rsm_cs}
\end{center}
\end{figure}
\begin{figure}
\begin{center}
\includegraphics[scale=0.45]{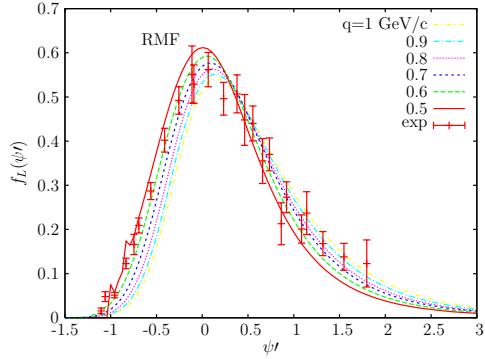}
\caption{
Longitudinal scaling function for $^{12}$C$(e,e')$ evaluated in the RMF model
for several values of the momentum transfer.
Experimental data from Ref.~\cite{Donnelly99b}.
}
\label{scaling_first_breaking}
\end{center}
\end{figure}

The extension of the RMF approach to inclusive processes such as
$(e,e')$ and $(\nu_\ell,\ell)$ requires to retain the contributions
from the inelastic channels.
One  approach is to use the real part of the
optical potential (rROP).
However orthogonalization is violated as
bound and ejected nucleon states are evaluated using different
potentials. The same happens in a second approach in which
FSI are turned off, known as the relativistic plane wave impulse
approximation (RPWIA). A third approach consists of
describing the scattered states as solutions of the Dirac equation
with the same real energy-independent RMF potential
considered for the initial bound nucleon
states.
This RMF model (same potential for bound and scattered nucleons)
preserves orthogonality, verifies current
conservation, and fulfills the dispersion relations.

In Fig.~\ref{rsm_cs}
we compare the three models of FSI, RPWIA, rROP
and RMF models, for inclusive $(e,e')$.  Results are presented for the
two current operators, CC1 and CC2.  Only the RMF model leads to a
significant tail extended to large values of $\omega$.
This points to the capability of the RMF model to
describe successfully inclusive electron scattering data in the QE
domain. Not only superscaling emerges from the calculations, as illustrated below, but also
the specific shape of the scaling function with a long tail extended
to high $\omega$.

\begin{figure}
\begin{center}
\includegraphics[scale=0.38]{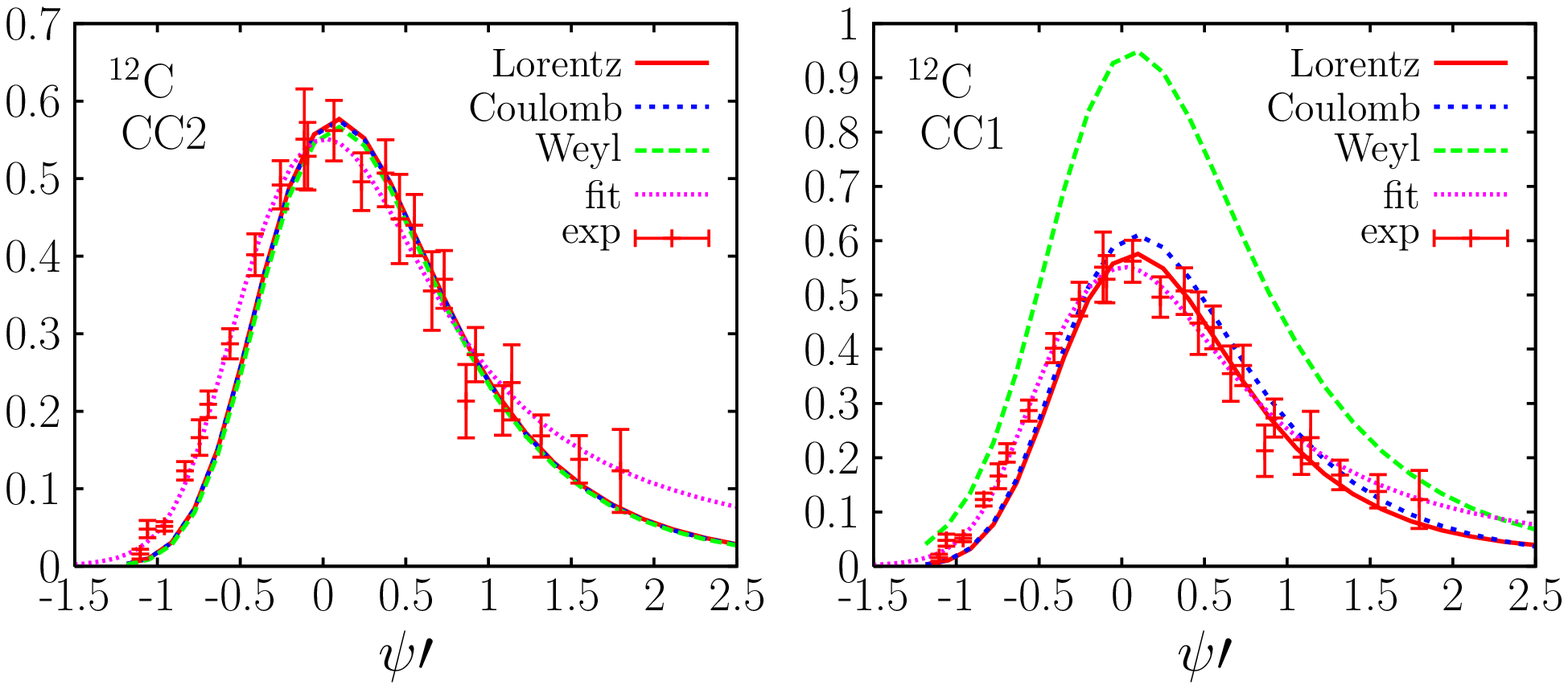} \\
\hspace{-0.5cm}
\includegraphics[scale=0.45]{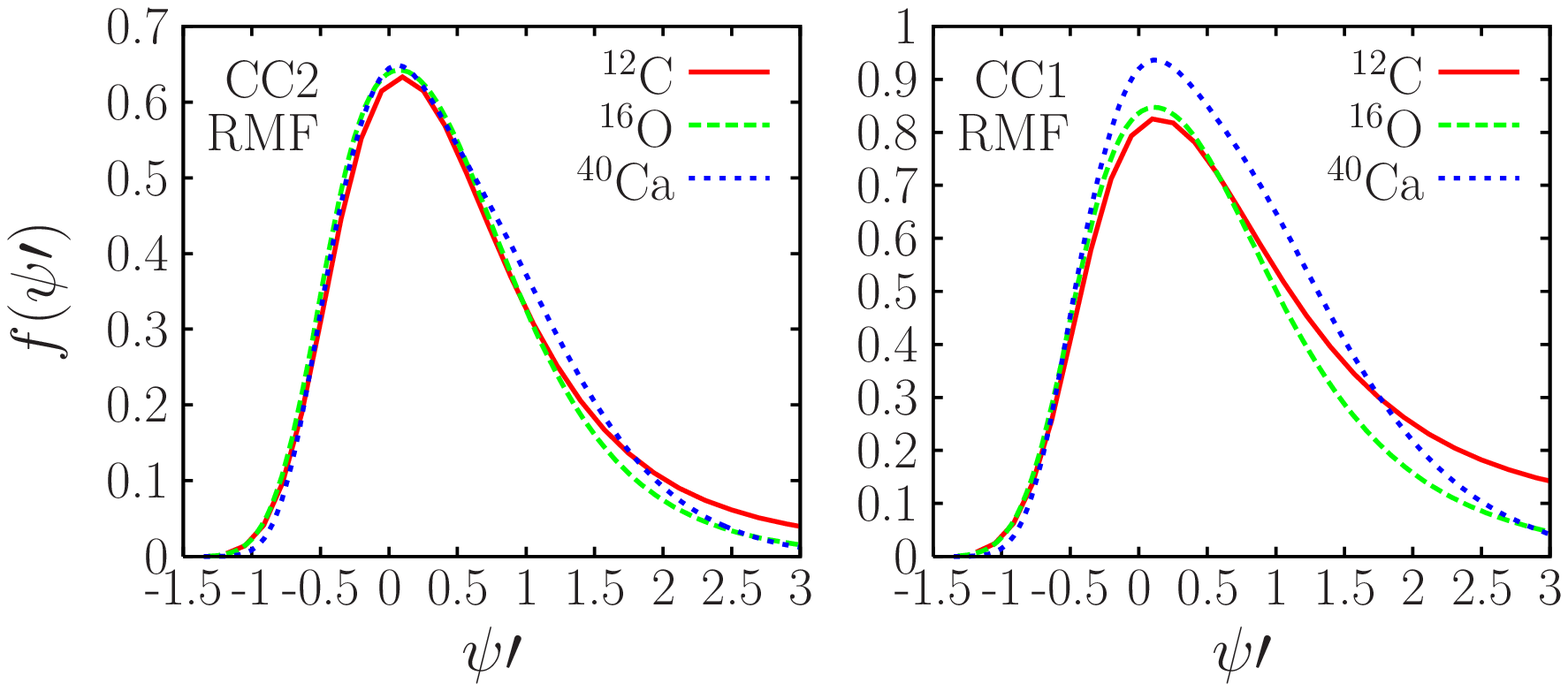}
\caption{Global scaling function corresponding to $(e,e')$ evaluated
  with the RMF model. Results are presented for the two current
  operators: CC2 (left panels) and CC1 (right panels). Off-shell
  effects are shown in the top panels where results are presented for
  $^{12}$C and three gauges: Lorentz, Coulomb and Weyl. Bottom panels
  refer to the analysis of second kind scaling comparing the results
  obtained for carbon, oxygen and calcium.}
\label{fL_gauge}
\end{center}
\end{figure}
\begin{figure}
\begin{center}
\includegraphics[scale=0.5]{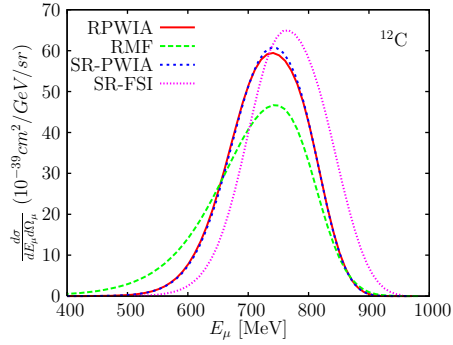}
\caption{Double differential cross section for charged-current
  neutrino scattering on $^{12}$C. Results correspond to fixed values
  of the neutrino energy and muon scattering angle:
  $\varepsilon_\nu=1$ GeV and $\theta_\mu=45^o$. Predictions of the
  RMF and RPWIA are compared with the results of a semirelativistic
  calculation (see text and Ref.~\cite{Amaro:2005dn} for details).}
\label{ja_quique}
\end{center}
\end{figure}
\begin{figure}
\begin{center}
\includegraphics[scale=0.5]{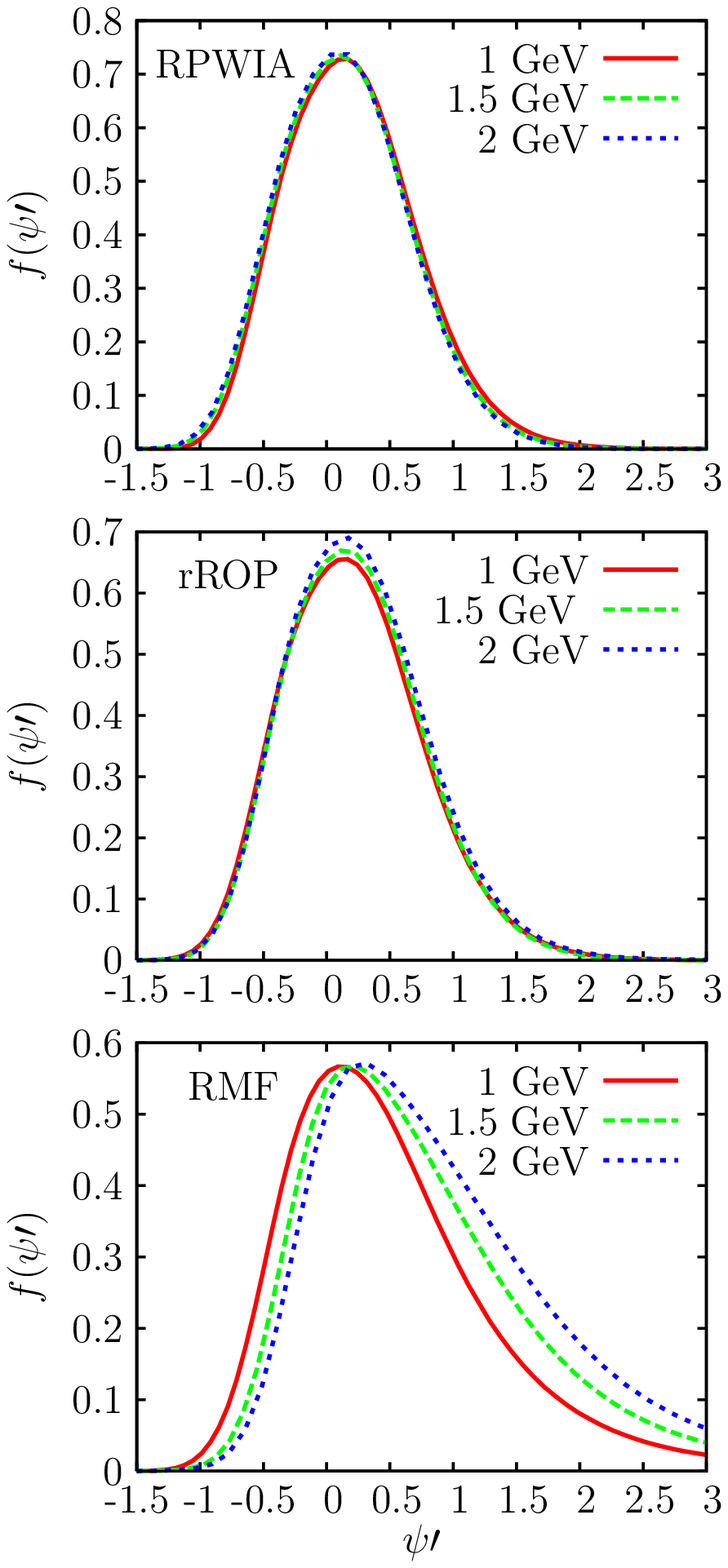}
\hspace{0.5cm}
\includegraphics[scale=0.5]{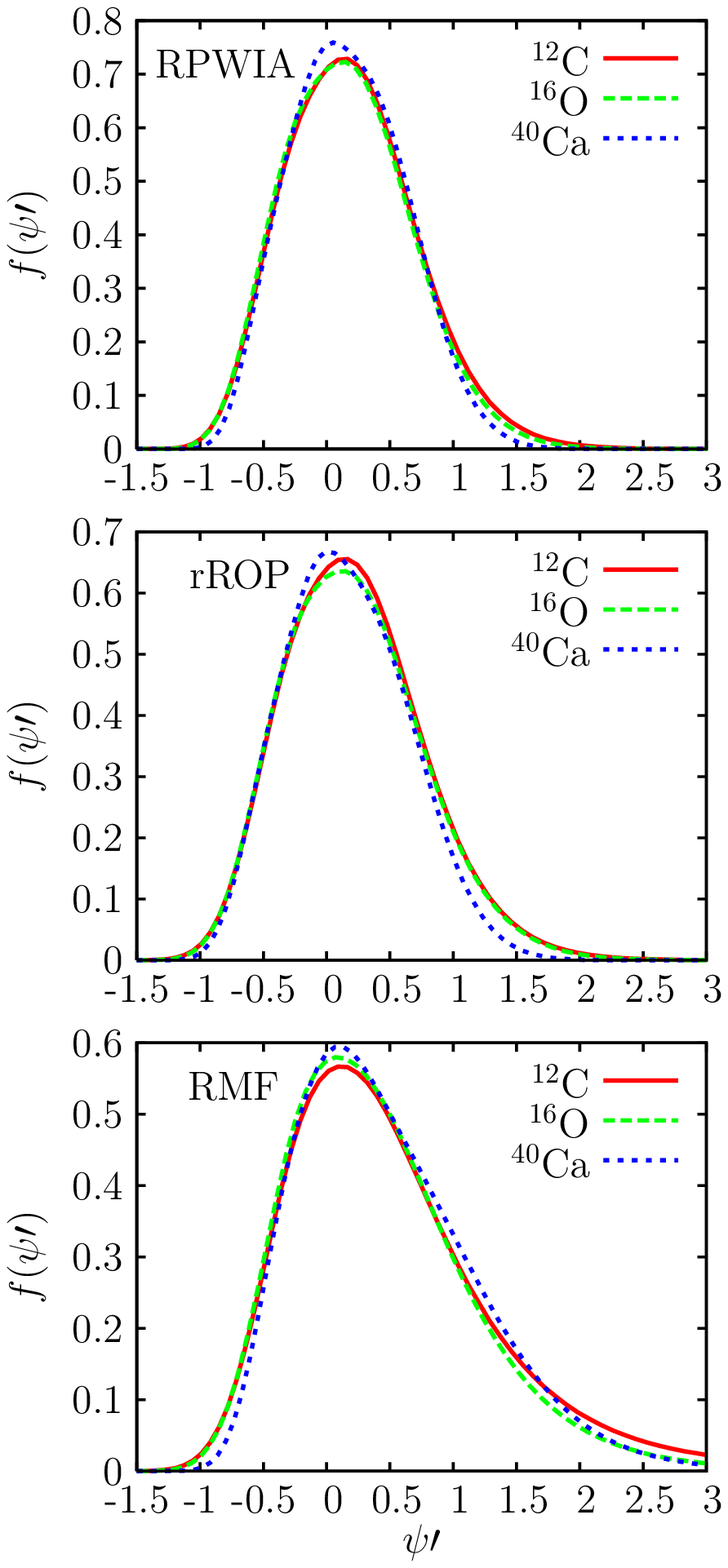}
\caption{Scaling function evaluated from $(\nu,\mu^-)$ on
  $^{12}$C. Results are shown for RPWIA (top panels), rROP (middle)
  and RMF (bottom). Scaling of the first kind is analyzed in the
  graphs on the left that compare results for different values of the
  energy beam. Scaling of the second kind is shown in the right panels
  that contain the scaling functions corresponding to $^{12}$C,
  $^{16}$O and $^{40}$Ca.}
\label{Fig2_neutrino}
\end{center}
\end{figure}
\begin{figure}
\begin{center}
\includegraphics[scale=0.5]{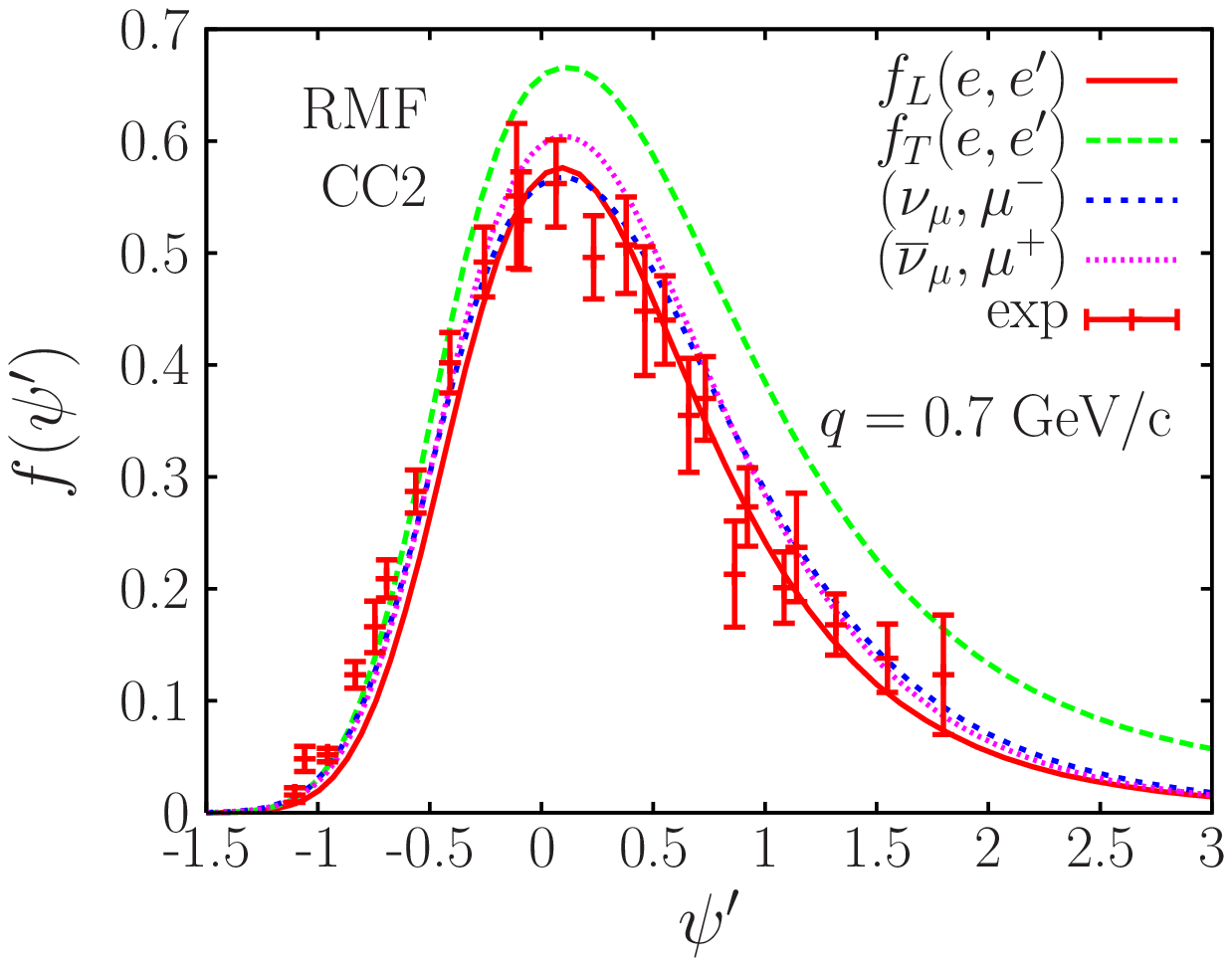}
\caption{(Color online) Longitudinal and transverse scaling functions
  for $(e,e')$ compared with $f(\psi')$ evaluated from $(\nu,\mu^-)$
  and $(\overline{\nu},\mu^+)$. All results correspond to the RMF
  approach using the CC2 current operator. The kinematics selected
  corresponds to fixed values of the incident lepton energy,
  $\varepsilon=1$ GeV, and momentum transfer, $q=0.7$ GeV/c. The
  averaged experimental function extracted from longitudinal electron
  scattering data is also shown~\cite{Maieron:2001it}. Figure taken
  from Ref.~\cite{Caballero:2007tz}.}
\label{Isospin_3new}
\end{center}
\end{figure}

To show how scaling of the first kind works,
in
Fig.~\ref{scaling_first_breaking} we compare  the longitudinal
scaling function evaluated with the RMF approach for different values
of $q$. A comparison
with data is also provided. Notice the shift in the RMF results to
larger positive $\psi$-values as the momentum transfer gets higher.
In spite of its merits,
the strong energy-independent scalar and vector potentials
involved in RMF lead to a very significant shift of the scaling
functions to higher $\omega$, and correspondingly, too severe a
decrease in the maxima. 

In Fig.~\ref{fL_gauge} (top panels) we analyze the off-shell effects
in the global RMF scaling function (including the contribution of both
the longitudinal and transverse channels).  We consider the two
current operators, CC1 and CC2, and show results for three different
gauges: Coulomb, Lorentz and Weyl
\cite{Caballero:2006wi,Caballero98a,Martinez:2003xs,Martinez:2002yx,Martinez:2002yw}. As
observed, the CC2 current leads to very similar results in the three
gauges whereas the Weyl prediction for the CC1 operator deviates very
significantly from the others. This gives us confidence in the use of
the CC2 current. The bottom panels of Fig.~\ref{fL_gauge} show the
scaling function for $^{12}$C, $^{16}$O and $^{40}$Ca for the two
current operators. As observed, scaling of second kind, i.e.,
independence of the scaling function on the nucleus, is highly
satisfied in the case of the CC2, while some discrepancies, mainly
connected with $^{40}$Ca, are present with the CC1 current. Again,
this reinforces our confidence in the use of the CC2 current operator.

The capability of the RMF model to provide a successful description of
the world QE $(e,e')$ data constitutes a benchmark in its extension to
the study of neutrino-nucleus reactions
\cite{Meucci:2011vd,Gonzalez-Jimenez14b,Caballero05,Caballero:2006wi,Amaro:2006tf,Amaro:2006if,Gonzalez-Jimenez:2013plb,Caballero:2007tz}. 
To show some illustrative
results, the differential cross sections for CCQE neutrino scattering
 $^{12}$C$(\nu_\mu,\mu^-)$  is compared in Fig.~\ref{ja_quique}
for different models. The energy of the
neutrino beam has been fixed to $\varepsilon_\nu=1$ GeV and the muon
scattering angle to $\theta_\mu=45^o$. We compare the RPWIA and RMF
predictions with  the semi-relativistic
model of \cite{Amaro:2005dn} with and without a WS potential (see
previous section). The
similarity between the RPWIA and SR-PWIA results  shows that,
within the plane wave approximation for the final state, the particular
description of the initial bound states as well as the current
operator (relativistic versus nonrelativistic) leads to almost
identical cross sections. On the contrary, the effects ascribed to
final state interactions, RMF versus SR-FSI (with WS potential),
produce cross sections that differ significantly. Not only the
strength in the maximum is very different, but also the shape of the
cross section with the long tail at smaller values of the final muon
energy being present in the RMF case.

In Fig.~\ref{Fig2_neutrino} we present the
neutrino scaling functions obtained
by dividing the $(\nu_\mu,\mu^-)$ cross section
by the weak single-nucleon cross sections \cite{Amaro05},
\begin{eqnarray}
  f^\nu(\psi) &=& \frac{k_F}{\overline{\sigma^{s.n.}_{\nu l}}}
      \frac{d^2\sigma}{dE_l d\cos\theta_l}
      \\
      \overline{\sigma^{s.n.}_{\nu l}} &=&
\sigma_0
\left\{
V_{CC} G_{CC}+
2{V}_{CL} G_{CL}
+{V}_{LL} G_{LL}+
{V}_{T} G_{T}
\pm
2{V}_{T'} G_{T'}
\right\}. 
\end{eqnarray}
where $\sigma_0$ is given in Eq. (\ref{eq:sig0}),
and the single nucleon factors $G_K$ are defined in Eq. (\ref{eq:RK}).
In the panels on the left we study scaling of
the first kind for the three models  RPWIA, rROP and
RMF. Panels on the right refer to scaling of second kind by showing
the results for carbon, oxygen and calcium. As noticed, both kinds of
scaling are fulfilled at high level except for the first kind within
the RMF model where a shift to higher values of the scaling variable
is clearly observed as the energy of the neutrino beam increases. This
was also the case for $(e,e')$ reactions. Thus we conclude that,
within the present models, scaling works at the same level of
precision for electron and neutrino scattering reactions with
nuclei.

In Fig.~\ref{Isospin_3new} we compare the scaling
functions corresponding to the two processes. In the case of electron
scattering we show separately the contributions ascribed to the two
channels, longitudinal and transverse, and compare with the results
for CC muon neutrinos (antineutrinos) processes. Notice that the
scaling function for neutrinos (antineutrinos), even being basically
transverse (the longitudinal channel is negligible), is more similar
to the longitudinal $(e,e')$ result than to the transverse one. This
is connected with the isoscalar/isovector contributions in electron
scattering processes compared with the pure isovector character of CC
neutrino reactions~\cite{Caballero:2007tz}. Moreover, the enhancement
shown by the $T$ contribution in $(e,e')$ is a consequence of the role
played by the lower components of the relativistic nucleon wave
functions. This effect is in accordance with the analysis of the
separate longitudinal/transverse data.

\subsection{Other models: RGF, SF, RPA, CRPA, GFMC}
     
\begin{figure}
\begin{center}
\includegraphics[scale=0.4]{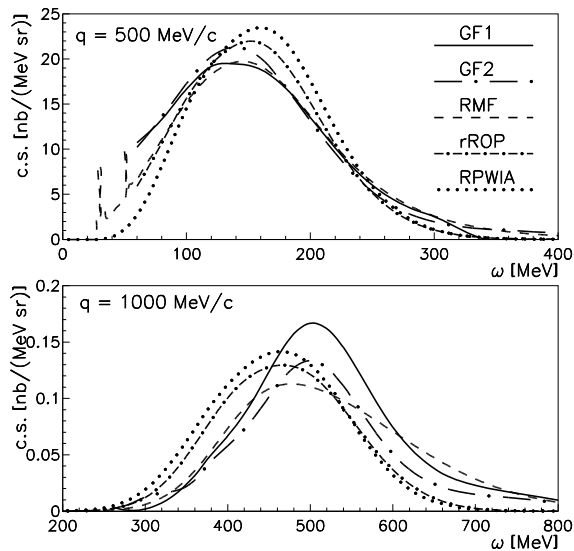}
\caption{Differential cross section of the $^{12}$C$(e,e')$ reaction
  at electron beam energy of 1 GeV. The GF predictions are compared
  with the RMF, rROP and RPWIA approaches. GF1 and GF2 refer to the
  RGF results obtained with EDAD1 and EDAD2 optical potentials. Figure
  taken from~\cite{Meucci:2009nm}.}
\label{Meucci}
\end{center}
\end{figure}  

An alternative, relativistic approach to FSI in $(e,e')$ and
$(\nu_\ell, \ell)$ reactions is provided by the Relativistic Green
Function (RGF) model. Here the description of the initial nucleon
states is analogous to the RMF model presented in the previous
section, but the scattered nucleon wave function is computed
with Green's function techniques.
The RGF method treats FSI
consistently in the inclusive and exclusive channels by
using a complex relativistic optical potential that describes
elastic nucleon scattering.
The flux reduction by the
imaginary part of the optical potential is redistributed into the other
channels, so the total flux is conserved in the
inclusive process.
The RGF has been applied to electron and
neutrino-nucleus scattering in 
Refs.~\cite{Meucci:2009nm,Giusti:2009ym,Meucci:2011pi,Meucci:2011vd,Gonzalez-Jimenez:2013xpa,Ivanov16b,Meucci:2003cv,Meucci03}.
As
an example we present in Fig.~\ref{Meucci} the inclusive
$^{12}$C$(e,e')$ cross section for fixed
 momentum transfers
and initial electron energy
$\varepsilon=1$ GeV. In each panel we compare the RGF predictions
with two different optical potentials  (GF1 and GF2),
with the results of the RMF,
rROP and RPWIA approaches.
As seen in the figure,
there are large differences between GF1 and GF2 cross sections depending 
on the kinematics
and both results  also deviate
significantly from the other models. Particularly interesting is that
the large asymmetry shown by the RMF prediction at $q=1$ GeV/c  is lost in
the other approaches.


\begin{figure}
\begin{center}
\includegraphics[scale=1.5]{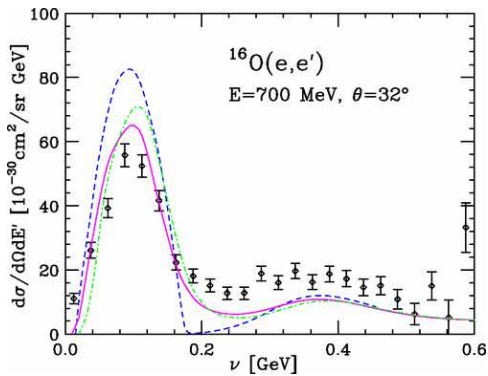}
\caption{Double differential cross section of the $^{16}$O$(e,e')$
  reaction at electron beam energy 700 MeV and electron scattering
  angle $32^o$. The SF calculation including FSI (solid line) is
  compared with the IA (no FSI) calculation (dot-dashed line) and the
  Fermi Gas (FG) model with $k_F=225$ MeV and a shift energy
  $\epsilon=25$ MeV. Data from Ref.~\cite{Anghinolfi:1996vm}. Figure
  taken from~\cite{Benhar:2005dj}.}
\label{Benhar}
\end{center}
\end{figure} 

\begin{figure}
\begin{center}
\includegraphics[scale=0.3]{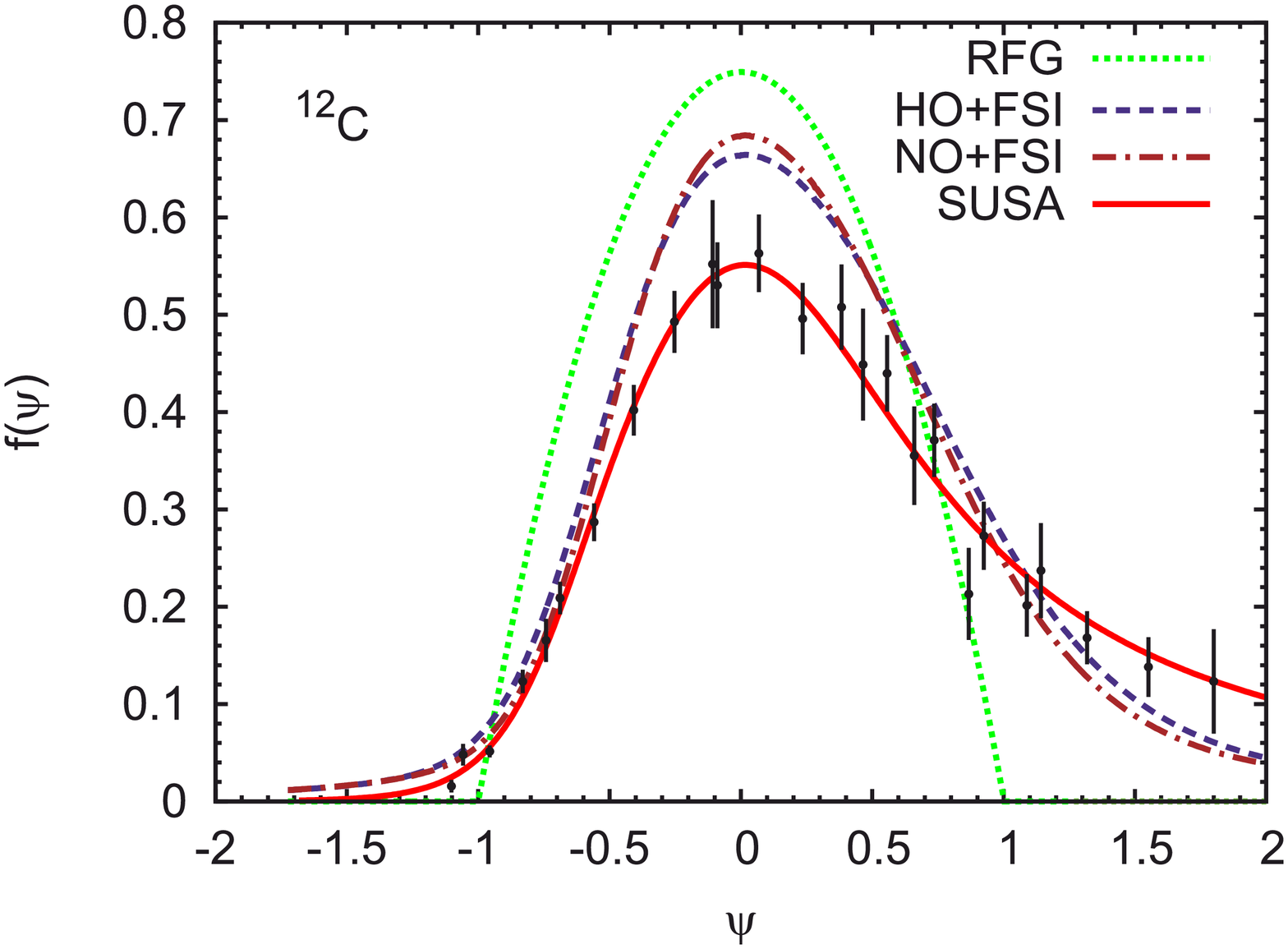}
\caption{Results for the superscaling function for $^{12}$C obtained
  using harmonic oscillator (HO) and natural orbitals (NO) approaches
  with FSI (see text for details). Comparison with the RFG and
  Superscaling (SUSA) predictions, as well as with the longitudinal
  experimental data. Figure taken from~\cite{M.V.Ivanov:2019jqp}.}
\label{Antonov}
\end{center}
\end{figure} 

In the  {\em spectral function} (SF) approach the inclusive scattering cross
section is factorized as the product of
the single-nucleon cross section and the
nuclear spectral function that gives the probability of removing a nucleon
 with certain values of momentum and
energy. Although factorization does not hold in general it
works well in the QE regime.
Contrary to the independent particle picture where the
bound nucleon shells correspond to discrete energy eigenvalues, the
formalism of the spectral function introduces dynamical correlations
induced by nucleon-nucleon (NN) force.
As a result,
the spectral function acquires tails that extend to large energy and
momentum. Note that the spectral function
formalism is non-relativistic. This can
introduce some doubts on its application to processes where the energy
and momentum transfers can be very high.

The existence of final state interactions between the ejected nucleon
and the residual nucleus has long been experimentally
established. However, the factorization assumption, implicit in the
spectral function formalism, is based on the Plane Wave Impulse
Approximation (PWIA). Different approaches have been used in the past
to incorporate FSI in the spectral function formalism. In most of the
cases the cross section (response functions) is finally written in
terms of the impulse approximation (IA) cross section convoluted with
a folding function that embodies FSI effects. A very detailed study on
this topic can be found in
\cite{Benhar:2005dj,Benhar:2006wy,Sick:1994vj,Vagnoni:2017hll,Rocco:2016ejr,Rocco16,Benhar:2015xga,Benhar:2013dq,Rocco16,Rocco:2018mwt}. We
present in Fig.\ref{Benhar} an illustrative example where it is
clearly shown that FSI produce a shift and a redistribution of the
strength leading to a quenching of the peak and to an enhancement of
the tail.

In the studies presented in 
Refs.~\cite{M.V.Ivanov:2019jqp,Ivanov:2013saa,Antonov:2011bi}
the spectral function
is constructed starting with the independent particle model and
introducing a Lorentzian function to describe the energy
dependence.  FSI are incorporated through a
time-independent optical potential~\cite{Ankowski:2007uy,Horikawa}.  In
Fig.~\ref{Antonov} we present some illustrative results for the
superscaling function evaluated from the double differential cross
section.
The bound nucleon wave functions are computed with the
harmonic oscillator (HO) and natural orbitals (NO) models.
These spectral function models also include FSI.
As noticed, the
role of FSI leads to a redistribution of the strength, with lower
values of the scaling function at the maximum and an asymmetric shape
around the peak position.
The scaling properties of the SF approach have also
been discussed in Ref.~\cite{Sobczyk:2017vdy}.

Other descriptions of the inclusive responses include the effects of
random phase approximation (RPA) in the local Fermi Gas and
Hartree-Fock (HF) models. In the former, RPA correlations are
introduced through a Landau Migdal residual interaction with pion and
rho exchange. This is the general strategy followed in a long series
of works performed by Martini {et al.} and Nieves and
collaborators  \cite{Martini:2009uj,Nieves:2011yp}. These calculations include contributions arising from
many-particle, many-hole (np-nh) excitations.
A fully relativistic model  of 
meson exchange currents (MEC) in 2p2h contributions is discussed in
 sect. \ref{sec:2p2h}.

A shell-model approach to RPA correlations is developed by the Gent
group. Bound and scattered nucleon wave functions are obtained with a
self-consistent Hartree-Fock model using a Skyrme-type nucleon-nucleon
interaction.  The HF model is later extended with collective
excitations described through a continuum random phase approximation
(CRPA).  Although inherently non-relativistic, the HF-CRPA model
incorporates some relativistic corrections
\cite{Amaro96,Jeschonnek:1997dm}, providing reliable results for the
scattering observables from very low momentum transfers, where
long-range correlations play an essential role, to moderate values of
$q$ \cite{Pandey:2016ee,Pandey15,VanCuyck:2016fab,Martini:2016eec,Jachowicz:2019eul}.

Studies on the
electromagnetic responses, as well as the
extension to neutrino-nucleus scattering processes
have also been performed with the Green's Function Monte Carlo
(GFMC) model
\cite{Lovato:2016gkq,Lovato:2015qka,Rocco:2016ejr,Lovato:2017cux,Pastore:2017uwc,Lynn:2017fxg,Lovato:2015oea,Carlson:2014vla,Bogner:2013pxa,Rocco:2017hmh}. This {\it ab initio} approach allows for a very
accurate description of the dynamics of constituent nucleons in
nuclei. In spite of some recent remarkable results concerning the
electromagnetic responses in QE inclusive $(e,e')$ reactions, the
computational complexity of such an approach is currently limited to
light nuclei ($^{12}$C). Not only the cost of calculation increases
exponentially with the number of nucleons, but also the need to
include relativistic kinematics and baryon resonance production
involve non trivial difficulties.

\subsection{Superscaling models: SuSA and SuSAv2}
\label{sec:susav2}

 \subsubsection*{SuperScaling Approach: a semiphenomelogical model}

\begin{figure}
\begin{center}
\vspace*{0.38cm}
\includegraphics[scale=0.4,clip,angle=0]{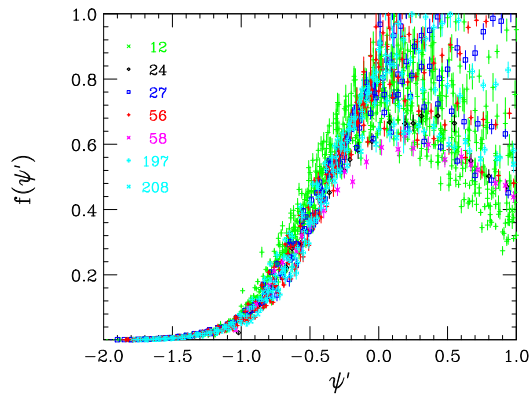}
\caption{Scaling function $f(\psi')$ as a function of $\psi'$ for
  different nuclei ($A\geq12$) and several kinematics covering from
  forward to very backward angles as well as from low to very high
  incident energies. The values of $A$ corresponding to different
  symbols are shown in the figure. Data taken from~\cite{Donnelly99b}.
}\label{fig2}
\end{center} 
\end{figure}

In \ref{section-rfg} we have seen that in the RFG all the response
functions of electron and neutrino scattering are proportional to the
same scaling function of the RFG, Eqs.  (\ref{rfge},\ref{eq:RK}).
This is in general not true in the real world as there are many
mechanisms violating scaling.  Starting with the experimental $(e,e')$
cross section data one can compute the experimental superscaling
function by dividing by the single nucleon cross section. In Fig.~\ref{fig2} data are plotted as a function of
the shifted variable, $\psi'$, defined with a shifted $\omega \rightarrow \omega - E_s$.  
As seen in the figure, scaling of the $(e,e')$ cross section is approximately fulfilled
in the region below the QE peak ($\psi'<0$), with a
narrow spread of data. The values of $E_s$ and $k_F$
were fitted to maximize scaling in this region.  On the
contrary, scaling is violated in the region above the QE peak, where
contributions breaking factorization play an important role.

\begin{figure}
\begin{center}
\includegraphics[scale=0.45,clip,angle=0]{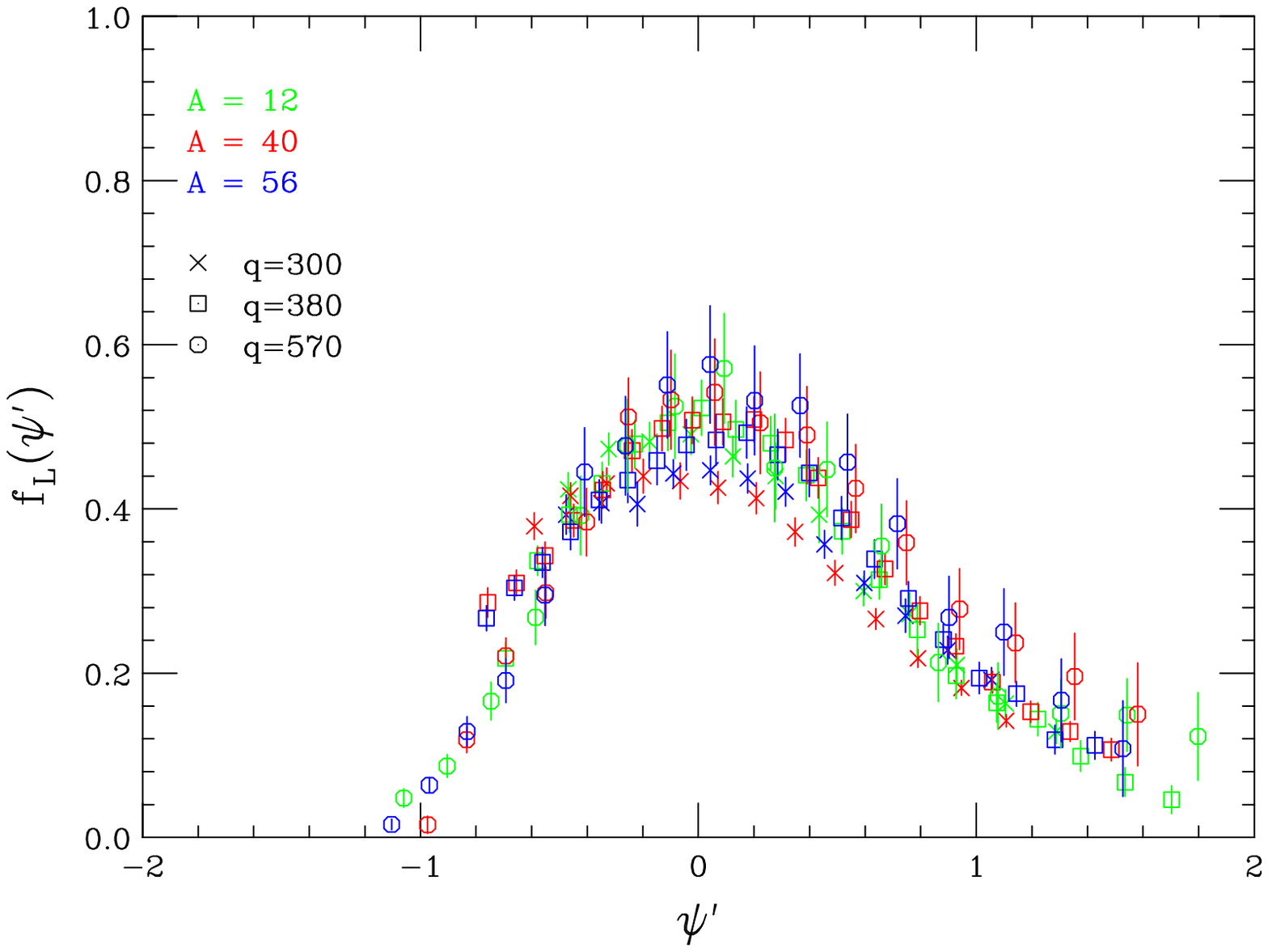}
\includegraphics[scale=0.45,clip,angle=0]{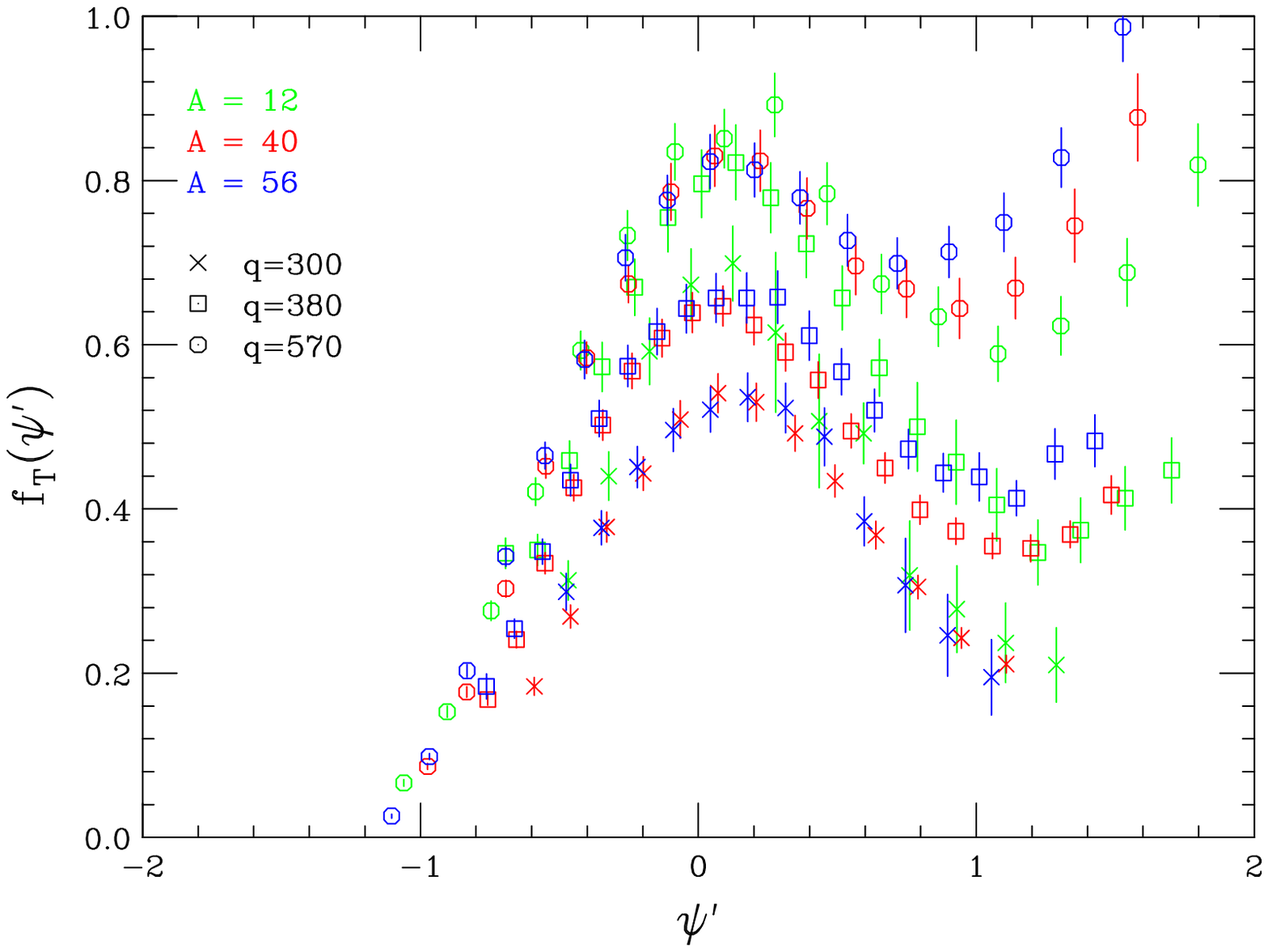}
\vspace{-0.3cm}
\caption{Scaling function, $f_L(\psi')$ and $f_T(\psi')$, from the
  longitudinal and transverse response, respectively, as a function of
  $\psi'$ for different nuclei ($A\geq12$) and for different values of
  $q$ (in MeV/c). Data taken from~\cite{Jourdan:1996aa}.
}\label{fig11}
\end{center} 
\end{figure}

This is clarified by studying the experimental longitudinal and
transverse response functions, that have been measured for several
nuclei. The corresponding longitudinal, $f_L(\psi')$, and transverse,
$f_T(\psi')$, experimental data, obtained dividing the responses by
the single nucleon factors, Eq. (\ref{eq:fT}), are shown in
Fig.~\ref{fig11}.  We see that the longitudinal response approximately
scales in the whole region of the QE peak, whereas the transverse data
do not scale, being scaling violations more prominent in the region
above the maximum of QE peak ($\psi'>0$).  Scaling violations at high
$\omega$ are produced by meson production and resonance excitations,
which are predominantly transverse.  Also
contributions beyond the impulse approximation such as two-particle
two-hole excitations produced by MEC, come into play in the transverse
channel (see Sect. \ref{sec:2p2h}).

In the SuperScaling Analysis (SuSA) one assumes that
$f_L=f_T$  if one
subtracts the contributions that violate scaling.
This is usually known as scaling of the zeroth kind.
Therefore the SuSA prescription is to identify
the universal scaling function with the experimental longitudinal
responses. This allows one
to compute all the response functions, for both electron and
neutrino scattering, by multiplying by the corresponding single nucleon factor
of the RFG. 
The analysis of the longitudinal response functions data
leads to the results already shown in Fig.~\ref{fig:SuSA} ,where
the solid line corresponds to a fit of data given by the following
parametrization of the SuSA scaling function
\begin{equation}
  f_{SuSA}(\psi')\equiv
  f_L(\psi')=\frac{p_1}{[1+p_2^2(\psi'-p_3)^2](1+e^{p_4\psi'})}
\end{equation}
with $p_1=2.9883$, $p_2=1.9438$, $p_3= 0.67310$ and $p_4=-3.8538$. As
already shown in Fig.~\ref{fig:SuSA}, the behavior of data leads to
the striking difference between $f_{SuSA}(\psi')$ and the prediction
provided by RFG, the former with a pronounced asymmetrical tail
extended towards positive values of $\psi'$, {\it i.e.},
$\omega>\omega_{QEP}$, and a maximum value
$f_{SuSA}(0)\sim 0.6$. This behavior sets a strong constraint to any
model aimed at describing lepton-nucleus scattering processes. In
fact, we have seen in the previous sections that most of the nuclear
models, even including FSI in different ways, fail to describe the
experimental scaling function, especially the asymmetry, while the RMF
is in accord with data reasonably well 
in a wide kinematical region \cite{Gonzalez-Jimenez14b}).

\subsubsection*{Extension of the Superscaling Approach
  from Relativistic Mean Field Theory: the SuSAv2 Model}

The motivation of the SuSAv2 model
is based on the RMF finding that 0th-kind scaling is
broken even in the impulse approximation, and generally $f_T(\psi') >
f_L (\psi')$.  This enhancement of the transverse response is an
expected property in order to reproduce the $(e,e')$ cross section
after adding the contributions of pion emission and MEC.
The SuSAv2 model extends the original SuSA by
incorporating the transverse scaling function from the RMF theory 
\cite{Gonzalez-Jimenez14b}.
The SuSAv2 model is based on the following four assumptions:
\begin{enumerate}
 \item $f_L$ superscales, {\it i.e}, it is independent of the
   momentum transfer and of the nuclear
   species. This is supported by experimental data.

    \item $f_T$ superscales.  Experimentally this is 
      approximately true  in the region $\psi<0$ for a wide
      range of $q$ ($400<q<4000$ MeV/c) \cite{Maieron:2001it}.
      We assume that superscaling can be extended to the whole range of
      $\psi$ once the contributions from non-QE processes are removed.

 \item The RMF model reproduces quite well the relationships between
   all scaling functions.  This
   assumption is supported by the fact that the RMF model is able to
   reproduce the longitudinal scaling function,
    and the fact that it naturally yields the
   enhancement of the transverse response, $f_T^{ee'}>f_L^{ee'}$.

 \item At very high $q$ the effects of FSI disappear and all scaling
   functions must approach the RPWIA results.

\end{enumerate}

Contrary to what is assumed in the SuSA model, where 
there is only one universal scaling function used
to build all nuclear responses, within SuSAv2 we use three RMF-based
{\it reference} scaling functions (which will be indicated with the
symbol $\widetilde f$): one for the transverse response, one for the
longitudinal isovector responses and another one to describe the
longitudinal isoscalar response in electron scattering.

\begin{figure}[htbp]
   \centering
         \includegraphics[width=0.3\textwidth,angle=270]{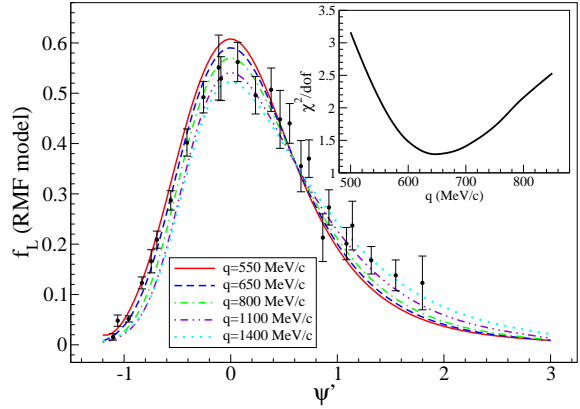}
     \caption{Longitudinal scaling function for $(e,e')$ computed
       within RMF.  The scaling functions have been shifted to place
       the maximum at $\psi'=0$.  In the inner smaller plot the
       reduced-$\chi^2$,
       taking into account  the experimental errors,
       is presented versus $q$. The minimum $\chi^2$ is around $q=650$
       MeV/c.  Data from Ref.~\cite{Donnelly99b}.}
     \label{fig:fLRMF}
\end{figure}

We employ the experimental scaling function $f_{L,\text{exp}}^{ee'}$
as guide in our choices for the {\it reference} ones.  In
Fig.~\ref{fig:fLRMF} we display the RMF longitudinal scaling function,
$f_L$, for several values of $q$. The
functions have been relocated by introducing an energy shift
\cite{Gonzalez-Jimenez14b} so that the maximum is at
$\psi'=0$. It appears that scaling of first kind is not perfect and
some $q$-dependence is observed.  Although all the curves are roughly
compatible with the experimental error bars, the scaling function that
produces the best fit to the data corresponds to $q\approx 650$
MeV/c. This is the result of a $\chi^2$-fit to the experimental
data of $f^{ee'}_{L,exp}$, as illustrated in the inner plot in
Fig.~\ref{fig:fLRMF}.

According to this result, we identify the reference scaling functions
with $f^{T=1,ee'}_L$, $f^{T=0,ee'}_L$ and $f^{T=1,ee'}_T$ evaluated
within the RMF model at $q=650$ MeV/c and relocated so that the
maximum is at $\psi'=0$:
\begin{eqnarray}
 \widetilde f_T &\equiv& f_T^{T=1,ee'}|^{RMF}_{q=650} \\
 \widetilde f_{L}^{T=1} &\equiv& f_L^{T=1,ee'}|^{RMF}_{q=650} \\
 \widetilde f_{L}^{T=0} &\equiv& f_L^{T=0,ee'}|^{RMF}_{q=650}\,,
\end{eqnarray}
where the isospin scaling functions, for $T=0,1$,
are defined by dividing the isospin response functions by the corresponding single nucleon contribution from the RFG
\begin{equation}
  f_K^{T,ee'}= k_F \frac{R^{T e.m.}_K}{G_K^{T e,e'}} .
\end{equation}
Thus, by construction, the $(e,e')$ longitudinal scaling function
built within SuSAv2 is $f_L|^{SuSAv2}=f_L|^{RMF}_{q=650}\approx
f_{L,exp}^{ee'}$.
We have parametrized these reference scaling functions by using a
skewed-Gumbel function \cite{Gonzalez-Jimenez14b}.

The neutrino scaling functions necessary to compute the nuclear
neutrino responses (see also Section~\ref{sec:genform}) are defined from the
reference ones as follows:
\begin{eqnarray}
\label{ratio1}
f_L^{VV,\nu(\overline{\nu})}(q) &\equiv& \widetilde f_{L}^{T=1} \\
\label{ratio2}
f_T^{VV,\nu(\overline{\nu})}(q) &=&
f_T^{AA,\nu(\overline{\nu})}(q) =
f_{T'}^{VA,\nu(\overline{\nu})}(q) = 
\widetilde f_{T}\,,
\\
f_{CC}^{AA,\nu(\overline{\nu})}(q) &=& 
f_{CL}^{AA,\nu(\overline{\nu})}(q) = 
f_{LL}^{AA,\nu(\overline{\nu})}(q) = 
\widetilde f_{L}^{T=1}\,.
\end{eqnarray}

\begin{figure}[htbp]
  \centering
         \includegraphics[width=0.3\textwidth,angle=270]{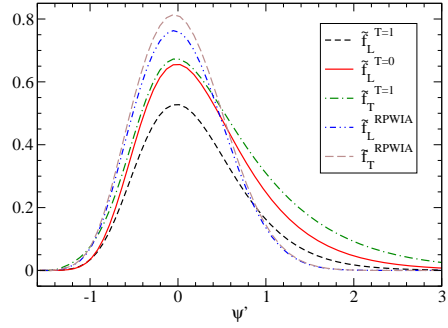}\vspace*{-0.4cm}
     \caption{Reference scaling functions for ($e,e')$ and CC neutrino
       induced reactions in the SuSAv2 model.}
     \label{refsusav2}
  \end{figure}

Finally, in order to implement the approaching of the RMF results to
the RPWIA ones at high kinematics, that is, the disappearance of FSI at high $q$, we build the SuSAv2 $L$ and $T$ scaling functions as
linear combinations of the RMF-based and RPWIA reference scaling
functions:
\begin{eqnarray}\label{fsusav2eq}
{\cal F}_L^{T=0,1} &\equiv& \cos^2 \chi(q) \widetilde f_L^{T=0,1} +
\sin^2 \chi(q) \widetilde f_L^{RPWIA}  \\
{\cal F}_T &\equiv& \cos^2 \chi(q) \widetilde f_T +
\sin^2 \chi(q) \widetilde f_T^{RPWIA}, 
\end{eqnarray}
where the "blending function" 
$\chi(q)$ is a $q$-dependent angle given by
\begin{equation}\label{transxi}
\displaystyle\chi(q)\equiv\frac{\pi}{2}\left(1-\left[1+e^{\left(\frac{q-q_0}{\omega_0}\right)}\right]^{-1}\right)
\end{equation}
with $q_0$ and $w_0$ the transition parameters between RMF and RPWIA
prescriptions which are given in Refs.~\cite{Gonzalez-Jimenez14b,
Megias16a}.
The reference RPWIA scaling functions, $\widetilde f_K^{RPWIA}$, are
evaluated at $q=1100$ MeV/c, where FSI effects are assumed to be
negligible, while the reference RMF scaling functions, $\widetilde
f_K$, are evaluated at $q=650$ MeV/c and are shown in
Fig.~\ref{refsusav2}. The explicit expressions of the RMF ($\widetilde
f_K$) and RPWIA ($\widetilde f_K^{RPWIA}$) scaling functions are given
in~\cite{Gonzalez-Jimenez14b}.
With this procedure we get a description of the responses based on RMF
behavior at low-intermediate $q$ values, while for higher momentum
transfers  the RPWIA trend is mimicked.  The transition between RMF and
RPWIA behaviors occurs at intermediate $q$-values, namely $\sim q_0$,
in a region of width $\sim w_0$.

The response functions are simply built as: \ba
 R_L^{ee'}(q,\omega) &=& \frac{1}{k_F}
    \left[ {\cal F}_{L}^{T=1}(\psi')  G_L^{T=1}(q,\omega)\right.\left.+  {\cal F}_{L}^{T=0}(\psi') G_L^{T=0}(q,\omega)\right]
             \label{RLee}\\
 R_T^{ee'}(q,\omega) &=& \frac{1}{k_F}
       {\cal F}_T(\psi') \left[G_T^{T=1}(q,\omega)\right.+ \left. G_T^{T=0}(q,\omega)\right]
      \label{RTee}
\ea
\ba
 R_L^{VV,\nu(\overline{\nu})}(q,\omega) = \frac{1}{k_F}
   {\cal F}_{L}^{T=1}(\psi') G_L^{VV}(q,\omega)\\
 R_{CC}^{AA,\nu(\overline{\nu})}(q,\omega)= \frac{1}{k_F}
    {\cal F}_{L}^{T=1}(\psi') G_{CC}^{AA}(q,\omega) \\
 R_{CL}^{AA,\nu(\overline{\nu})}(q,\omega) = \frac{1}{k_F}
    {\cal F}_{L}^{T=1}(\psi') G_{CL}^{AA}(q,\omega) \\
 R_{LL}^{AA,\nu(\overline{\nu})}(q,\omega) = \frac{1}{k_F}
    {\cal F}_{L}^{T=1}(\psi') G_{LL}^{AA}(q,\omega)
%
\ea
\ba
 R_T^{\nu(\overline{\nu})}(q,\omega) &=& \frac{1}{k_F}
             {\cal F}_T(\psi') \left[G_T^{VV}(q,\omega)\right.+ \left. G_T^{AA}(q,\omega)\right]
             \label{RTnu}\\
 R_{T'}^{\nu(\overline{\nu})}(q,\omega) &=& \frac{1}{k_F}
     {\cal F}_T(\psi')G_{T'}^{VA}(q,\omega).
    \label{RT'nu}
\ea

Furthermore, in order to reproduce the peak position of RMF and RPWIA
scaling functions within SuSAv2 we consider a $q$-dependent energy
shift, namely, $E_{shift}(q)$. This quantity modifies the scaling
variable
$\psi(q,\omega)\longrightarrow\psi'=\psi(q,\omega-E_{shift})$.
In
particular, we build this function $E_{shift}(q)$ from the results of
the RMF and RPWIA models presented in~\cite{Gonzalez-Jimenez14b}.  The
choice of $E_{shift}(q)$ depending on the particular $q$-domain region
considered is solely based on the behavior of the experimental cross
sections and their comparison with our theoretical predictions (see
results in Section~\ref{sec:susav2-results}).

In Section~\ref{sec:susav2-results} we will show that the SuSAv2 model
describes more accurately the neutrino induced reactions than the
semiphenomenological SuSA approach. This result mainly comes from the
$q$-dependence on the energy shift, the enhancement on the transverse
response via RMF prescriptions as well as the possibility of
separating scaling functions into isoscalar and isovector
contributions.

\section{Relativistic model for CC MEC and 2p2h responses}
\label{sec:2p2h} 

In this section we present a fully relativistic model of meson
exchange currents (MEC) to describe the inclusive neutrino scattering
in the two-nucleon emission channel (2p-2h).  The present approach was
developed in the RFG model of Ref.~\cite{Simo:2016ikv}, where the
difficulties of the relativistic reaction are overcome thanks to the
plane waves used for the single-nucleon states. However this channel
is cumbersome on the computational side, if compared to the analytical
simplicity of the 1p-1h responses.

The 2p-2h hadronic tensor is given by summation (integration) over all
the 2p-2h excitations of the RFG with two holes $\nh_1$, $\nh_2$, and
two particles $\np'_1$, $\np'_2$, in the final state, with $h_i < k_F$
and $p'_i> k_F$
 \begin{eqnarray}
W^{\mu\nu}_{\rm 2p2h}
=\frac{V}{(2\pi)^9}\int
d^3p'_1
d^3p'_2
d^3h_1
d^3h_2
&&
\frac{m_N^4}{E_1E_2E'_1E'_2}
w^{\mu\nu}(\np'_1,\np'_2,\nh_1,\nh_2)\;
\delta(E'_1+E'_2-E_1-E_2-\omega)
\nonumber\\
&&
\times 
\Theta(p'_1,h_1)\Theta(p'_2,h_2)
\delta(\np'_1+\np'_2-\nq-\nh_1-\nh_2) \, ,
\label{amaro-hadronic12}
\end{eqnarray}
where the Pauli blocking function $\Theta$ is defined as 
the product of step-functions 
\begin{equation}
\kern -8mm
\Theta(p',h) \equiv
\theta(p'-k_F)
\theta(k_F-h) .
\end{equation}
The 2p-2h equivalent to the single-nucleon hadronic tensor inside the
integral is called $w^{\mu\nu}(\np'_1,\np'_2,\nh_1,\nh_2)$, and
describes two-nucleon transitions with given initial and final
momenta, summed up over spin and isospin,
\begin{equation}
w^{\mu\nu}(\np'_1,\np'_2,\nh_1,\nh_2) = \frac{1}{4}
\sum_{s_1s_2s'_1s'_2}
\sum_{t_1t_2t'_1t'_2}
j^{\mu}(1',2',1,2)^*_A
j^{\nu}(1',2',1,2)_A \, .
\label{amaro-elementary}
\end{equation}
Here the 2p-2h current function, in short-hand notation for the full
spin-isospin-momentum dependence, is
\begin{equation} 
j^\mu(1',2',1,2)
\equiv j^\mu(\np'_1s'_1t'_1,\np'_2s'_2t'_2,\nh_1s_1t_1,\nh_2s_2t_2)
\end{equation}
and $j^\mu(1',2',1,2)_A$ refers to the 
direct minus exchange matrix elements
\begin{equation} \label{amaro-anti}
j^{\mu}(1',2',1,2)_A
\equiv j^{\mu}(1',2',1,2)-
j^{\mu}(1',2',2,1) \,.
\end{equation}
 The factor $1/4$ in Eq.~(\ref{amaro-elementary}) accounts for the
 antisymmetry of the two-body wave function, to avoid double counting
 in the number of final 2p-2h states.  The exchange $1\leftrightarrow
 2$ in the second term implies implicitly the exchange of momenta,
 spin and isospin quantum numbers.

Momentum conservation allows to integrate over $\np'_2$ to compute the
2p-2h response functions as a 9-dimensional integral
\begin{eqnarray}
R^{K}_{\rm 2p2h}
&=&
\frac{V}{(2\pi)^9}\int
d^3p'_1
d^3h_1
d^3h_2
\frac{m_N^4}{E_1E_2E'_1E'_2}
\Theta(p'_1,h_1)\Theta(p'_2,h_2)
r^{K}(\np'_1,\np'_2,\nh_1,\nh_2)\;
\delta(E'_1+E'_2-E_1-E_2-\omega) ,
\label{amaro-hadronic}
\end{eqnarray}
where $\bf p'_2= h_1+h_2+q-p'_1$.  The five elementary response
functions for a 2p2h excitation, $r^K$, are defined in terms of the
elementary hadronic tensor $w^{\mu\nu}$, for the five indices
$K=CC,CL,LL,T,T'$.

Due to azimuthal symmetry of nucleon emission around the $z$ axis,
defined by $\nq$, we can fix the azimuthal angle of particle 1', by
setting $\phi'_1=0$, and multiplying by a factor $2\pi$. Finally, the
Dirac delta-function enables us to integrate over $p'_1$, and so the
integral in Eq.~(\ref{amaro-hadronic}) can be reduced to seven
dimensions.  The details of the integration method can be found in
Ref.~\cite{Simo:2014wka}.

\begin{figure}[t]
\centering
\includegraphics[width=10cm,bb=110 310 500 690]{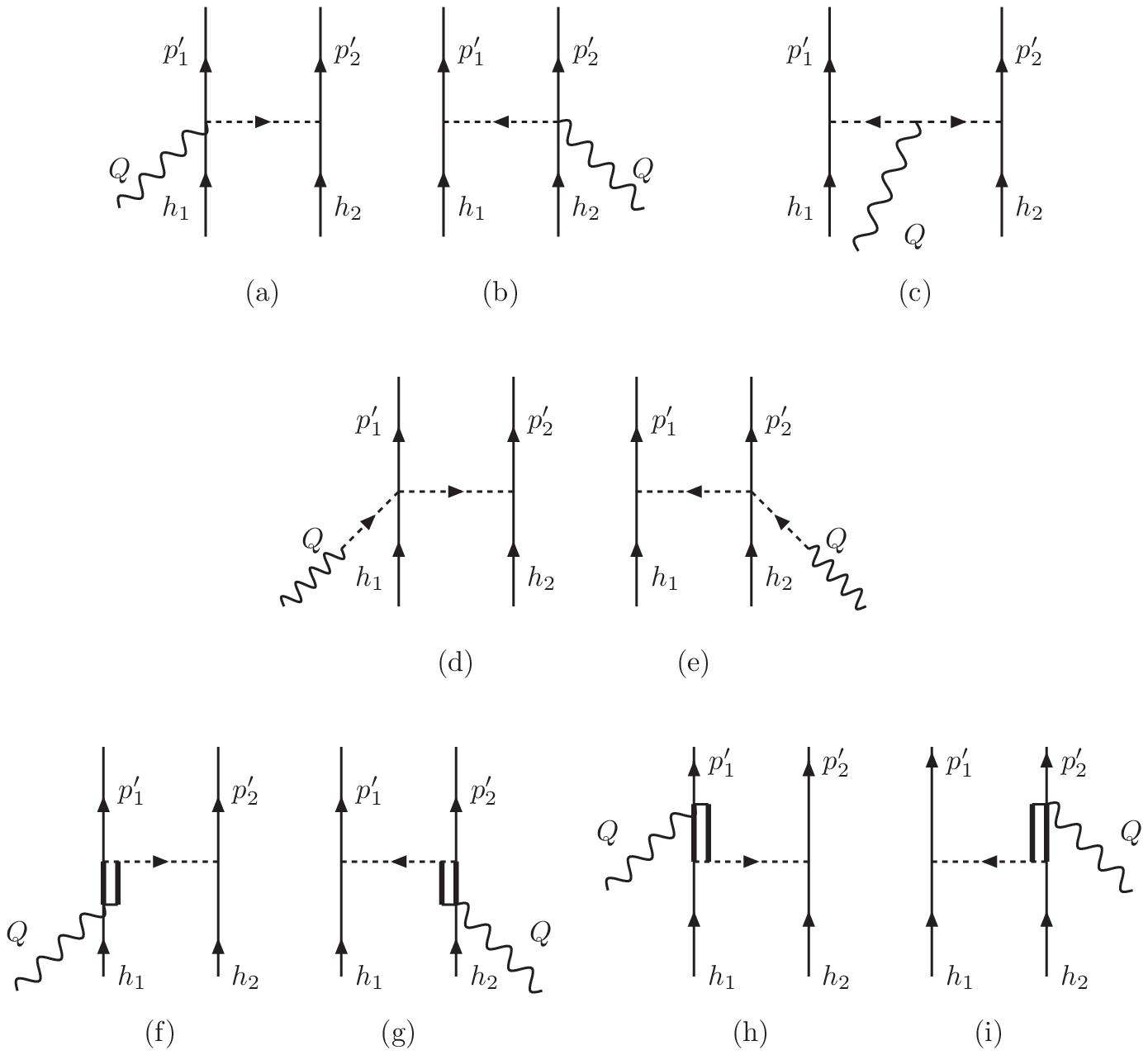}
\caption{Feynman diagrams for the electroweak MEC model discussed in 
this review \cite{Simo:2016ikv}.  }\label{fig_feynman}
\end{figure}

To specify a model for the two-body current matrix elements
$j^{\mu}(1',2',1,2)$, we consider  MEC containing the Feynman
diagrams depicted in Fig.~\ref{fig_feynman}.  The different
contributions have been taken from the pion production model of
\cite{Hernandez07}.

Our MEC are given as the sum of four two-body currents: 
\begin{equation}
j^\mu(1',2,1,2) = 
j_{\rm sea}^\mu +j_{\rm \pi}^\mu + j_{\rm pole}^\mu + j_{\rm \Delta}^\mu,
\end{equation}
corresponding in Fig. \ref{fig_feynman} to the seagull (diagrams
a,b), pion in flight (c), pion-pole (d,e) and $\Delta(1232)$
excitation (f,g,h,i).  Their expressions are given by
 \begin{eqnarray}
 j^\mu_{\rm sea}&=&
  \left[I_V^{\pm}\right]_{1'2',12}
\frac{f^2}{m^2_\pi}
V_{\pi NN}^{s'_1s_1}(\np'_1,\nh_1) 
 \nonumber\\
 &\times& 
 \bar{u}_{s^\prime_2}(\np^\prime_2)
 \left[ F^V_1(Q^2)\gamma_5 \gamma^\mu
 + \frac{F_\rho\left(k_{2}^2\right)}{g_A}\,\gamma^\mu
   \right] u_{s_2}(\nh_2)
   +
   (1\leftrightarrow2) \,
\label{seacur}
\\
 j^\mu_{\pi}&=& \left[I_V^{\pm}\right]_{1'2',12}
 \frac{f^2}{m^2_\pi}
 F^V_1(Q^2)
V_{\pi NN}^{s'_1s_1}(\np'_1,\nh_1) 
V_{\pi NN}^{s'_2s_2}(\np'_2,\nh_2) 
\left(k^\mu_{1}-k^\mu_{2}\right)
\label{picur}
\\
j^\mu_{\rm pole}
&=&
\left[I_V^{\pm}\right]_{1'2',12}
\frac{f^2}{m^2_\pi}\,
\frac{F_\rho\left(k_{1}^2\right)}{g_A}
\frac{
Q^\mu
\bar{u}_{s^\prime_1}(\np^\prime_1)\Qbar u_{s_1}(\nh_1)
 }{Q^2-m^2_\pi}
V_{\pi NN}^{s'_2s_2}(\np'_2,\nh_2) 
\nonumber\\
&&
+(1\leftrightarrow2)
\label{polecur}
\\
j^\mu_{\Delta}
&=&
\frac{f^* f}{m^2_\pi}\,
V_{\pi NN}^{s'_2s_2}(\np'_2,\nh_2) 
\bar{u}_{s^\prime_1}(\np^\prime_1)
\left\{ 
\left[U_{\rm F}^{\pm}\right]_{1'2',12}
 k^\alpha_{2}
G_{\alpha\beta}(h_1+Q)
\Gamma^{\beta\mu}(h_1,Q)
\right.
\nonumber\\
&& +
\left.
\left[U_{\rm B}^{\pm}\right]_{1'2',12}\; 
k^\beta_{2}
\hat{\Gamma}^{\mu\alpha}(p^\prime_1,Q)
G_{\alpha\beta}(p^\prime_1-Q)
\right\}
u_{s_1}(\nh_1)
+(1\leftrightarrow2)
\label{deltacur}.
\end{eqnarray}
In these equations we use the following notation.  The momentum
transfer to $i$th nucleon is $k_i^\mu = p'_i{}^\mu-h_i^\mu$.  The $\pi
NN$ interaction and the pion propagator are contained in the following
spin-dependent function:
\begin{equation}
V_{\pi NN}^{s'_1s_1}(\np'_1,\nh_1) \equiv 
\frac{\bar{u}_{s^\prime_1}(\np^\prime_1)\,\gamma_5
 \kbar_{1} \, u_{s_1}(\nh_1)}{k^2_{1}-m^2_\pi} .
\end{equation}
We have also defined the  isospin operators 
\begin{eqnarray}
I_V^{\pm}  &=& (I_V)_x\pm i (I_V)_y
\\
\Ivec_V  & =&  i \tauvec(1) \times\tauvec(2) ,
\end{eqnarray}
where the $+ (-)$ sign refers to neutrino (antineutrino) scattering.
In the $\Delta$ current, the forward, $U_{\rm F}^{\pm}= U_{Fx}\pm i
U_{Fy}$, and backward, $U_{\rm B}^{\pm}= U_{Bx}\pm i U_{By}$, isospin
transition operators are obtained from the cartesian components of the
operators
\begin{eqnarray}
U_{\rm Fj}&=&\sqrt{\frac32}
\sum_i\left(T_i T_j^\dagger\right)\otimes
\tau_i
\,,\ \ \ \ 
U_{\rm Bj}
=
\sqrt{\frac32}
\sum_i\left(T_{j}\, T^\dagger_i\right)\otimes
\tau_i,
\label{backward}
\end{eqnarray}
where $\vec{T}$ is an isovector transition operator from isospin
$\frac32$ to $\frac12$, normalized as
\begin{equation}
T_iT_j^{\dagger}= \frac23\delta_{ij} -\frac{i}{3}\epsilon_{ijk}\tau_k.
\end{equation}

The coupling constants in the MEC are $f=1$, $g_A=1.26$, and
$f^*=2.13$.  The electroweak form factors $F_1^V$ and $F_{\rho}$ in
the seagull and pionic currents are taken from Ref.~\cite{Hernandez07}.  To take into account the finite size of the
hadrons, we apply phenomenological strong form factors (not written
explicitly in the MEC) in all the $NN\pi$ and $N\Delta \pi$ vertices
\begin{eqnarray}
F_{\pi NN} (k) &=& \frac{\Lambda_\pi^2-m_\pi^2}{  \Lambda_\pi^2-k^2} 
\,,\ \ \ \ 
F_{\pi N\Delta} (k) 
= \frac{\Lambda_\Delta^2}{  \Lambda_\Delta^2-k^2} \,,
\end{eqnarray}
where the cutoff constants are $\Lambda_\pi=1.3$ GeV and
$\Lambda_\Delta= 1.15$ GeV.  These values of the strong form factors
in the MEC are similar to past works on the two nucleon emission
responses for electron scattering both in the non-relativistic
\cite{Alberico:1983zg}, and relativistic Fermi gas model
\cite{Dekker:1994yc}. In the formalism presented here, which is an
extension of the Torino model \cite{DePace:2003spn}, we use still the
same strong form factors for consistency with previous modelling.

In our approach the weak $N\rightarrow\Delta$ transition vertex tensor
in the forward current, $\Gamma^{\beta\mu}(P, Q)$ is given by
\begin{equation} \label{amaro-vertex}
\Gamma^{\beta\mu}(P,Q)=
\frac{C^V_3}{m_N}
\left(g^{\beta\mu}\Qbar-Q^\beta\gamma^\mu\right)\gamma_5
+ C^A_5 g^{\beta\mu} .
\end{equation}
We have kept only the $C_3^V$ and $C_5^A$ form factors and neglected
the smaller contributions of the others.  They are taken from
\cite{Hernandez07}.  The additional terms have been neglected
because they are of order $1/m_N^2$ or negligible for the intermediate
energy kinematics of interest for the quasielastic regime.  Moreover
the vector part of the vertex (\ref{amaro-vertex}) is equivalent to
the $\Delta$ operator used in ref \cite{DePace:2003spn} for electron
scattering.
  
The corresponding vertex tensor entering in the backward current is
defined by
\begin{equation}
\hat{\Gamma}^{\mu\alpha}(P^\prime, Q)=\gamma^0
\left[\Gamma^{\alpha\mu}(P^\prime,-Q)\right]^{\dagger}
\gamma^0 \, .
\end{equation}

The $\Delta$-propagator takes into account the finite decay width of
the $\Delta\,(1232)$ by the prescription
\begin{equation}\label{delta_prop}
 G_{\alpha\beta}(P)= \frac{{\cal P}_{\alpha\beta}(P)}{P^2-
 M^2_\Delta+i M_\Delta \Gamma_\Delta+
 \frac{\Gamma^2_{\Delta}}{4}} \, ,
\end{equation}
where the width $\Gamma_\Delta(P^2)$ depends only on the $\Delta$
invariant mass. The explicit dependence is taken from
\cite{Dekker:1994yc}.

The projector ${\cal P}_{\alpha\beta}(P)$ over spin-$\frac32$
particles is given by
\begin{eqnarray}
{\cal P}_{\alpha\beta}(P) = -(\Pbar+M_\Delta)
\left[g_{\alpha\beta}-\frac13\gamma_\alpha\gamma_\beta-
\frac23\frac{P_\alpha P_\beta}{M^2_\Delta}
+ \frac13\frac{P_\alpha\gamma_\beta- P_\beta\gamma_\alpha}{M_\Delta}\right].
\end{eqnarray}

In the present approach of MEC we do not consider the nucleon pole
(NP) terms, also called nucleon correlation current, with an
intermediate nucleon propagation. 
These contributions have been
evaluated in some works for electron scattering
\cite{Alberico:1983zg,Amaro:2010iu} and for neutrino scattering
\cite{Martini:2009uj,Martini:2010ex,Nieves:2011pp}.  But
these approaches are problematic from the theoretical point of view
due to the presence of double poles in the hadronic tensor, that need
to be regularized in some way not free from ambiguities.  Alternative
approaches (for instance \cite{Amaro:1998rr}) include these contributions in the nuclear wave function,
in form of Jastrow correlations.  A
third possibility is to include the nuclear correlation effects, at
least partially, by using a phenomenological scaling function
extracted from $(e,e')$ data, as discussed in Sect. III.

Note also that our MEC do not include heavier mesons such as
$\rho$-exchange, which is expected to provide a smaller contribution
than the leading pion-exchange for the energies of interest.

The computation of the 2p-2h responses in the RFG model presents two
main difficulties: the evaluation of the 2p-2h hadronic tensor by
performing the spin and isospin traces, and the evaluation of the 7D
integral over the phase space.  The first problem in fact can be faced
with analytical results for the 2p-2h tensor
$w^{\mu,\nu}(\np'_1,\np'_2,\nh_1,\nh_2)$, but hundreds of thousands
terms arise already in the electron scattering case
\cite{DePace:2003spn} and many more are present for neutrino
scattering.  We then follow the alternative and easier approach of
computing numerically the Dirac matrix elements of the currents and
then summing over spins.

The second problem consists on computing efficiently the 7D integral
of the MEC. This can be done in several ways. The integration in the
Laboratory system was studied in Ref.~\cite{Simo:2014wka} in the
context of the phase space integral
\begin{eqnarray}
F(q,\omega)
&\equiv&
\int
d^3p'_1
d^3h_1
d^3h_2
\frac{m_N^4}{E_1E_2E'_1E'_2}
\Theta(p'_1,h_1)\Theta(p'_2,h_2)
\,
\delta(E'_1+E'_2-E_1-E_2-\omega) .
\label{amaro-phase-function}
\end{eqnarray}
The study of this universal function of the RFG was done in detail in
Ref.~\cite{Simo:2014wka} for relativistic and non relativistic
kinematics.

\begin{figure}
\includegraphics[width=5cm, bb=130 280 410 780]{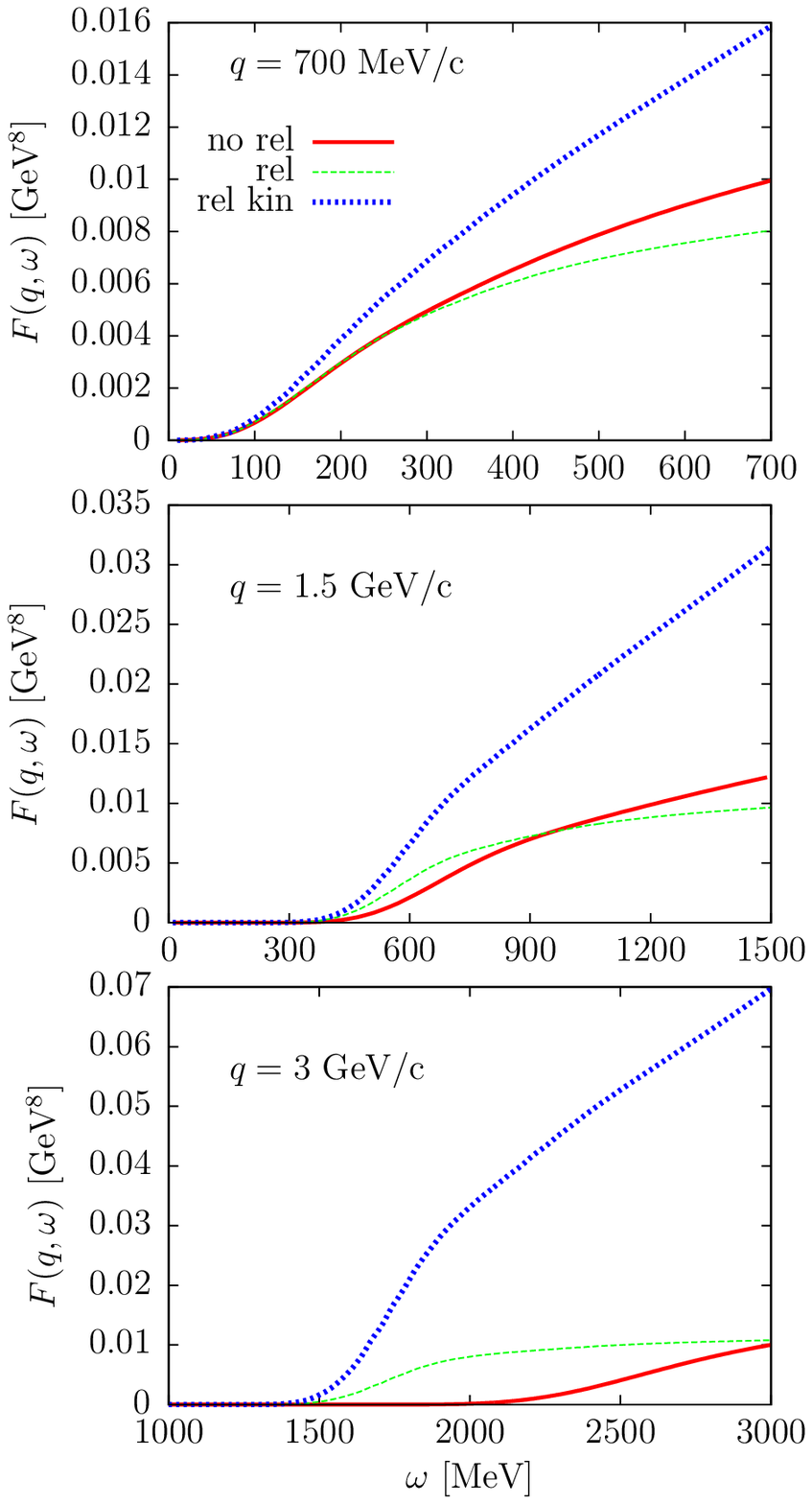}
\caption{
\label{amaro-fasico}  
Effect of implementing relativistic kinematics in a non-relativistic
calculation of $F(q,\omega)$. No rel: non-relativistic result.  Rel
kin: relativistic kinematics only without the relativistic factors
$m_N/E$. Rel: fully relativistic result.  \cite{Simo:2014wka}.  }
\end{figure}

One important conclusion of that study is presented in
Fig. \ref{amaro-fasico}, where we compare the relativistic and non
relativistic phase space functions. The latter uses non-relativistic
kinematics in the delta of energies and has no boost factors
$m_N/E$ inside the integral. As expected the relativistic effect
increases with the momentum transfer. We also show that using
relativistic kinematics without the boost factors worsens the results.
This shows that this procedure of ``relativizing'' a non-relativistic
MEC model using relativistic kinematics is not advisable.

In Ref.~\cite{Simo:2014esa} we followed the alternative approach of
computing the 7D integral going to the center of mass frame of the
final two-nucleon system.  There analytical results were found for the
phase space function in the non-Pauli blocking regime. The angular
distribution of the final particles in the phase space integral in the
CM system was found to be isotropic (independent on the emission
angle), as it is assumed in the Monte Carlo analyses of two nucleon
emission for neutrino scattering \cite{Sobczyk:2012ms}.  However,
these distributions are going to change after including the effects of
the MEC, because the angular distribution for fixed momenta, $\nh_1$
and $\nh_2$, is multiplied by the 2p-2h hadronic tensor. Further
studies are needed to correct the angular distribution from the
isotropic assumption, to reduce the systematic uncertainties of
neutrino event generators in the 2p-2h channel.

The 2p-2h inclusive cross section requires to compute two
electromagnetic response functions, $R^{T,L}$, for $(e,e')$ reactions
and five weak response functions, $R^{CC,CL,LL,T,T'}$, for
$(\nu_\mu,\mu^-)$.  All these responses were computed and analyzed in
Ref.~\cite{Simo:2016ikv}.

The numerical results of our model are in agreement with the
calculations of \cite{DePace:2003spn} and with the much more recent
 calculation of \cite{Rocco:2018mwt}, which uses the same
theoretical MEC model. These results for the response functions have
been parametrized in the kinematic range relevant for electron
scattering \cite{Megias16a}, and for the current neutrino quasielastic
scattering experiments \cite{Megias:2016nu}.  The parametrization is
convenient because in neutrino scattering there is an additional
integration over the incident neutrino flux to obtain the cross
section.  The parametrization is available from the authors; it
has been already used in other analyses \cite{Butkevich:2017mnc} and
has been implemented in Monte Carlo event generator GENIE
\cite{Dolan:2019bxf}.

\begin{figure*}[ht]
\begin{center}
\includegraphics[scale=0.6,bb=30 450 590 780]{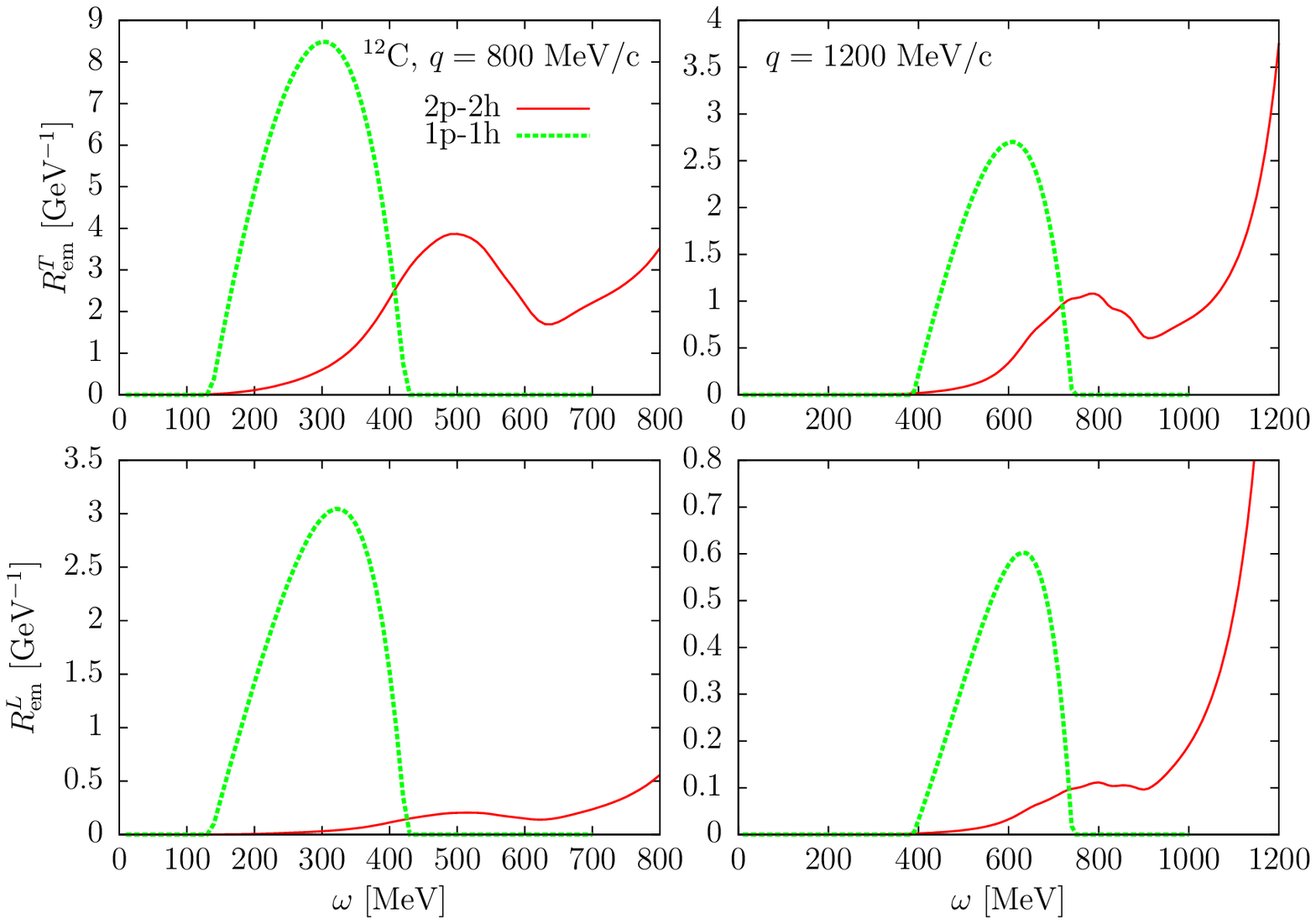}
\caption{Comparison between 1p-1h and 2p-2h response functions for
  electron scattering off $^{12}$C for two values of the momentum
  transfer.  The Fermi momentum of the RFG is $k_F=228$ MeV/c
  \cite{Simo:2016ikv}.  }
\label{electron-responses}
\end{center}
\end{figure*}

\begin{figure*}[ht]
\begin{center}
\includegraphics[scale=0.6,bb=30 270 590 780]{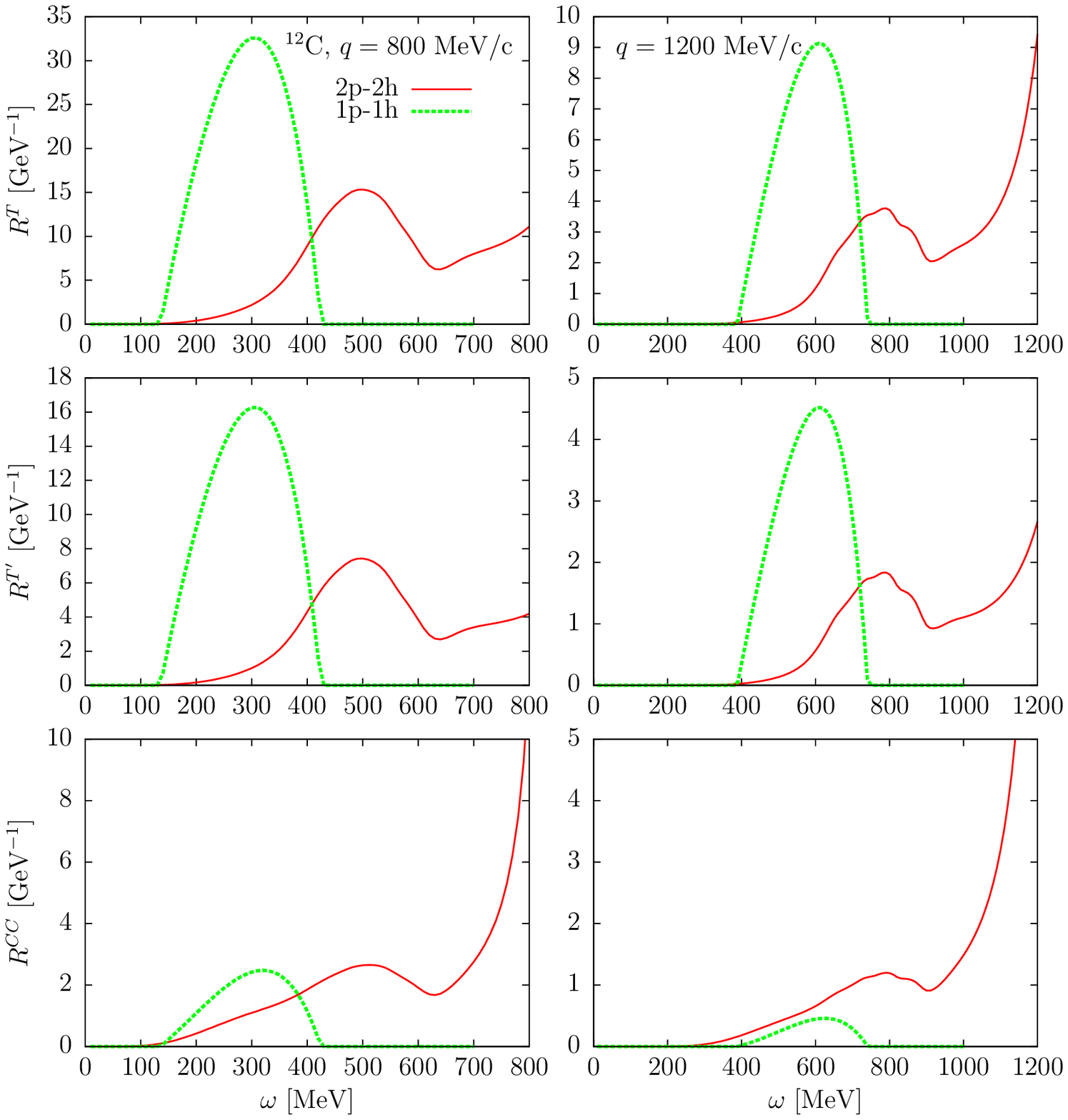}
\caption{Comparison between 1p-1h and 2p-2h response functions for CC
  neutrino scattering off $^{12}$C for two values of the momentum
  transfer.  The Fermi momentum of the RFG is $k_F=228$ MeV/c
  \cite{Simo:2016ikv}. }
\label{neutrino-responses}
\end{center}
\end{figure*}

To illustrate the qualitative predictions of the RFG model, in Figs.
\ref{electron-responses} and \ref{neutrino-responses} we compare the
1p-1h and 2p-2h neutrino responses for two values of the momentum
transfer for electron and neutrino scattering, respectively.  The
1p-1h responses only contain the one-body (OB) current. For these
values of $q$ large MEC effects are found. The 2p-2h responses present
a maximum due to the presence of the $\Delta$ propagator in the
$\Delta$ current.  The MEC effects are similar in the $T$ and $T'$
responses.  The 2p-2h contribution to $(e,e')$ is predominantly
transverse, due to dominance of the $\Delta$ excitation in the
transverse vector current.  The MEC effects in the $CC$ response are
relatively much larger than in the transverse ones. This comes from a
large longitudinal contribution of the axial MEC.  However, each
response function appears in the cross section multiplied by a
kinematic factor, which slightly alters the relative contribution to
the cross section.  In particular, the three responses $R^{CC}$,
$R^{CL}$ and $R^{LL}$ largely cancel each other, yielding a small
charge/longitudinal cross section. In figs.  \ref{electron-responses}
and \ref{neutrino-responses} only the real part of the denominator of
the $\Delta$ propagator has been considered in the calculation of the
MEC, a prescription taken also in Refs.~\cite{DePace:2003spn,
  Simo:2016ikv}.

\begin{figure}
\begin{center}
\includegraphics[scale=0.8,bb=200 600 420 790]{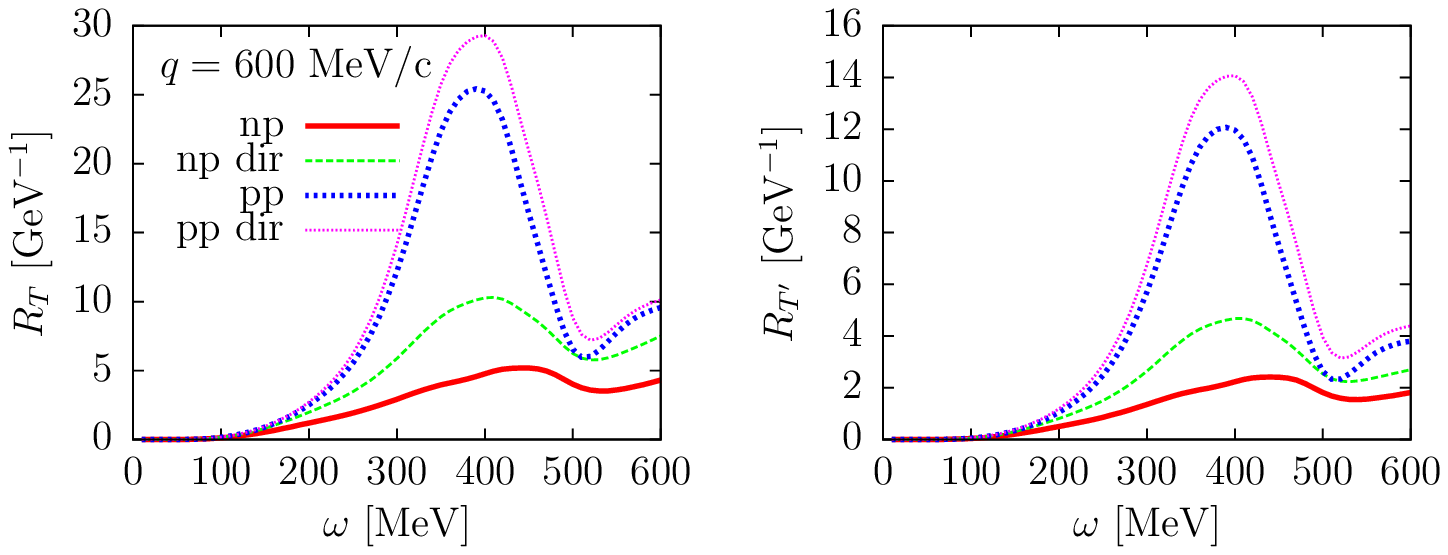}
\vspace{-1.2cm}
\caption{Separate pp and np contributions to the neutrino $T$ and $T'$
  2p-2h response functions of $^{12}$C, for $q=600$ MeV/c, compared to
  the direct contributions obtained by neglecting the direct-exchange
  interference. The value of the Fermi momentum $k_F=228$ MeV/c
  \cite{RuizSimo:2016ikw}. }
\label{amaro-pnseparation}
\end{center}
\end{figure}

In our model the elementary 2p-2h response functions contain
direct-direct, exchange-exchange and direct-exchange contributions due
to the antisymmetry of the 2p-2h states.  The exchange-exchange term
is identical to the direct-direct for the inclusive reactions
considered here. Therefore one can write the elementary response
functions as direct minus direct-exchange contribution. For instance
for the $T$ responses of neutrino scattering in the pp emission
channel we have
\begin{eqnarray}
r^{T}_{pp}
&=& 
4\sum_{\mu=1}^2
 \sum_{s_1s_2s'_1s'_2}
\left\{ 
\left| 
J^\mu_{pp}(1'2';12)
\right|^2
\right.
-{\rm Re}\;
J^\mu_{pp}(1'2';12)^*
J^\mu_{pp}(2'1';12)
\Big\} ,
\label{wPP}
\end{eqnarray} 
where $J^\mu_{pp}(1'2';12)$ is the matrix element for pp emission with
neutrinos.  The first term in Eq. (\ref{wPP}) is the ``direct''
contribution, and the second one is the ``exchange'' contribution,
actually being the interference between the direct and exchange matrix
elements.  There are similar expressions for the remaining response
functions (see Ref.~\cite{Simo:2016ikv} for details).

In other models of 2p-2h based on the Fermi gas, the exchange
interference contributions are disregarded.  Specifically they are
neglected in the models of Lyon \cite{Martini:2010ex, Martini:2009uj}
and Valencia \cite{Nieves:2011pp,Gran:2013kda}, because they involve
higher dimensional integrals than the direct terms, which can be
reduced to low dimensions within some assumptions.

To quantify the importance of the exchange, this contribution was
separated in our model in Ref.~\cite{Simo:2016ikv}. It was found to
be typically about 25\% of the 2p-2h inclusive cross section.  Later,
it was separated in the individual channels of neutron-proton (np) and
proton-proton (pp) emission in Refs.~\cite{RuizSimo:2016ikw,
  Simo:2016imi}.  An example is given in Fig.
\ref{amaro-pnseparation}, where we see the transverse, T and T',
response functions for np and pp emission with neutrinos. In the np
case the exchange interference reduces the strength by 50\%.  This
example shows that the ratio pp/np critically depends on the
treatment of the exchange interference.

In other works \cite{RuizSimo:2017onb, RuizSimo:2017hlc} we have
developed some assumptions which simplify and reduce the number of
nested integrations to calculate the inclusive 2p-2h responses. In the
first of them, Ref.~\cite{RuizSimo:2017onb}, we used the "frozen
nucleon approximation" to reduce the 7-dimensional integral to just a
single integration, by assuming the initial nucleons to be at
rest. The strong dependence on the kinematics near the $\Delta$ peak
arising in the frozen approximation was avoided by using a smeared
$\Delta$ propagator that was parametrized for the relevant values of
the momentum transfer, in order to make these approximate results  
consistent with those calculated with the full integration
procedure.  In the second work, Ref.~\cite{RuizSimo:2017hlc}, we took an approach similar 
to that of the IFIC-Valencia
model \cite{Nieves:2011pp, Gran:2013kda}: we developed a convolution
model of the elementary 2p-2h responses, given in Eq. (\ref{amaro-elementary}), 
with two 1p-1h responses (Lindhard functions)
sharing the energy and momentum transfer from the electroweak probe.
The importance of these works is two-fold: from the computational
point of view, reducing the time to evaluate the inclusive 2p-2h
responses makes these models suitable to be incorporated in the Monte
Carlo event generators; additionally, with the second work
\cite{RuizSimo:2017hlc}, we also can incorporate the direct-exchange
terms, that appear in Eq. (\ref{amaro-elementary}), in the
calculation of the inclusive 2p-2h response functions in the same or
similar fashion as it is done in the Valencia model
\cite{Nieves:2011pp}.  But this is even more than what is included in
the original Valencia model \cite{Nieves:2011pp} in this respect,
because in this last model those terms were not considered because
their calculation was extremely difficult with the technique of the
Cutkosky rules.

\section{Pion production in the resonance region}
\label{sec:pion}
          
There are several reasons that make pion production relevant for
present and future accelerator-based neutrino oscillation experiments.
Neutral current $\pi^0$ production is an important background in the
electron neutrino and antineutrino appearance analyses due to the
misidentification of the photons from the $\pi^0$ decay with the
electron or positron signal.  Also, the final state for events with a
pion produced in the primary vertex may mimic that of a QE one i) if
the pion is not detected, and ii) if due to FSI the pion is absorbed.
Finally, in the few-GeV energy region, the number of events observed
in the far and near detectors arising from neutrino-induced pion
production competes with those from the quasielastic process. Since
pions can be detected, for example, in coincidence with the final
lepton and possibly other hadrons, a good understanding of the pion
production mechanisms could help to increase the statistics of events
that are selected for the reconstruction of the neutrino energy.
         
\begin{figure}[htbp]
\centering  
(a)
\includegraphics[width=.25\textwidth,angle=270]{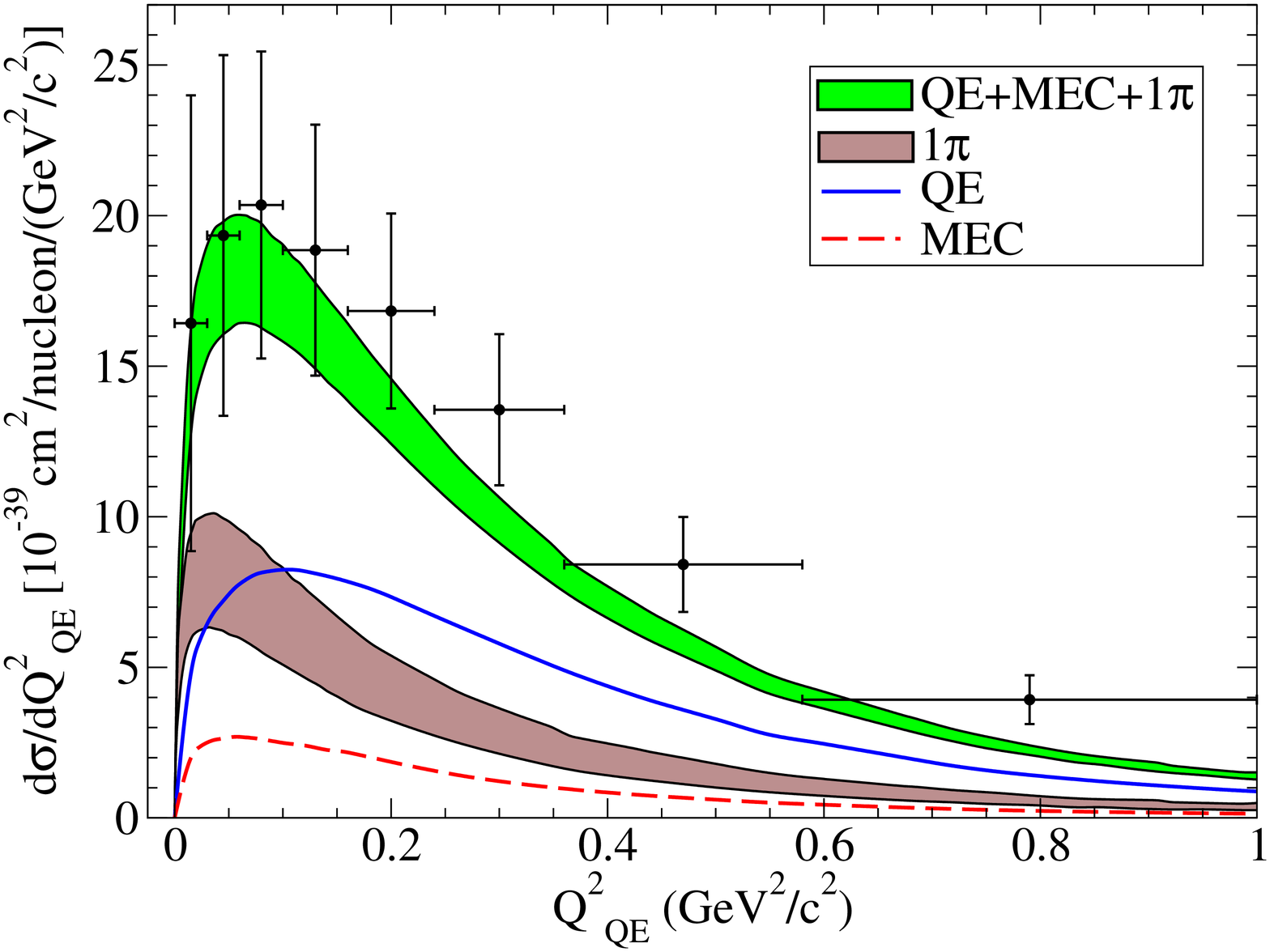}
\hspace{0.5cm}
(b)
\includegraphics[width=.25\textwidth,angle=270]{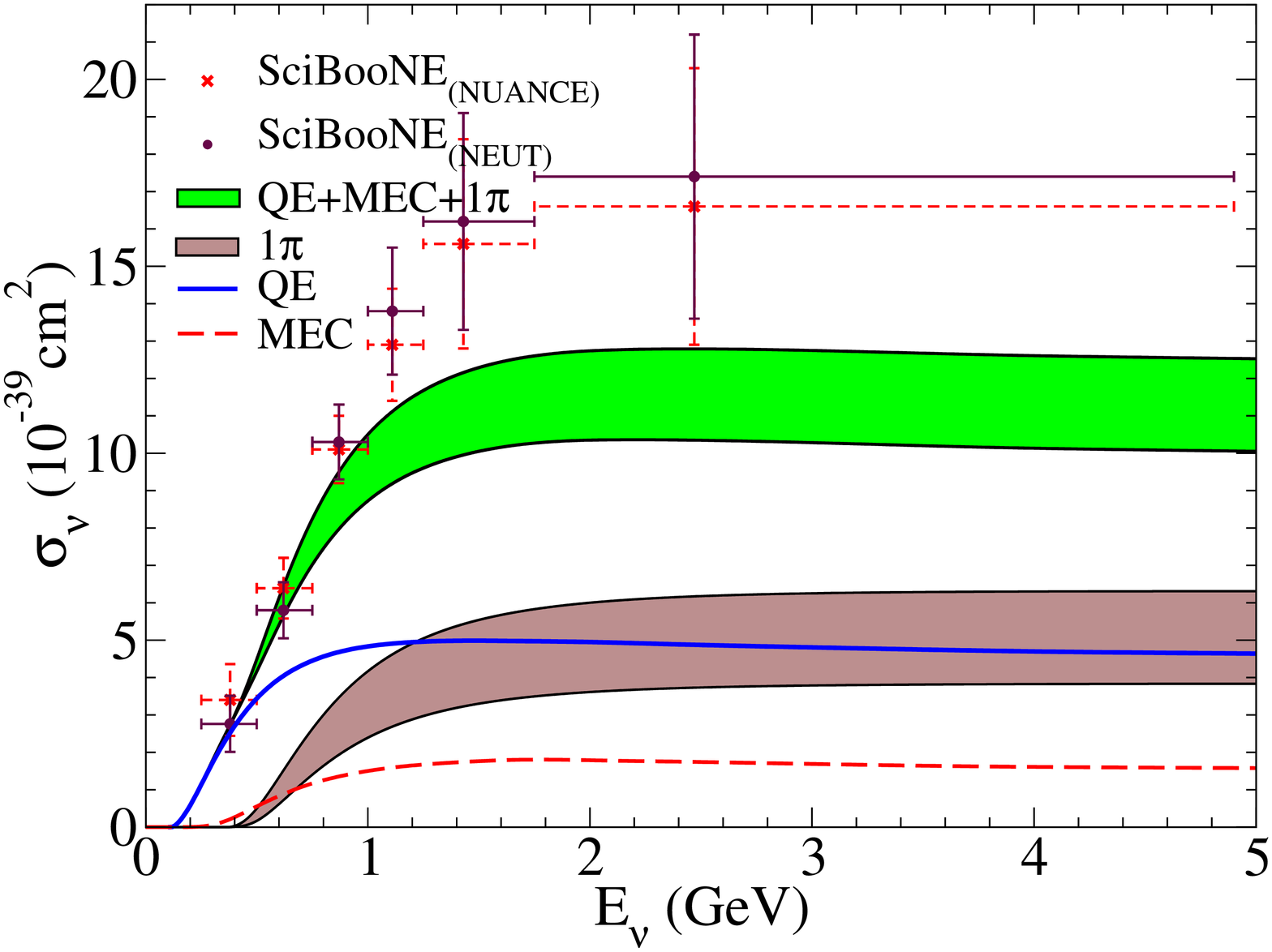}
\caption{ (a) We show the CC-inclusive T2K flux-folded
  $\nu_e$-$^{12}$C $Q^2_{QE}$ differential cross section per
  nucleon. (b) The CC $\nu_\mu$ total cross section on C$_8$H$_8$ is
  presented. Experimental data are from~\cite{T2Kinclusive14} (a)
  and~\cite{SciBooNE11} (b). Theoretical predictions for QE, non-QE
  (1$\pi$) and the 2p2h MEC are shown
  separately. 
  Plots updated from Ref.~\cite{Ivanov16}.}\label{fig:T2KQE-SciB}
\end{figure}

To illustrate the relative magnitude of QE and non-QE processes, in
Fig.~\ref{fig:T2KQE-SciB}(a) we show the $\nu_e$-$^{12}$C inclusive
$Q^2_{QE}$\footnote{$Q^2_{QE}$ is the reconstructed four-momentum transfer, obtained assuming an initial state nucleon at rest with a constant binding energy.} differential cross section averaged with the T2K flux, and
in Fig.~\ref{fig:T2KQE-SciB}(b) the $\nu_\mu$ total cross section on
C$_8$H$_8$ target.  
Three different contributions are shown: 
i) QE scattering computed within the SuSAv2 model (Sect.~\ref{sec:susav2}), 
ii) the 2p2h MEC (Sect.~\ref{sec:2p2h}), and 
iii) an inelastic contribution containing mainly one-pion production (labelled as $1\pi$), that will be described below
\footnote{Notice that Figs.~\ref{fig:T2KQE-SciB}(a) and (b) are updated versions of those originally shown in Ref.~\cite{Ivanov16}.
Here the 2p2h MEC contains the vector-vector, vector-axial and axial-axial contributions while in Ref.~\cite{Ivanov16} only its vector-vector contribution was available.}.
One observes that for the $\nu_e$ T2K flux the QE
and $1\pi$ contributions are similar in magnitude. This should not be
surprising since, even though the T2K flux peaks at around $E_\nu=0.5$
GeV, it has a long tail extending up to 10 GeV~\cite{T2Kinclusive14}.
For the total cross section, Fig.~\ref{fig:T2KQE-SciB}(b), the QE
contribution dominates up to $E_\nu\approx1$ GeV. For higher neutrino energies the QE
and $1\pi$ contributions are comparable.

The motivation of the non-QE SuSAv2 model used in
Fig.~\ref{fig:T2KQE-SciB} is to extend the superscaling arguments,
previously applied in the QE domain, to the $\Delta$ resonance
region~\cite{Amaro05,Maieron09}.  Hence, the underlying idea is that
pion production in the $\Delta$ region is indeed strongly dominated by
the excitation of the $\Delta$ resonance.  The first step is to define
an experimental scaling function in this inelastic region,
$f^{non-QE}$. For that we subtract the QE and 2p2h MEC contributions
to the inclusive electron scattering experimental cross section:
\ba
    \left(\frac{d^2\sigma}{d\Omega d\omega}\right)_{non-QE}\equiv \left(\frac{d^2\sigma}{d\Omega d\omega}\right)_{exp} - \left(\frac{d^2\sigma}{d\Omega d\omega}\right)^{1p1h}_{QE,SuSAv2} - \left(\frac{d^2\sigma}{d\Omega d\omega}\right)^{2p2h}_{MEC}\,.
\ea
The result is called non-QE cross section. Then the superscaling function is constructed as
\ba
     f^{non-QE}(\psi_\Delta) = k_F\frac{\left(\frac{d^2\sigma}{d\Omega d\omega}\right)_{non-QE}}{\sigma_M(v_LG_L^\Delta+v_TG_T^\Delta)}\label{fnonQE}
\ea
where $G_{L,T}^\Delta$ are the single-nucleon functions for the
$\gamma N\Delta$ vertex and $\psi_\Delta$ is the $\Delta$ scaling
function (the explicit expressions can be found in
Refs.~\cite{Amaro99,Amaro05,Maieron09}).

\begin{figure}[htbp]
\centering  
(a)
\includegraphics[width=.45\textwidth,angle=0]{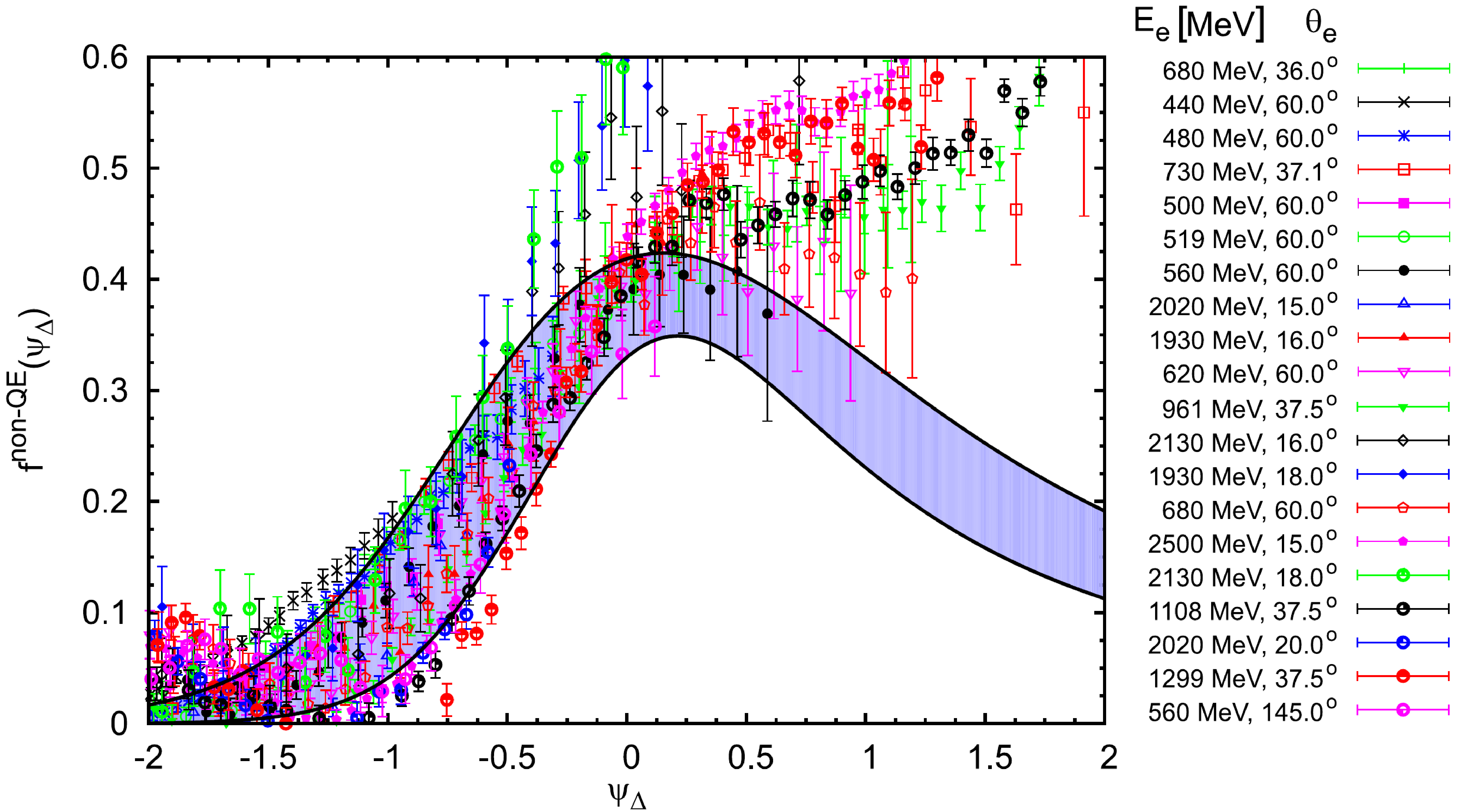}
\hspace{0.5cm}
(b)
\includegraphics[width=.36\textwidth,angle=0]{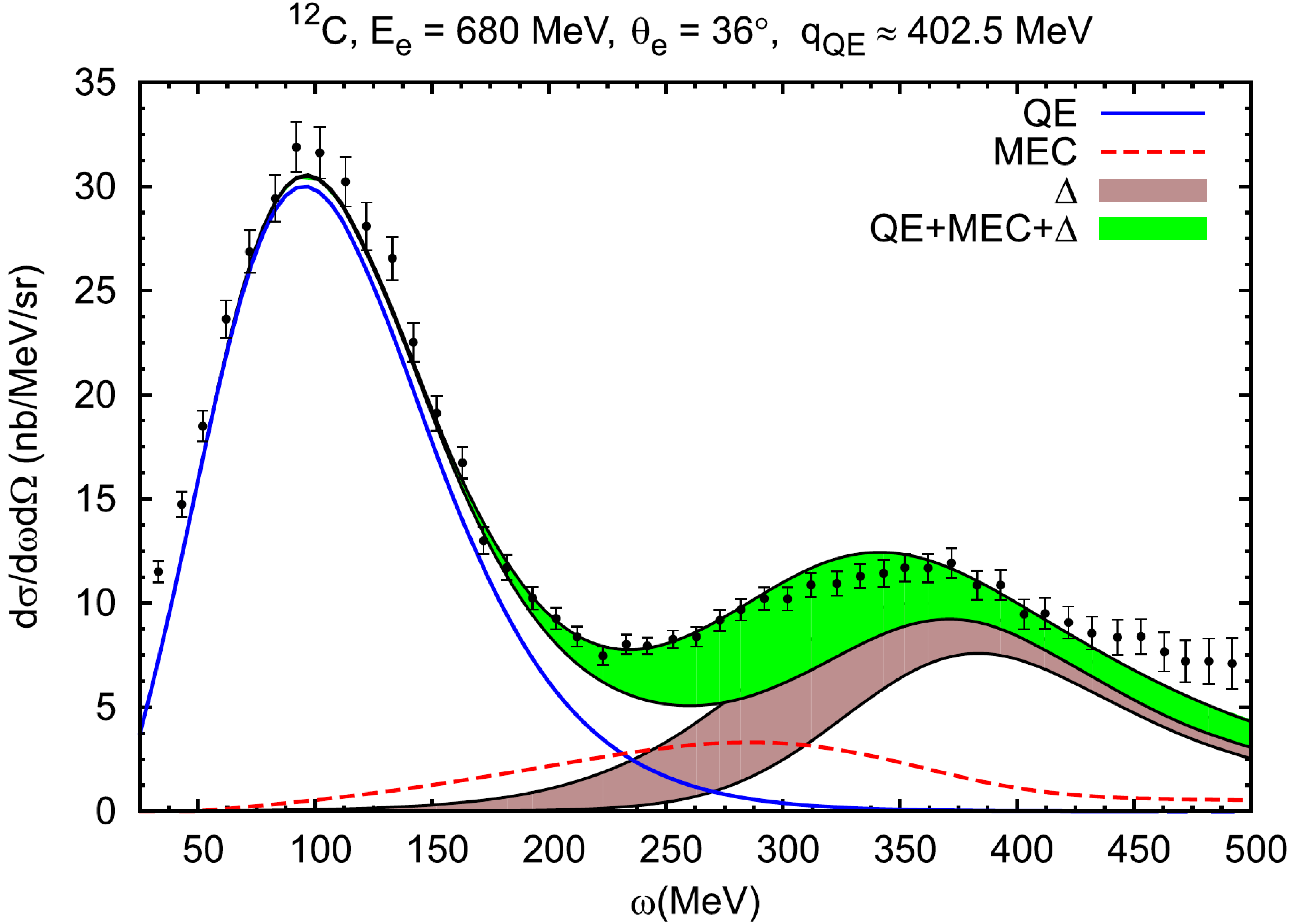}
\caption{ (a) We show the $f^{non-QE}(\psi_\Delta)$, the experimental
  data can be found in~\cite{QESarxiv}. (b) We compare inclusive
  electron-$^{12}$C data~\cite{Barreau:1983ht} with the predictions of
  the QE and non-QE SuSAv2, and the 2p2h MEC contribution. These plots
  are taken from Ref.~\cite{Ivanov16}.}\label{fig:nonQE}
\end{figure}

The function $f^{non-QE}(\psi_\Delta)$ is represented in
Fig.~\ref{fig:nonQE}(a) for a large set of inclusive electron-$^{12}$C
scattering data.  One observes that in the region $\psi_\Delta<0$
scaling occurs but it is not as good as in the QE case. For that
reason, the scaling function is parametrized as a broad band that
accounts for the spread of the data.  This band can be understood as
the systematic error attached to the model.  Notice that in the region
$\psi_\Delta>0$, as expected, scaling breaks because contributions
from processes beyond the Delta production start to be important.
Thus, the predictions of this model beyond the Delta peak should be
taken with caution. In Fig.~\ref{fig:nonQE}(b) we show that the model
works well for inclusive electron scattering (see more
in~\cite{Ivanov16}).

Inclusive models like superscaling-based approaches, or the model of
Ref.~\cite{Martini:2009uj}, do not provide any information about the
final state hadrons.  For that, one normally uses theoretical
approaches that focus on one particular reaction channel.  The
inclusive signal should be recovered by adding all channels allowed
for the given energy and momentum transferred and integrating over the
hadronic variables. This implies, therefore, the non-trivial task of
modelling all those possible reaction channels.

In an effort to improve our current knowledge on the neutrino-induced
pion production, in recent years the
MiniBooNE~\cite{MBNCpi010,MBCCpionC11,MBCCpion011},
MINERvA~\cite{MINERvACCpi015,MINERvACCpi15,MINERvACCpi16,MINERvAnuCC17,MINERvAdiff17}
and T2K~\cite{T2KCC1pi17} collaborations have reported (pion detected)
differential cross sections for NC and, mainly, CC neutrino-induced
pion production on different nuclear targets. 
It is important to stress that up to date, no model is able to simultaneously
reproduce and explain the MiniBooNE, MINERvA and T2K data sets~\cite{Sobczyk15,Mosel17,Nikolakopoulos18,Gonzalez-Jimenez18}. 
This problem is sometimes referred as the ``pion puzzle'', and 
its solution is crucial for future oscillation experiments like DUNE, in which pion production plays a major role. 
From the theoretical
side, we aim at providing support to improve our understanding of
these and forthcoming data in a more consistent way.
Accordingly, in this section we focus on the single-pion production
(SPP) process, which is the dominant one in the resonance region and
is typically characterized by invariant masses~\footnote{The invariant
  mass $W$ is defined as $W^2 = s = (Q+P_i)^2 = (K_\pi+P_f)^2$, with
  the four vectors defined in Fig.~\ref{fig:SPPkin}(a).} going from
the pion threshold to $W<1.8-2$~GeV.  The strategy shared by most
approaches is based on the idea that the lepton interacts with only
one nucleon in the nucleus.  This allows one to decompose this complex
problem into different pieces: $1$) the elementary vertex, $2$) the
nuclear framework, including the description of the initial and final
nuclear states, and $3$) the final-state interactions.  In what
follows we review some of the features of this complex problem. For
further reading we refer to the recent review
articles~\cite{Alvarez-Ruso14,Katori17,Nakamura17,Alvarez-Ruso17}.

\subsection{Elementary vertex}

One finds the following reaction channels for SPP off a free nucleon
induced by neutral current interactions (EM or WNC): \ba \gamma + p
\rightarrow \pi^{+} + n\,,\,\,\,\,\,\,& \gamma + n \rightarrow \pi^{-}
+ p\,,\non\\ \gamma + p \rightarrow \pi^0 + p\,,\,\,\,\,\,\,& \gamma +
n \rightarrow \pi^0 + n\,.  \ea $\gamma$ represents a real
(photoproduction) or virtual (electroproduction) photon. For WNC
interaction, the virtual photon is replaced by a $Z^0$ boson.
In the case of neutrino ($W^+$ exchanged) and antineutrino ($W^-$) CC
interactions, one has: \ba W^+ +& p \rightarrow \pi^{+} +
p\,,\,\,\,\,\,\,& W^- + n \rightarrow \pi^{-} + n\,,\non\\ W^+ +& n
\rightarrow \pi^0 + p\,,\,\,\,\,\,\,& W^- + p \rightarrow \pi^0 +
n\,,\\ W^+ +& n \rightarrow \pi^+ + n\,,\,\,\,\,\,\,& W^- + p
\rightarrow \pi^- + p\,.\non \ea

\begin{figure}[htbp]
\centering  
(a)
\includegraphics[width=.35\textwidth,angle=0]{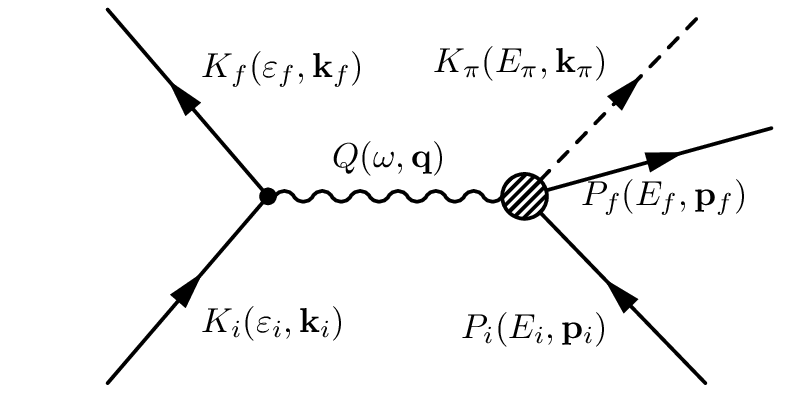}
\hspace{0.5cm}
(b)
\includegraphics[width=.35\textwidth,angle=0]{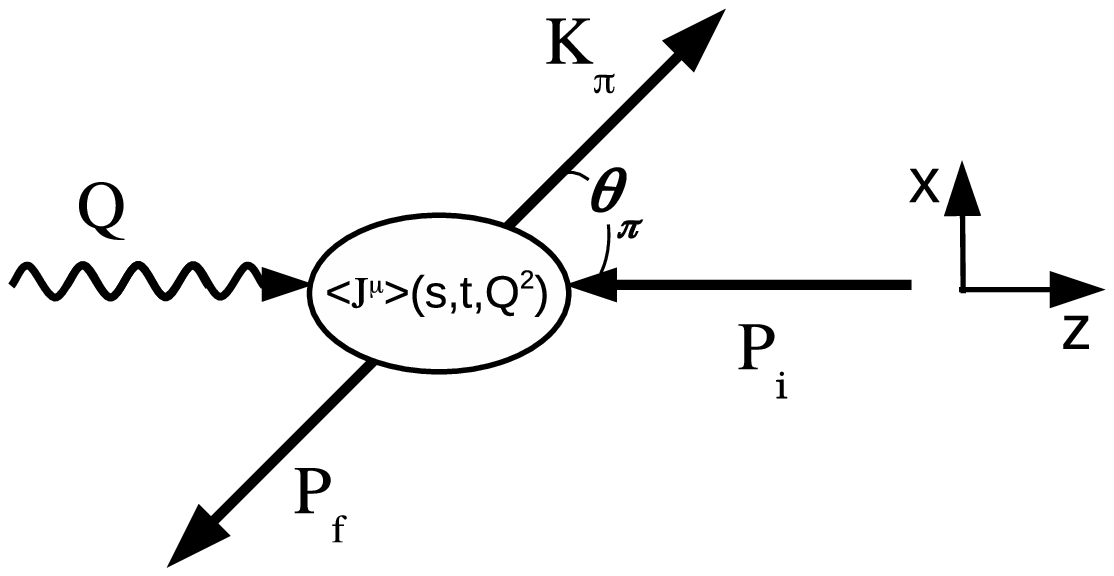}
\caption{(a) Feynman diagram representing the single-pion production
  off a free nucleon.  (b) Representation of the hadronic subprocess
  in the center of momentum frame. }\label{fig:SPPkin}
\end{figure}

The process is depicted in Fig.~\ref{fig:SPPkin}(a), and it will be
discussed in what follows.\\

\paragraph{Cross Section} 

The differential cross section for SPP off the nucleon in a reference frame of collinear incident particles is: 
\ba \frac{d\sigma}{d\nk_f d\nk_\pi} =
\frac{K}{\phi} \frac{{\cal R}_X}{(2\pi)^5} \eta_{\mu\nu}\, h^{\mu\nu}
\delta(\varepsilon_i + E_i - \varepsilon_f - E_\pi -
E_f)\,, \label{CScov} \ea with \ba K = \frac{1}{2\varepsilon_i\,
  \varepsilon_f} \frac{1}{2E_i\,E_f} \frac{1}{2E_\pi}\,,\,\,\,\,\,\,
\phi = \left| \frac{\nk_i}{\varepsilon_i}-\frac{\np_i}{E_i} \right|\,.
\ea This expression is valid for EM, CC and WNC interactions. The
factor ${\cal R}_X$, where the subscript $X$ refers to the type of
interaction, include the boson propagator as well as the coupling
constants at the leptonic and hadronic vertices: \ba {\cal R}_{EM} =
\left(\frac{4\pi\alpha}{Q^2}\right)^2 \,,\,\,\,\,\, {\cal R}_{CC} =
\left(G_F\cos\theta_c\right)^2\,,\,\,\,\, {\cal R}_{WNC} = G_F^2\,.
  \label{R_X}
\ea
$\eta_{\mu\nu}$ is the leptonic tensor discussed in
Sect.~\ref{sec:genform}.  We define the hadronic tensor as 
\ba
h^{\mu\nu} = 2m_{N,i}m_{N,f}\overline\sum_{S_iS_f} (J^\mu)^*J^\nu.  
\ea 
with the hadronic current \ba J^\mu = \bar u(\np_f,S_f) {\cal O}^\mu_{1\pi}
u(\np_i,S_i).  \ea $u(\np_{i,f},S_{i,f})$ represent the free Dirac
spinors for the initial and final state nucleons with spin projection
$S_i$ and $S_f$.~\footnote{The convention for the plane wave and Dirac
  spinor is
  $\psi(\nx,\np)=\sqrt{\dfrac{m_N}{VE}}\,u(\np,s)\,e^{i\nx\cdot\np}$,
  with $u(\np,s)=\sqrt{\dfrac{m_N+E}{2m_N}}
  \bma\chi_s\\ \frac{\nsigma\cdot\np}{E+m_N}\chi_s\ema$, and $\chi_\half
  = \bma 1\\ 0\ema$ and $\chi_{-\half} = \bma 0\\ 1\ema$.}  ${\cal
  O}^\mu_{1\pi}$ is the transition operator between the initial
one-nucleon state and the final one-nucleon one-pion ($N \pi$) state.

We can use the energy-conservation delta function to integrate over
$k_\pi$. This leads to \ba
\frac{d^5\sigma}{d\varepsilon_fd\Omega_fd\Omega_\pi} = \frac{K}{\phi}
\frac{\varepsilon_fk_f E_\pi k_\pi}{ f_{rec}^{(k_\pi)} } \frac{{\cal
    R}_X}{(2\pi)^5} \eta_{\mu\nu}\, h^{\mu\nu}\,, \label{CSkf} \ea
where we have introduced the nucleon recoil function \ba
f_{rec}^{(k_\pi)}=\left|1+ \frac{E_\pi}{E_f}
\left(1-\frac{\nk_\pi\cdot(\nq+\np_i)}{k_\pi^2}\right) \right|\,.  \ea

Alternatively, if we want to study distributions as a function of the
pion energy, we can integrate over $k_f$.  In this case one obtains
\ba \frac{d^5\sigma}{d\Omega_fdE_\pi d\Omega_\pi} = \frac{K}{\phi}
\frac{\varepsilon_fk_f E_\pi k_\pi}{ f_{rec}^{(k_f)}} \frac{{\cal
    R}_X}{(2\pi)^5} \eta_{\mu\nu}\, h^{\mu\nu}\,, \label{CSkpi} \ea
with \ba f_{rec}^{(k_f)} = \left|1+ \frac{\varepsilon_f}{E_f}
\left(1-\frac{\nk_f\cdot(\nk_i+\np_i-\nk_\pi)}{k_f^2}\right)
\right|\,.  \ea
Notice that by choosing a reference frame in which $\nk_f$ has no
$\hy$ component, the cross sections in eqs.~(\ref{CSkf}) and
(\ref{CSkpi}) will not depend on the azimuthal angle
$\phi_f$. Alternatively, if one chooses a reference frame in which
$\nk_\pi$ has no $\hy$ component, the cross sections will be
independent on the azimuthal angle $\phi_\pi$.

We will focus now on the description of the hadronic current.  In the
center of momentum system (CMS) of the hadronic subprocess,
Fig.~\ref{fig:SPPkin}(b), the hadronic current depends on only three
independent variables.
 
A usual choice for these are the Lorentz invariants $Q^2$, $s$ and
$t$, with \ba t =(Q-K_\pi)^2 = m_\pi^2 - |Q^2| - 2 ( \omega E_\pi -q
k_\pi\cos\theta_\pi )\,, \label{eq:textreme} \ea where
\begin{align}
  \omega = \frac{s-m_N^2-|Q^2|}{2W}\,,\,\,
   E_\pi = \frac{s+m_\pi^2-m_N^2}{2W}\,.\label{eq:Epi}
\end{align}
The maximum and minimum $t$ values allowed by energy-momentum
conservation correspond to the values $\cos\theta_\pi=-1$ and
$\cos\theta_\pi=1$.
Thus, a possible strategy is to compute the hadronic current
$J^\mu(Q^2,s,t)$ in the CMS and then bring it back to the lab frame
where the experiments are performed. This would allow one to save
important computational time when sophisticated models are employed in
the calculation of the hadronic current.\\

\paragraph{Nucleon Resonances}

SPP in the resonance region is dominated by the excitation and
de-excitation of nucleon resonances.  The properties of the resonances
(spin, isospin, mass, decay width, branching ratios, etc.) lying in
the region $W<2$ GeV are, in general, known with great precision
thanks to partial waves analyses of (mainly) elastic pion-nucleon
scattering data~\cite{Tiator11}.  Information on their electroweak
properties, on the other hand, can be extracted from analysis of
photon- and lepton-induced single-pion production data.

\begin{figure}[htbp]
\centering  
\includegraphics[width=.5\textwidth,angle=0]{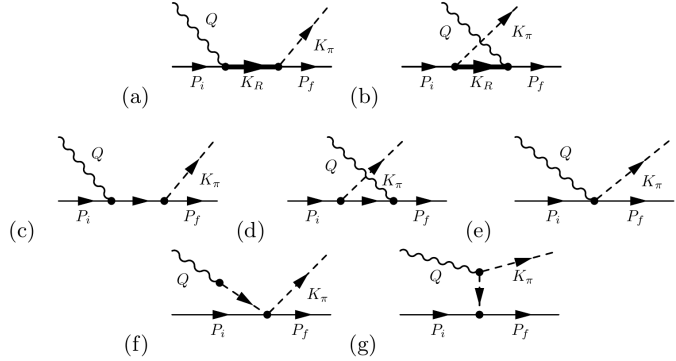}
\caption{Diagrams contributing to the SPP. Resonance contribution:
  Direct (a) and cross (b) channel terms. Background contributions:
  Direct (c) and cross (d) nucleon pole, contact term (e), pion pole
  (f), and pion-in-flight (g) term.}\label{fig:Res-Bg}
\end{figure}

The Feynman diagram representing the electroweak resonant mechanism is
illustrated in Fig.~\ref{fig:Res-Bg}(a). The cross resonance term,
which is sometimes considered to maintain crossing
symmetry~\cite{Fernandez-Ramirez06,Fernandez-Ramirez08a}, is shown in
Fig.~\ref{fig:Res-Bg}(b).  In these diagrams we can identify three
pieces: the electroweak vertex $\Gamma_{QNR}^{\mu\nu}$, the propagator
of the resonance $S_{\mu\nu}$, and the hadronic vertex $\Gamma_{R\pi
  N}^\mu$. The amplitude for these diagrams is given by the
contraction of them `sandwiched' between the nucleon spinors. For
example, for a spin-$3/2$ resonance, such as the Delta, one finds: \ba
J^\mu_{(a)} &\propto& \bar u(\np_f,s_f)\, \Gamma_{R\pi N}^\alpha\,
S_{\alpha\beta}\, \Gamma_{QNR}^{\beta\mu}
u(\np_i,s_i)\,,\non\\ J^\mu_{(b)} &\propto& \bar u(\np_f,s_f)\,
\bar\Gamma_{QNR}^{\alpha\mu}\, S_{\alpha\beta}\, \Gamma_{R\pi N}^\beta
u(\np_i,s_i)\,, \ea with
$\bar\Gamma_{QNR}^{\alpha\mu}(P_f,Q)\equiv\gamma^0
[\Gamma_{QNR}^{\alpha\mu}(P_f,-Q) ]^\dagger\gamma^0$. The three pieces
are described below for the case of a spin-$3/2$ resonance, and within
a Rarita-Schwinger formalism~\cite{Rarita41}.
\begin{enumerate}
 \item {\it The electroweak vertex} is characterized by the properties
   of the resonance (spin, isospin, and parity) and the electroweak
   $Q^2$-dependent form factors. It is parametrized
   by~\cite{Llewellyn-Smith72}: \ba \Gamma_{QNR}^{\mu\nu} =
   \left(\Gamma_{QNR,V}^{\mu\nu} + \Gamma_{QNR,A}^{\mu\nu}
   \right)\widetilde{\gamma}^5\,, \ea with $\widetilde{\gamma}^5 =
   \munit$ for even parity resonances, and $\widetilde{\gamma}^5 =
   \gamma^5$ for odd ones.  The vector contribution is 
   \ba
   \Gamma_{QNR,V}^{\mu\nu} = \Bigg[ \frac{C_3^V}{m_N}
     (g^{\mu\nu}\slashed{Q} - Q^\mu\gamma^\nu) + \frac{C_4^V}{m_N^2}
     (g^{\mu\nu}Q\cdot K_{R} - Q^\mu K_{R}^\nu) + \frac{C_5^V}{m_N^2}
     (g^{\mu\nu}Q\cdot P - Q^\mu P^\nu) + C_6^Vg^{\mu\nu}
     \Bigg]\gamma^5\,, \ea and the axial one is \ba
   \Gamma_{QNR,A}^{\mu\nu} = \frac{C_3^A}{m_N} (g^{\mu\nu}\Qslash -
   Q^\mu\gamma^\nu) + \frac{C_4^A}{m_N^2} (g^{\mu\nu}Q\cdot K_{R} -
   Q^\mu K_{R}^\nu) + C_5^A g^{\mu\nu} + \frac{C_6^A}{m_N^2}
   Q^{\mu}Q^{\nu}\,.  \ea $K_R^\mu$ stands for $K_s^\mu=P^\mu+Q^\mu$
   and $K_u^\mu=P^\mu-K_\pi^\mu$ for the direct and cross channels
   respectively.
The resonance form factors $C^{V,A}_i$ account for the internal
structure, whose dynamics is explored with a resolution $Q^2$.

The vector form factors can be extracted from helicity amplitude
analyses of single-pion electro-production
data~\cite{Tiator11,Lalakulich06,Hernandez08,Leitner09}.
In the axial sector, however, the situation is quite different.  While
the couplings may be guided by the PCAC hypothesis (see e.g. Appendix
C in Ref.~\cite{Leitner09}), the rather scarce experimental data does
not allow to constrain the $Q^2$ dependence, which is usually taken as
dipole shapes with sets of parameters inspired by the vector and QE
cases.
 \item {\it The propagator} is \ba S_{\alpha\beta} =
   \frac{-(\Kslash_{R} +m_{R})}{ K_{R}^2 - m_{R}^2 + im_{R}
     \Gamma_{\text{{\tiny width}}}(W) } \Big( g_{\alpha\beta}
   -\frac{1}{3}\gamma_\alpha\gamma_\beta - \frac{2}{3
     m_{R}^2}K_{R,\alpha} K_{R,\beta} - \frac{2}{3
     m_{R}}(\gamma_\alpha K_{R,\beta} -K_{R,\alpha}\gamma_\beta )
   \Big)\,,\label{eq:Sdelta} \ea with $\Gamma_{\text{{\tiny
         width}}}(W)$ the resonance decay width (see for instance
   Ref.~\cite{Leitner09}).
 \item The {\it decay vertex} is \ba \Gamma_{R\pi N}^\alpha =
   \frac{\sqrt{2}f_{\pi NR}}{m_\pi}K_\pi^\alpha\widetilde\gamma^5\,,
   \ea with $f_{\pi NR}$ the strong coupling
   constant~\cite{Leitner09,Gonzalez-Jimenez17}.

In Ref.~\cite{Pascalutsa99}, it is proposed an alternative description
of the $\Delta \pi N$ vertex that remove the `unwanted' coupling to
the spin-$1/2$ component of the delta propagator in
eq.~(\ref{eq:Sdelta}), some results comparing the two vertices were
shown in Ref.~\cite{Praet09}.
\end{enumerate} 

More details on nucleon resonances in neutrino-induced processes can
be found
in~\cite{Lalakulich06,Hernandez07,Hernandez08,Hernandez13,Gonzalez-Jimenez17,LeitnerThesis,Rafi-Alam16}
and references therein.\\

\paragraph{Models}

Resonance contributions alone are not sufficient to reproduce the
experimentally observed photo-, electro- and neutrino-production data.
Background contributions arising from different mechanisms are
necessary, e.g. those in Fig.~\ref{fig:Res-Bg}(c)-(g).  This is
especially evident at the pion threshold, far from the Delta peak,
where this background is fully determined by chiral
symmetry~\cite{Hilt13}.  In Ref.~\cite{Hernandez07}, the leading order
terms of a chiral perturbation theory Lagrangian (ChPT) for the
$\pi$-nucleon system were computed~\footnote{Recently, in
  Refs.~\cite{Yao18,Yao19} it has been presented a study on the weak
  pion production in ChPT up to next-to-next-to-leading order, using
  pions, nucleons, and Delta degrees of freedom.}, the diagrams
contributing are shown in Fig.~\ref{fig:Res-Bg}(c)-(g), while the
expressions for their amplitudes can be found
in~\cite{Hernandez07,Sobczyk13,Lalakulich10,Gonzalez-Jimenez17}.

Strictly speaking, the applicability of the background
model~\cite{Hernandez07} is limited to the pion production threshold,
however, it has been shown that the incorporation of form factors and
the Delta terms, Fig.~\ref{fig:Res-Bg}(a) and (b), results in an
effective model that provides a reasonably good agreement with
photon-, electron-, and neutrino-induced SPP data in the region
$W<1.4$ GeV (see~\cite{Hernandez07,Hernandez13,Lalakulich10,Sobczyk13}
and references therein).
This approach has been employed by different
groups~\cite{Lalakulich10,Sobczyk13,Rafi-Alam16,Gonzalez-Jimenez17,Kabirnezhad18,Graczyk19}
in the description of neutrino-induced SPP.  Its strength lies in its
relative simplicity and that it is easily exportable to different
nuclear frameworks.

To summarize the different models for SPP in the resonance region,
that have been applied to electron and neutrino interactions, we use
the classification given in Refs.~\cite{Nakamura15,Nakamura17}. The
first kind of models are those that include only resonance amplitudes,
e.g. Ref.~\cite{Lalakulich06}. Models of the second kind combine
resonances and background
contributions~\cite{Fogli79,Rein81,Hernandez07,Leitner09,Serot12,Kabirnezhad18}.
First and second kind, they both have in common that do not respect
the unitarity condition of the amplitude.
In the third kind we have the dynamical coupled-channels (DCC)
model~\cite{Kamano12,Kamano13,Nakamura15,Nakamura19}, which in words of the
authors~\cite{Nakamura17} is an extension of the Sato and Lee (SL)
model~\cite{Sato03,Matsui05}. The DCC model goes a few steps further
than the first and second kind: i) it accounts for the re-scattering
between the hadrons, ii) it includes coupled channels that open for
increasing energy transfer, such as $\eta N$, $K \Lambda$, $K \Sigma$
and $\pi\pi N$, and iii) the unitarity condition is fulfilled by
construction.  Similar degree of sophistication is achieved by the
MAID unitary isobar model~\cite{Drechsel07}, but unfortunately, to our
knowledge, this has not been consistently extended to neutrino-induced
reactions yet.

It is worth mentioning that unitarity was partially restored in the
model of Ref.~\cite{Hernandez07} by incorporating the relative phases
between the background diagrams and the dominant partial wave of the
Delta pole~\cite{Alvarez-Ruso16}.  On top of that, in
Ref.~\cite{Hernandez17} an additional contact term was added, aiming
at improving the description of the $\nu_\mu n\rightarrow\mu^-n\pi$
channel and, at the same time, removing spurious spin-$1/2$
contributions in the Delta diagrams.
Recently, in Ref.~\cite{Sobczyk19}, this model, the DCC and the SL
models were compared for a variety of reaction channels and
kinematical scenarios. The authors concluded that, in the region
$W<1.4$ GeV, the simplest model~\cite{Hernandez17} is able to
reproduce the bulk results from the other more sophisticated ones.

It should be stressed that, for the axial sector, all models rely, in
one way or another, on the neutrino-deuterium and neutrino-hydrogen
pion-production data from the old bubble chamber BNL and ANL
experiments~\cite{CC-ANL82,WNC-ANL80,CC-BNL86}. On top of poor
statistics and large systematic errors, the interpretation of these
data has the additional challenge of describing the reaction on a
deuterium target~\cite{Alvarez-Ruso99,Wu15}.  Therefore, it is clear
that new data on neutrino-hydrogen SPP would be of great value to move
forward on a firmer basis. \\

\paragraph{SPP in the high-$W$ region}\label{Hybrid-model}

All the models mentioned describe the amplitude using lowest order
interaction terms.  For increasing $W$, higher-order contributions are
needed, and hence, the predictions of these low-energy models (LEM)
beyond $W=1.4-1.5$ GeV should be taken with care.  The DCC model can
provide reliable predictions for somewhat larger invariant masses
($W\lesssim2$ GeV) thanks to the incorporation of the relevant
resonances up to $W\approx2$ GeV along with the unitarization of the
amplitude.

The procedure of adding higher order contributions soon becomes
unfeasible.  Regge phenomenology~\cite{Perl74} is
an alternative approach that provides the correct $s$-dependence of
the hadronic amplitude in the high-$s$ small-$|t|$ limit.  The method,
first proposed in~\cite{Levy73} and further developed
in~\cite{Guidal97}, boils down to replacing the $t$-channels
meson-exchange diagrams by the corresponding Regge amplitudes, which
include an entire family of hadrons.  In the context
of~\cite{Guidal97}, the $s$- and $u$-channel Born diagrams are kept in
order to maintain CVC.  As mentioned, Regge phenomenology only
provides the $s$-dependence of the amplitude, but not the $t$- or
$Q^2$-dependence.  The $t$-dependence of the Regge amplitude is
determined by the background terms of the LEM. The idea behind this is
that the physical region of the reaction is not too far from the
nearest pole of the Regge trajectory \cite{Omnes77,Perl74}.
This Regge-based $t$-channel approach~\cite{Guidal97} compares well to
photoproduction and low-$Q^2$ electroproduction data, but it fails to
reproduce the higher $Q^2$ data and the transverse component of the
electroproduction cross section. To remedy this, in
Ref.~\cite{Kaskulov10}, this Regge-based model was supplemented with
hadronic form factors, that are included in the $s$- and $u$-channel
Born terms, and effectively account for the contributions of nucleon
resonances in the high-$Q^2$ sector.
 
\begin{figure}[htbp]
\centering  
\includegraphics[width=.2\textwidth,angle=270]{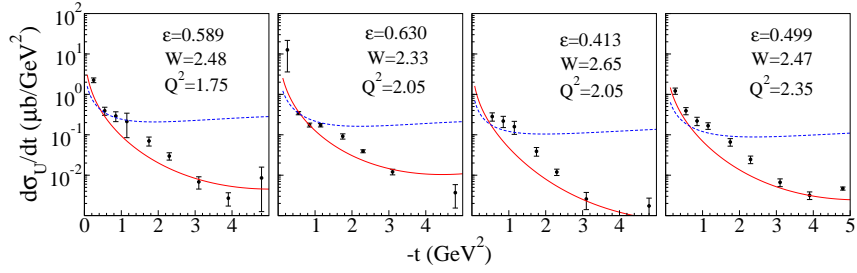}
\caption{Exclusive $p(e,e'\pi^+)n$ differential cross section
  ($d\sigma_U/dt=d\sigma_T/dt+\varepsilon d\sigma_L/dt$). The
  low-energy model (dashed line) overshoots the data, the high-energy
  model (solid line) reproduces its trend better. Data are
  from~\cite{Park13}. Figure is adapted from
  Ref.~\cite{Gonzalez-Jimenez17}.}\label{fig:ChPT-vs-HEM}
\end{figure}

An extension to neutrino-induced reactions of the models of
Refs.~\cite{Guidal97,Kaskulov10} was recently developed
in~\cite{Gonzalez-Jimenez17}.  The model presented in
Ref.~\cite{Gonzalez-Jimenez17} consists in a low-energy model (LEM)
and a high-energy model (HEM) that are combined by a smooth
$W$-dependent transition function.  This results in a {\it hybrid
  model} that  can be safely applied over a large $W$ region.  In
particular, the LEM is built up from the coherent sum of background
terms~\cite{Hernandez07} and the $s$- and $u$-channel diagrams for the
$P_{33}(1232)$ (Delta), $D_{13}$ (1520), $S_{11}$ (1535) and $P_{11}$
(1440) resonances. The parametrization of the Olsson phases from
Ref.~\cite{Alvarez-Ruso17}, as well as the corresponding Delta form
factors, were used.
The HEM is obtained by Reggeizing the background terms~\cite{Guidal97}
and introducing hadronic form factors~\cite{Kaskulov10,Vrancx14}. This
HEM is compared with the LEM and with exclusive $p(e,e'\pi^+)n$ data
in Fig.~\ref{fig:ChPT-vs-HEM}. The results show that the LEM clearly
overshoots the data while the HEM provides a better description of
them.
Finally, the $s$- and $u$-channel resonance diagrams are regularized
by using phenomenological cut-off form factors, whose role is just to
eliminate their contributions far from the resonance peak $W\approx
m_R$.
More details can be found in~\cite{Gonzalez-Jimenez17}.

Event generators, such as NuWro~\cite{Sobczyk05} and
GiBUU~\cite{Buss12,Lalakulich12}, use DIS-based
formalisms~\cite{Bodek:1980ar,Bodek02} and PYTHIA hadronization
routines~\cite{Pythia6} to describe neutrino-induced SPP in the high
energy region ($W>1.5-2$ GeV).  The transition from the resonance
region to the high-$W$ regime is done by a transition function that
smoothly, for increasing $W$ values, switches on (off) the
high-(low-)energy components of the model\footnote{The results of the
  hybrid model and NuWro for neutrino-induced SPP on the nucleon were
  compared in Ref.~\cite{Gonzalez-Jimenez17}.}.
The validity of these DIS-based approaches is not clear at small
$Q^2$, where it concentrates most of the strength of
SPP~\cite{Kaskulov10}. This could be tested with the wide collection
of electron-nucleon exclusive SPP data in the high-$W$
region~\cite{Brauel76a,Brauel76b,Brauel77,Brauel79,Horn06,Tadevosyan07,Airapetian08,Horn08,Block08,Qian10,Park13}. \\

\paragraph{Some results}

\begin{figure}[htbp]
\centering  
\includegraphics[width=15cm,bb=80 528 566 672]{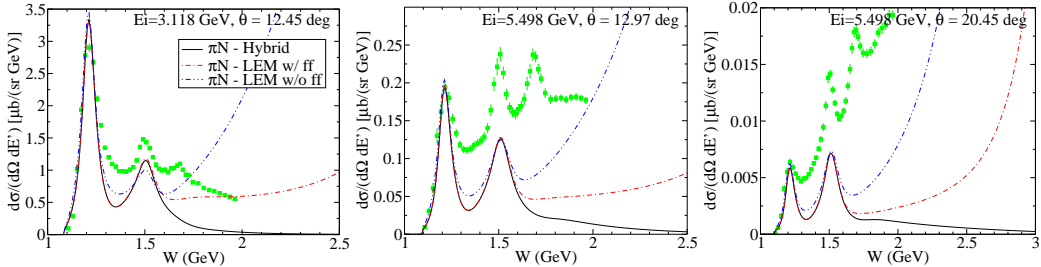}
\vspace{-0.5cm}
\caption{Inclusive electron-proton scattering cross sections
  data~\cite{JLab-database} are compared with the SPP results of the
  low-energy (LEM) model (with and without resonance form factors) and
  the hybrid model of
  Ref.~\cite{Gonzalez-Jimenez17}.}\label{fig:IncluLEM}
\end{figure}

In Fig.~\ref{fig:IncluLEM} we show inclusive electron-proton
scattering cross sections for three different energies and scattering
angles.
The curves labelled as ``LEM w/o ff'' and ``LEM w/ ff'' correspond to
the LEM without and with these resonance form factors,
respectively. In both cases the LEM background~\cite{Hernandez07} is
used.
The model labelled as ``Hybrid'' includes the resonance form factors
and the transition to the Regge-based model~\cite{Gonzalez-Jimenez17}.
The theoretical predictions include only the single-pion production
channel, therefore, they should underestimate the inclusive data.  By
construction, the three models provide essentially the same results
for $W<1.3$ GeV. The need of cut-off resonance form factors and a
high-energy model becomes evident if one moves towards higher $W$.

\begin{figure}[htbp]
\centering  
\includegraphics[width=15cm,bb=80 528 566 672]{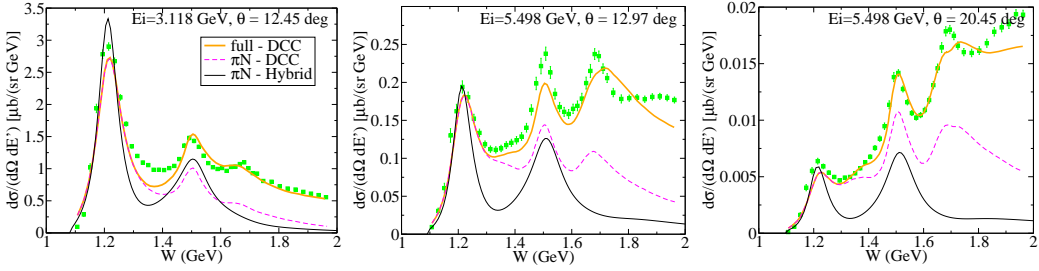}
\vspace{-0.5cm}
\caption{Inclusive electron-proton scattering cross sections
  data~\cite{JLab-database} are compared with the SPP results from the
  DCC and Hybrid models. The inclusive DCC results (full-DCC) are also
  shown. The results of the DCC model have been taken from
  Ref.~\cite{Nakamura15}.}\label{fig:IncluDCC}
\end{figure}

The predictions of the hybrid and DCC model~\cite{Nakamura15} for SPP
are shown in Fig.~\ref{fig:IncluDCC}, for the same process and
kinematics as in Fig.~\ref{fig:IncluLEM}. The inclusive DCC
predictions, which include other reaction channels such as two-pion
production, is also shown.
Up to the Delta peak, the DCC and hybrid models, and the data are
comparable.  Above the Delta peak, it is evident that the basic
ingredients in the hybrid model are not enough to reproduce the
strength of the SPP channel predicted by the DCC model. In particular,
the hybrid model underpredicts the `dip' between the Delta peak and
the second resonance region.  This may be related with the lack of
other processes, such as $\rho$- and $\omega$-exchanged contributions,
the lack of pion-nucleon re-scattering, and the use of too strong
cut-off resonance form factors.  The third resonance region (third
peak in the spectra) is also underpredicted by the hybrid model. This
could be improved by including the relevant resonances in that region.

\begin{figure}[htbp]
\centering  
\includegraphics[width=.2\textwidth,angle=270]{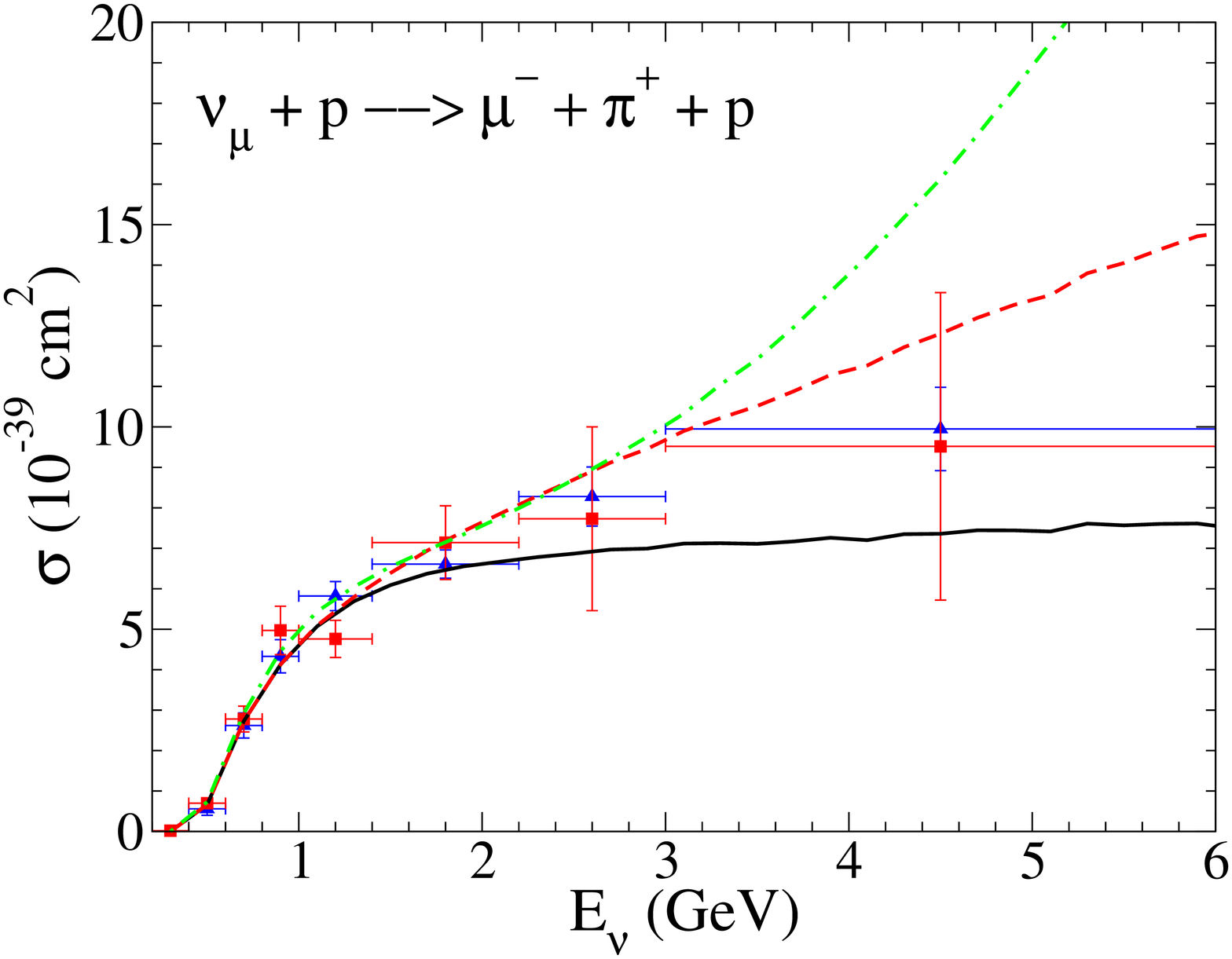}
\includegraphics[width=.2\textwidth,angle=270]{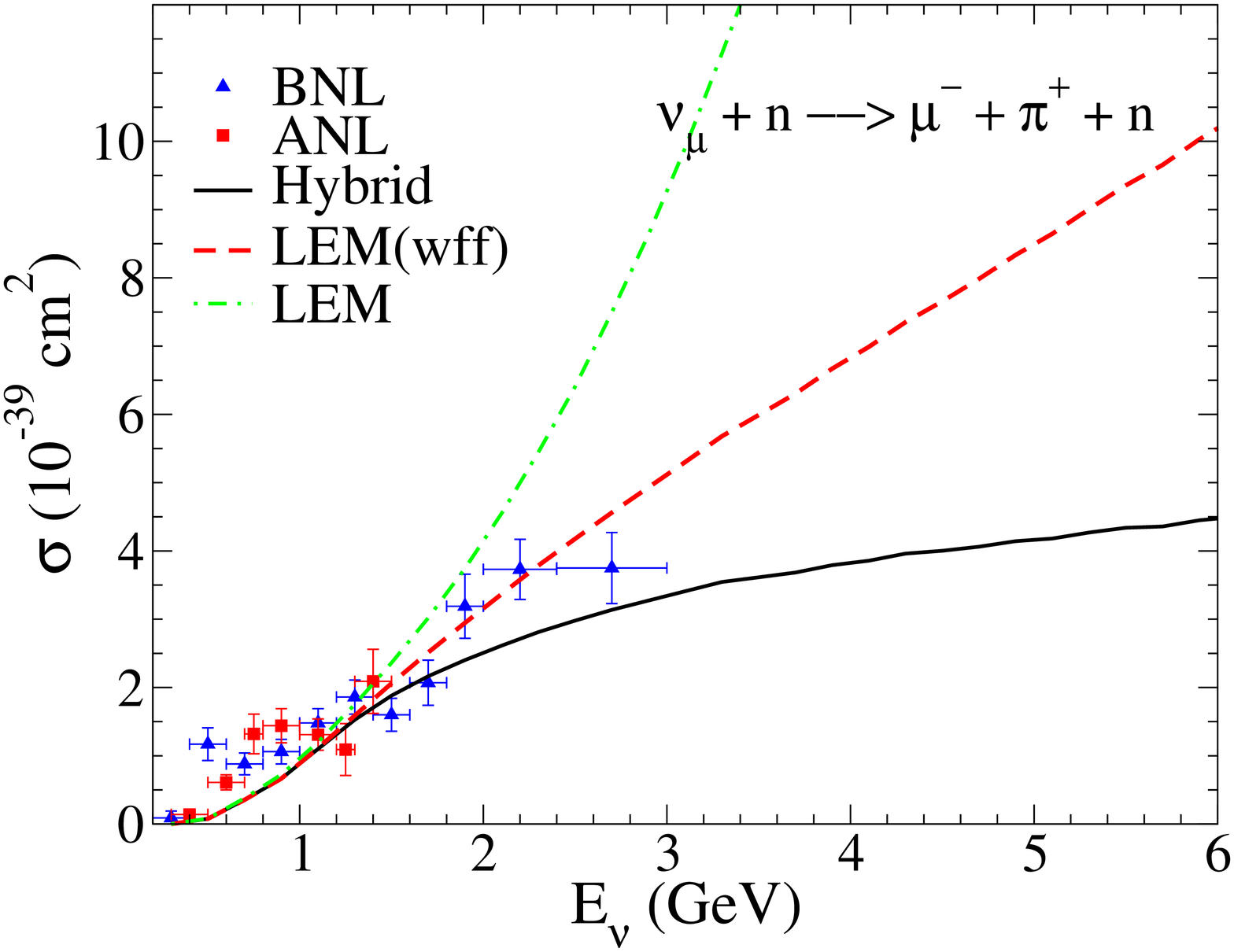}
\includegraphics[width=.2\textwidth,angle=270]{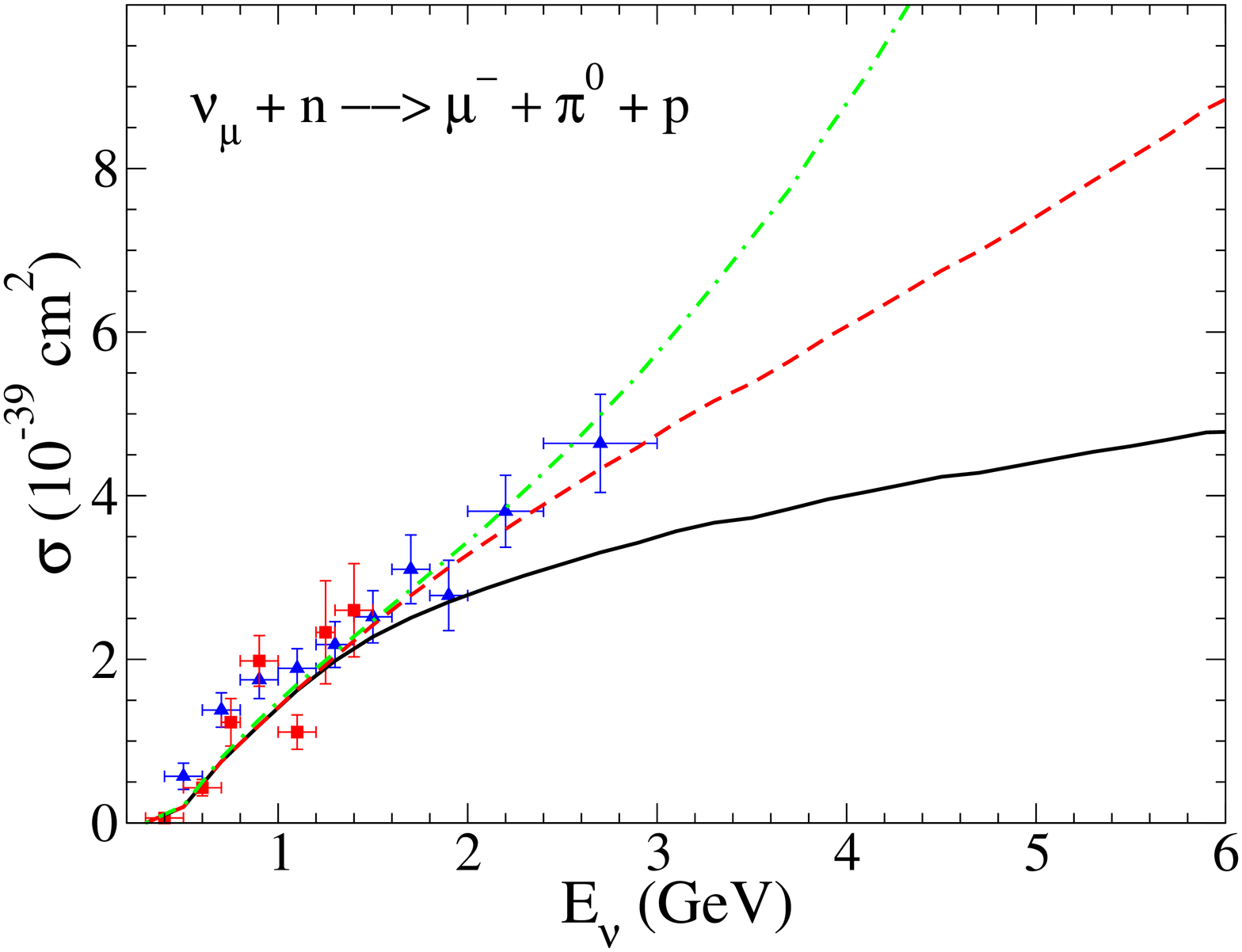}
\caption{Total cross sections for CC neutrino-induced SPP. The recent
  analysis~\cite{Wilkinson14} of the ANL and BNL
  data~\cite{CC-ANL82,CC-BNL86} is compared with the low-energy and
  hybrid models (data without cuts in the invariant mass). Deuteron
  effects are not considered. The figure was adapted
  from~\cite{Gonzalez-Jimenez17}.}\label{fig:totBNLANL}
\end{figure}

Finally, in Fig.~\ref{fig:totBNLANL} we compare the ANL and BNL data
for CC neutrino-induced SPP with the low energy and hybrid models. In
the region $E_\nu>2$ GeV the differences between the three models are
important. Since the predictions of the LEMs cannot be considered
realistic in this region, these results point to the need of a
HEM. Also, the apparent agreement of the LEMs with data in the region
$E_\nu>1-1.5$ GeV should be interpreted with care.

\subsection{Nuclear framework}

The elementary reaction described above for SPP can be incorporated in
a nuclear framework using the impulse approximation, i.e., considering
that the electroweak boson interacts with only one nucleon in the
nucleus.
The simplest choice is the global relativistic Fermi gas  model
which, in spite of its simplicity, is still employed by MC generators
in the analyses of
data~\cite{MBNCpi010,MBCCpionC11,MBCCpion011,MINERvACCpi15,MINERvACCpi16,T2KCC1pi17}.
The global RFG is a fully relativistic model, it respects the
fundamental symmetries of a relativistic quantum field theory and it
is able to capture the gross features of the nuclear response. Though,
it is quite far from the level of precision required in the new era of
the neutrino-oscillation programme~\cite{Alvarez-Ruso17}.
The local Fermi gas (lFG) or its relativistic version (lRFG) are also
widespread~\cite{Ahmad06,Szczerbinska07,Zhang12,Hernandez13,Sobczyk13,Martini14}.
To achieve a more realistic description of the nuclear dynamics,
different groups add different ingredients on top of the `bare' local
Fermi gas (see~\cite{Alvarez-Ruso14,Katori17} for a review).
The factorization approach has also been used for neutrino-induced
pion production~\cite{Benhar:2005dj,Rocco19}. This factorization of
the elementary vertex and the nuclear dynamics, the latter contained in
the nuclear spectral function, makes it very appealing to be used in
MC event generators.
Finally, mean-field based models are the state of the art regarding
the description of the nuclear dynamics within a quantum mechanical
framework that allows for the treatment of inclusive and exclusive
processes.  In what follows we review the lepton-induced SPP on the
nucleus in the context of a the relativistic mean-field
model~\cite{Fernandez-Ramirez08b,Praet09,Gonzalez-Jimenez18,Nikolakopoulos18,Gonzalez-Jimenez19a}.\\

\paragraph{What Mean-Field models can offer to the neutrino interaction community}

Mean-field models are able to capture a good part of the nuclear
dynamics by describing the ground state nucleus as a set of
independent-particle nucleon wave functions that are solutions of the
mean-field equations. In previous sections we already discussed these
models applied to electron and neutrino scattering reactions. Here we
simply summarize the basic points of interest for the discussion that
follows.

As known, the case of inclusive processes can be efficiently modelled
by describing the final states as scattering solutions of the
corresponding wave equation: the Dirac or Schr\"odinger equation for the
nucleons and the Klein-Gordon equation for the pions.  Since the flux has
to be conserved, one can use mean-field potentials with only real
part~\cite{Maieron03,Kim03,Caballero05,Meucci09,Kim07,Butkevich07,Gonzalez-Jimenez19a}
or full complex optical potentials, in which the flux lost
(transferred to inelastic channels) due to the imaginary term is
recovered by a summation over those channels, as done in the
Relativistic Green Function (RGF)
model~\cite{Capuzzi91,Meucci09,Ivanov16b}.
On the contrary, in the exclusive case one needs to account for the
flux moved to the inelastic channels (absorption, multi-particle
emission, charge exchange, etc.). This can be accounted for by using
phenomenological complex optical potentials~\cite{Udias93,Udias01}
(see discussion in Sect.\ref{sec:RMF}) fitted to elastic
nucleon-nucleus scattering~\cite{Cooper93,Cooper09} or elastic
pion-nucleus scattering~\cite{Nagl91} data, and correcting the results
with spectroscopic factors.

In neutrino experiments, however, fully exclusive conditions are never
satisfied: the neutrino energy is unknown and the limited
acceptances of the detectors make it impossible to detect the complete
final state.
Therefore, Monte Carlo (MC) neutrino event generators have to deal
with inclusive and semi-inclusive events and, by combining the
available experimental information with the nuclear theory,
reconstruct the neutrino energy for every selected event.
Modelling all possible semi-inclusive scenarios in a consistent way
means to solve an extremely complicated many-body coupled-channel
problem. A situation still far from being resolved.
In spite of that, one can try to minimize the systematic errors that
propagate to oscillation analyses by improving the theoretical nuclear
models implemented in the MC's.

In MC event generators there are two clearly separated steps: i) the
elementary vertex, and ii) the propagation in the nuclear medium of
the created hadrons, using a classical cascade model that generates
the complete final state.
Due to this factorization, `elementary vertex $\times$ hadron
propagation', the inclusive cross section will not be affected by the
cascade process. Therefore, the primary model (the one that describes
the elementary vertex) should be able to provide a good inclusive
response.  In summary, the primary model puts 100\% of the strength in
the elastic channel, then the cascade takes care of splitting the flux
into the different inelastic channels.
It is, therefore, preferable to use primary models that provide
information not only on the final lepton (inclusive models) but also
on the hadrons, so that they can be used as the `seed' for the cascade
in a more consistent way.
The mean-field models discussed here can satisfy those requirement:
full hadronic information and good inclusive results.\\

\paragraph{Cross section} 

\begin{figure}[htbp]
  \centering
      \includegraphics[width=0.45\textwidth,angle=0]{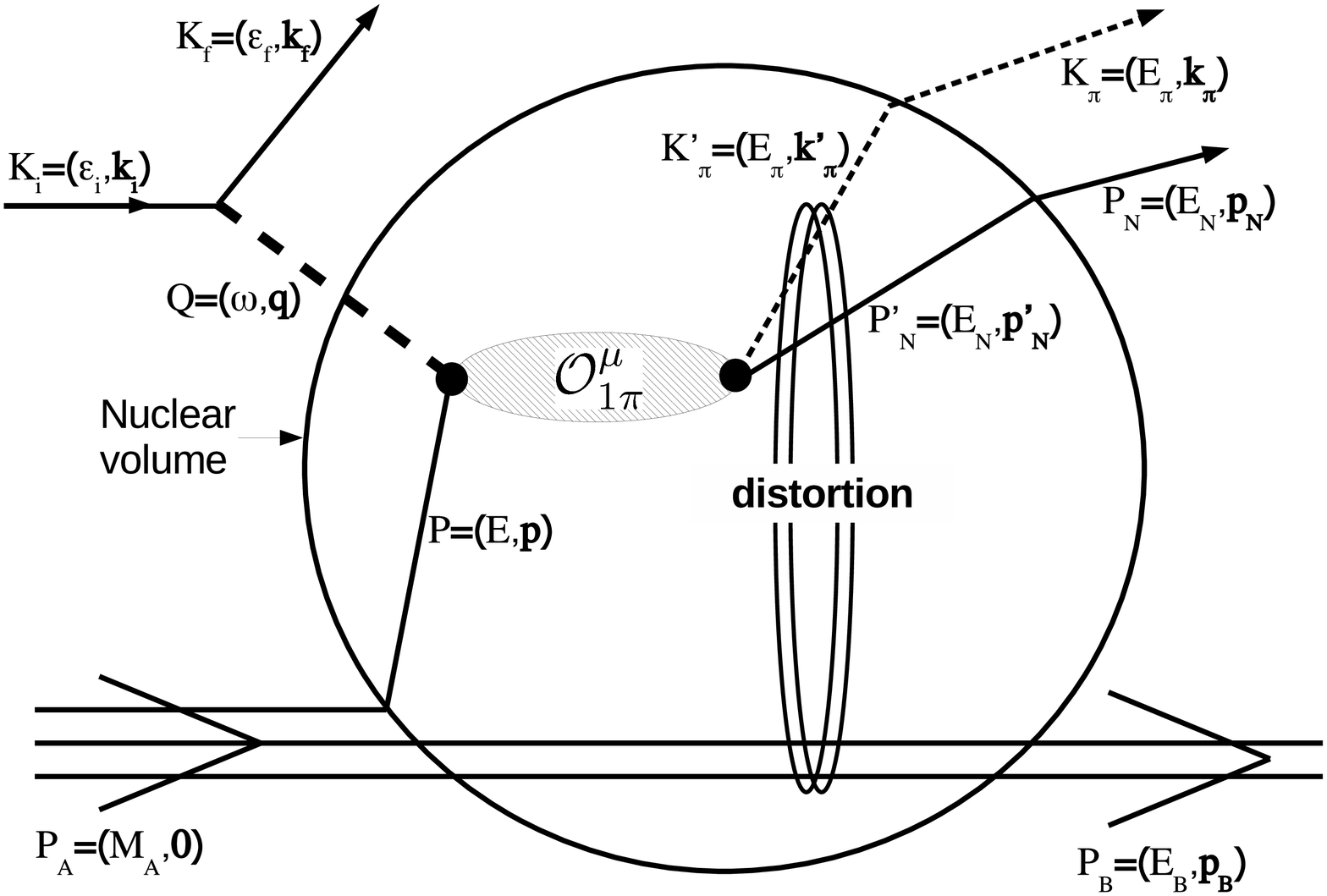}
      \vspace{-0.5cm}
  \caption{Representation of the electroweak SPP on a nucleus within
    the impulse approximation.  An incoming lepton $K_i$ scatters on a
    nucleus $P_A$. A single boson $Q$ is exchanged.  The final state
    consists in a scattered lepton $K_f$, the residual system $P_{B}$,
    and a knockout nucleon and pion with asymptotic four momentum
    $P_N$ and $K_\pi$, respectively.  Inside the nucleus the momenta
    of the initial and final nucleons ($\np$ and $\np'_N$) and the
    pion ($\nk'_\pi$) are given by probability distributions. See text
    for details. }
  \label{fig:SPP-nucleus}
\end{figure}

Given the momenta $\nk_i$ and $\np_A$ of the incoming lepton and
target nucleus, the SPP scattering process is completely determined by 9
independent variables~\cite{Gonzalez-Jimenez19b}. The nucleon wave
functions are computed in the center of mass of the target nucleus,
therefore, it is natural to work in the laboratory reference frame
where the target nucleus is at rest ($\np_A=0$).
Hence, we use the laboratory variables $\varepsilon_f$, $\theta_f$,
$\phi_f$, $E_\pi$, $\theta_\pi$, $\phi_\pi$, $\theta_N$, $\phi_N$, and
the missing energy $E_m$, as independent variables (see
Fig.~\ref{fig:SPP-nucleus}). In terms of these, the differential cross
section for the process in Fig.~\ref{fig:SPP-nucleus}
reads~\cite{Gonzalez-Jimenez19b}
\ba
 \frac{d^{9}\sigma}{d \varepsilon_f\Omega_f d E_\pi d\Omega_\pi d\Omega_N dE_m} = 
	\frac{{\cal R}_X}{(2\pi)^{8}}\, \frac{k_f p_N E_N E_\pi k_\pi }{2\varepsilon_i\, f_{rec}}\ \eta_{\mu\nu}h_{(\kappa)}^{\mu\nu} \rho_B(E_m).\label{XS}
\ea
The function $f_{rec}$ accounts for the recoil of the residual nucleus
and is given by~\footnote{We note that there is a typo in the
  expression of $f_{rec}$ given in Ref.~\cite{Gonzalez-Jimenez18}.}
\ba
  f_{rec} = \left|1 + \frac{E_N}{E_{B}}\left(1 +\frac{\np_N\cdot(\nk_\pi-\nq)}{p_N^2}\right) \right|\,.
\ea
The energy of the outgoing nucleon $E_N$ and the residual nucleus
$E_B$ are obtained from the independent
variables~\cite{Gonzalez-Jimenez19b}.
The factor ${\cal R}_X$ was defined in Eq.~\ref{R_X} and the leptonic
tensor $\eta_{\mu\nu}$ in Sect. II.  The hadronic
tensor for one nucleon knock out from a $\kappa$ shell is given by
\ba
  h_{(\kappa)}^{\mu\nu} = \frac{1}{2j+1}\sum_{m_j}\sum_{s_N} (J^\mu)^\dagger J^\nu\,,\label{hmunu} 
\ea
where $\sum_{s_N=\pm1/2}$ is the sum over the spin projections of the knockout nucleon and $\frac{1}{2j+1}\sum_{m_j}$ is the
 average over the bound nucleons of a $\kappa$ shell, being $j$  the total angular
momentum and $m_j$ its third component.

The fact that the residual nucleus may be in an excited state is
introduced by the function $\rho_B(E_m)$, that represents the density
of final states for the residual nucleus. Its mass, $M_B$, is
determined from the energy-conservation relation $M_B = M_A + E_m -
m_N$.  Within a pure shell model
\ba
 \rho_B(E_m)=\sum_{\kappa}\delta(E_m-E_m^{\kappa})\,,
\ea
where $E_m^{\kappa}$ is a fixed value for each shell. \\

\paragraph{Hadron current}

In this framework, the most general hadronic current (for given
$\kappa$, $m_j$ and $s_N$) in momentum space reads
\begin{equation}     \label{J-general}
J^\mu =
\frac{1}{  ( 2\pi )^{3/2} } \int d\np_N' \int d\np \,
    \psib^{s_N}(\np_N',\np_N) \phib(\nk'_\pi,\nk_\pi) \,
         {\cal  O}_{1\pi}^\mu(Q,P,K'_\pi,P'_N)
    \psi_\kappa^{m_j}(\np),
  \end{equation}
    where $K'_\pi =
    Q+P-P'_N$.  $\psi_\kappa^{m_j}(\np)$ is the Dirac spinor of the
    bound nucleon (see Ref.~\cite{Caballero98a} for details) and $P$
    represents its four momentum.
$\psi^{s_N}(\np_N',\np_N)$ and $\phi(\nk'_\pi,\nk_\pi)$ represent the
    wave functions of the final nucleon and pion (see
    Fig.~\ref{fig:SPP-nucleus}).~\footnote{The definition of the wave function including distortion effects can be found, for example, in Refs.~\cite{Udias:1993zs,Udias93,Martinez06} for the nucleon, and in Ref.~\cite{Nagl91} for the pion.}
In deriving Eqs.~(\ref{XS}) and (\ref{J-general}), it has been
considered that all the particles have well defined energy in all
steps of the process, so that the energy of the bound nucleon $E$,
which is needed to build the operator ${\cal O}_{1\pi}^\mu$, can be
determined by energy conservation $\omega + E = E_N + E_\pi$.  Note
that the transition operator ${\cal O}_{1\pi}^\mu(Q,P,K'_\pi,P'_N)$
has to be evaluated for every $\np$ and $\np'_N$.  All this, together
with the large phase space for this process (4 particles in the final
state), represents a non trivial problem from the computational point
of view.

In what follows we describe some approximations that simplify the
problem and have allowed us to obtain numerical results that can be
compared with cross section data:
\begin{enumerate}
 \item Using the asymptotic values for the four momenta of the final
   nucleon and pion in the operator ${\cal O}_{1\pi}^\mu$, i.e.: \ba
   {\cal O}_{1\pi}^\mu(Q,P,K'_\pi,P'_N) \longrightarrow {\cal
     O}_{1\pi}^\mu(Q,P,K_\pi,P_N)\,\, \text{with }\,
   P=P_N+K_\pi-Q\,.\label{J-approx-asy} \ea In this way, ${\cal
     O}_{1\pi}^\mu$ can be evaluated out of the integrals in
   Eq.~(\ref{J-general}).
 \item Describing the pion as a plane wave. In this case,
   Eq.~(\ref{J-general}) reduces to: \ba J^\mu =
   \frac{1}{\sqrt{2E_\pi}}\int{d\np}\, \psib^{s_N}(\np_N',\np_N) {\cal
     O}_{1\pi}^\mu(Q,P,K_\pi,P'_N)
   \psi_\kappa^{m_j}(\np), \label{J-approx1} \ea with
   $P'_N=Q+P-K_\pi$.  Notice that if the
   approximation~(\ref{J-approx-asy}) is not employed, the operator
   should be calculated for every $\np$ value.
 \item Describing the pion and the outgoing nucleon as plane
   waves. Thus, Eq.~(\ref{J-approx1}) reads: \ba J^\mu =
   (2\pi)^{\frac{3}{2}} \sqrt{\frac{m_N}{2E_\pi E_N}}\, \ubarN {\cal
     O}_{1\pi}^\mu(Q,P,K_\pi,P_N)\,
   \psi_\kappa^{m_j}(\np),\label{J-approx2} \ea with
   $P=P_N+K_\pi-Q$. In this case, the momenta of all particles are
   well defined, so that the approximation~(\ref{J-approx-asy}) does
   not apply.
 \item Describing the initial and final nucleon, and the pion as plane
   waves.  The hadronic current~(\ref{J-approx2}) results \ba J^\mu =
   (2\pi)^3 \delta^3(\np_N+\nk_\pi-\nq-\np)\, \sqrt{\frac{m_N^2}{2E_\pi
       E_N E}}\, \ubarN {\cal O}_{1\pi}^\mu(Q,P,K_\pi,P_N)\,
   \uu\,,\label{J-pw} \ea the replacement $\frac{1}{2j+1}\sum_{m_j}
   \rightarrow \frac{1}{2}\sum_s$ in Eq.~(\ref{hmunu}) is assumed.
   This is the hadronic current in the RFG model.
From Eqs.~(\ref{XS})~and~(\ref{J-pw}), it is straightforward to obtain
numerical results within the RFG; additionally, the free nucleon case
can be recovered as the limit $p_F\rightarrow0$, with $p_F$ the Fermi
momentum.
\end{enumerate}

\begin{figure}[htbp]
  \centering
      \includegraphics[width=0.235\textwidth,angle=270]{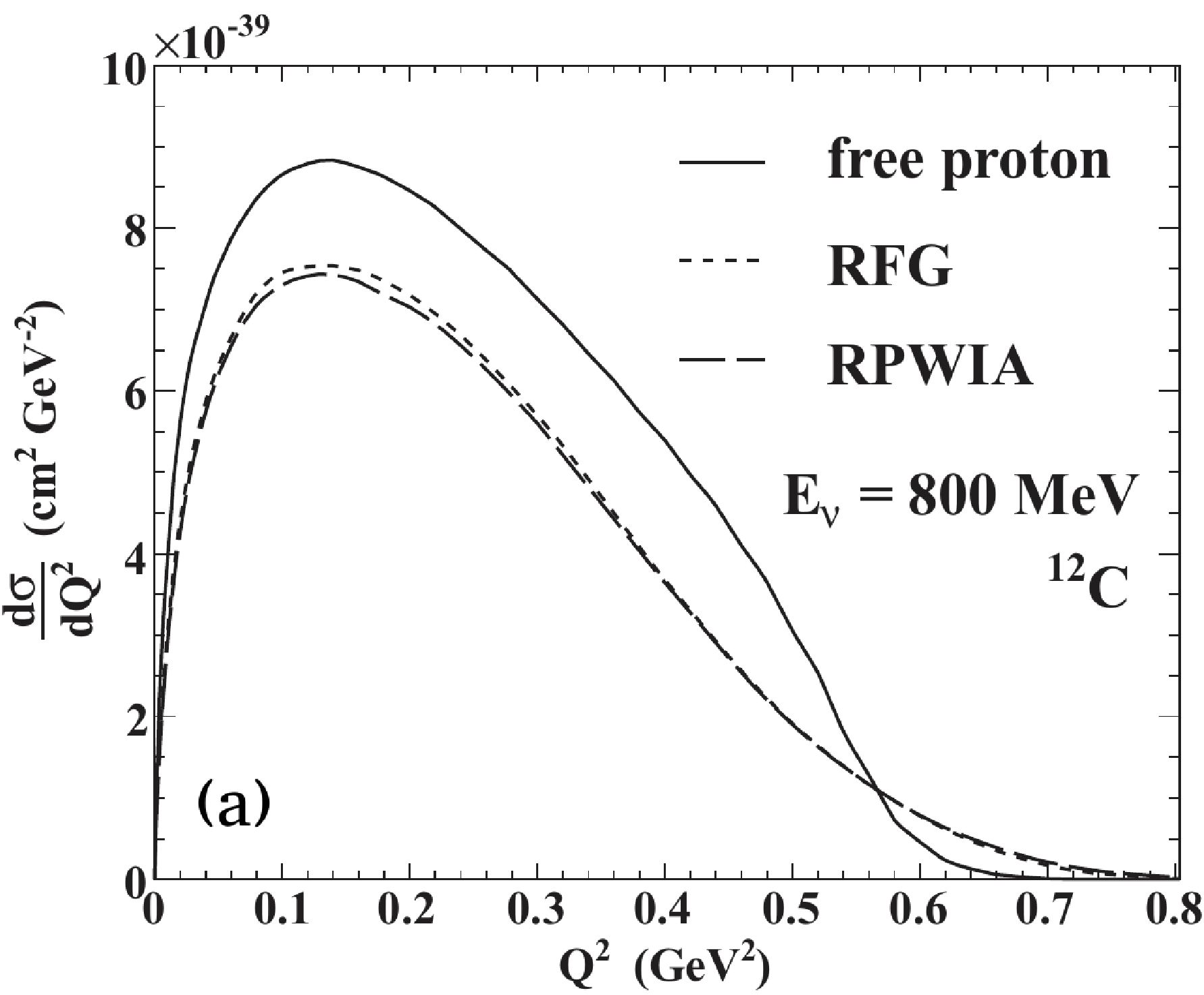}
      \hspace{0.5cm}
      \includegraphics[width=0.23\textwidth,angle=270]{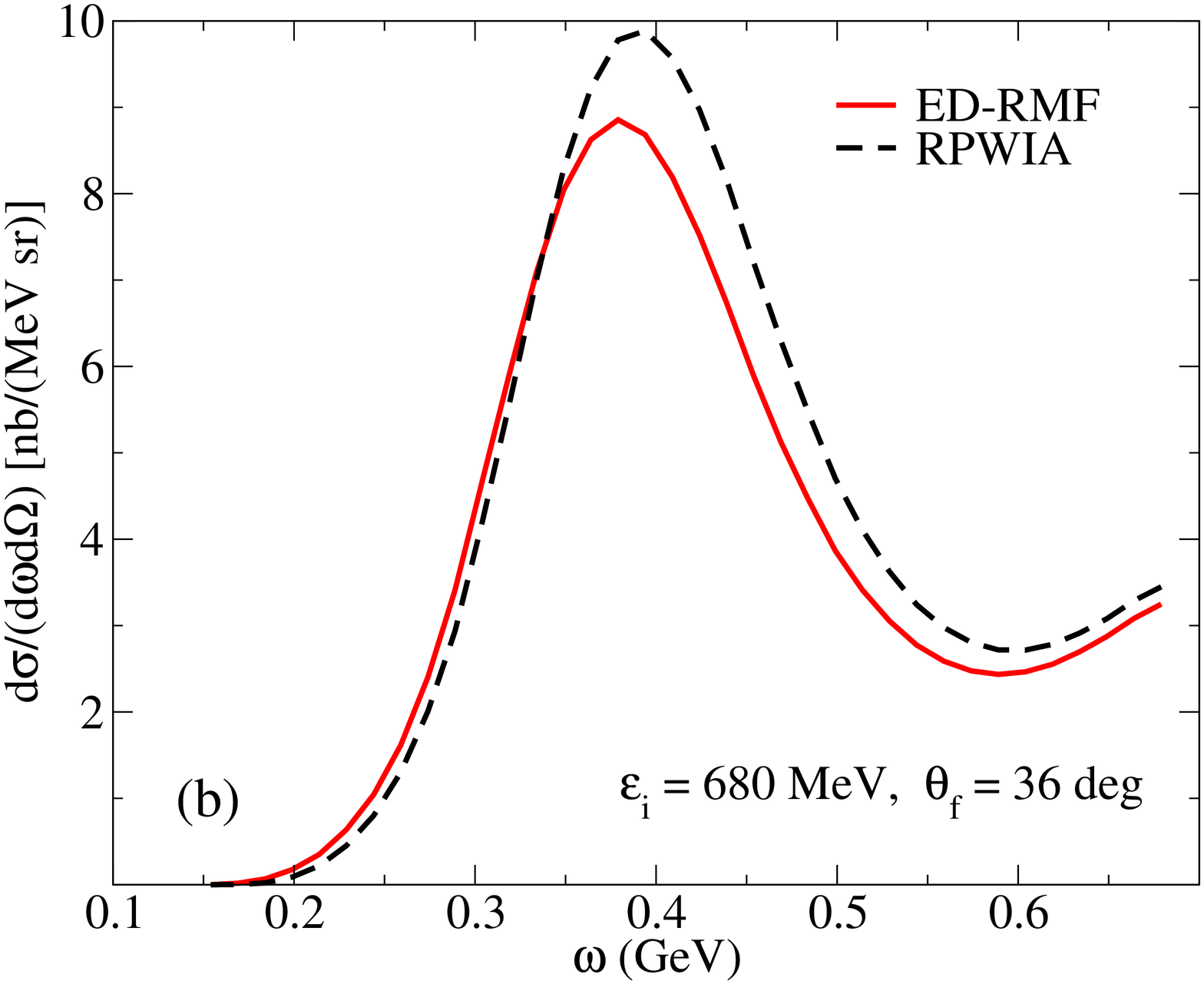}
  \caption{(a) $Q^2$ differential cross section for the process
    $\nu_\mu + ^{12}C \rightarrow \mu^- + ^{11}B + p + \pi^+$ computed
    with RPWIA, RFG and considering a free proton target; figure taken
    from Ref.~\cite{Praet09}. (b) Electron-induced SPP double
    differential cross section evaluated under the approximations
    described in Eqs.~\ref{J-approx1} (ED-RMF) and \ref{J-approx2}
    (RPWIA).  }
  \label{fig:SPP-approx}
\end{figure}

The impact of the approximations described above will depend on the
particular case under study, but in general we can say that the more
exclusive conditions (smaller phase space) and the further we are from
the maximum of the distributions, the greater the dependence on these
nuclear effects.  In Fig.~\ref{fig:SPP-approx}(a), we show the $Q^2$
differential cross section (per nucleon) for the process $\nu_\mu +
{}^{12}C \rightarrow \mu^- + {}^{11}B + p + \pi^+$ computed with three
different approaches: i) considering a free proton target, ii) the
RFG, and iii) the RPWIA.  This comparison allows us to estimate the
effect of Fermi motion and binding energy as well as the impact of
using a distribution instead of a Dirac delta in the initial state
(compare Eqs.~(\ref{J-approx2}) and (\ref{J-pw})).  One observes that
the free-nucleon model clearly departs from the RPWIA and RFG results,
while the latter are almost indistinguishable. As explained in the
original paper~\cite{Praet09}, this is the case for sufficiently high
energy and momentum transfer and for a RFG model with an appropriate
binding-energy correction.

In Fig.~\ref{fig:SPP-approx}(b) we show the inclusive ${}^{12}C(e,e')$
electron scattering cross section computed with two approaches based
on Eqs.~(\ref{J-approx1})~and~(\ref{J-approx2}), i.e.: the pion is
described as a plane wave and the initial nucleon is computed within
the RMF model, the final nucleon is i) a plane wave
(Eq.~(\ref{J-approx2}), labelled as RPWIA), or ii) it is a continuum
relativistic distorted wave (Eq.~(\ref{J-approx1}), labelled as
ED-RMF).
The results show that in spite of the large phase space that
contribute (integrals over the nucleon and pion variables have been
performed), the nuclear effects, such as Pauli blocking and
distortion, that are incorporated in the ED-RMF model and not in the
RPWIA, remain quite important (see Ref.~\cite{Gonzalez-Jimenez19a} for
further details).
 
The approximation defined in Eq.~(\ref{J-approx-asy}) was considered
in the ED-RMF calculations shown in Fig.~\ref{fig:SPP-approx}(b), and
in those of Ref.~\cite{Gonzalez-Jimenez19a}. The impact of this
approximation has been studied in the past for different reactions, in
Ref.~\cite{Tiator84} for pion photoproduction and in
Ref.~\cite{Leitner09b} for coherent pion production.  Its effect in
lepton-induced SPP within the RMF framework will be the subject of
future work.

Finally, describing the pion as a plane wave (Eq.~\ref{J-approx1}) is
clearly an oversimplification of the problem, especially if we aim at
describing semi-inclusive results in which the pion is detected.  This
is further discussed in Sect.~\ref{sec:pionsFSI}.\\

\paragraph{In-medium modification of the Delta resonance} 

The properties of the resonances are modified in the nuclear
medium. Due to the dominant role of the Delta resonance, the in-medium
modification of its decay width is one of the main nuclear effects in
neutrino-induced SPP~\cite{Kim96,Singh98}.

The procedure developed in Refs.~\cite{Oset87,Nieves93,Gil97} consists
in modifying the Delta decay width by the complex part of the Delta
self-energy calculated in the nuclear medium, as well as accounting
for Pauli blocking that reduces the phase space available for the
Delta to decay. It has been discussed in many
references~\cite{Hernandez13,Sobczyk13,Lalakulich13b,Martini14,Gonzalez-Jimenez16a},
so we do not repeat it here.
The uncertainties and inconsistencies of using this method, which was
developed within a Fermi gas model, in the framework of our
relativistic mean-field model were discussed in
Ref.~\cite{Gonzalez-Jimenez18,Nikolakopoulos18}. Its effect in the
cross section is shown in Fig.~\ref{fig:pions-nudata} as a red band,
that represents the uncertainty attached to this nuclear effect.

Other studies on the medium modification of the Delta width have been
presented in Refs.~\cite{Cugnon88,Arve94,Lee96,Kim97,Ghosh17}.

\subsection{Pion final-state interactions}\label{sec:pionsFSI}

In Refs.~\cite{Gonzalez-Jimenez18,Nikolakopoulos18} theoretical SPP
cross sections were compared with different sets of 1$\pi$-detected
data reported by the
MiniBooNE~\cite{MBNCpi010,MBCCpionC11,MBCCpion011},
MINERvA~\cite{MINERvACCpi015,MINERvACCpi15,MINERvACCpi16,MINERvAnuCC17}
and T2K~\cite{T2KCC1pi17} experiments.  Due to the semi-inclusive
nature of these experiments, a meaningful comparison with these data
requires to account for the pion final-state interactions.  The pions
produced in the primary vertex may suffer elastic rescattering, charge
exchange, be absorbed, create new pions, etc.  Indeed, the pion that
reaches the detector may not be the one produced in the primary vertex
but another one originated in a secondary interaction.

\begin{figure}[htbp]
  \centering
      \includegraphics[width=0.22\textwidth,angle=270]{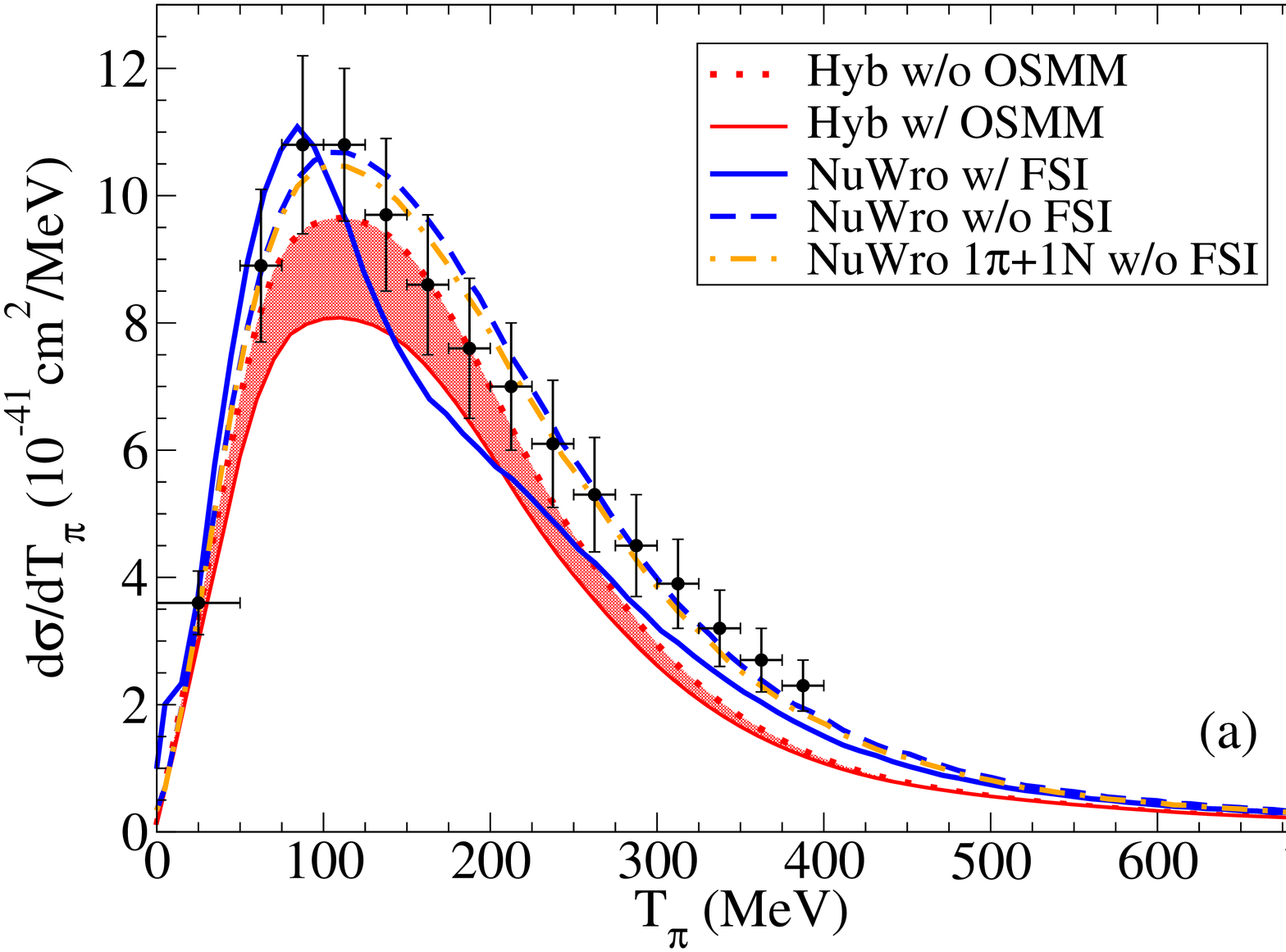}
      \includegraphics[width=0.22\textwidth,angle=270]{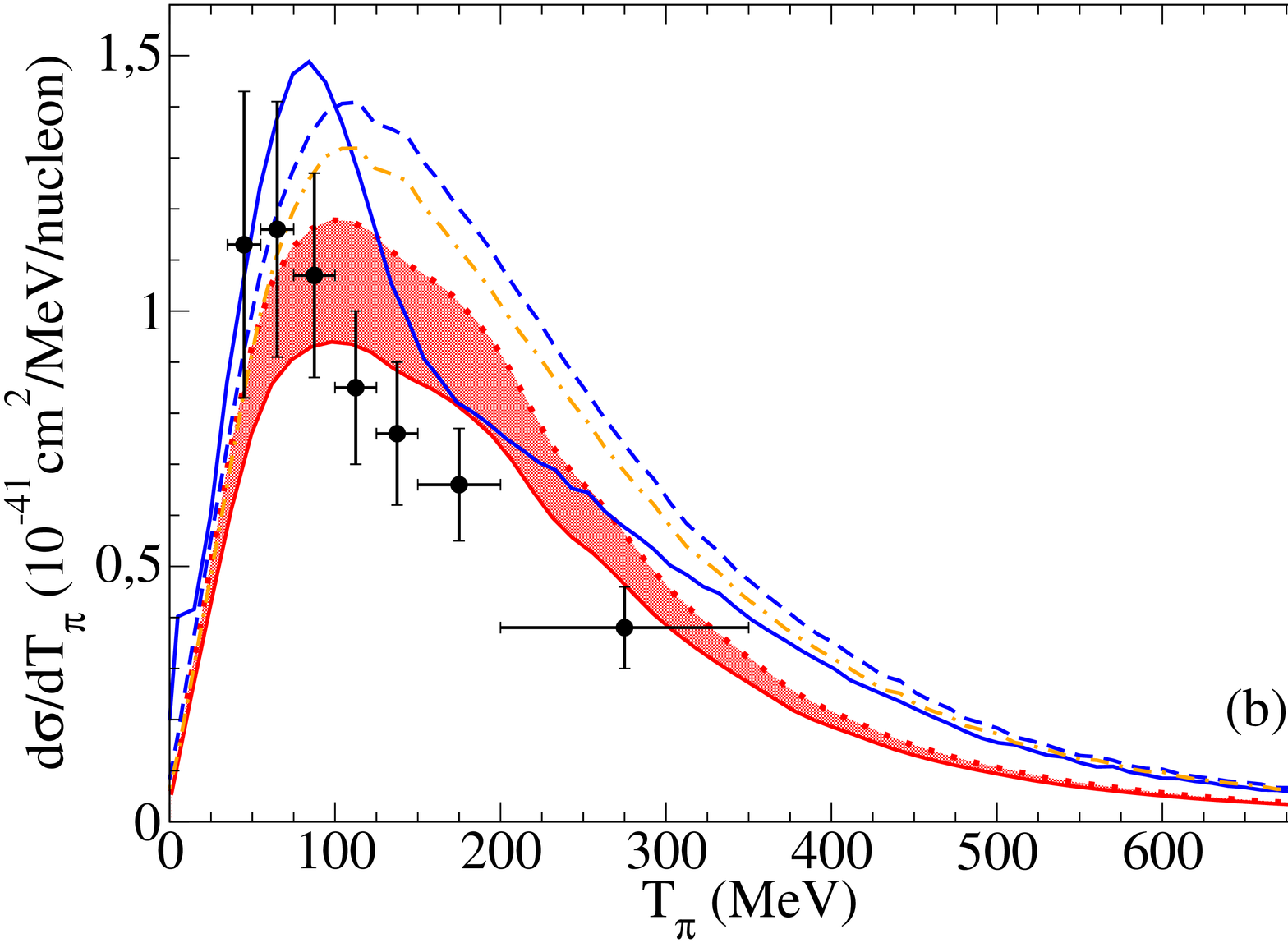}
      \includegraphics[width=0.22\textwidth,angle=270]{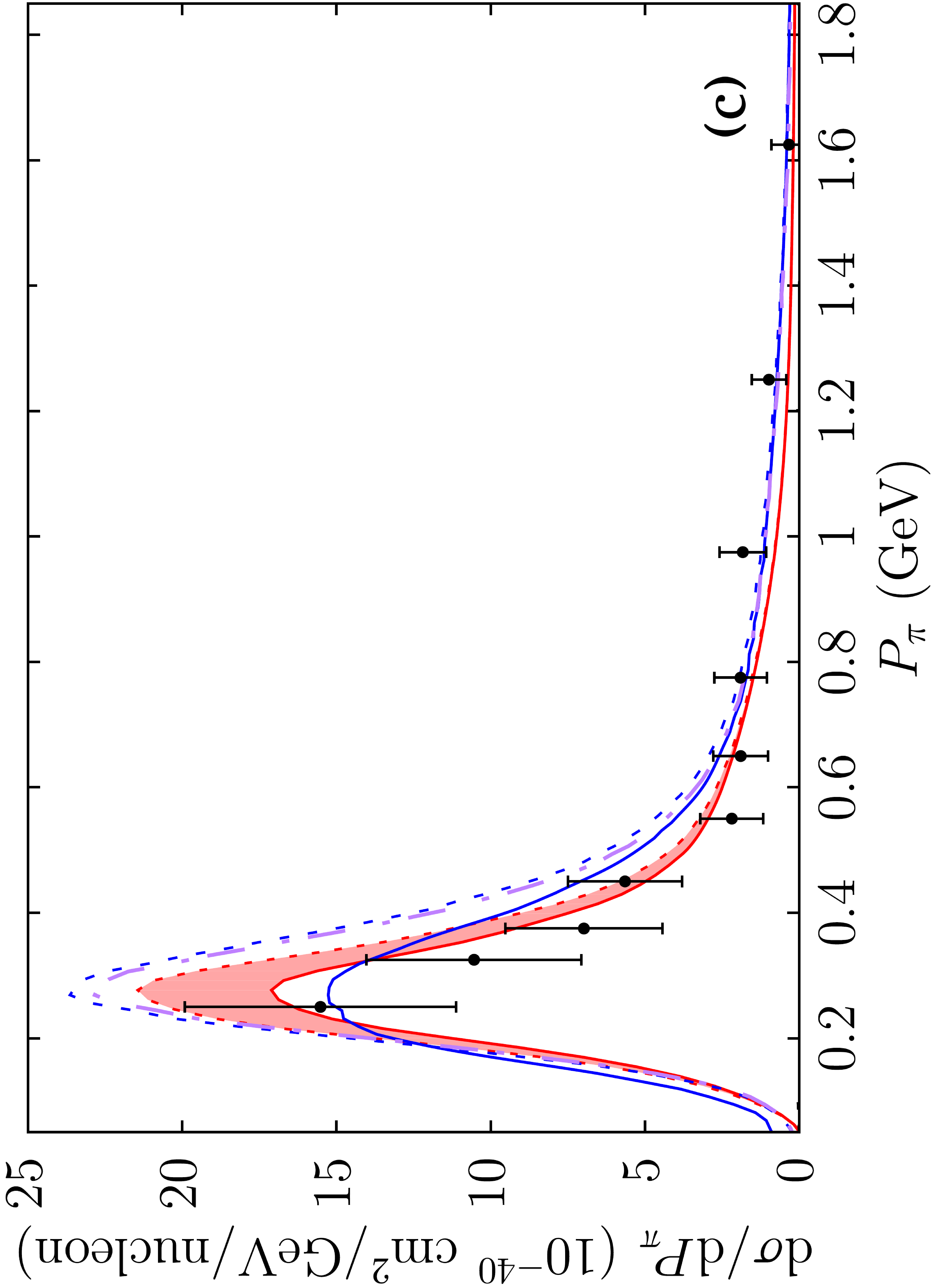}\\
      \vspace{0.3cm}
      \includegraphics[width=0.225\textwidth,angle=270]{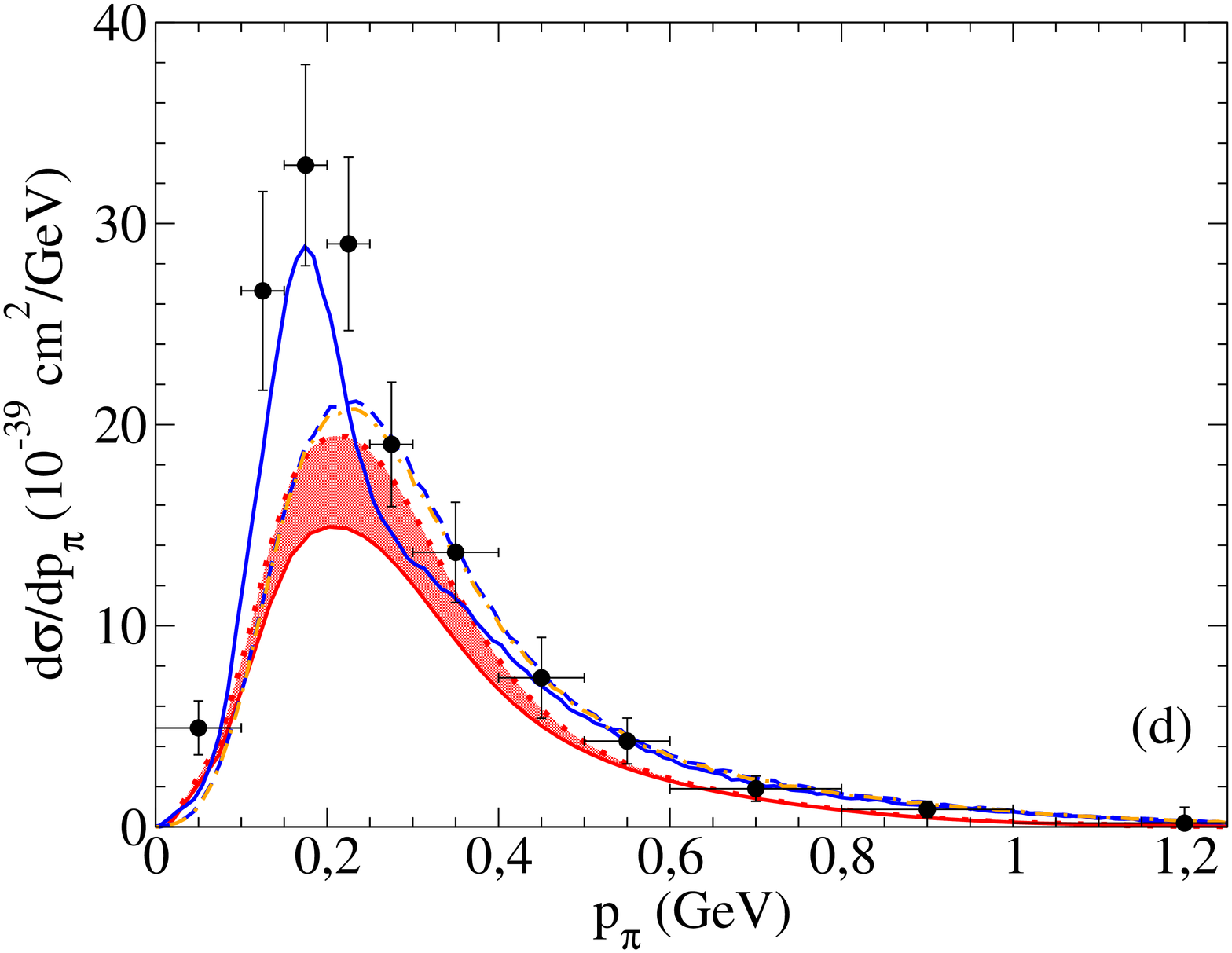}
      \hspace{0.3cm}
      \includegraphics[width=0.22\textwidth,angle=270]{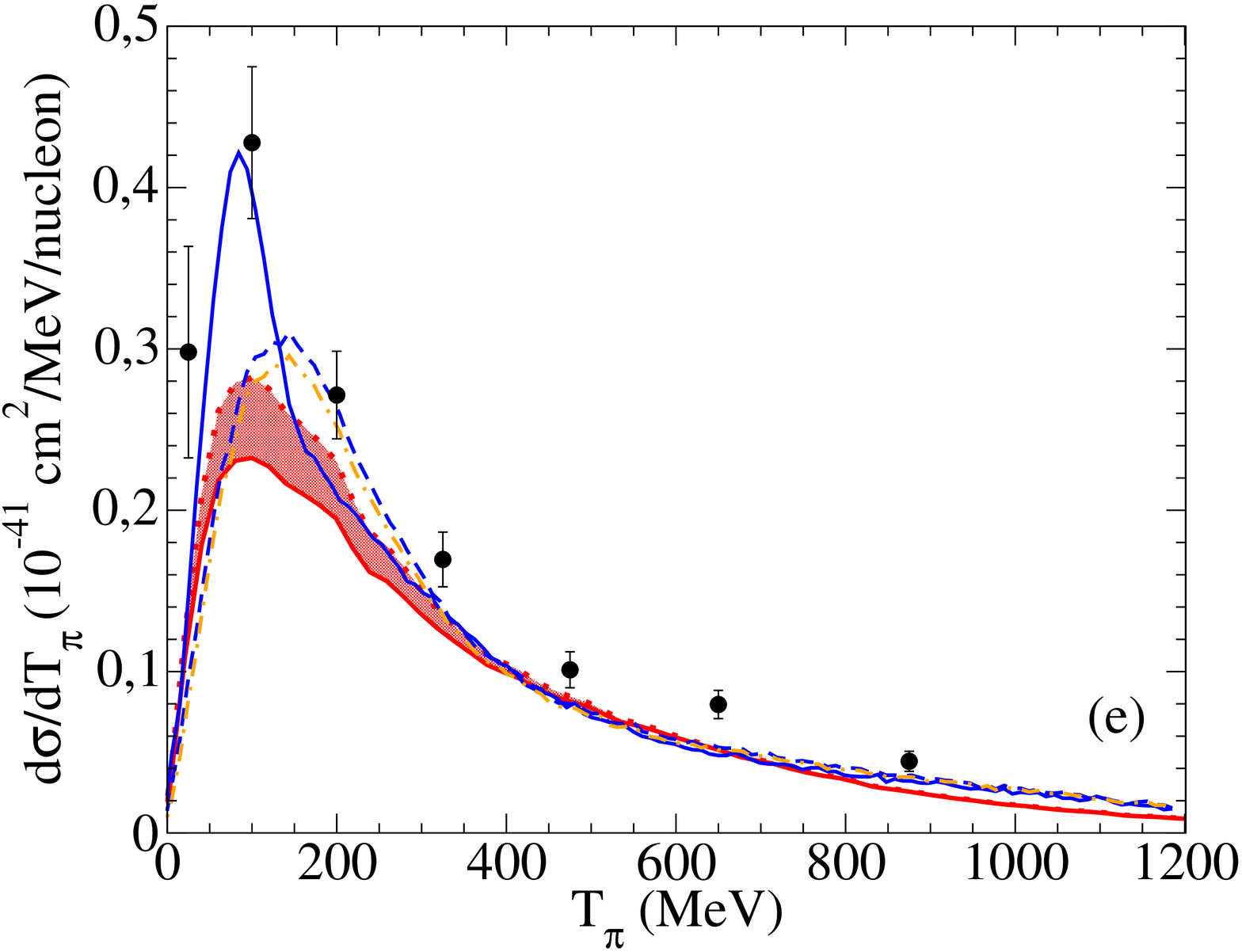}
  \caption{CC $\nu_\mu$-induced one pion production on the
    nucleus. Single differential cross sections are represented as a
    function of the kinetic energy $T_\pi$ or the momentum $p_\pi$ of
    the pion.  Panels (a), (b) and (c): MiniBooNE~\cite{MBCCpionC11},
    MINERvA~\cite{MINERvACCpi15,MINERvACCpi16} and
    T2K~\cite{T2KCC1pi17} 1$\pi^+$ production on CH$_2$, CH and
    H$_2$O, respectively.  Panels (d) and (e):
    MiniBooNE~\cite{MBCCpion011} and MINERvA~\cite{MINERvACCpi015}
    1$\pi^0$ production on CH$_2$ and CH, respectively.  Figures
    adapted from Refs.~\cite{Gonzalez-Jimenez18,Nikolakopoulos18}. }
  \label{fig:pions-nudata}
\end{figure}

In Fig.~\ref{fig:pions-nudata}, we present a selection of the results
presented in Refs.~\cite{Gonzalez-Jimenez18,Nikolakopoulos18}.
On the one hand, we show the predictions of the hybrid-RPWIA model. In
this approach, the hybrid model described in Sect.~\ref{Hybrid-model}
was used for the current operator, and the RPWIA was used for the
nuclear part (Eq.~\ref{J-approx1}). Thus, final-state interactions
(FSI) are neglected.
On the other hand, we show the results of the MC neutrino event
generator NuWro~\cite{Golan12}. In NuWro the elementary
vertex is described in a somewhat simpler way: the Delta resonance and
a background, which is an extrapolation of the DIS contribution to the
low-$W$ region~\cite{Sobczyk05}. On the contrary, NuWro accounts for
FSI using a sophisticated cascade model~\cite{Salcedo88,Golan12} (see
Ref.~\cite{Niewczas19} for a recent study in which the model is
updated and benchmarked), so it gives us a quantitative estimate of
the effect of FSI on the pion distributions.
In the calculations shown in Fig.~\ref{fig:pions-nudata}, red bands
correspond to the Hybrid-RPWIA with (lower line) and without (upper
line) medium modification of the Delta width. The blue-solid lines are
NuWro results using the same definition of the signal as in the
experiment. The blue-dashed lines are NuWro but without FSI. The
orange dash-dotted lines are NuWro results without FSI and selecting
those events in which only one pion and one nucleon exit the
nucleus. These latter results correspond to the elementary SPP process
as predicted by NuWro, and they could be compared with the
hybrid-RPWIA results.

Finally, the effect of the pion FSI in Fig.~\ref{fig:pions-nudata} can
be understood attending to the mechanisms that are implemented in the
cascade, i.e.: secondary interactions slow down the pions, which shift
the strength towards lower $T_\pi$ regions; pion absorption reduces
the total strength; charge-exchange reactions moves flux from one
channel to another, but since the 1$\pi^+$ channel is dominant in CC
$\nu_\mu$-induced pion production, the net effect is to reduce
(increase) the magnitude of the 1$\pi^+$ (1$\pi^0$) distributions.

In Refs.~\cite{Mosel17,Hernandez13,Sobczyk15} other theoretical predictions were compared with MiniBooNE, MINERvA and/or T2K pion-detected data.

\section{Higher Inelastic, DIS and its implementation in the scaling formalism}
\label{sec:DIS}

\subsection{Higher lying 
resonance contributions and
general models} 
Beyond the excitation of
$\Delta(1232)$ resonance, other higher lying resonances ($W>1.4$ GeV)
have to be taken into account in a description based on resonance
production followed by their subsequent decays, either on specific
channels ($\pi N$, $\pi\pi N$, $\eta N$, $K Y$...), or summing over
all of them for inclusive $(e,e^\prime)$ or $(\nu_l,l)$ reactions at
higher energy transfers. This region of invariant masses $W>1.4$ GeV
is usually called the second resonance region
\cite{Lalakulich:2006sw}.  The most relevant and theoretically studied
resonances in this region are the three isospin 1/2 ($N^\ast$) states
known as $P_{11}(1440)$, $D_{13}(1520)$ and $S_{11}(1535)$, where the
nomenclature $L_{2I\,2J}$ is used to classify them, with $I$, $J$ and
$L$ being the isospin, spin and partial wave of the resonances,
respectively.

Most of the information about the electromagnetic and weak resonance
transition vector form factors comes from the studies of pion electro
and photo-production data off nucleons by including these resonances
in the analyses \cite{Burkert:2004sk, Burkert:2002zz,Aznauryan:2004jd,
  Tiator:2003uu}. The electromagnetic nucleon-to-resonance form
factors can be usually obtained from the helicity amplitudes, in the
same fashion as the MAID model \cite{Drechsel:1998hk}. The
corresponding weak vector transition form factors can then be related
to the electromagnetic ones by assuming that the weak vector charged
current belongs to the same triplet of current operators as the
isovector component of the electromagnetic current (for a very
detailed description on how to work out this, please see the
appendices of Ref.~\cite{Leitner09}).  Normally, the transition
axial form factors are much more difficult to determine because of the
lack of guidance from the electromagnetic interactions. The usual
trick is to use PCAC to relate two of the axial form factors, and the
coupling of one of them at $q^2=0$ can be determined from the partial
decay width of the particular resonance into the $\pi N$ channel
\cite{Lalakulich:2006sw, Lalakulich:2005ak}.  This is done in the same
fashion as when we relate the nucleon pseudo-scalar and axial form
factors, while the axial coupling is obtained by assuming pion-pole
dominance and by relating it to the strong $\pi NN$ coupling through
the Goldberger-Treiman relation.

The aforementioned procedure is general and can be applied to the
neutrino-production of many resonances in the range of invariant
masses between 1 and 2 GeV. It has been applied, for instance, in
Refs.~\cite{Lalakulich:2006sw, Lalakulich:2005cs,Leitner:2006ww,
  Leitner:2006ww,Leitner09} to the study of electron and
neutrino-nucleus scattering in the resonance region, including the
$\Delta$ region as well.  For more recent works where the inclusion of
higher lying baryon resonances is taken into account for the study of
single pion production driven by weak and neutral currents, the reader
is referred to Ref.~\cite{Alam:2015gaa}. Other works where the
excitation of the $N^\ast(1535)$ and $N^\ast(1650)$ resonances is
taken into account for the calculation of $\eta$-meson production
cross sections off nucleons can be found in
\cite{Alam:2015zla,Alam:2013xoa}.

One of the most relevant works in the analysis of single pion
neutrino-production is that of Rein and Sehgal
\cite{Rein81}. They included all the relevant resonances below 2
GeV for the analysis of both charged and neutral current single pion
production data by then.  In this latter work, however, the helicity
amplitudes are not phenomenologically obtained, but theoretically
calculated within the relativistic harmonic oscillator quark model of
Feynman, Kislinger and Ravndal \cite{Feynman:1971wr}.  The Rein-Sehgal
model is widely used in some of the Monte Carlo event generators used
for the analyses and simulations of single pion production events,
namely GENIE, Neut and Nuance \cite{Stowell:2019zsh,
  Andreopoulos:2009rq, Andreopoulos:2015wxa, Hayato:2009zz,
  Casper:2002sd}. This model has been also extended to
account for the final lepton mass and spin in lepton polarization
studies in neutrino-nucleon scattering \cite{Kuzmin:2003ji,
  Kuzmin:2004ya}.

It is also worth mentioning that some specific channels, such as the
two pion production $\pi\pi N$ one, have been very scarcely studied
from the experimental and theoretical side. In
Ref.~\cite{Hernandez08} it is shown that the inclusion of the
Roper resonance, $N^\ast(1440)$, is relevant for this reaction channel
near its production threshold.

Up to now we have only mentioned the excitation of non-strange
resonances. However the weak charged current can induce
strangeness-changing transitions $|\Delta S|=1$ as well, although
these are highly suppressed with respect to their
strangeness-preserving $|\Delta S|=0$ counterparts because the
amplitudes for the latter are proportional to the cosine of the
Cabibbo angle, $\cos\theta_c\sim0.974$, while those for
strangeness-changing are proportional to the sine of the same angle,
which is much smaller, $\sin\theta_c \sim 0.226$. This kind of
processes normally requires the inclusion of strange baryon resonance
degrees of freedom, normally the $\Sigma^\ast(1385)$, which belongs to
the same SU(3) multiplet as the $\Delta(1232)$. Some relevant 
and quite recent works
studying the production of strange particles induced by electrons,
neutrinos and anti-neutrinos, either
 through the strangeness-preserving or 
 the strangeness-changing part of
the weak charged current, some of them studying
polarization observables and second-class currents effects,
 can be found in Refs.~\cite{Singh:2006xp,RafiAlam:2010kf,Adera:2010zz,Alam:2011xq, 
RafiAlam:2019rft,Alam:2012ry,Fatima:2018wsy,Fatima:2018tzs,
Akbar:2017qsf}.

\subsection{Inelastic contribution
within the scaling formalism and the SuSAv2 approach}\label{SuSAv2-inelastic}

Within the scaling formalism, in Ref.~\cite{Barbaro:2003ie}, an
extension of the RFG model to the inelastic region was carried out by
allowing the reached final invariant masses $W_X$ to be larger than in the
QE case ($m_N+m_\pi \leq W_X \leq m_N+\omega-E_s$, with $E_s$ the
nucleon separation energy), and then performing an integration over all the allowed
invariant masses with a "spectral function" (the inelastic RFG
scaling function in terms of the inelastic scaling variable $\psi_X$ - see Appendix~\ref{scaling-appendix})
accounting for the energy-momentum distribution of nucleons in the RFG
description of the initial nuclear state. To apply this formalism it is
necessary to resort to phenomenological parameterizations
of the inelastic single-nucleon structure functions, extracted mainly
from DIS $e-N$ scattering, but also incorporating resonances in the
$W<2$ GeV region \cite{Bodek:1980ar,Bodek:1981wr,Bodek:1979rx,Stein:1975yy}.  
More recent and complete parameterizations of these inelastic single-nucleon structure
functions extracted from inelastic electron-proton and
electron-deuteron scattering experiments can be found in Refs.~\cite{Christy10, Bosted08}.

\begin{figure}[htbp]
\begin{center}
\vspace{-0.35cm}\hspace*{-0.195cm}\includegraphics[scale=0.3, angle=270]{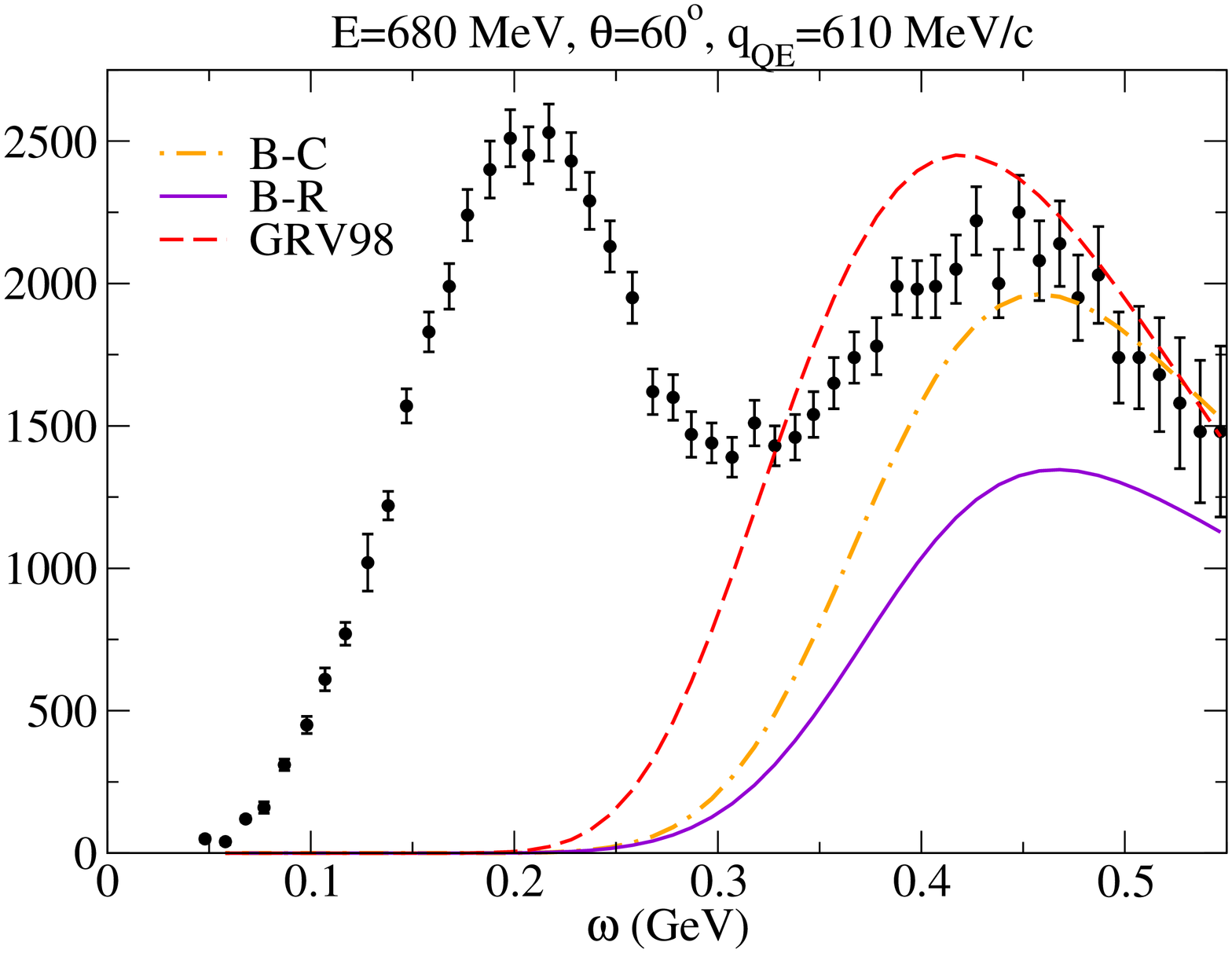}\hspace*{-0.9cm}\includegraphics[scale=0.3, angle=270]{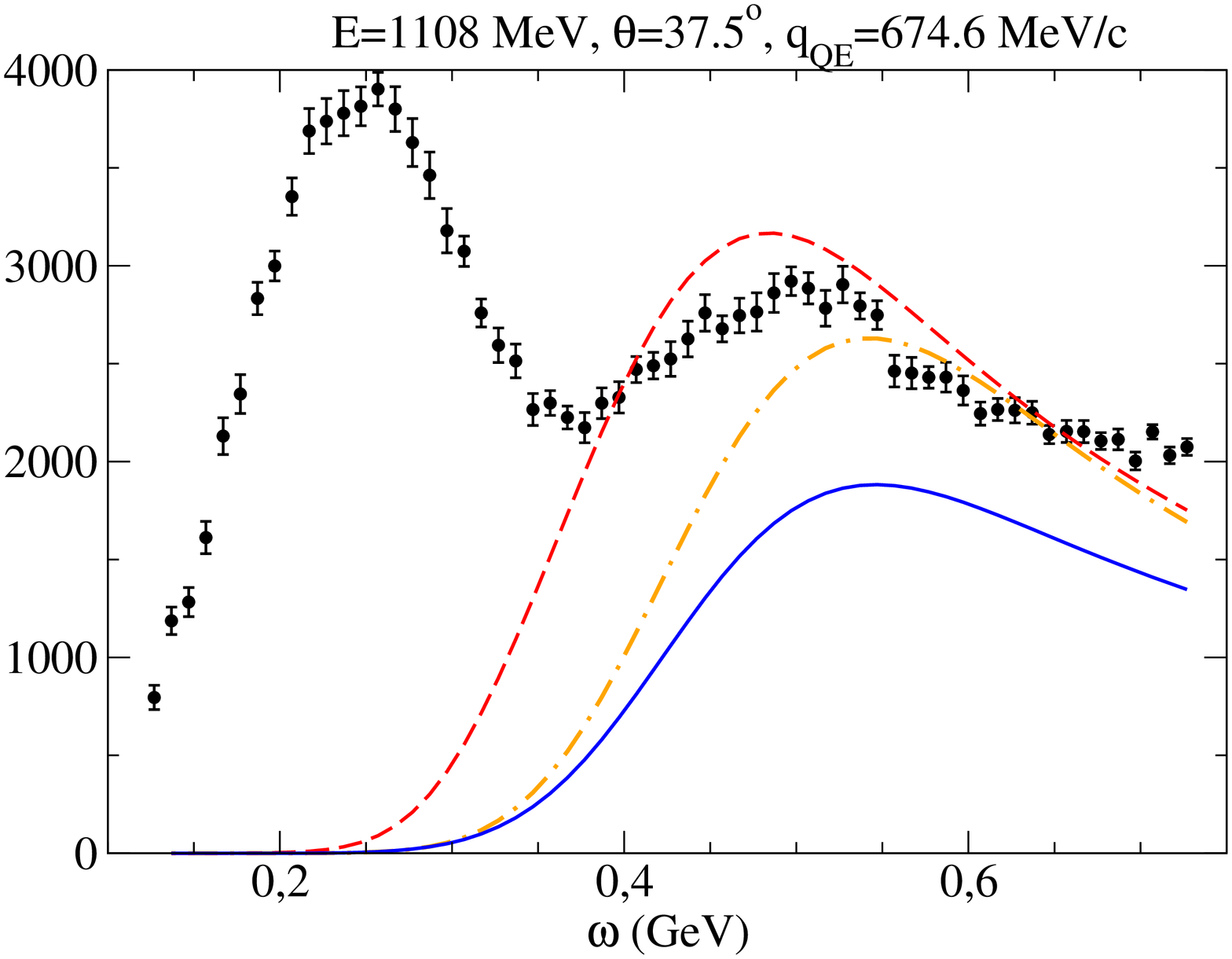}\\
\end{center}\vspace{-0.8cm}
\caption{Comparison of inclusive $^{12}$C($e,e'$) double differential cross sections and predictions for the inelastic regime of the Bodek-Ritchie parametrization (solid lines), Bosted-Christy parametrization (dot-dashed lines) and GRV98 PDFs (dashed lines) at different kinematics (incident electron beam and scattering angle) in terms of the energy transferred to the nucleus ($\omega$). Experimental data taken from~\cite{QESarchive,QESarxiv}. The $y$-axis represents $d^2\sigma/d\Omega/d\omega$ in nb/GeV/sr. The value of $q$ at the QE peak ($q_{QE}$) is shown as reference.
}\label{inelasticsf}
\end{figure}

This formalism has been also applied to the SuSAv2 approach to extend the description to the full inelastic regime, just replacing the RFG scaling functions by the QE SuSAv2 ones and considering the corresponding inelastic scaling variable ($\psi_X$) aforementioned. The inelastic nuclear response functions are thus defined as: 
\begin{eqnarray}
 R^{K}_{inel}(\kappa,\tau) &=& \frac{{\cal N}}{ \eta_F^3\kappa}
\xi_F \int_{\mu_{X}^{min}}^{\mu_{X}^{max}} d\mu_X \mu_X
f^{model}(\psi_X^\prime) U^{K}\; .
\label{eq:rlt_funiv}
\end{eqnarray}\vspace*{0.075cm}
Then, we can identify the inelastic scaling function which englobes all the nuclear dependence of the interaction, $f^{model}$, and employ the one arising from the specific model considered (RFG, SuSAv2, etc), in a similar way as done for the QE regime. The previous formalism allows for a complete description of the inelastic cross sections on lepton-nucleus interactions where the nuclear effects can be included by means of the inelastic RFG or SuSAv2 scaling functions ($f^{model}$). The procedure in the SuSAv2-inelastic approach is similar to that applied for the QE regime before, keeping the same functional form for the RMF and RPWIA scaling functions but using a different scaling variable ($\psi_X^\prime$) and a different $q_0$ transition parameter for the blending function introduced in Eq.~\eqref{transxi}. We make use of the Bosted and Christy parametrization for the single-nucleon inelastic structure functions
\cite{Christy10, Bosted08} which describes DIS, resonant and non-resonant
regions, providing a good description of the resonant structures in
($e,e'$) cross sections and covering a wide kinematic region. As shown
in Fig.~\ref{inelasticsf} and in~\cite{Megias:2017PhD}, the use of other choices such as the
Bodek-Ritchie~\cite{Bodek:1980ar,Bodek:1981wr} parametrization and
models based on Parton
Distribution Functions 
(PDF)\footnote{For a general
discussion on electron
and charged current or neutral
current neutrino/antineutrino
DIS off nucleons,
the reader is referred to 
Ref \cite{bilenky_book}.}, such as GRV98~\cite{Glueck},
leads to large discrepancies with ($e,e'$) data. Although the phenomenological structure functions ($F_1, F_2$) employed in the SuSAv2-inelastic model were obtained via electromagnetic interactions, the extension to the weak sector is also possible under some assumptions. The description of the inelastic regime for weak interactions implies an additional structure function, $F_3$, related to the parity violating contribution associated to the $V-A$ interference. An accurate determination of this weak function is hard to achieve from neutrino experiments as well as from parity-violating electron scattering~\cite{perezf3,nakamuraf3} due to the large uncertainties associated to the cross section measurements. Nevertheless, as mentioned before, within the quark-parton model, it can be established a relationship among the electromagnetic and weak structure functions and between $F_2$ and $F_3$~\cite{Bodek:1980ar,Bodek:1981wr,Thomas1}. This is based on the assumption that the corresponding structure functions $W_i$ can be written in terms of quark $\mathcal{Q}$ and antiquark $\overline{\mathcal{Q}}$ distributions~\cite{haiderqcd,hobbsqcd}
\begin{eqnarray}\vspace*{-0.295cm}
 F_2=\nu W_2=\mathcal{Q}+\overline{\mathcal{Q}} \quad ;\quad
 F_3=x\nu W_3=\mathcal{Q}-\overline{\mathcal{Q}}\, \quad\rightarrow\quad
x\nu W_3&=&\nu W_2-2\overline{\mathcal{Q}}\label{f3f2}\,.
\end{eqnarray}
For electron scattering, the isoscalar $F_2$ structure function of the nucleon,
defined as the average of the proton and neutron structure functions,
is given (at leading order in $\alpha_s$ and for three flavors) by
\vspace*{-0.025cm}
\begin{equation}
F_2^{eN} = \frac12 \left( F_2^{ep} + F_2^{en} \right)
= \frac{5x}{18}
  \left(u + \overline{u} + d + \overline{d} \right)+\frac{x}{9}\left( s + \overline{s}
  \right)\, .
\label{eq:F2eN}
\end{equation}
The quark distributions are defined to be those in the proton and the factors $5/18$ and $1/9$ arise from the squares of the quark charges. For neutrino scattering, the corresponding $F_2$ structure function is given by
\begin{eqnarray}
F_2^{\nu N} &=& x (u + \overline{u} + d + \overline{d} + s + \overline{s})\, ,
\label{eq:F2nuN}
\end{eqnarray}
where quark charges are not considered. In the moderate and large-$x$ region, where strange quarks
are suppressed, the weak and electromagnetic $F_2$ structure functions approximately satisfy
\begin{equation}
F_2^{eN}\ \approx\ \frac{5x}{18} 
\left( u+\overline{u}+d+\overline{d} \right) \
      \approx\ \frac{5}{18}\, F_2^{\nu N}\ .
\end{equation}
Under this assumption, which has been analyzed in connection with experimental results~\cite{Kimprl,Seligman,haiderqcd,GRM99}, one can readily obtain the weak structure functions from the existing parametrization of the electromagnetic structure functions and the antiquark distribution\footnote{The extension of the SuSAv2-inelastic model to the weak sector is under way.} .
More specific details about the SuSAv2-inelastic model can be found in~\cite{Megias:2017PhD}.

In a different approach~\cite{Maieron09}, 
the scaling and super-scaling properties of the non-QE experimental
data have been analysed within the framework of several theoretical models. To this end,
the QE contribution was subtracted from the inclusive
$(e,e^\prime)$ experimental data to obtain a non-QE cross section. The QE contribution was calculated
by assuming a universal
phenomenological longitudinal scaling function fitted in Refs.~\cite{Donnelly99a,Donnelly99b} and
assuming the QE transverse and longitudinal scaling functions  to be equal ($f^{\rm
  QE}_L(\psi_{\rm QE}) =f^{\rm QE}_T(\psi_{\rm QE})$). The non-QE cross section so obtained 
  was then translated into a non-QE scaling
function, $f^{non-QE}(\psi_\Delta)$, by dividing by the single-hadron
$N\rightarrow \Delta$ cross section, whose explicit expressions can be
found in Refs.~\cite{Amaro99,Amaro:2004bs}.  The
residual pseudo-data were mainly compared with two  theoretical
models: the first one was basically that of Ref.~\cite{Barbaro:2003ie}, 
already discussed in the previous paragraph,
but with updated inelastic single-nucleon structure functions taken
from \cite{Christy10, Bosted08}; the other model used to
analyze the data was that of Ref.~\cite{Amaro99}, which is based
on the assumption of the $\Delta$ dominance in the inelastic region,
but taking into account the finite $\Delta$ width.  One of the main
conclusions of this work \cite{Maieron09} is that, besides the
importance of non-impulsive mechanisms (2p-2h, correlations...) to be
incorporated consistently in any model, there were kinematics much
better described by the inelastic model of Ref.~\cite{Barbaro:2003ie}
than by the $\Delta$-dominance model of \cite{Amaro99}, thus
indicating the importance of higher inelastic contributions beyond the
$\Delta$ at those kinematics (mainly corresponding to larger incident
electron energies and medium/large scattering angles, thus linked to
higher momentum transfers as well).

   \section{The SuSAv2-MEC model}\label{sec:susav2-results} 
    
      In Section~\ref{sec:susav2} a detailed description of the SuSAv2
      model was given, that, as known, incorporates the predictions
      from the RMF theory and a transition to the RPWIA model at high
      values of the momentum transfer. This transition between the RMF
      and RPWIA regimes is governed by a blending function whose
      explicit expression was described in Secion~\ref{sec:susav2}
      (see also ~\cite{Gonzalez-Jimenez14b,Megias16a} for details). In
      order to apply the SuSAv2 model not only to the QE regime but
      also to the full inelastic spectrum, it is required a good
      control of the transition parameters ($q_0,\omega_0$) that
      determine the relative strength of the RMF and RPWIA responses,
      and how the transition between them evolves as the transfer
      momentum varies.  Accordingly, the transition parameter, $q_0$,
      is expected to increase with $q$ in such a way that the RMF
      contribution will be dominant at low kinematics whereas the
      RPWIA one starts to be relevant at higher energies.  The
      particular procedure to determine the $q_0$-behavior with $q$ is
      in accordance to the best fit to a large amount of $(e,e')$
      measurements on $^{12}$C in a wide kinematical region, covering
      from low to high $q$-values ($q$: $239-3432$ MeV/c). For this
      analysis, $^{12}$C is employed as reference target due to the
      ample variety of existing data for electron scattering as well
      as its relevance for neutrino oscillation experiments. The
      method applied to determine the RMF/RPWIA transition in the
      SuSAv2 model in both QE and inelastic regimes is based on a
      reduced-$\chi^2$ analysis of the data sets
      (see~\cite{Megias:2016nu} for details).

In this section, the so-called SuSAv2-MEC model, which adds the 2p2h-MEC contributions
described in Section~\ref{sec:2p2h} to the SuSAv2 approach, is used to describe the CC0$\pi$ channel in neutrino interactions
and the full regime in ($e,e'$) reactions due to the extension of the
RMF-based SuSAv2 approach to the inelastic regime. Work is in progress
to apply this extension to the weak sector.

\subsection{Comparison with electron scattering data on $^{12}C$}

   In this Section, we compare the SuSAv2-MEC predictions with inclusive $^{12}$C ($e,e'$)
   experimental data in a wide kinematical region. The QE regime is
   described in terms of the SuSAv2 model described in
   Section~\ref{sec:susav2}, which has been extended to include the
   complete inelastic spectrum --- resonant, nonresonant and deep
   inelastic scattering --- as described
   in~\cite{Megias:2017PhD,Megias16a,Maieron:2001it}. We also discuss
   the impact of 2p-2h meson-exchange currents following the fully
   relativistic procedure described in Section~\ref{sec:2p2h}. Our
   predictions are also compared with Rosenbluth separated cross
   section data in terms of longitudinal and transverse
   contributions. Finally, the extension of the SuSAv2-MEC formalism
   to other nuclei is also addressed. The capability of the model to
   describe electron scattering data with accuracy gives us confidence in the validity of
   its subsequent extension and application to recent
   neutrino oscillation experiments.

\begin{figure}
\begin{center}\vspace{-0.89cm}
\hspace*{-0.195cm}\includegraphics[scale=0.194, angle=270]{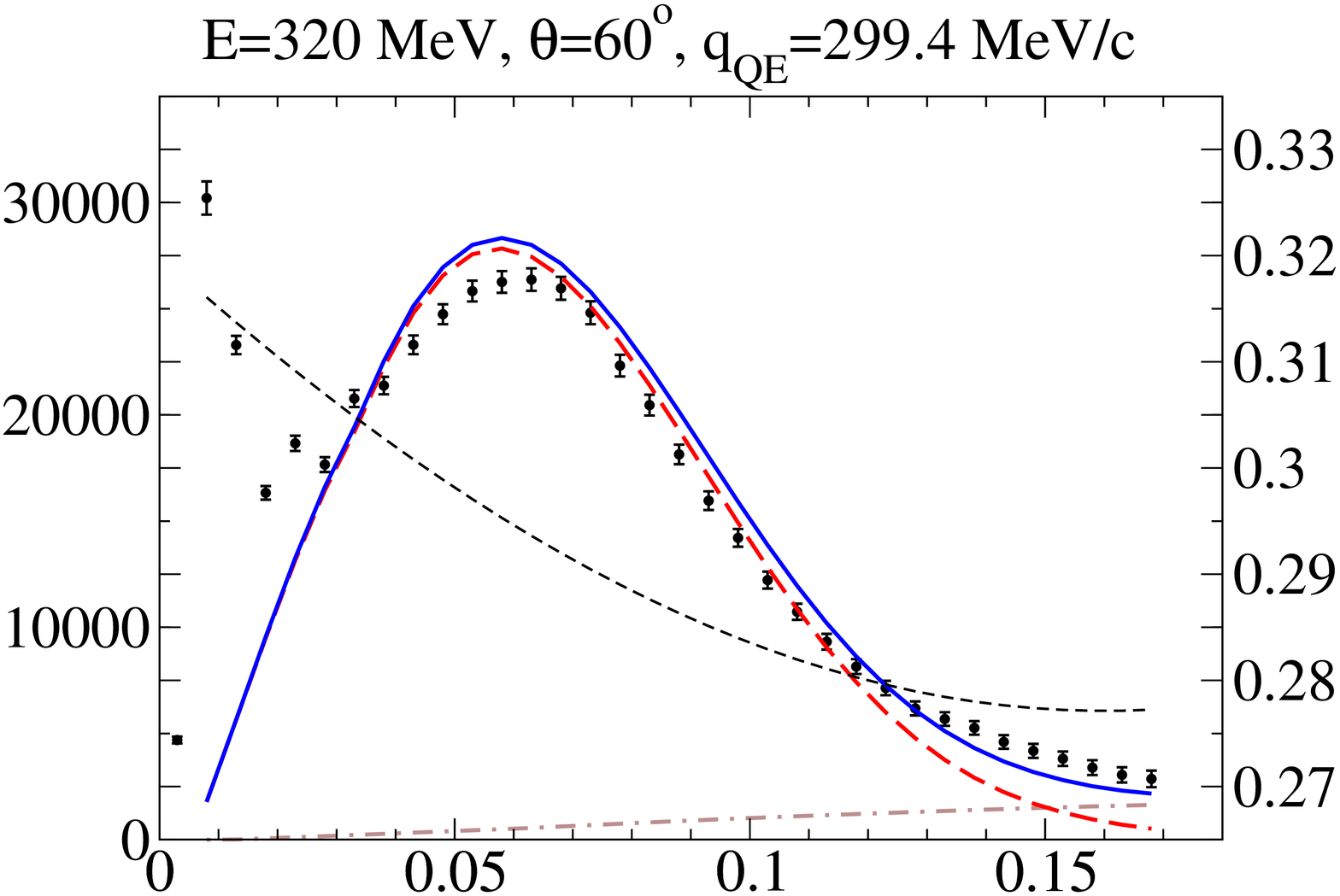}
\hspace*{-0.195cm}\includegraphics[scale=0.194, angle=270]{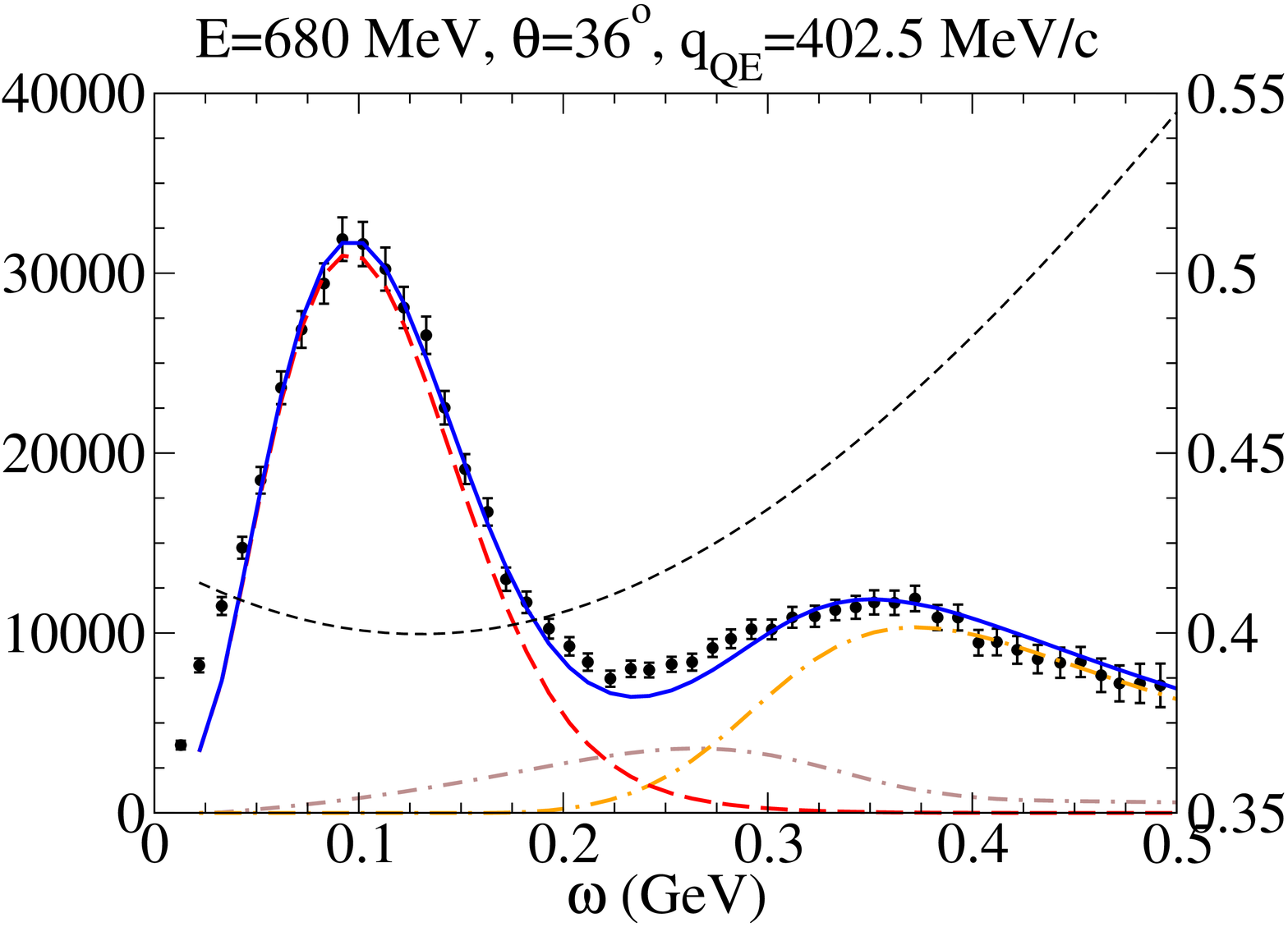}
\hspace*{-0.195cm}\includegraphics[scale=0.194, angle=270]{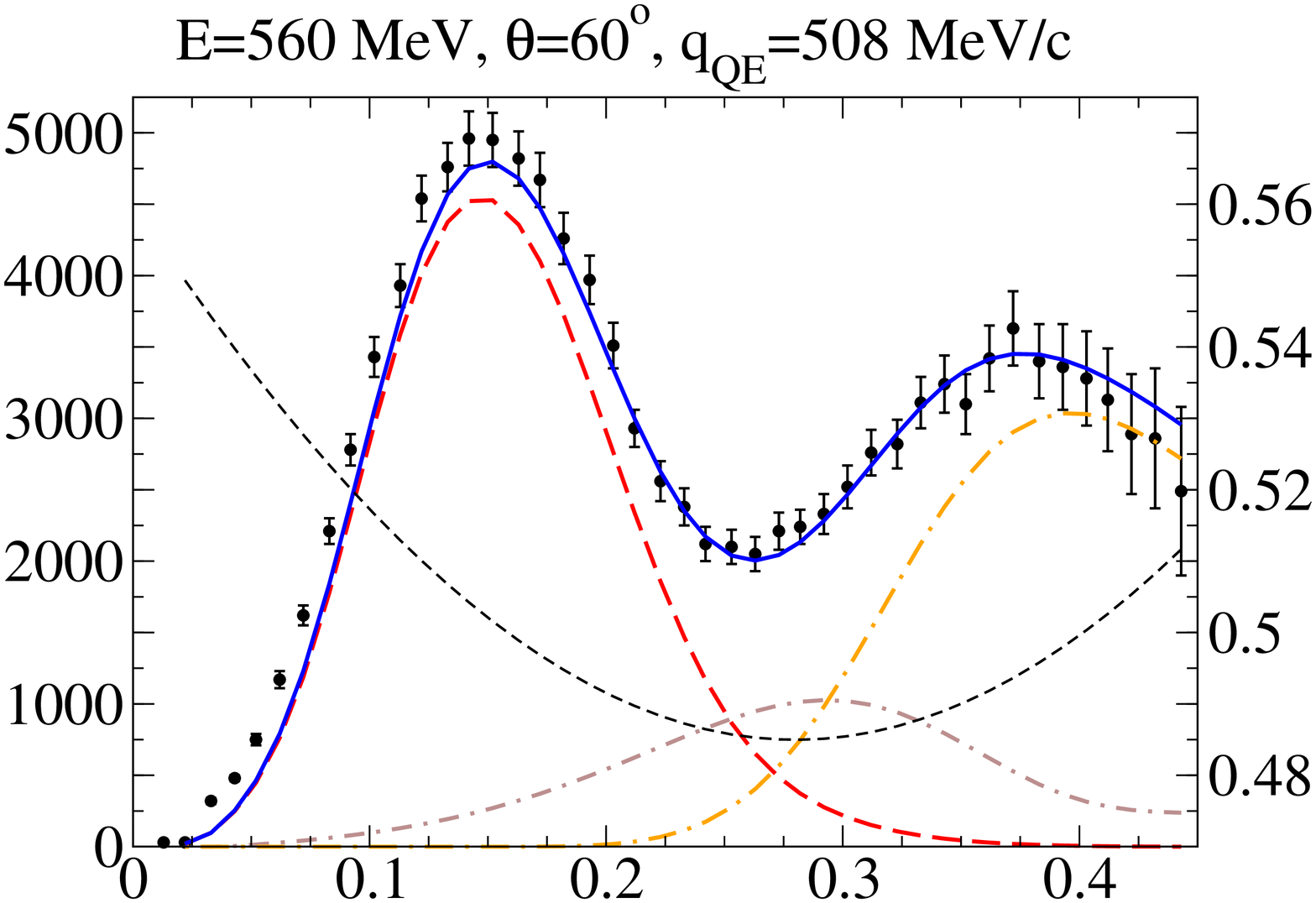}\\
\hspace*{-0.195cm}\includegraphics[scale=0.194, angle=270]{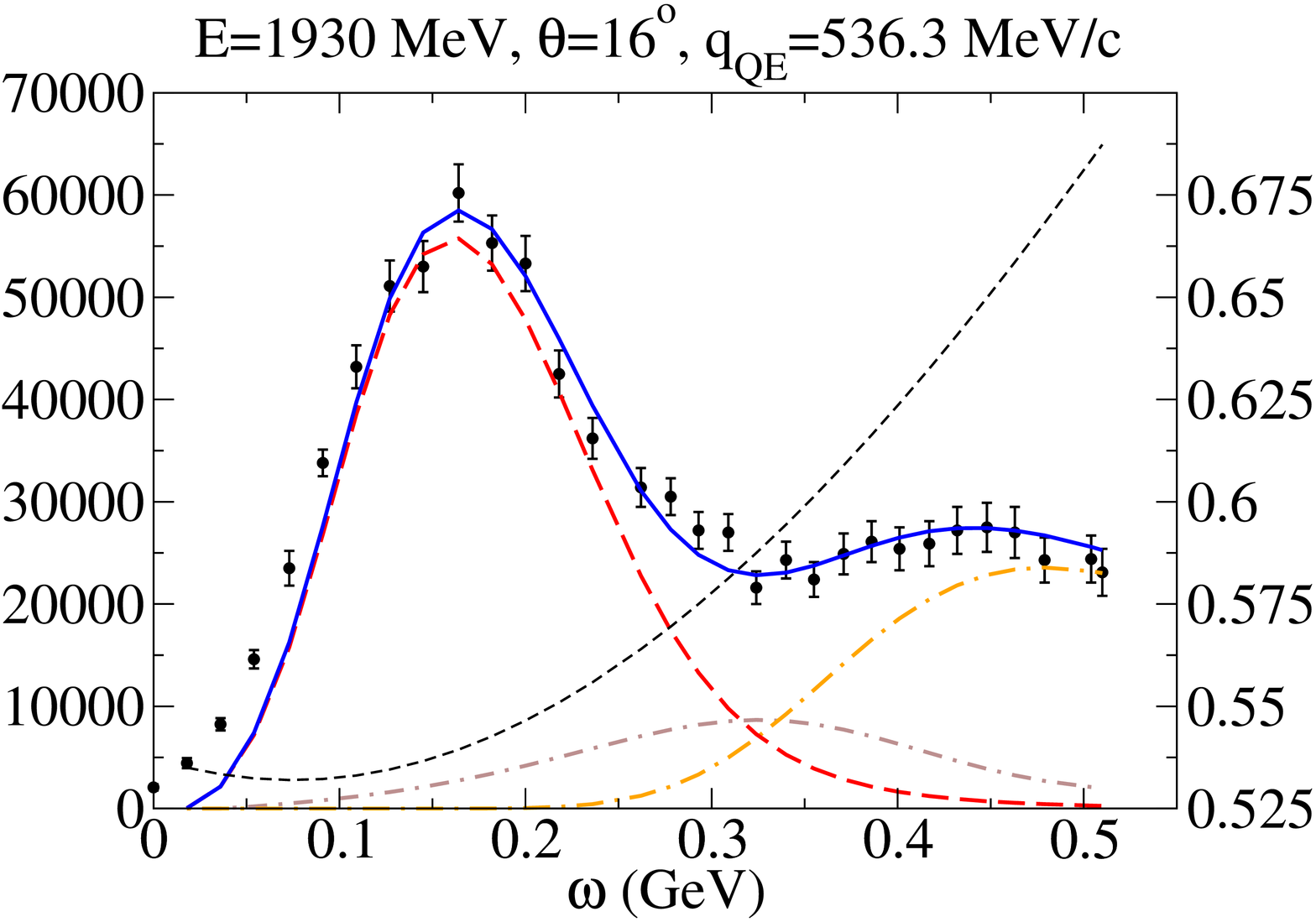}
\hspace*{-0.195cm}\includegraphics[scale=0.194, angle=270]{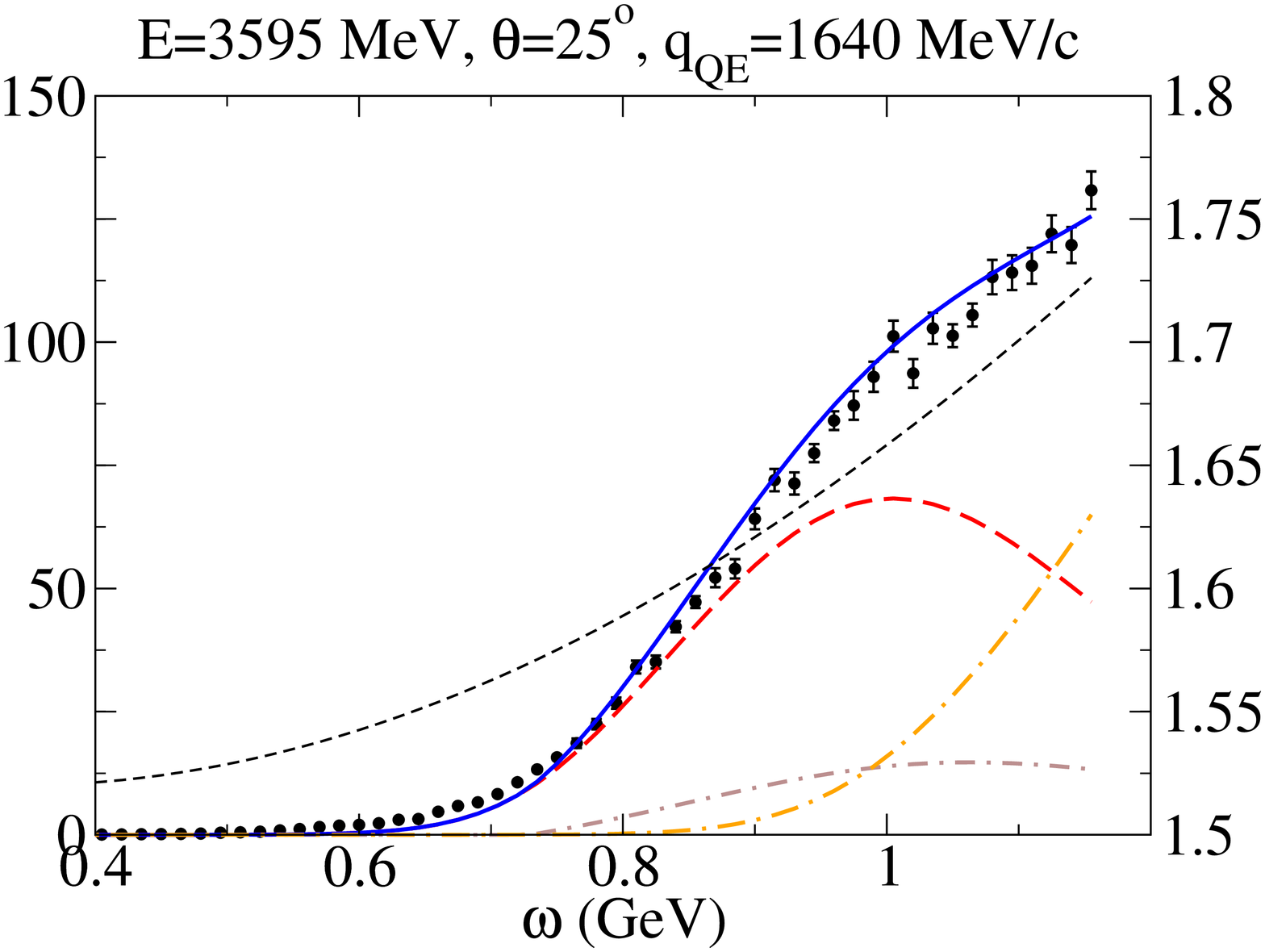}
\hspace*{-0.195cm}\includegraphics[scale=0.194, angle=270]{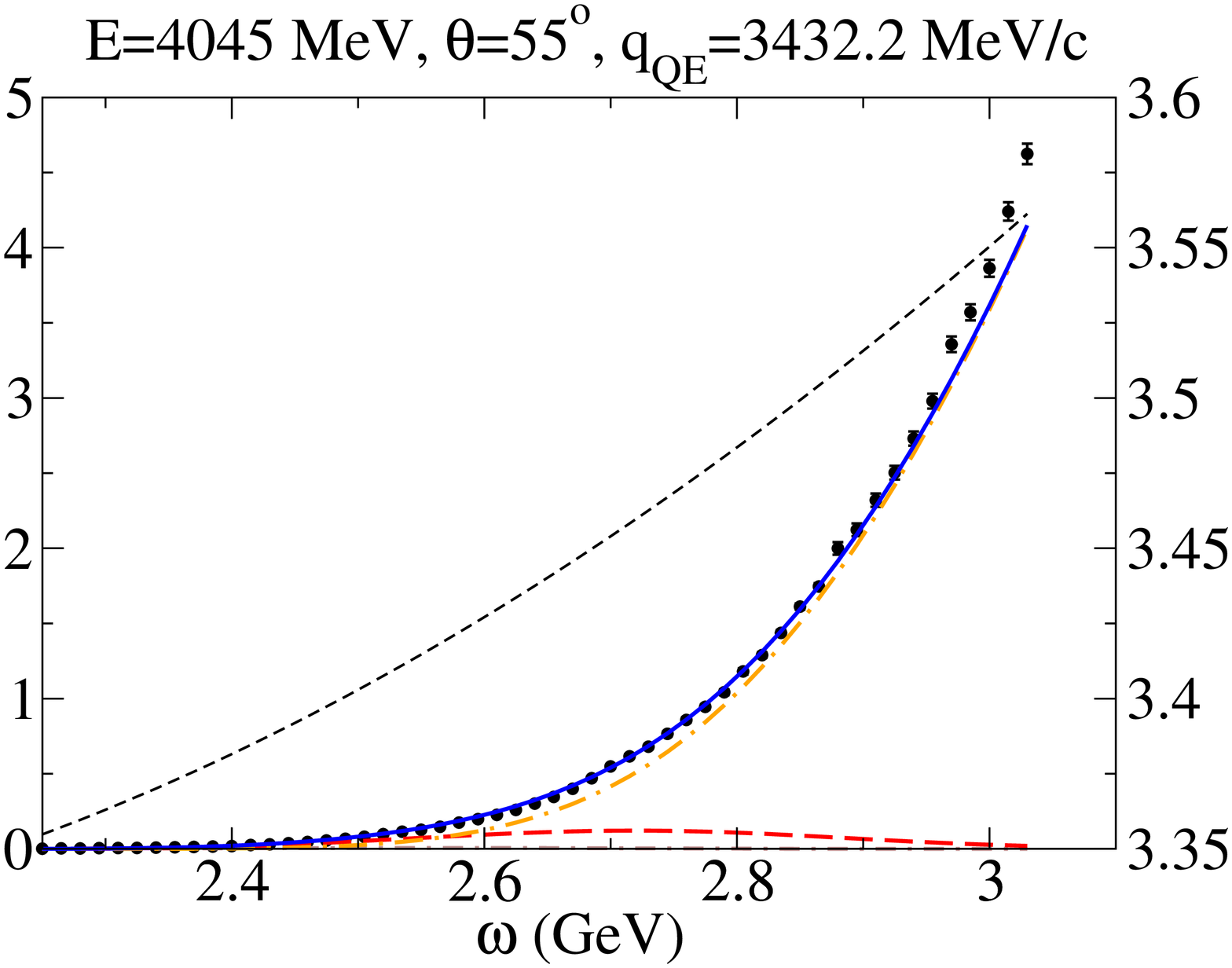}
\begin{center}
\vspace{-0.65cm}
\end{center}
\end{center}\vspace{-0.295cm}
\caption{ Comparison of inclusive $^{12}$C(e,e') cross sections and
  predictions of the QE-SuSAv2 model (long-dashed red line), 2p-2h MEC
  model (dot-dashed brown line) and inelastic-SuSAv2 model (long
  dot-dashed orange line). The sum of the three contributions is
  represented with a solid blue line. The $q$-dependence with $\omega$
  is also shown (short-dashed black line). The y-axis on the left
  represents $d^2\sigma/d\Omega/d\omega$ in nb/GeV/sr, whereas the one
  on the right represents the $q$ value in GeV/c. Experimental data taken from~\cite{QESarchive,QESarxiv}.}\label{ee_lowq2}
\end{figure}

In what follows we present the double differential inclusive
$^{12}$C($e,e'$) cross section versus the energy transferred to the
nucleus ($\omega$), confronting our SuSAv2-MEC predictions with the
available experimental data~\cite{Benhar:2006wy,QESarxiv}. Results are
shown in Fig.~\ref{ee_lowq2}: in each panel we show the three separate
contributions to the inclusive cross section, namely, quasielastic,
2p-2h MEC and inelastic.  We adopt the Bosted and Christy
parametrization for the single-nucleon inelastic structure functions
\cite{Christy10, Bosted08} which describes DIS, resonant and non-resonant
regions, providing a good description of the resonant structures in
($e,e'$) cross sections and covering a wide kinematic region. For the QE
regime, we employ the electromagnetic form factors of the extended
Gari-Krumpelmann (GKex) model~\cite{Lomon1,Lomon2,Crawford1} which
improves the commonly used Galster parametrization for $|Q^2|>1$
GeV$^2$ (see \cite{Megias:2017PhD} for details). The sensitivity of
the QE results to the different parametrizations of the nucleon form
factors has been discussed in \cite{Megias:2013aa} and it will be
addressed in Section~\ref{section-ff}. Additionally, for the Fermi
momentum we employ the values given in~\cite{Maieron03}, namely
$k_F=228$ MeV/c for $^{12}$C.

The comparisons are carried out for kinematics ranging
from low-intermediate energies to the highly-inelastic regime. Each
panel corresponds to fixed values of the incident electron energy
($E$) and the scattering angle ($\theta$). To ease the discussion,
the  panels have been ordered according to the
corresponding value for the momentum transfer at the quasielastic
peak, denoted as $q_{QE}$. This gives us the value of $q$ where the
maximum in the QE peak appears. However, it is important to point out
that as $\omega$ varies, $q$ also varies. For completeness, we also
include in each panel a curve that shows how the momentum transfer
changes with $\omega$. Results illustrate that at very forward angles
the value of $q$ increases with the energy transfer, whereas this
trend tends to reverse at backward angles. Thus for electrons
scattered backwards, the $q$-values corresponding to the inelastic
process are smaller than those ascribed to the QE regime. However, in
this situation the cross section is clearly dominated by the QE
peak. On the contrary, at very forward kinematics the inelastic
process takes place at larger values of $q$. Thus, the two regimes, QE
and inelastic, overlap strongly, the inelastic processes being the
main ones responsible for the large cross sections observed at
increasing values of $\omega$. Finally, for intermediate scattering
angles the behavior of $q$ exhibits a region where it decreases
(QE-dominated process), whereas for higher $\omega$ (inelastic regime)
the behavior of $q$ reverses and starts to go up. In these situations
the QE peak, although significantly overlapped with the inelastic
contributions, is clearly visible even for very high electron
energies.

The systematic analysis presented in Fig.~\ref{ee_lowq2} demonstrates
that the present SuSAv2-MEC model provides a very successful
description of the whole set of $(e,e')$ data, validating the
reliability of our predictions. The positions, widths and heights of
the QE peak are nicely reproduced by the model taking into account not
only the QE domain but also the contributions given by the 2p-2h MEC
terms (around $\sim10-15\%$). Notice also that the "dip" region (between the QE and $\Delta$ peaks)
is successfully reproduced by the theory. A more detailed analysis of the
SuSAv2-MEC model with regard to ($e,e'$) data can be found
on~\cite{Megias16a,Megias:2017PhD}.

Some comments concerning the "dip'' region are in order. This is the region where the QE and
the inelastic contributions overlap the most and where FSI effects
that modify in a significant way the tail of the QE curve at large
$\omega$-values can introduce an important impact. Moreover, the role
of the 2p-2h MEC effects is essential because its maximum contribution
occurs in this region. Thus, only a realistic calculation of these
ingredients beyond the IA can describe successfully the behavior of
the cross section.

To conclude, the accordance between theory and data in the inelastic
regime, where a wide variety of effects are taken into account, also
gives us a great confidence in the reliability of the
calculations. The inelastic part of the cross section is dominated by
the $\Delta$-peak that mainly contributes to the transverse response
function. At low electron scattering angles the longitudinal QE
response function dominates the cross section and the inelastic
contribution is smaller (as will be shown in
Section~\ref{LTsepsect}). The opposite holds at large scattering
angles, where the $\Delta$-peak contribution is important. On the
other hand, for increasing values of the transferred momentum the
peaks corresponding to the $\Delta$ and QE domains become closer, and
their overlap increases significantly. This general behaviour is
clearly shown by our predictions compared with data. In those
kinematical situations where inelastic processes are expected to be
important, our results for the QE peak are clearly below the data
which is compensated by the larger inelastic contribution. On the
contrary, when the inelastic contributions are expected to be small,
the QE theoretical predictions get closer to data. Note also the
excellent agreement in some very high energy situations (bottom panels in
Fig.~\ref{ee_lowq2}), even being aware of the limitations and
particular difficulties in order to obtain phenomenological fits of
the inelastic structure functions, and the reduced cross sections at
these kinematics.
    
\subsection{Separate $L/T$ analysis}\label{LTsepsect}

\begin{figure}
\begin{center}\vspace{-0.599cm}
\hspace*{-0.24cm}\includegraphics[scale=0.22, angle=270]{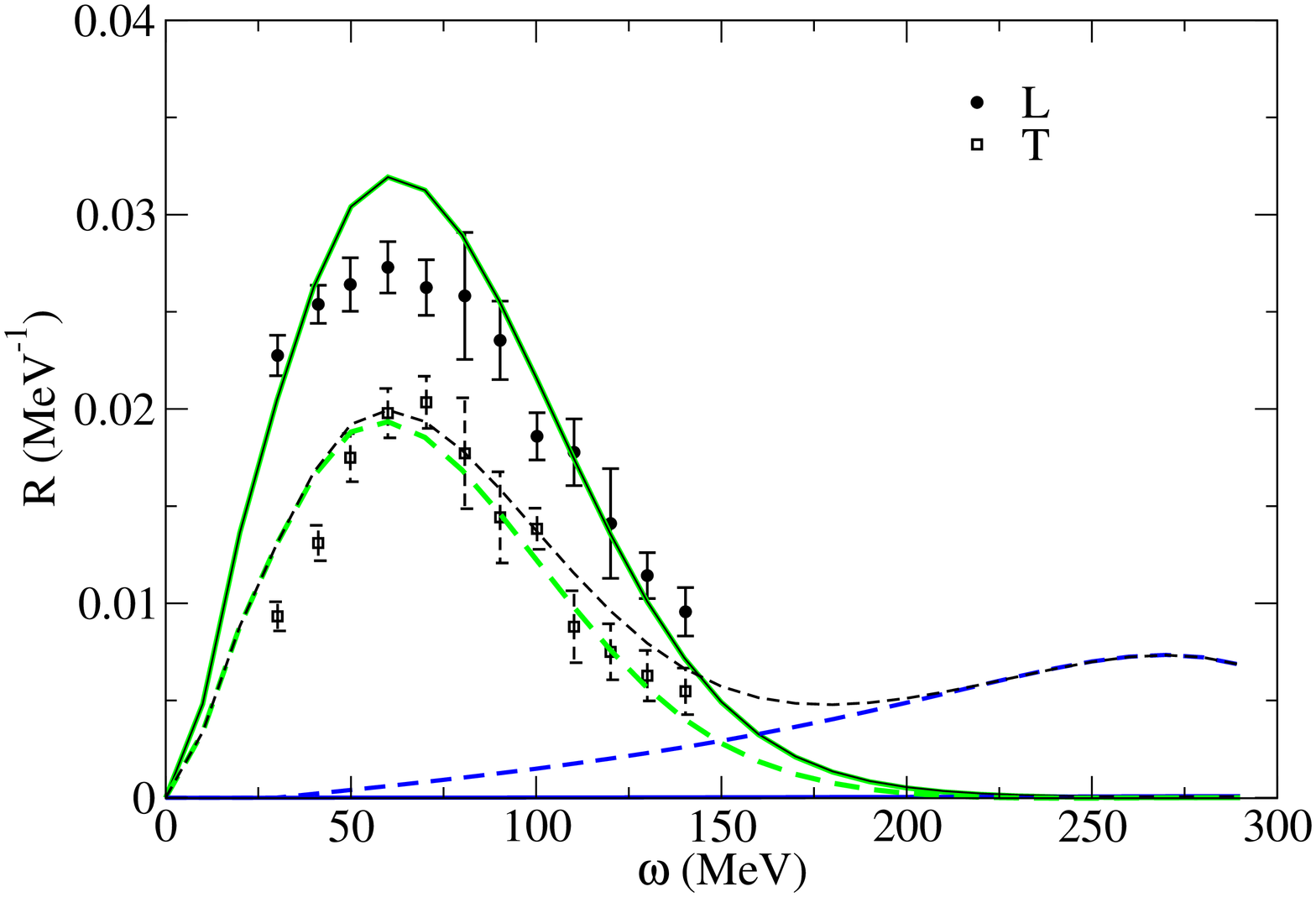}
\hspace*{-0.49cm}
\includegraphics[scale=0.22, angle=270]{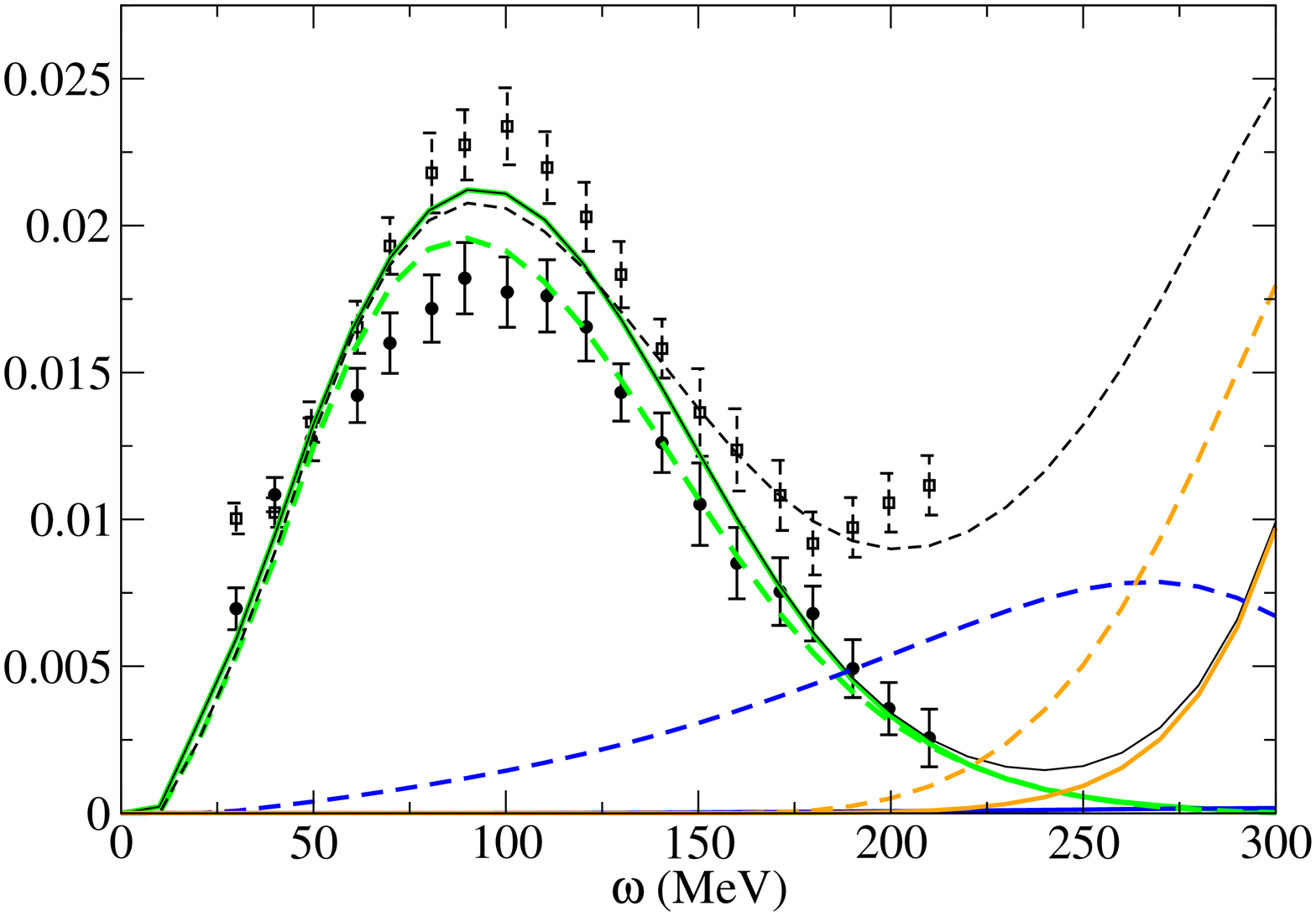}
\hspace*{-0.49cm}
\includegraphics[scale=0.22, angle=270]{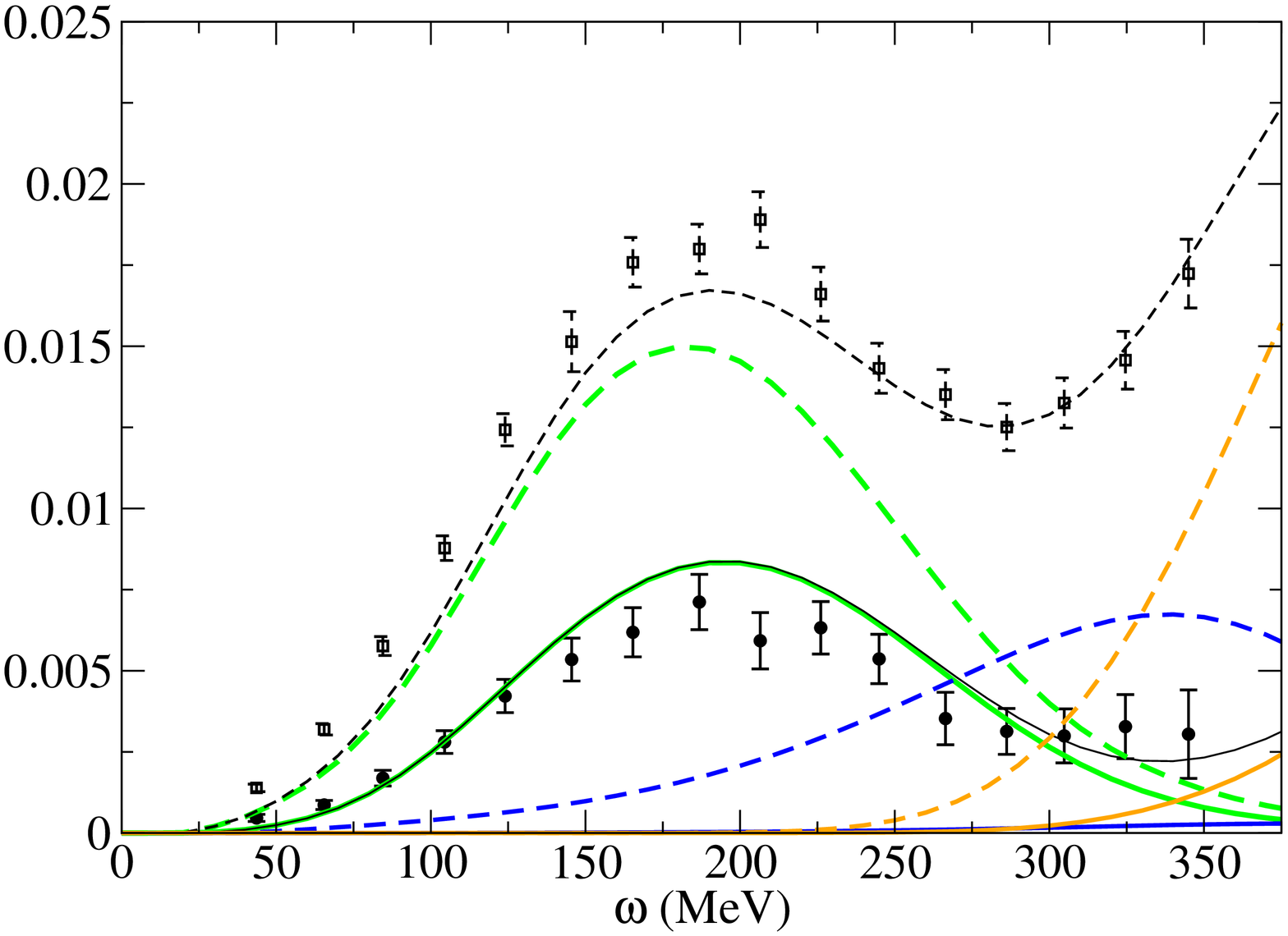}
\end{center}\vspace{-0.59cm}
\caption{Analysis of the longitudinal (solid lines) and transverse
  responses (dashed lines) in $(e,e')$ scattering at $q=300$ MeV/c
  (left panel), $q=380$ MeV/c (middle panel) and $q=570$ MeV/c (right
  panel). QE, MEC and inelastic contributions are shown, respectively,
  as green, blue and orange lines.  The total response is shown by the
  black lines. Data taken from~\cite{Jourdan:1996aa}.  }\label{ltsep}
\end{figure}

The separate analysis of the longitudinal and transverse response
functions of $^{12}$C is presented in Fig.~\ref{ltsep}. The SuSAv2-MEC
predictions are compared with data taken from Jourdan~\cite{Jourdan:1996aa} based
on a Rosenbluth separation of the $(e,e')$ world data\footnote{Notice that there still exist some debate on this L/T separation. See for instance~\cite{Morgenstern:2001jt}.}.  In each case we
isolate the contributions corresponding to the QE, inelastic and 2p-2h
MEC sectors. Three kinematical situations corresponding to fixed
values of the momentum transfer have been considered in
Fig.~\ref{ltsep}: $q=300$ MeV/c, 380 MeV/c  and
570 MeV/c. As observed, the longitudinal channel is totally
dominated by the QE contribution. Only at very large values of
$\omega$ does the inelastic process enter giving rise to a minor
response, whereas the effects due to 2p-2h MEC are negligible. This
result is in accordance with previous
work~\cite{Jourdan:1996aa,Amaroee,Meucci03,Pandey15}, and it
clearly shows that the longitudinal response is basically due to one-nucleon knockout. 
On the contrary, the transverse sector shows an important
sensitivity to MEC and inelastic processes. Note that the inelastic
transverse response gives rise to the high tail shown by data at large
$\omega$-values, whereas the 2p-2h MEC can modify significantly the
transverse response in the dip region as well as in the maximum of the
QE peak. It is also worth mentioning that the natural enhancement in
the transverse response arising from the RMF model included in the SuSAv2 approach is necessary in
order to reproduce the separate $L/T$ data, rejecting the idea of 0-th
kind scaling.

From results in Fig.~\ref{ltsep} we observe that the model leads to a
reasonable agreement with data in both channels, although some
discrepancies also emerge. In particular, the longitudinal prediction at
$q=300$ MeV/c ($q=380$ MeV/c) overestimates the data by $\sim12\%$
($\sim15\%$).  It is important to point out that SuSAv2 is based on
the existence of the scaling phenomenon for $(e,e')$ data, and this is
completely fulfilled when the value of $q$ is large enough ($q\geq
400$ MeV/c). Therefore, the extension of the superscaling approach to
low $q$-values is not well established even though a good agreement at
low kinematics has been achieved in the previous section. Furthermore,
the minor discrepancies observed may also be due to the specific
Rosenbluth separation method used in~\cite{Jourdan:1996aa}, which
introduces some level of model dependence through $y$-scaling
assumptions and the treatment of radiative corrections.

\subsection{Extension of the SuSAv2-MEC model to other nuclei}\label{susav2othernuc}

Once analyzed the capability of the SuSAv2 model to reproduce the
$^{12}$C($e,e'$) measurements, we extend the previous formalism to the
analysis of electron scattering data on other nuclei. For this
purpose, no differences in the scaling functions are assumed for the
different nuclei except for the values used for the Fermi momentum and
energy shift (see~\cite{Maieron:2001it} for details). The use of the
same scaling functions for different nuclear systems is consistent
with the property of scaling of second type, {\it i.e}, independence
of the scaling function from the nucleus, and it also follows from the
theoretical predictions provided by the RMF and RPWIA models on which
SuSAv2 relies. This has been studied in detail in previous works (see
\cite{Caballero:2006wi,Gonzalez-Jimenez:2013plb,Caballero:2007tz,Meucci:2011vd,Caballero05})
where the electromagnetic and weak scaling functions evaluated with
the RMF and RPWIA approaches have been compared for $^{12}$C, $^{16}$O
and $^{40}$Ca. Other theoretical analyses for these nuclei can also be
found in~\cite{Pandey15} and~\cite{Giusti:ee} in the framework of
the RGF and CRPA models, respectively.  In Figure~\ref{Fig2_neutrino} we
compared the general RMF scaling functions for these nuclei, which
exhibit no remarkable differences in terms of the nuclear
species. Therefore, we apply the reference scaling functions for
$^{12}$C to the analysis of QE and inelastic regimes in other nuclei.

In the case of $^{16}$O, the $k_F$-value selected ($k_F=230$ MeV/c) is
consistent with the general trend observed in \cite{Maieron:2001it},
{\it i.e.}, an increase of the Fermi momentum with the nuclear
density. Additionally, an offset of -4 MeV is applied with regard to
the $^{12}$C $E_{shift}$-values. The choice of the $k_F$- and
$E_{shift}$-values provides a consistent analysis of the superscaling
behavior in the deep scaling region below the QE peak, as shown
in~\cite{Megias:2017PhD}, and more importantly, it also provides a
good agreement with electron scattering data.

\begin{figure}\vspace*{-0.409cm}
      \begin{center}
\hspace*{-0.09cm}\includegraphics[width=4.9cm,angle=270]{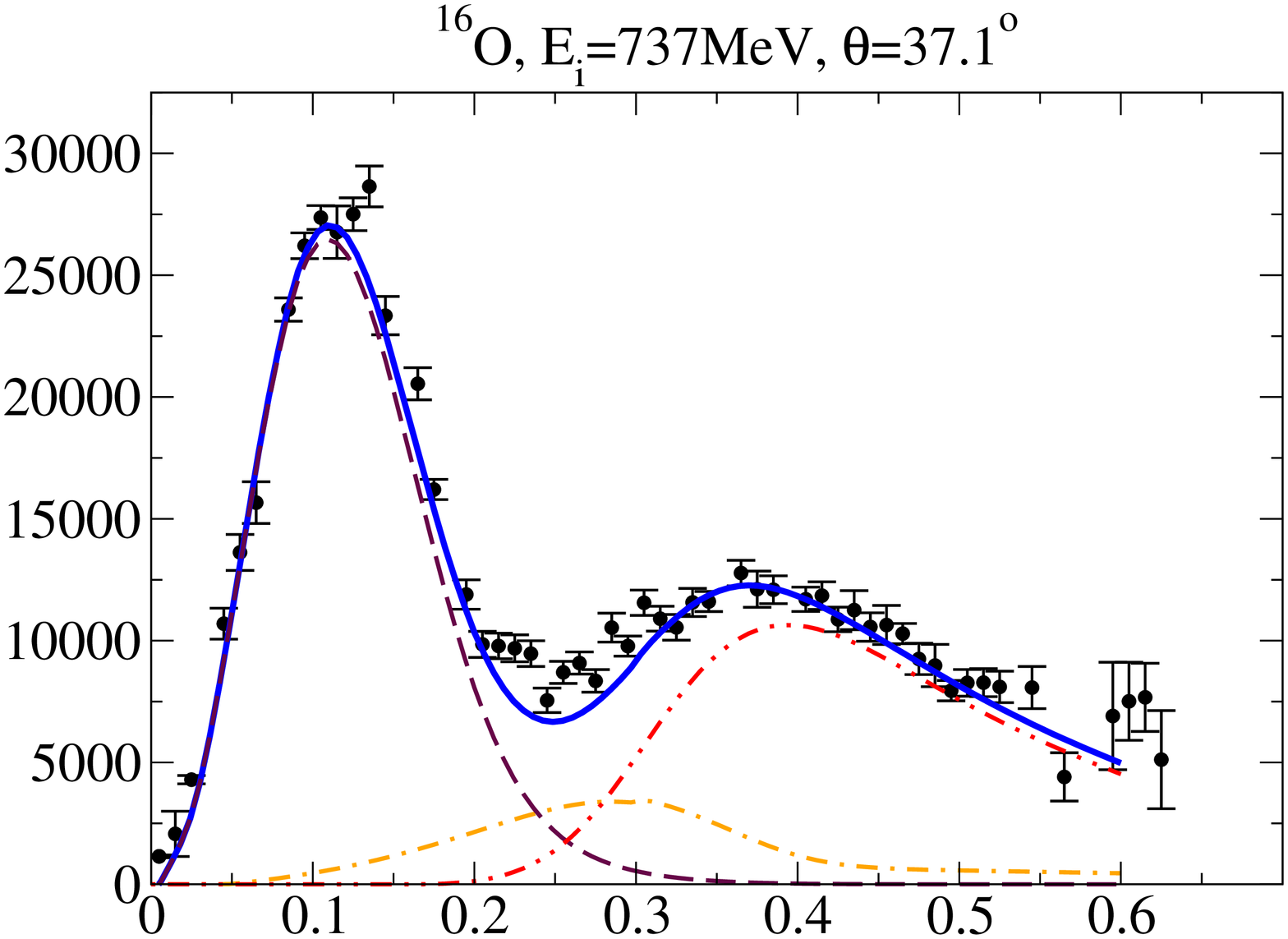}\hspace*{-0.09cm}
\hspace*{-0.49cm}\includegraphics[width=4.9cm,angle=270]{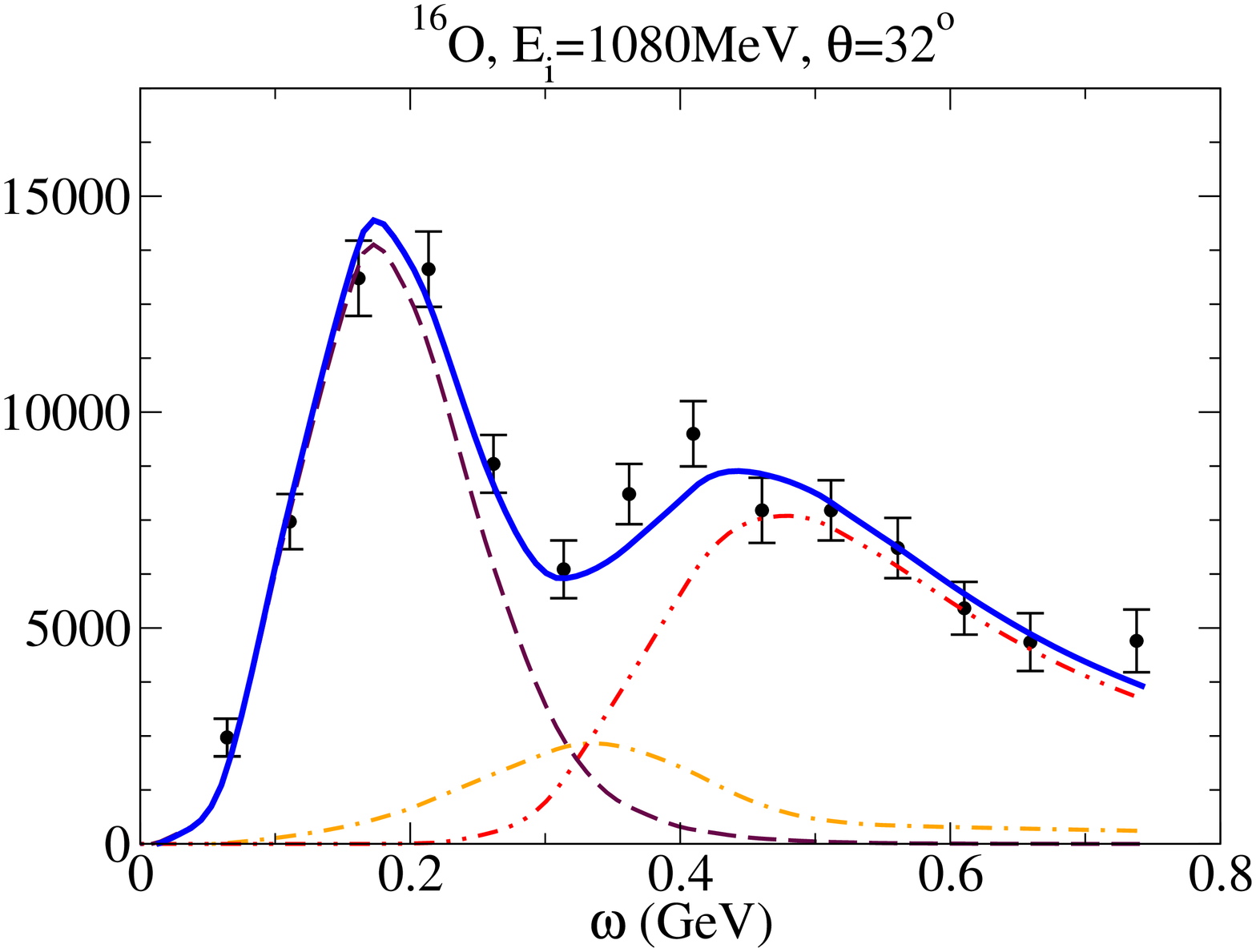}\hspace*{-0.529cm}
\includegraphics[width=4.9cm,angle=270]{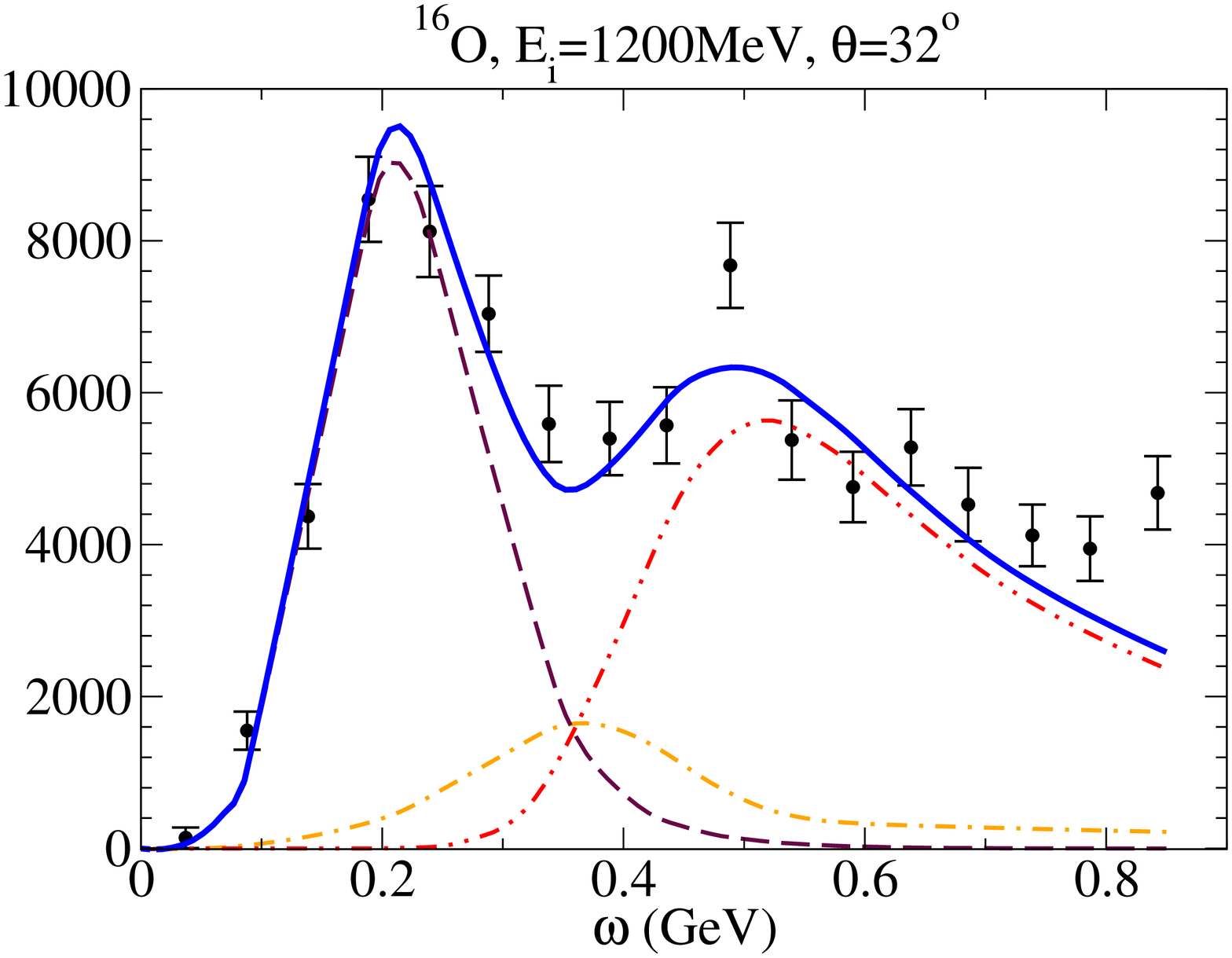}\hspace*{-0.529cm}\\
\vspace{-0.5cm}
\hspace*{-0.15cm}\includegraphics[scale=0.23,angle=270]{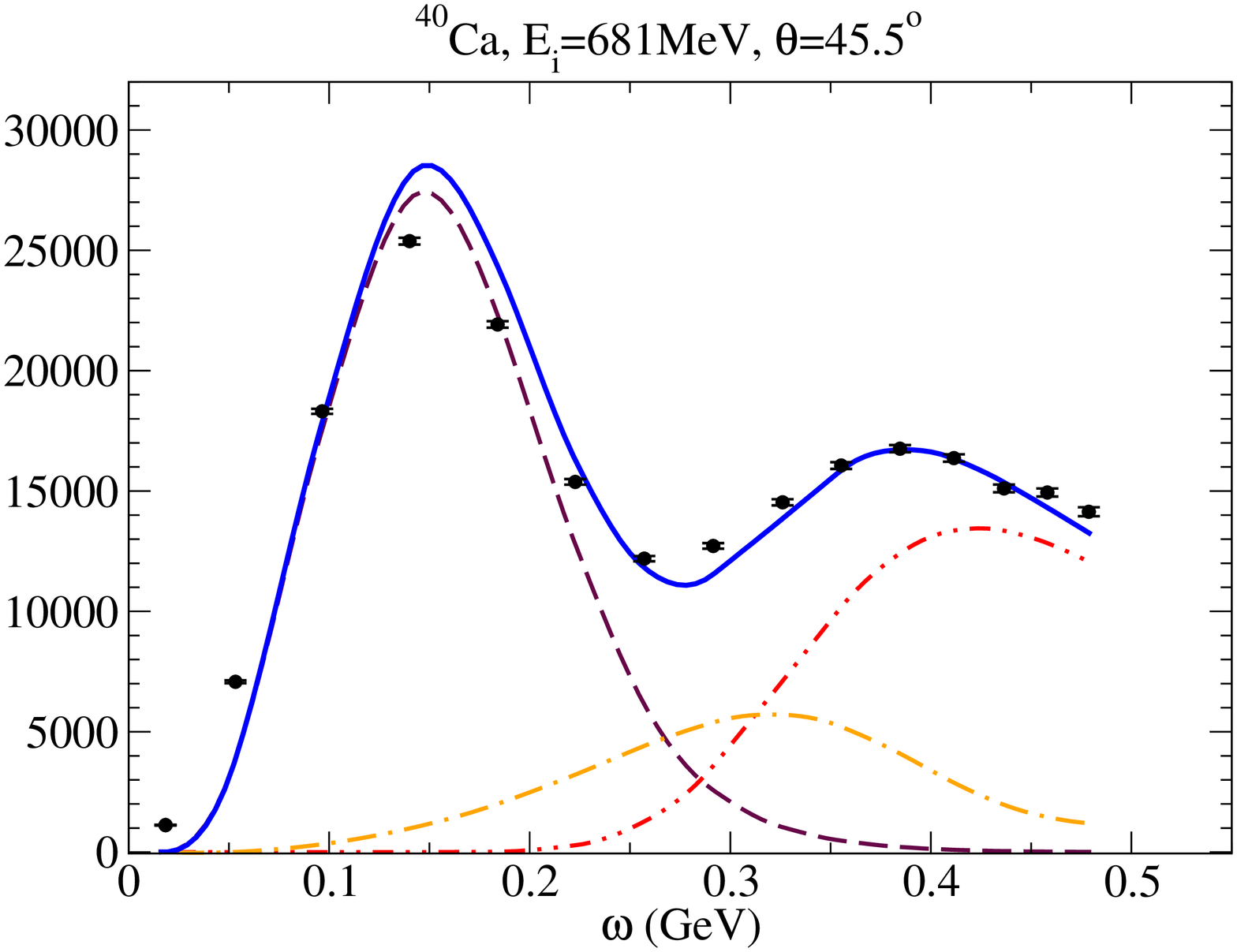}\hspace*{-0.47cm}
\hspace*{-0.15cm}\includegraphics[scale=0.23,angle=270]{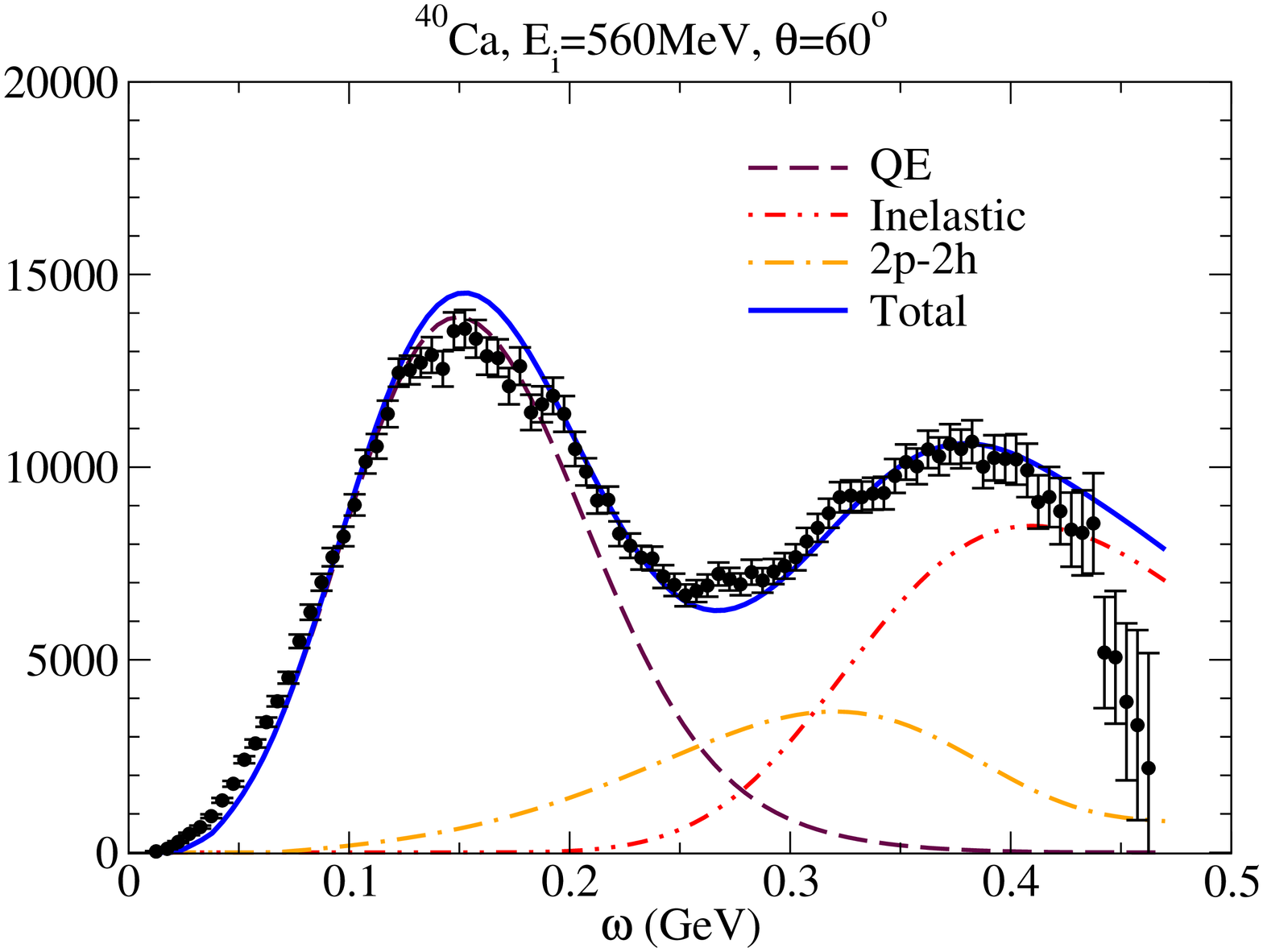}\hspace*{-0.7cm}
\includegraphics[scale=0.23,angle=270]{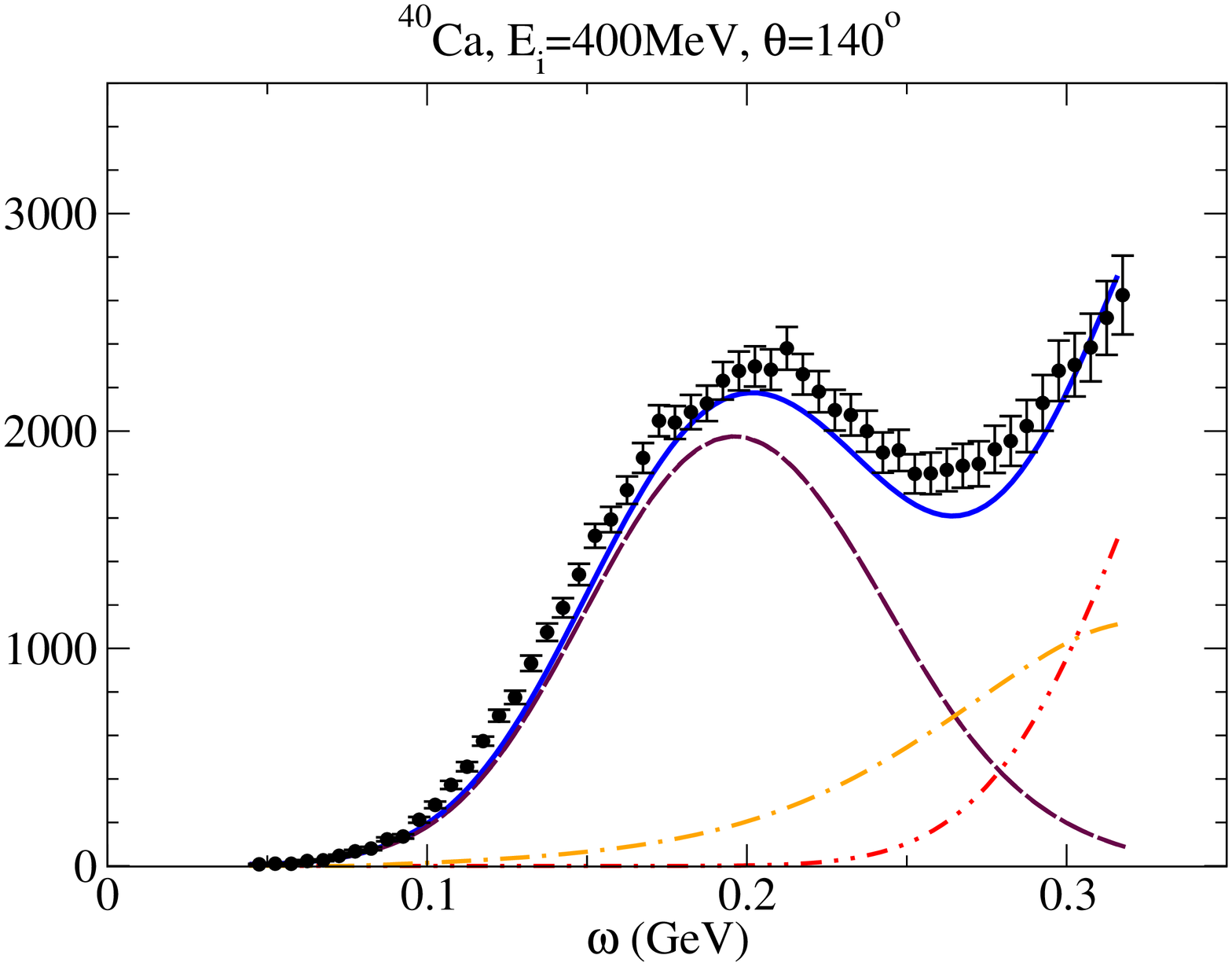}\vspace*{-0.394cm}
\end{center}
\caption{\label{fig:fig1ox} Comparison of inclusive $^{16}$O$(e,e')$ (top panels) and $^{40}$Ca$(e,e')$ (bottom panels)
  cross sections and predictions of the SuSAv2-MEC model. The separate
  contributions of the pure QE response (dashed line), the 2p-2h MEC
  (dot-dashed), inelastic (double-dot dashed) are displayed. The sum
  of the three contributions is represented with a solid blue
  line. The $y$ axis represents $d^2\sigma/d\Omega/d\omega$ in
  nb/GeV/sr.  Data from Refs.~\cite{Anghinolfi:1996vm,O'Connell:1987ag,Williamson:1997,Meziani:1984is}.}
  \end{figure}

Thus, in accordance with the previous analysis, we show in
Fig.~\ref{fig:fig1ox} (top panels) the predictions of the SuSAv2-MEC model for three
different kinematical situations, corresponding to the available
($e,e')$ data on $^{16}$O. In all the cases we present the separate
contributions for the QE, 2p-2h MEC and inelastic regimes. The 2p-2h
MEC responses are extrapolated from the exact calculation performed
for $^{12}$C assuming the scaling law $R_{2p2h}\sim Ak_F^2$ deduced in
Ref.~\cite{Amaro:2017eah}. 
The inclusive
cross sections are given versus the transferred energy ($\omega$), and
each panel corresponds to fixed values of the incident electron energy
($E_i$) and the scattering angle ($\theta$). 

As observed, the SuSAv2-MEC predictions are in very good accordance
with data for all kinematical situations. 

For completeness, we also present in the bottom panels of Figure~\ref{fig:fig1ox} the
calculations for the heavier target $^{40}$Ca ($k_F=241$ MeV/c,
$E_{shift}=28$ MeV within the RFG formalism~\cite{Maieron:2001it}) where the
comparison with data is again very precise from forward to very
backward angles. The analysis of these results is relevant because of
the similarity with $^{40}$Ar, a target of interest for recent and
forthcoming neutrino oscillation experiments. Note also that the 2p-2h
MEC contributions are more prominent for $^{40}$Ca than for $^{12}$C
and $^{16}$O with respect to the QE regime due to the different $k_F$
dependence of the QE and 2p-2h MEC contributions, $A/k_F$ and
$Ak_F^2$, respectively~\cite{Barbaro:2018kxa}.

    \subsection{Comparison with recent JLab data}
    
\begin{figure}   \vspace*{-0.495cm}        
\includegraphics[scale=0.2084, angle=270]{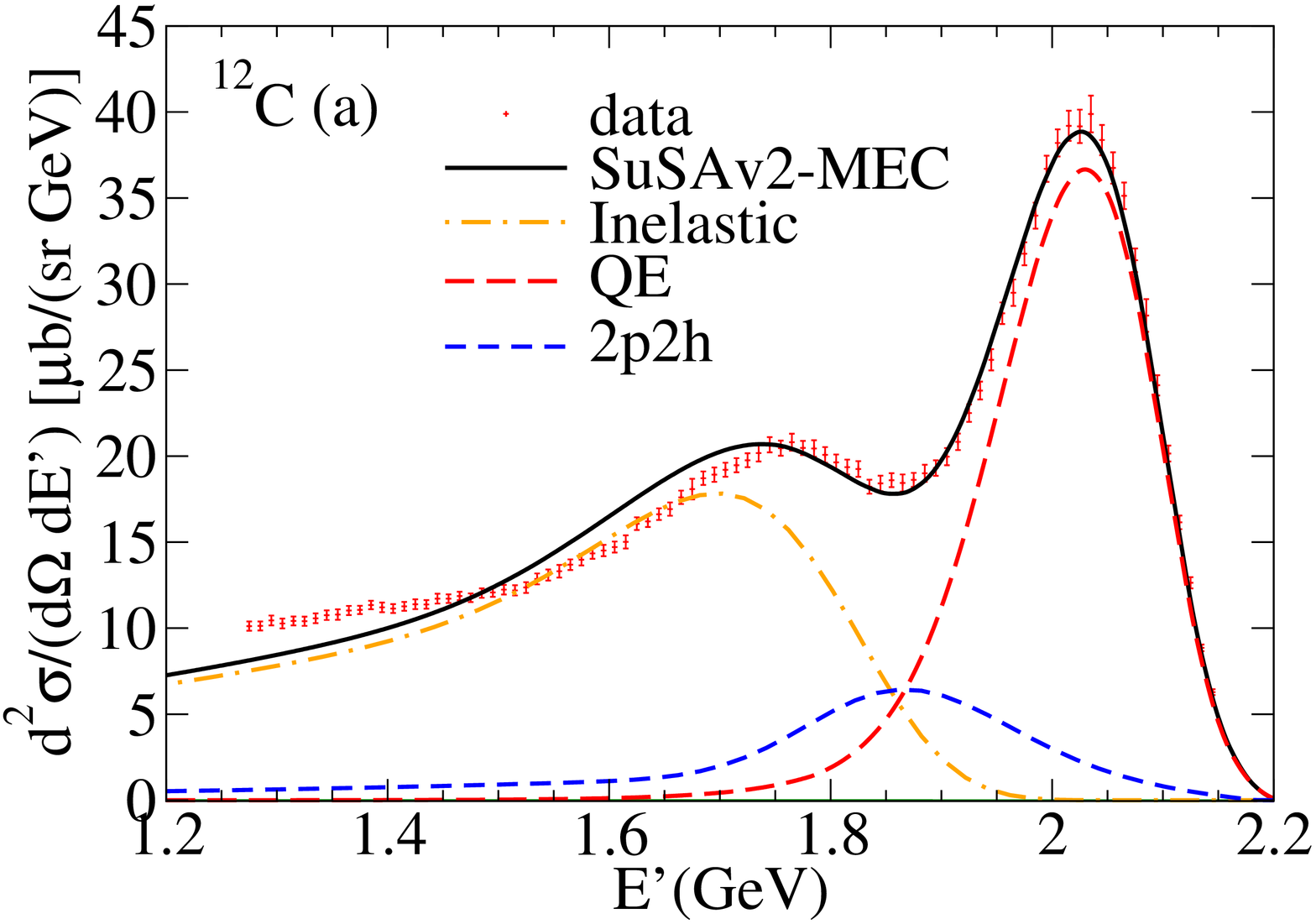}\hspace*{-0.39cm} 
\includegraphics[scale=0.2084, angle=270]{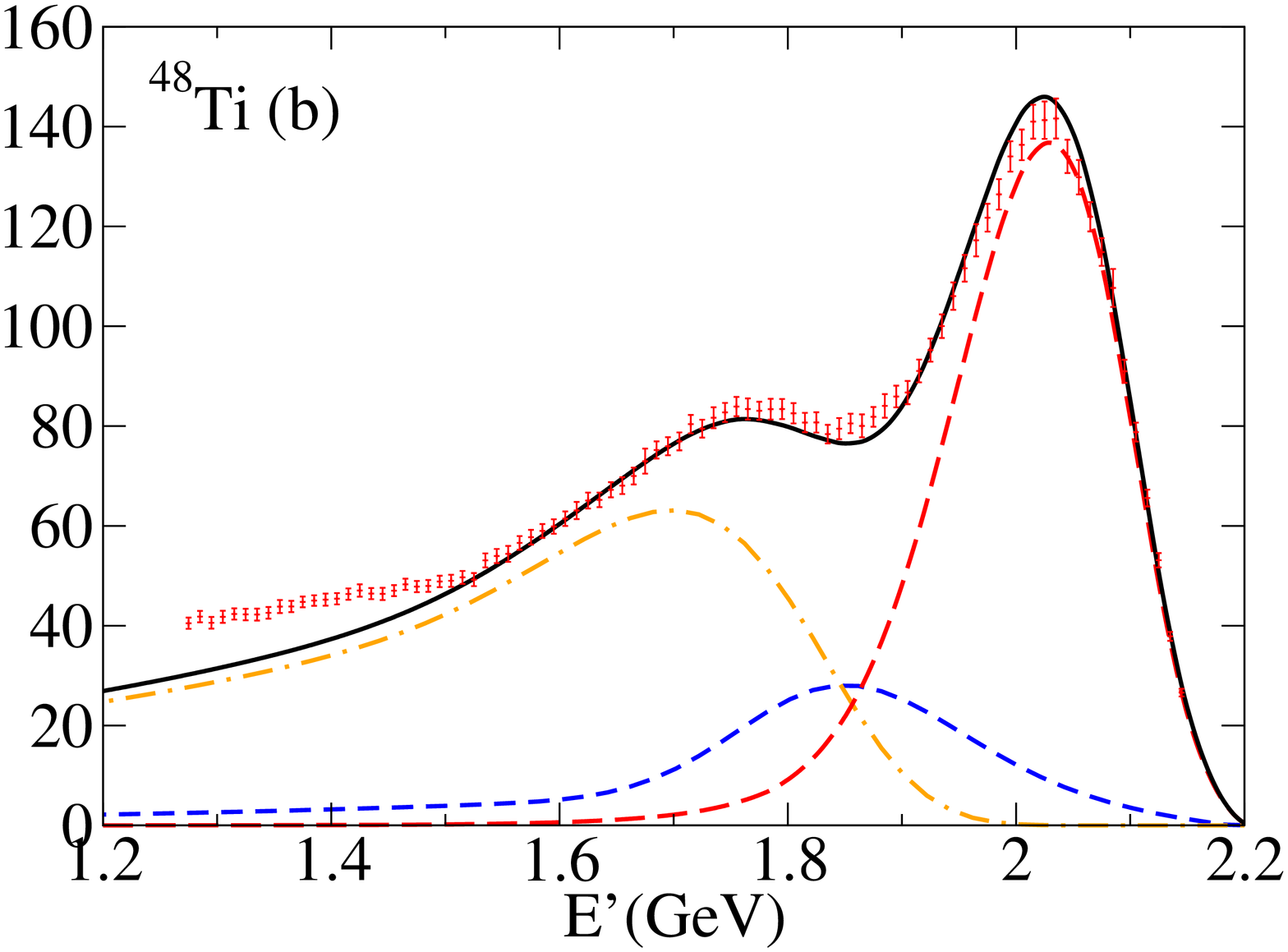}\hspace*{-0.39cm} 
\includegraphics[scale=0.2084, angle=270]{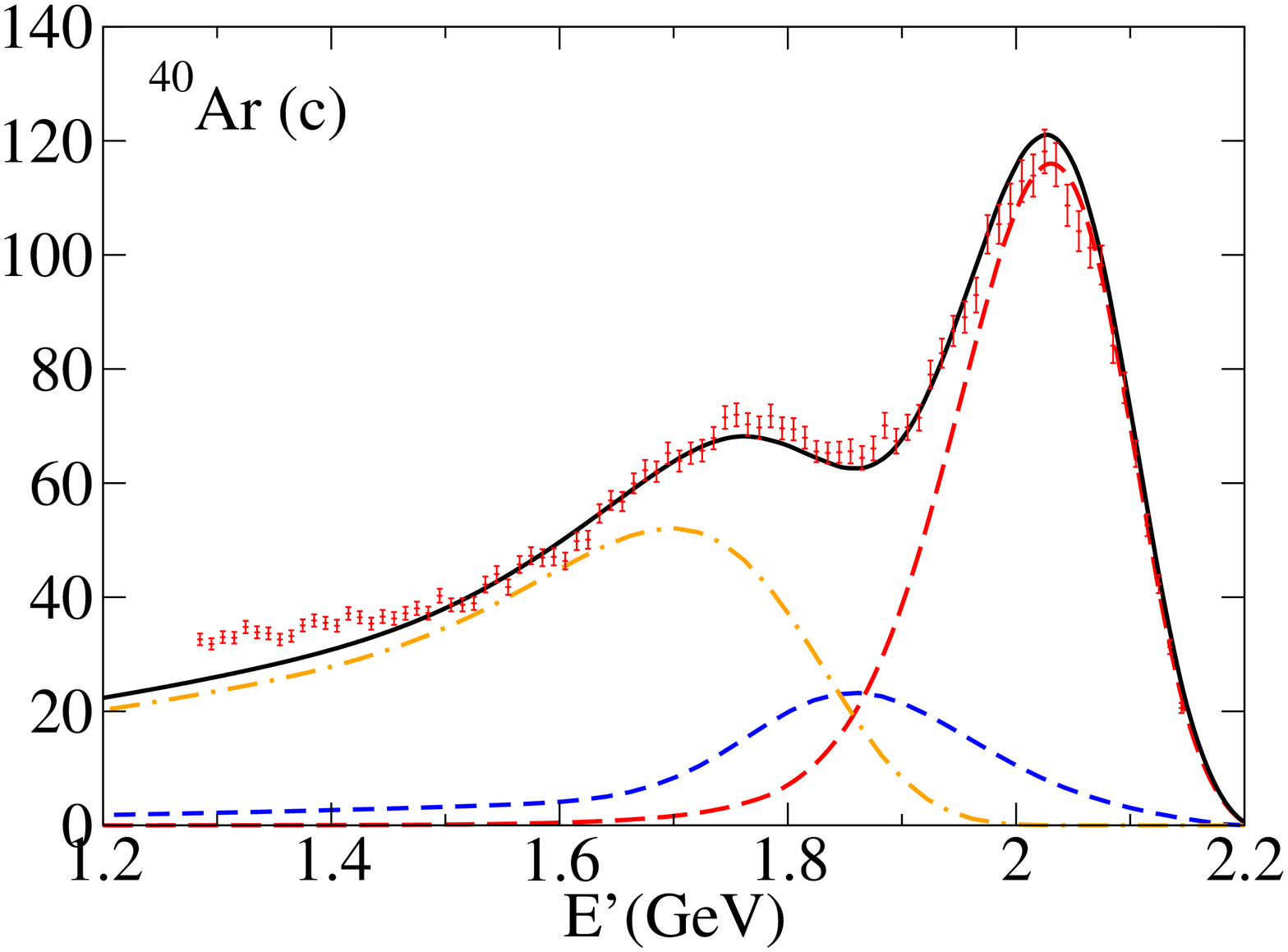}\vspace*{-0.295cm}
		\caption{The $(e,e')$ double differential cross
                  section of carbon, titanium and argon from
                  \cite{Dai:2018xhi,Dai:2018gch}, compared with the
                  SuSAv2-MEC prediction. For completeness, the
                  separate QE, 2p2h and inelastic contributions are
                  also shown. The beam energy is $E$=2.222 GeV and the
                  scattering angle $\theta$=15.541
                  deg.}\label{fig:fig1}\vspace*{-0.2cm}
\end{figure}

\begin{figure*}[ht]\vspace*{-0.2cm}
\hspace*{-0.14cm}\includegraphics[scale=0.68, angle=0]{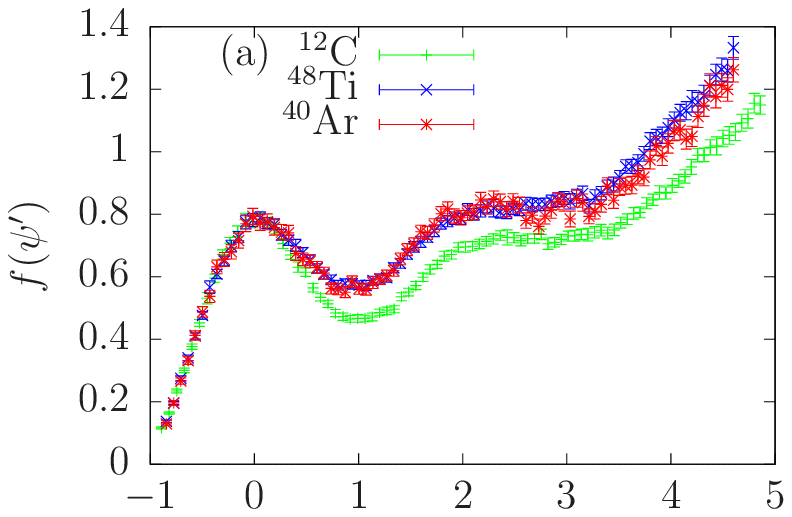}
\\
\includegraphics[scale=0.68, angle=0]{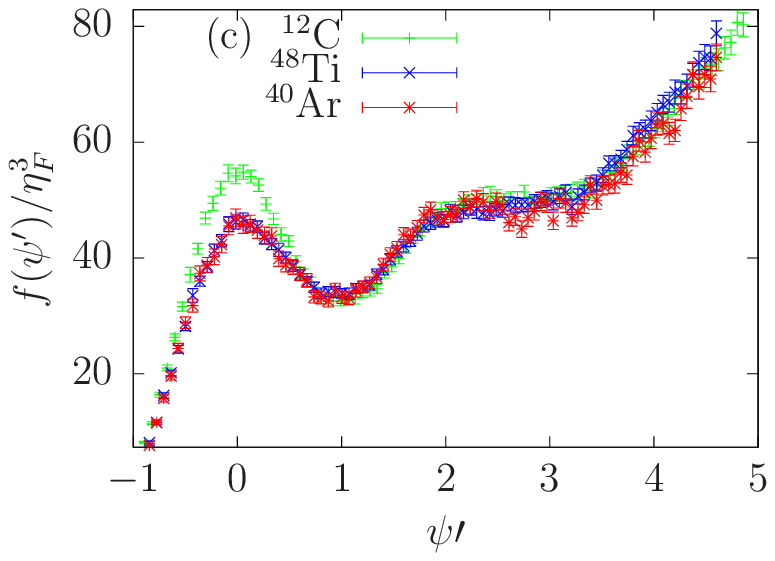}
	\caption{The superscaling function $f$ (top) and $f$ divided by $\eta_F^3$ (bottom) extracted from
          the JLab data \cite{Dai:2018xhi,Dai:2018gch}.}\vspace*{-0.8cm}
	\label{fig:figJLabscaling}
\end{figure*}

In addition to the previous analyses, we have also tested the validity
of the SuSAv2-MEC model and the scaling rules applied when
extrapolating to different nuclei through the analysis of the recent
inclusive electron scattering JLab data~\cite{Dai:2018gch} on
three different targets (C, Ar and Ti)~\cite{Barbaro:2019vsr}. As observed in Fig.~\ref{fig:fig1}, the
agreement is very good over the full energy spectrum, with some
discrepancy seen only in the deep inelastic region. The 2p2h response,
peaked in the dip region between the QE and $\Delta$-resonance peak,
is essential to reproduce the data. We also analyze in
Fig.~\ref{fig:figJLabscaling} the general behavior of
the data in terms of scaling of second kind:
if this kind of scaling was perfectly fulfilled the superscaling function extracted from the data should be independent of $k_F$. 
Deviations from this behaviour indicate a violation on second kind scaling. In Fig.~\ref{fig:figJLabscaling} we show that the 2p2h
response scales very differently from the quasielastic one, in full
accord with what was predicted by the model. As observed, scaling of
second kind only works in the QE peak, then it breaks down because
non-impulsive contributions (2p2h) and inelastic channels come into
play, consistently with previous
analyses~\cite{Donnelly99a,Donnelly99b}.  We also present (bottom panel) the
superscaling function $f(\psi^\prime)$ but divided by $\eta_F^3$,
where $\eta_F\equiv k_F/m_N$. As shown, scaling is highly broken in
the QE peak (results for carbon are significantly higher), however,
results collapse into a single curve within the dip region. This
result confirms our previous study in \cite{Amaro:2017eah}, where we
predicted that 2p2h response scales as $k_F^3$. It is important to
point out that the minimum in the cross section shown in
Fig.~\ref{fig:figJLabscaling} corresponds to the maximum in the 2p2h
contribution (see Fig.~\ref{fig:fig1}). Although contributions from
the QE and inelastic domains also enter in the dip region, and this
may at some level break the scaling behavior, the results in the bottom panel of
Fig.~\ref{fig:figJLabscaling} strongly reinforce our confidence in
the validity of our 2p2h-MEC model. It is also
interesting to note that the same type of scaling, i.e,
$f(\psi^\prime)\sim\eta_F^3$, seems to work reasonably well not only in
the dip region but also in the resonance and DIS domains. Further
studies on the origin of this behavior are underway.  Overall,
these results represent a valuable test of the applicability of the
model to neutrino scattering processes on different nuclei. A more
detailed analysis of these results can be found
in~\cite{Barbaro:2019vsr}.

    \subsection{Comparison with neutrino scattering data}
  In this Section we present our theoretical predictions compared with
  charged-current neutrino scattering data from different
  collaborations, mainly T2K~\cite{Abe:2016tmq,T2Kwater} and MINERvA~\cite{Patrick:2018gvi,PhysRevD.99.012004}, but we will also analyze MiniBooNE
  kinematics~\cite{AguilarArevalo:2010zc,AguilarArevalo:2013hm}. The SuSAv2-MEC model, that has already proven to
  describe accurately $(e,e')$ data (see previous section), is here
  applied to the analysis of recent neutrino data with the aim of
  showing its capability to describe successfully a large variety of
  experimental measurements covering a wide range of kinematics. Our
  study is mainly restricted to the ``quasielastic-like'' regime where
  the impulse approximation used to describe the one-nucleon knockout
  process in addition to the effects linked to the 2p-2h
  meson-exchange currents play a major role. 
  
 Other models~\cite{Nieves:2011pp,Nieves:2011yp,Martini:2010ex,Martini12PRD,Gallmeister16,Meucci,Meucci15,Rocco16,Rocco:2018mwt,Golan12,Lovato:2015qka,Pandey:2016ee,Butkevich:2017mnc,Ivanov:2013saa,PhysRevD.99.093001,Martini:2011wp,Nieves:2013nubar,Martini:2013sha,Mosel:2014lja,Meucci:2014bva,MartiniCP,Mosel,Dolan:2018sbb} have been developed to describe such 1p1h and 2p2h processes for CC neutrino reactions, also providing appropriate descriptions of the experimental measurements. It is important to note that although a similar comparison with inclusive neutrino measurements can be reached, these models rely on different assumptions and descriptions of the nuclear dynamics. Further comparisons of these models with more exclusive measurements could help to determine the goodness of the different approaches.
  
\begin{figure}\vspace{0.08cm}
	\begin{center}\vspace{0.80cm}
			\hspace*{-0.94cm}\includegraphics[scale=0.178, angle=270]{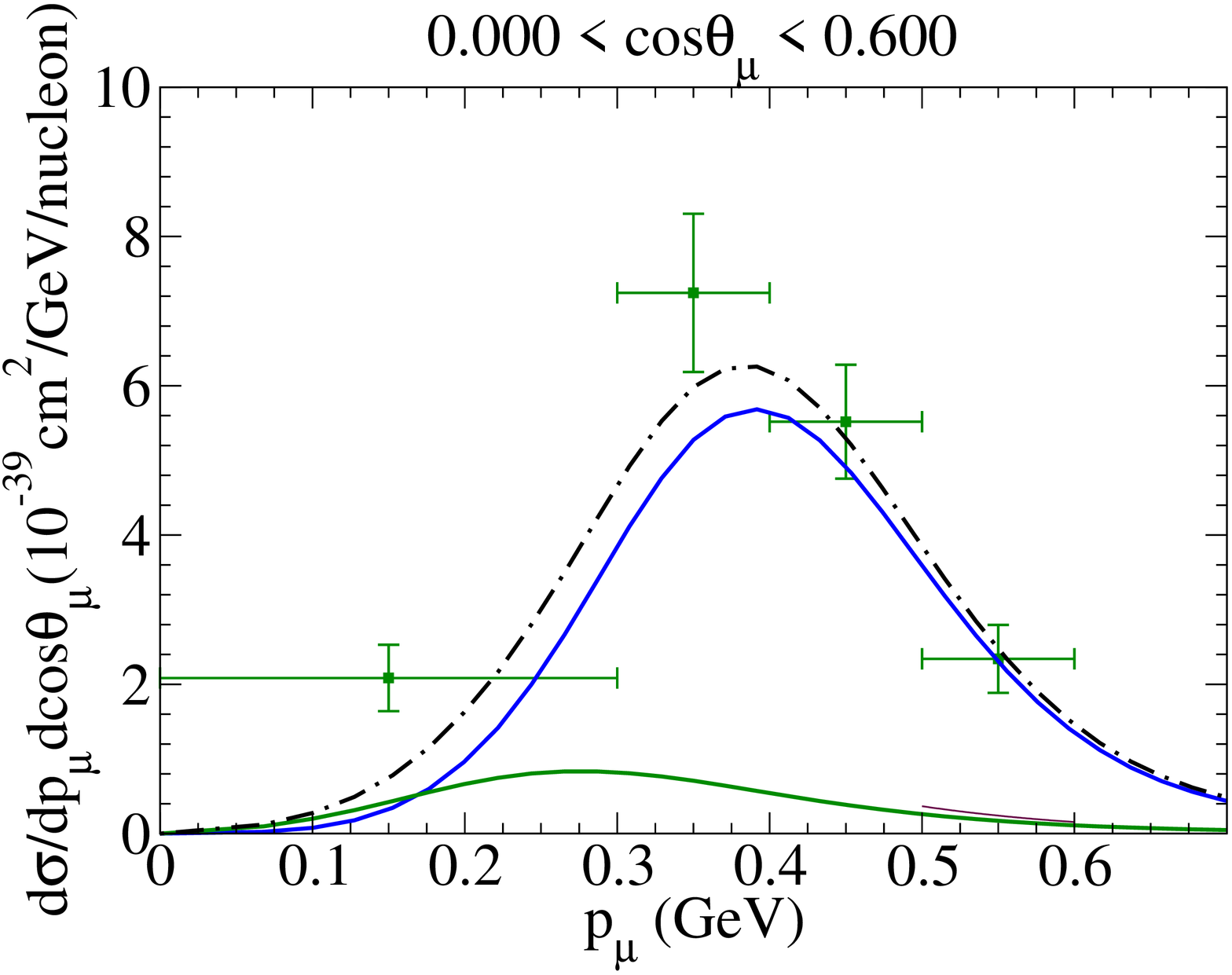}\hspace*{-0.395cm}%
		\includegraphics[scale=0.178, angle=270]{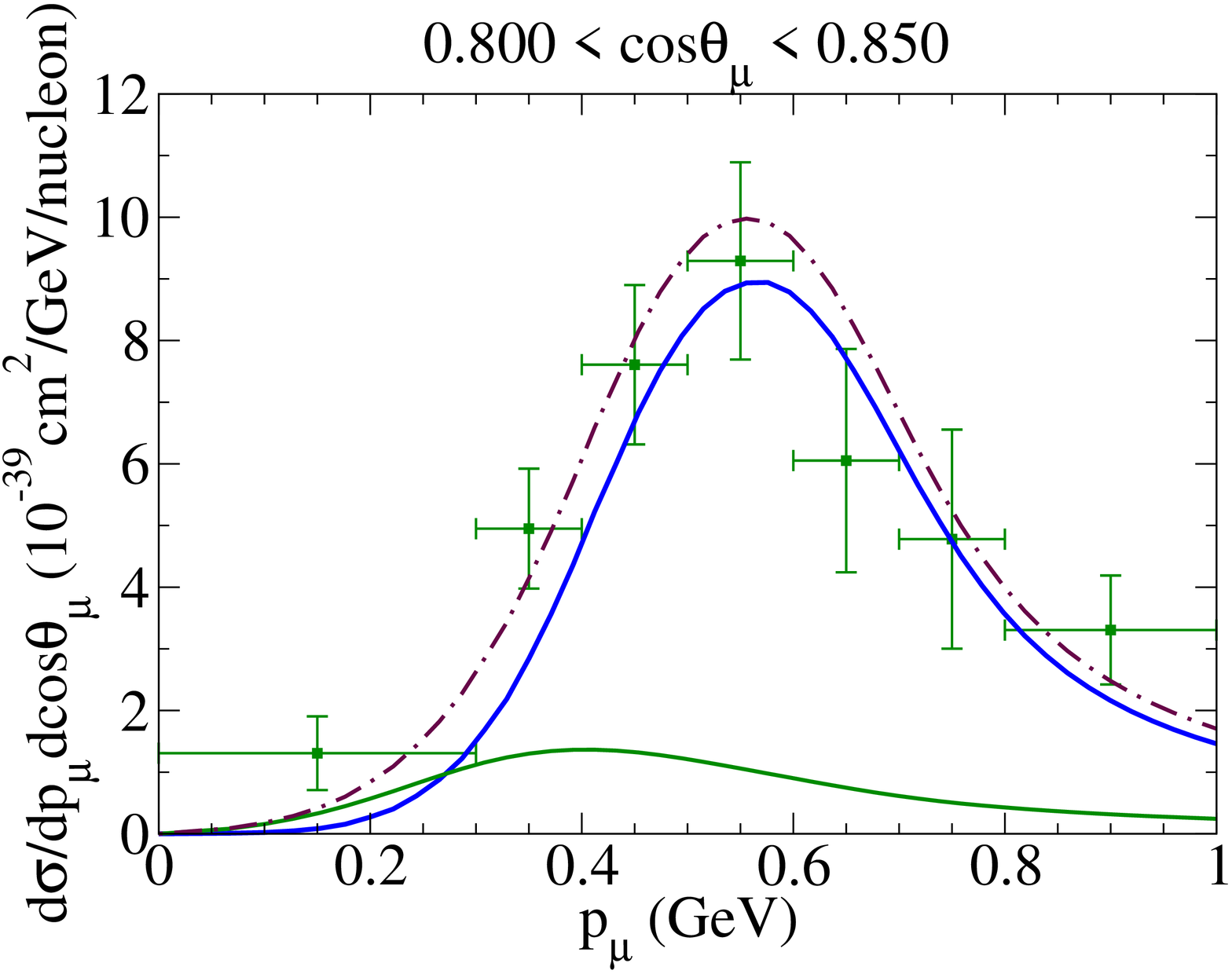}\hspace*{-0.395cm}
		\includegraphics[scale=0.178, angle=270]{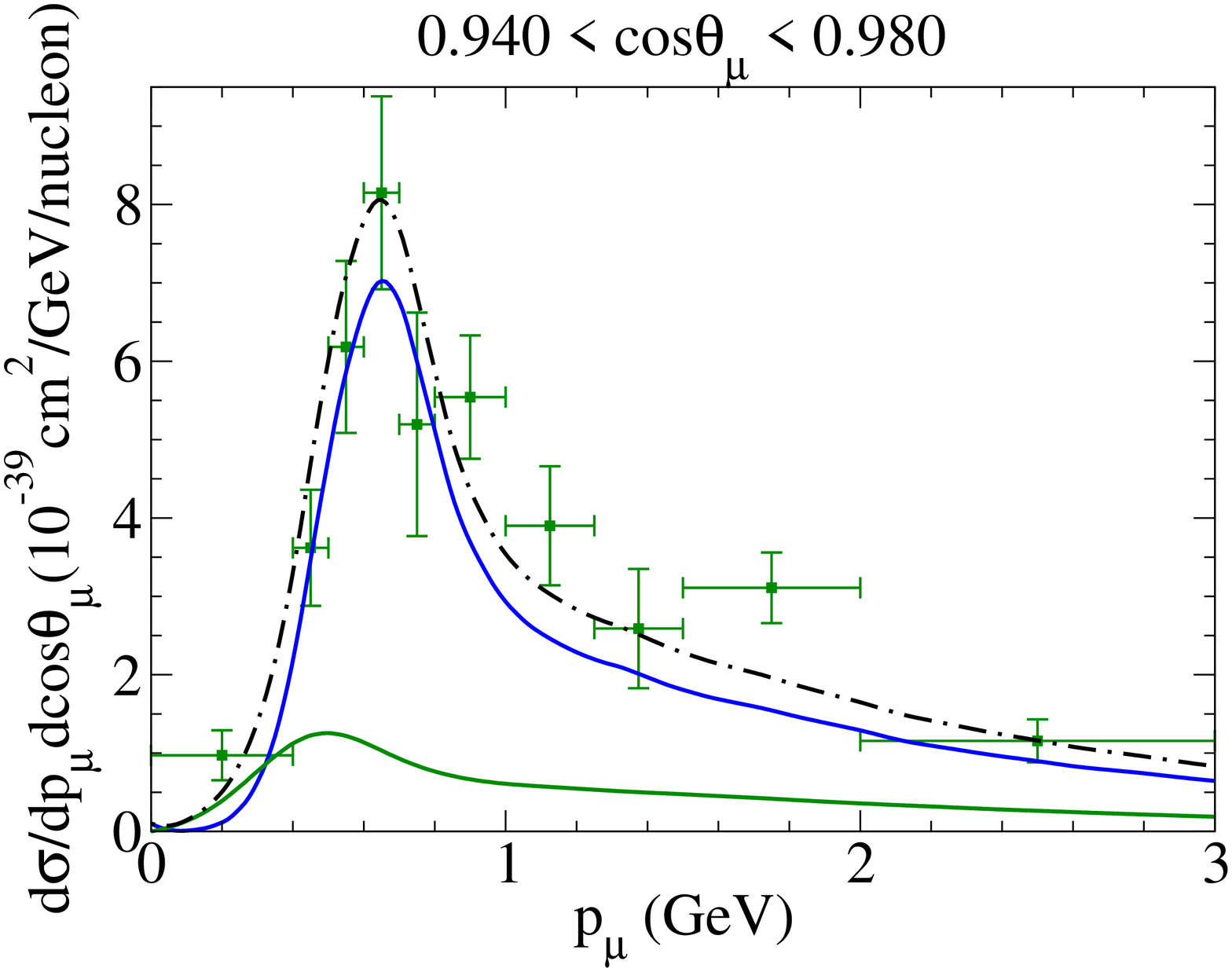}\hspace*{-0.395cm}%
		\includegraphics[scale=0.178, angle=270]{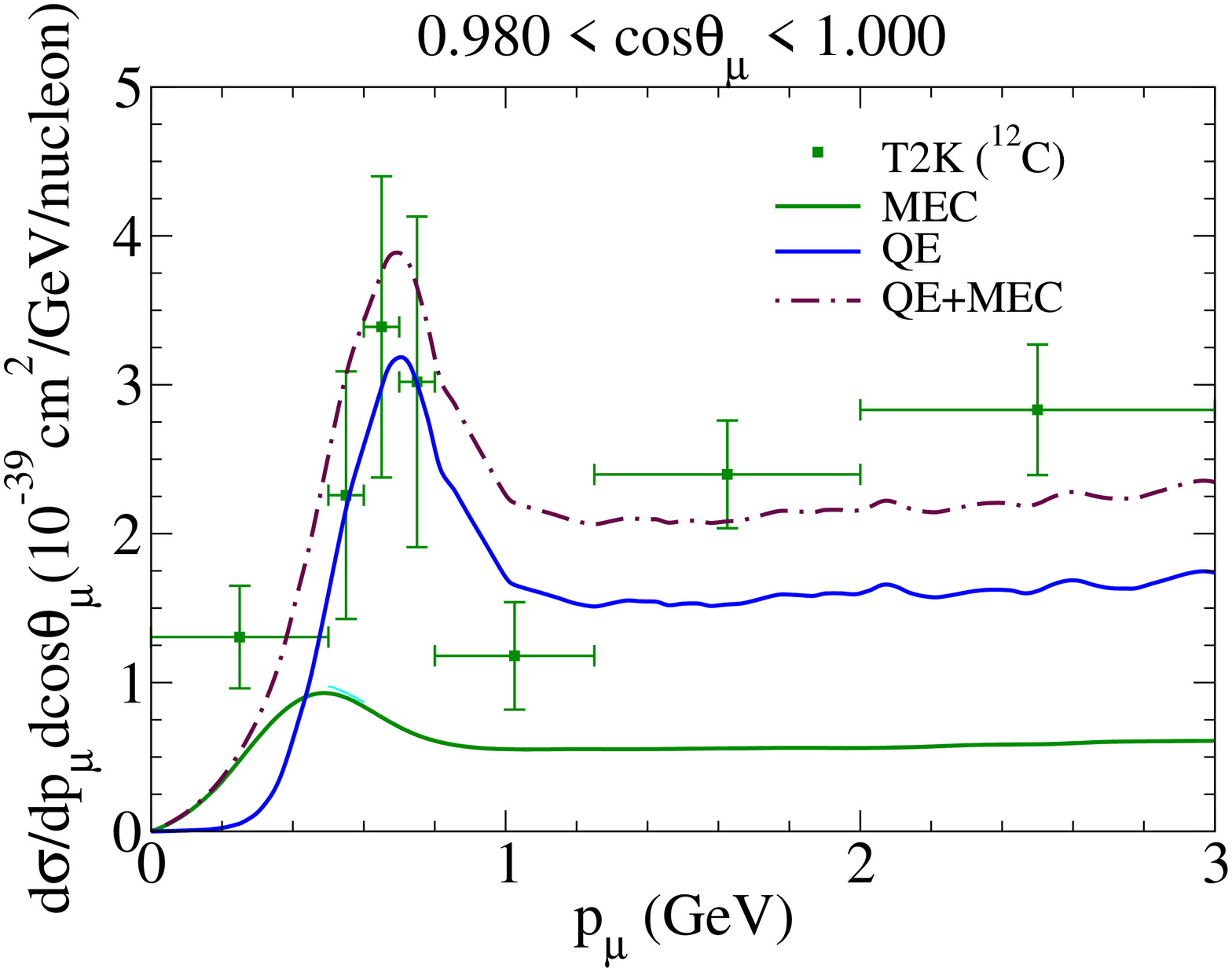}\\
		\hspace{-0.984cm}\includegraphics[scale=0.178, angle=270]{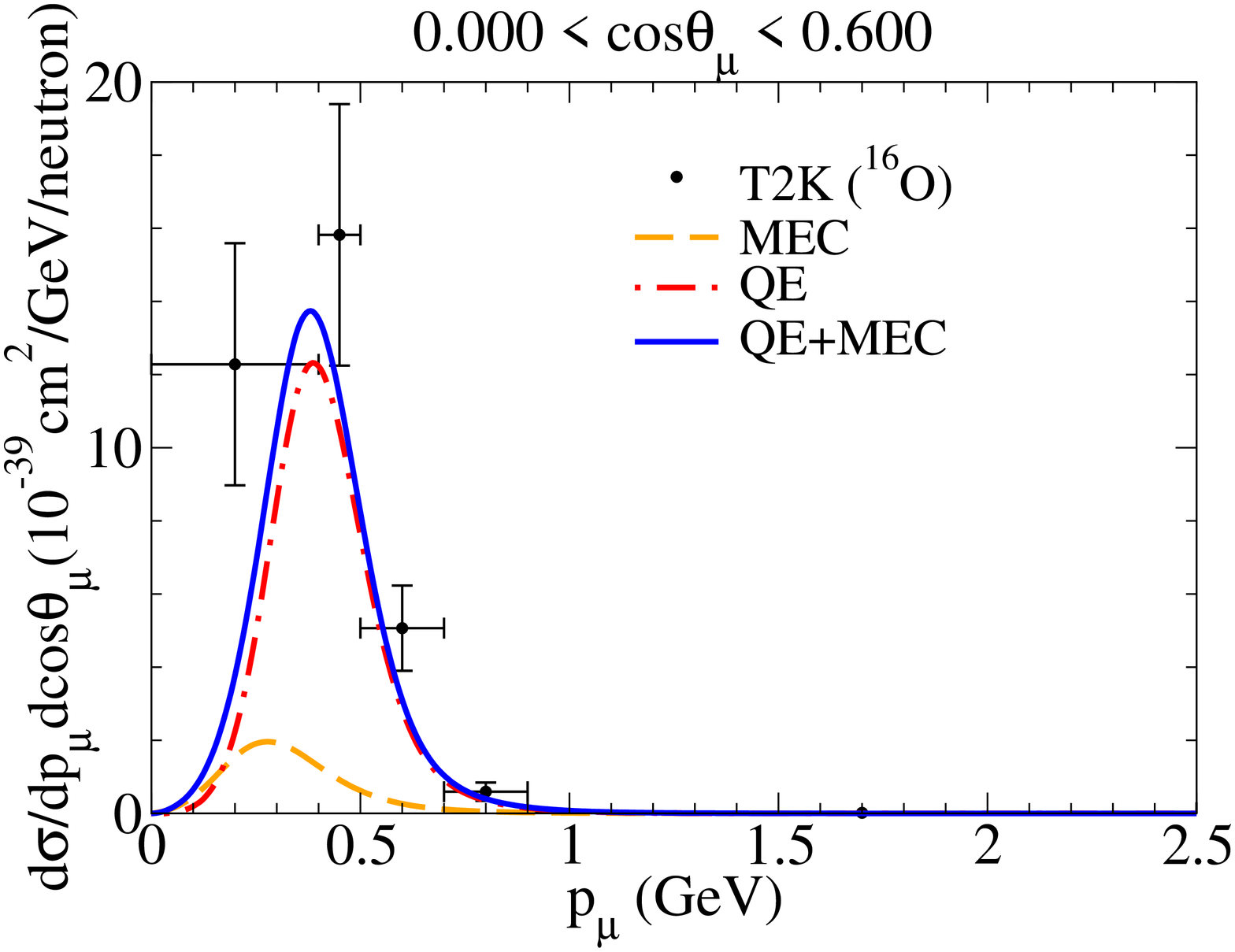}\hspace{-0.4cm}\includegraphics[scale=0.178, angle=270]{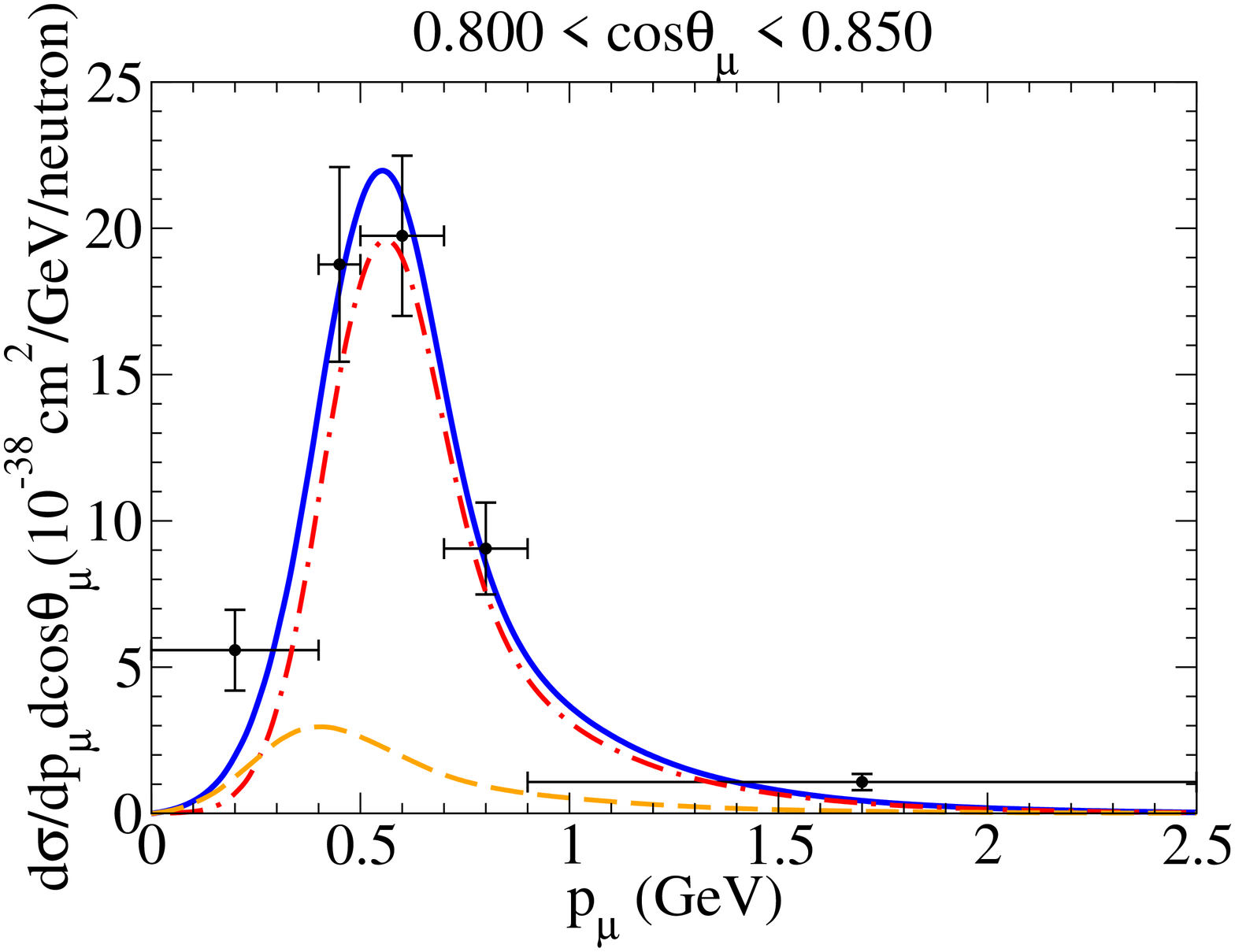}\hspace{-0.4cm}\includegraphics[scale=0.178, angle=270]{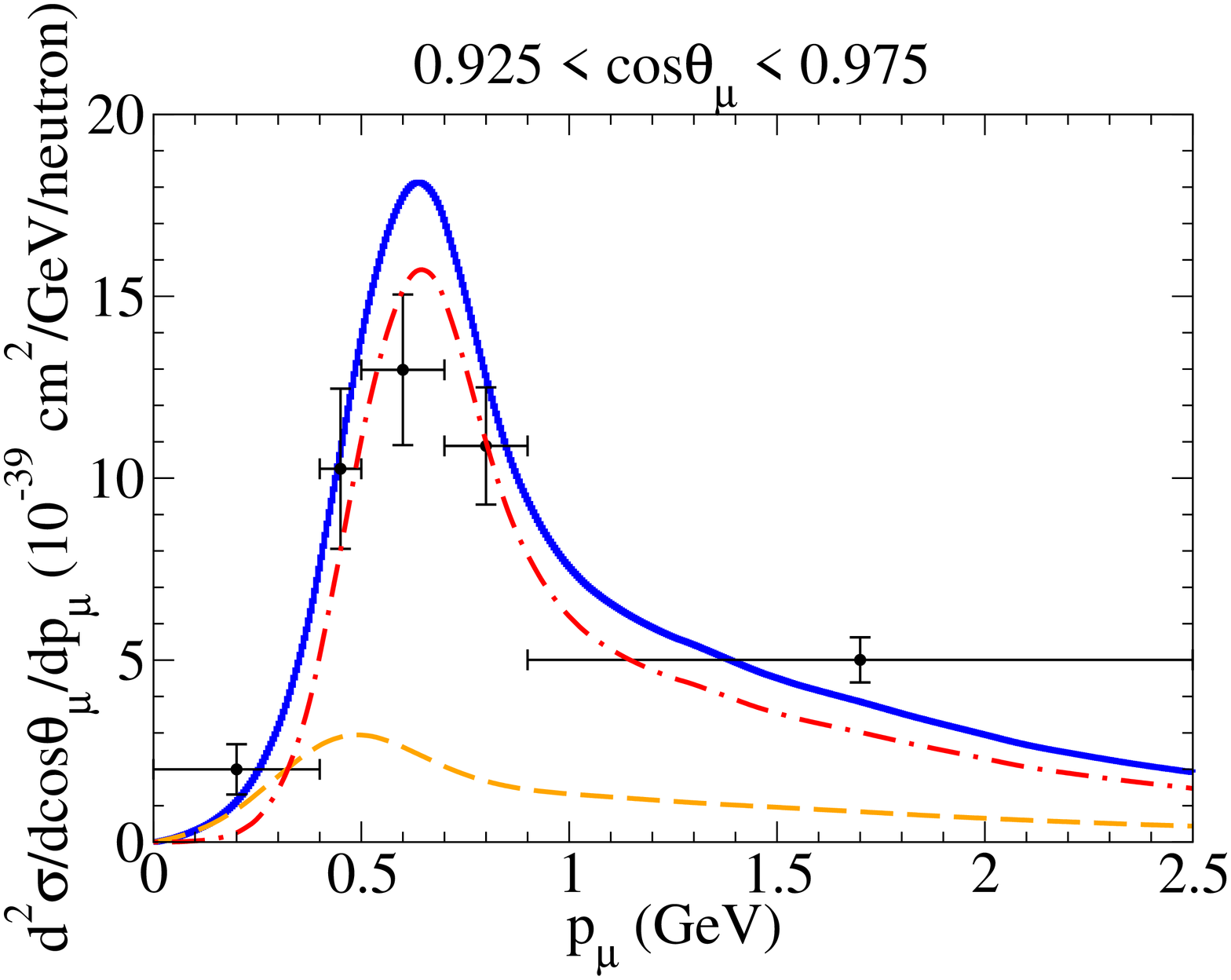}\hspace{-0.4cm}\includegraphics[scale=0.178, angle=270]{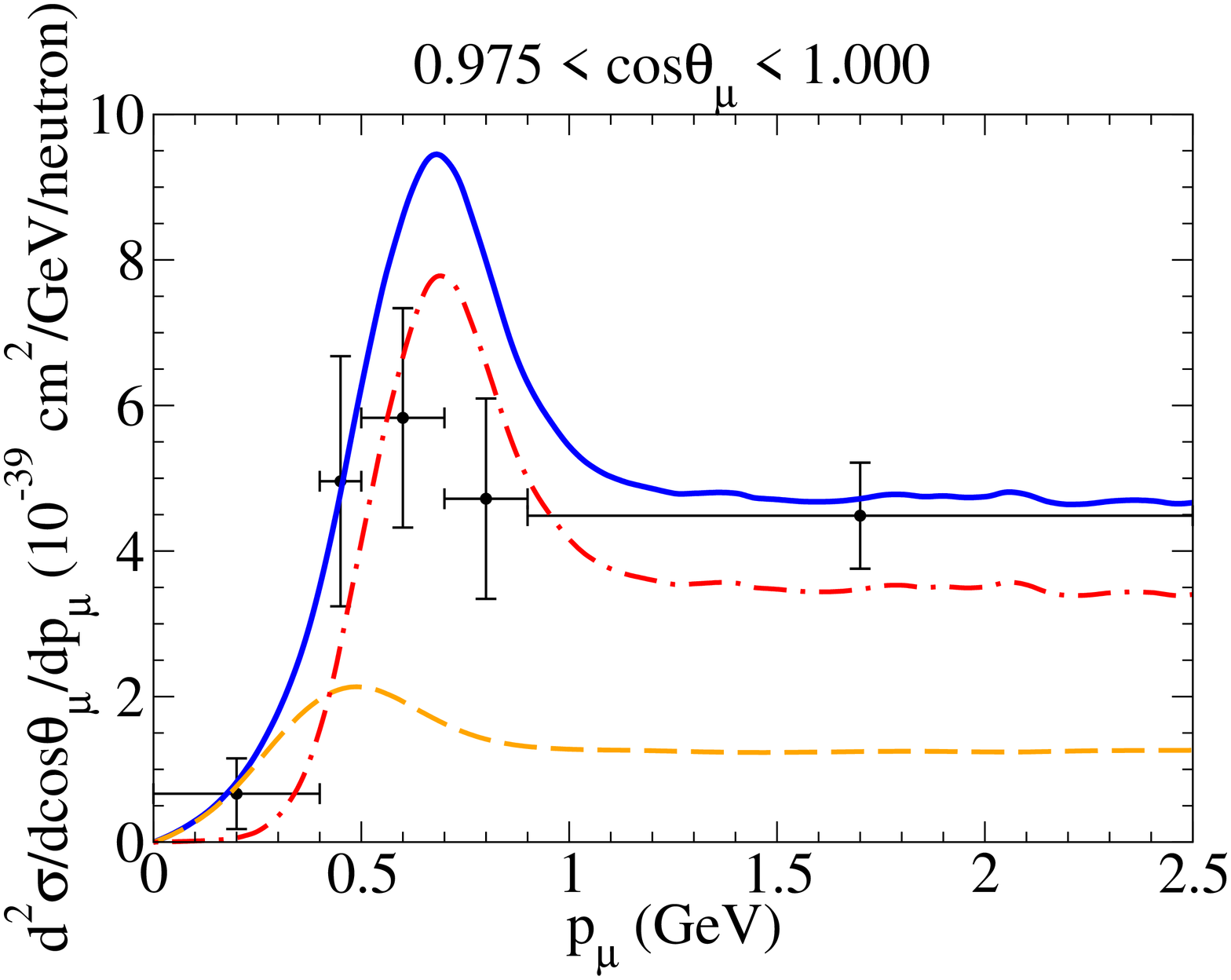}
	\end{center}\vspace*{-0.25cm}
	\caption{The T2K flux-integrated CCQE and 2p2h
          double-differential cross-section for neutrino for
          scattering on $^{12}$C (top panels) and $^{16}$O (bottom panels), within the SuSAv2-MEC model at T2K
          kinematics in units of 10$^{-39}$ cm$^2$/GeV per nucleon (for carbon) and per nucleon target (for oxygen).
          The CC$0\pi$ T2K data are from Ref.~\cite{Abe:2016tmq} and~\cite{T2Kwater}.
	\label{fig:T2K_d2snew}}
\end{figure}

\begin{figure}
	\begin{center}
		\includegraphics[scale=0.16, angle=270]{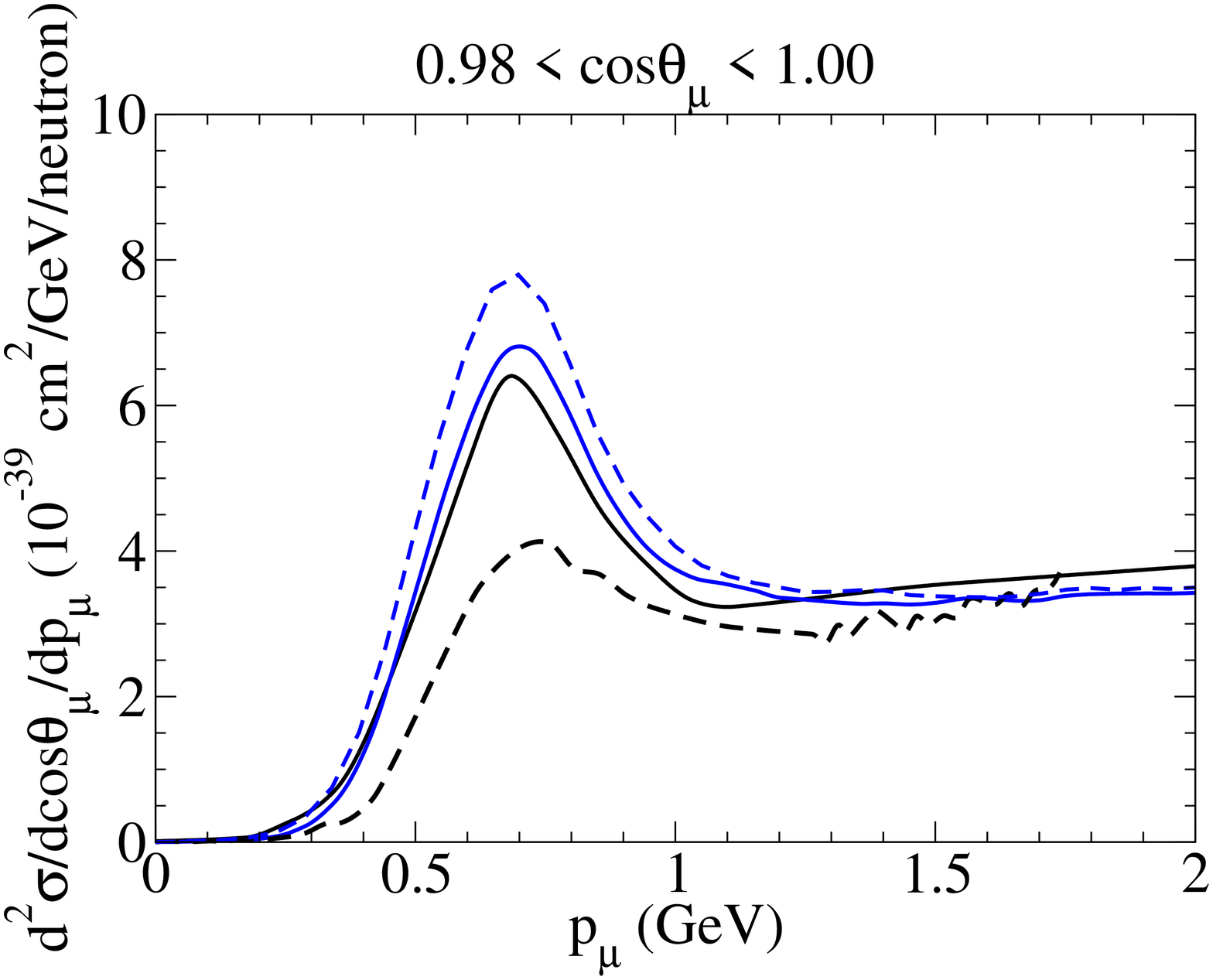}\hspace*{-0.15cm}\includegraphics[scale=0.16, angle=270]{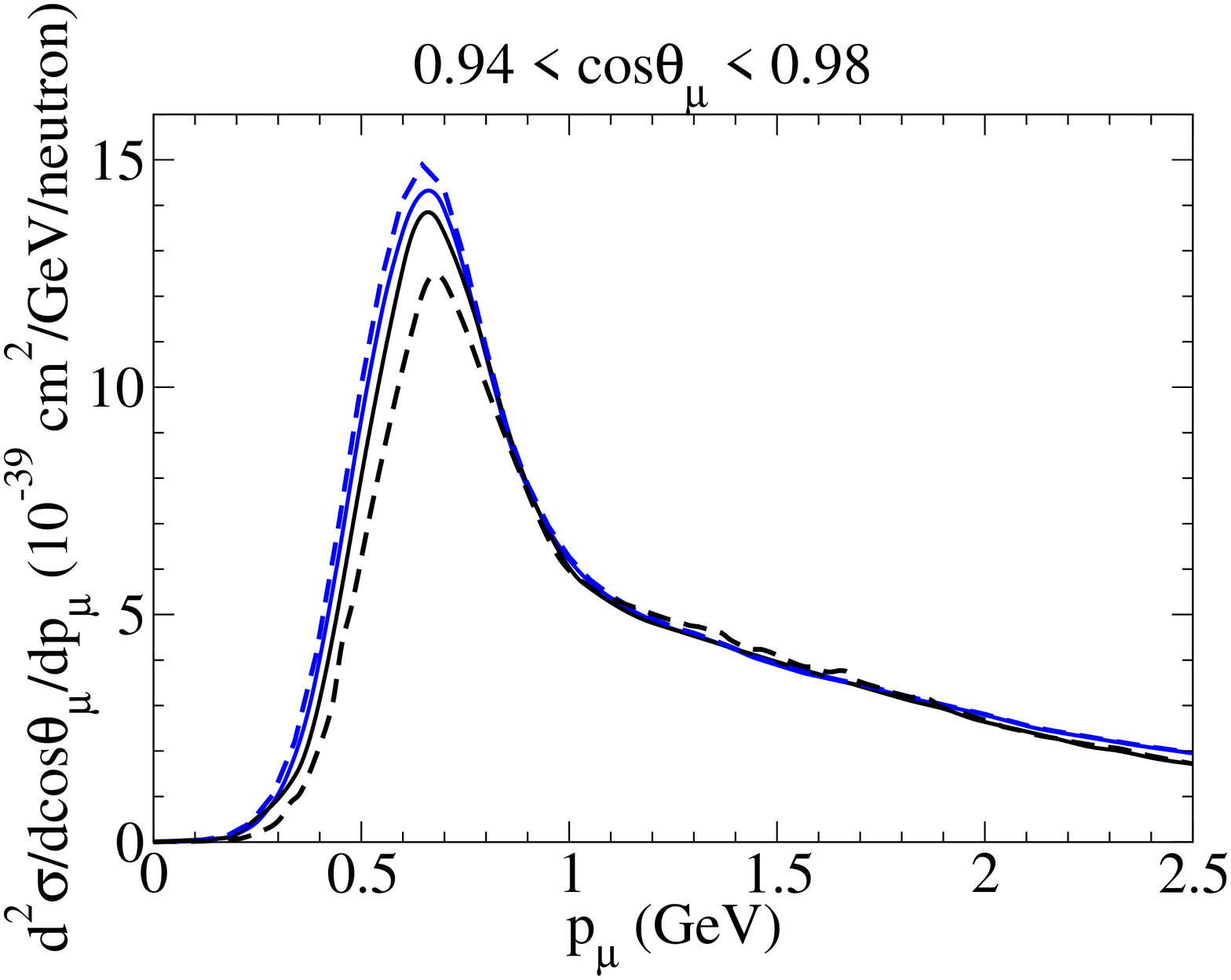}
		\includegraphics[scale=0.16, angle=270]{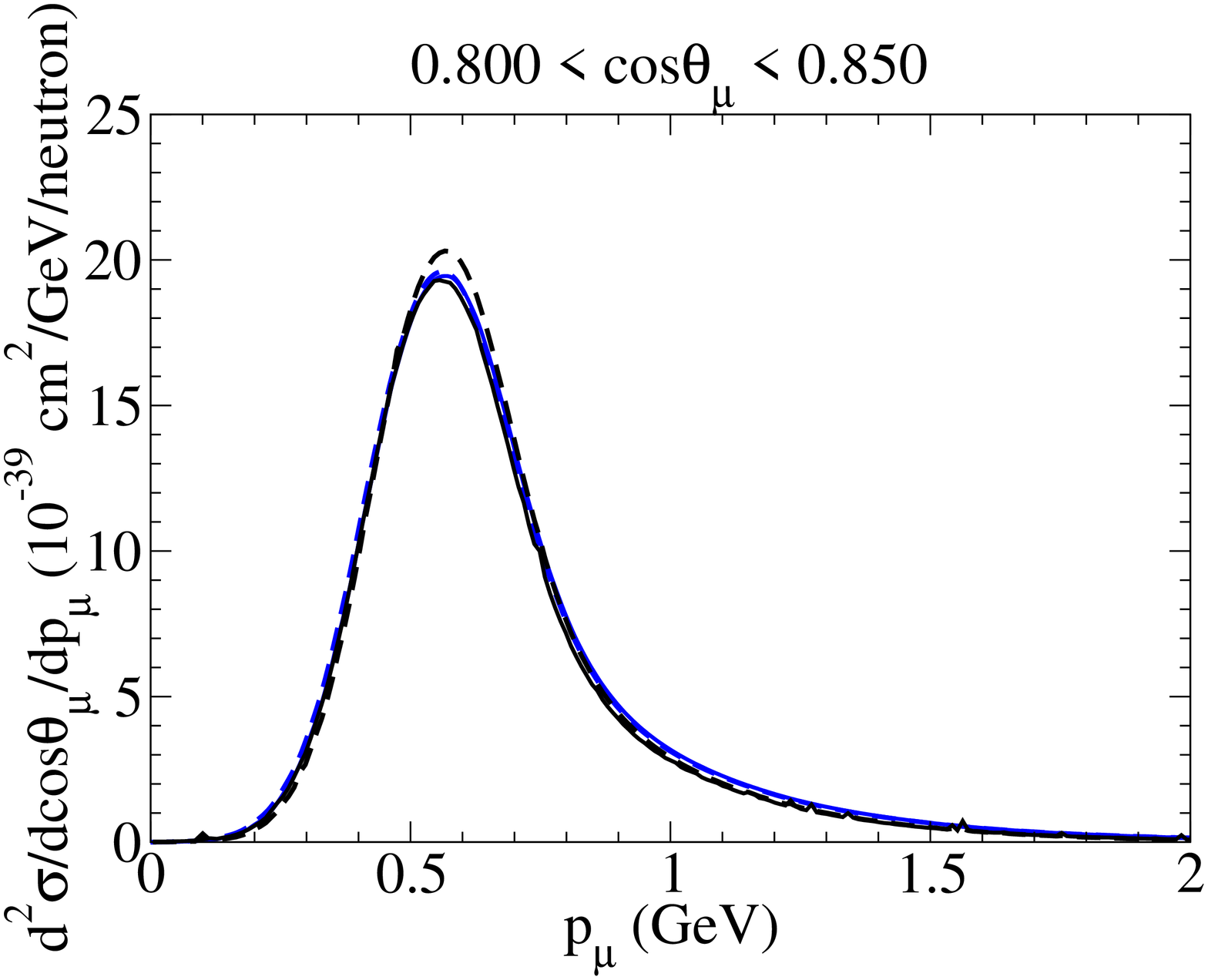}\hspace*{-0.15cm}\includegraphics[scale=0.16, angle=270]{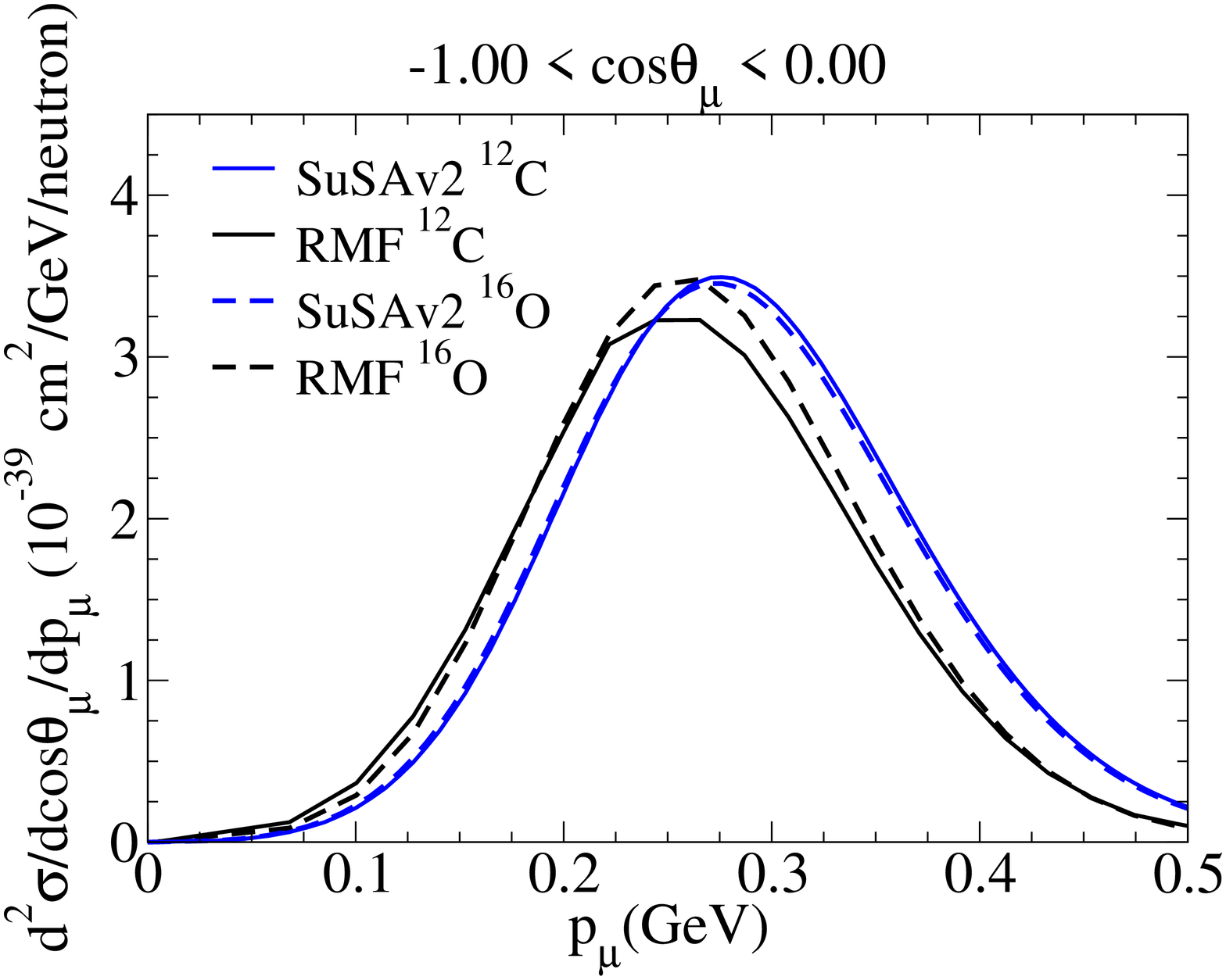}
		\begin{center}
		\end{center}
	\end{center}\vspace*{-0.79cm}
	\caption{Comparison of double differential cross sections on
          $^{12}$C (solid lines) and $^{16}$O (dashed lines) at T2K
          kinematics within the SuSAv2 and RMF models. Results are
          displayed from forward to backward angles.}\label{rmfv2}
\end{figure}

  In Fig.~\ref{fig:T2K_d2snew} we show the T2K CC0$\pi$ (anti)neutrino
  data on $^{12}$C and $^{16}$O in comparison with the
  SuSAv2-MEC model. The CC0$\pi$ scattering is defined, equivalently
  to the CCQE-like one, as the process where no pions are detected in
  the final state. As already explained in previous sections,
  quasielastic (QE) scattering and multi-nucleon excitations dominated
  by 2p-2h MEC contributions should be included in the analysis
  together with other minor effects such as pion-absorption processes
  in the nucleus that can mimic a CCQE-like event. However, as will
  be shown in Section~\ref{impl-section}, the pion-absorption effects
  are not particularly relevant at T2K
  kinematics. Note also that these two main mechanisms, QE and 2p2h, have in
  general a different dependence upon the nuclear species, namely they
  scale differently with the nuclear density\cite{Amaro:2017eah}, as
  previously analyzed. As observed in Fig.~\ref{fig:T2K_d2snew},
  the SuSAv2-MEC model is capable of reproducing the T2K data for both
  nuclei~\cite{Megias:2018oxygen}, which can help to disentangle how nuclear effects enter in
  the analysis of the T2K experiment as well as to reduce
  nuclear-medium uncertainties. 
  It is interesting to point out the results for the most forward angles, {\it i.e.,}
  the panel on the right-bottom corner. Notice that the QE and 2p-2h
  MEC contributions are stabilized to values different from zero for
  increasing muon momenta as a consequence of the high energy tail of
  the T2K neutrino flux. This is at variance with all remaining
  situations where the cross sections decrease significantly as the
  muon momentum $p_\mu$ goes up. It is also important to notice that the slight overestimation of the SuSAv2 predictions in the peaked region at very forward angles is clearly related to the limitations of the model at very low $q$ and $\omega$ values where RMF scaling violations due to nuclear-medium effects are not properly included in the SuSAv2 approach. This drawback of the model will be solved in the forthcoming ED-RMF approach~\cite{Gonzalez-Jimenez19a,Gonzalez-Jimenez20}. This point is also stressed in Fig.~\ref{rmfv2} where large differences between RMF predictions for $^{12}$C and $^{16}$O are observed due to different nuclear-medium effects and binding energies, while these effects are not properly accounted for in the current SuSAv2 approach where the results for $^{12}$C and $^{16}$O are pretty similar.  In this sense, it is important to mention the recent exploration of the C/O
  differences carried out by the T2K Collaboration~\cite{Abe:2020uub}, that has allowed to analyze the validity of various models to extrapolate between carbon and oxygen nuclear targets, as is required in T2K oscillation analyses, showing particular model separation for very forward-going muons. As also observed in~\cite{Abe:2020uub}, the SuSAv2 approach overestimates these new data at very forward angles while the RMF model corrects this drawback, yielding a good agreement at low kinematics.
  
\begin{figure*}[!ht]\vspace{-0.128cm}
		\hspace*{-0.295cm}\includegraphics[scale=0.192, angle=270]{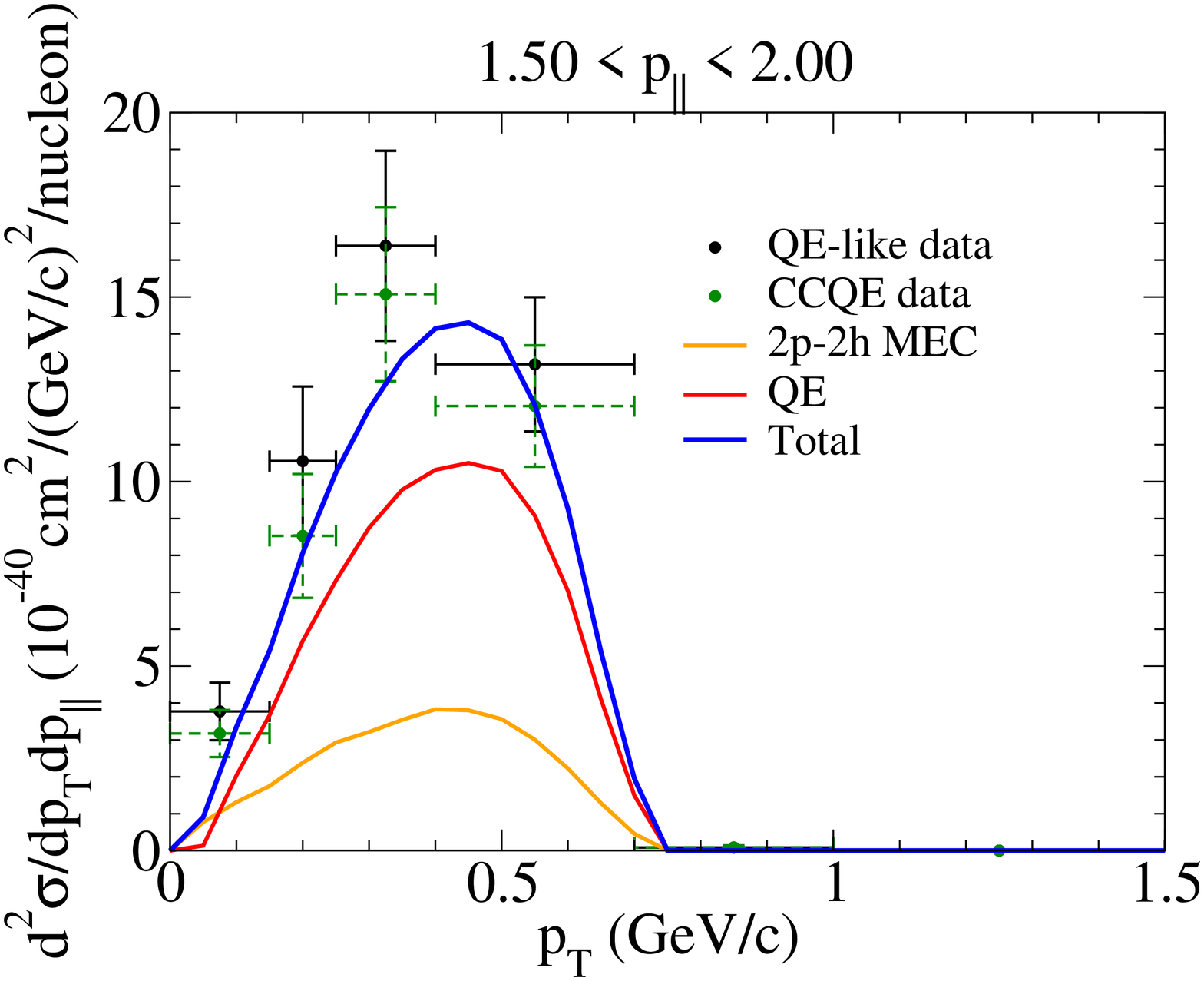}\hspace*{-0.584cm}%
		\hspace*{-0.295cm}\includegraphics[scale=0.192, angle=270]{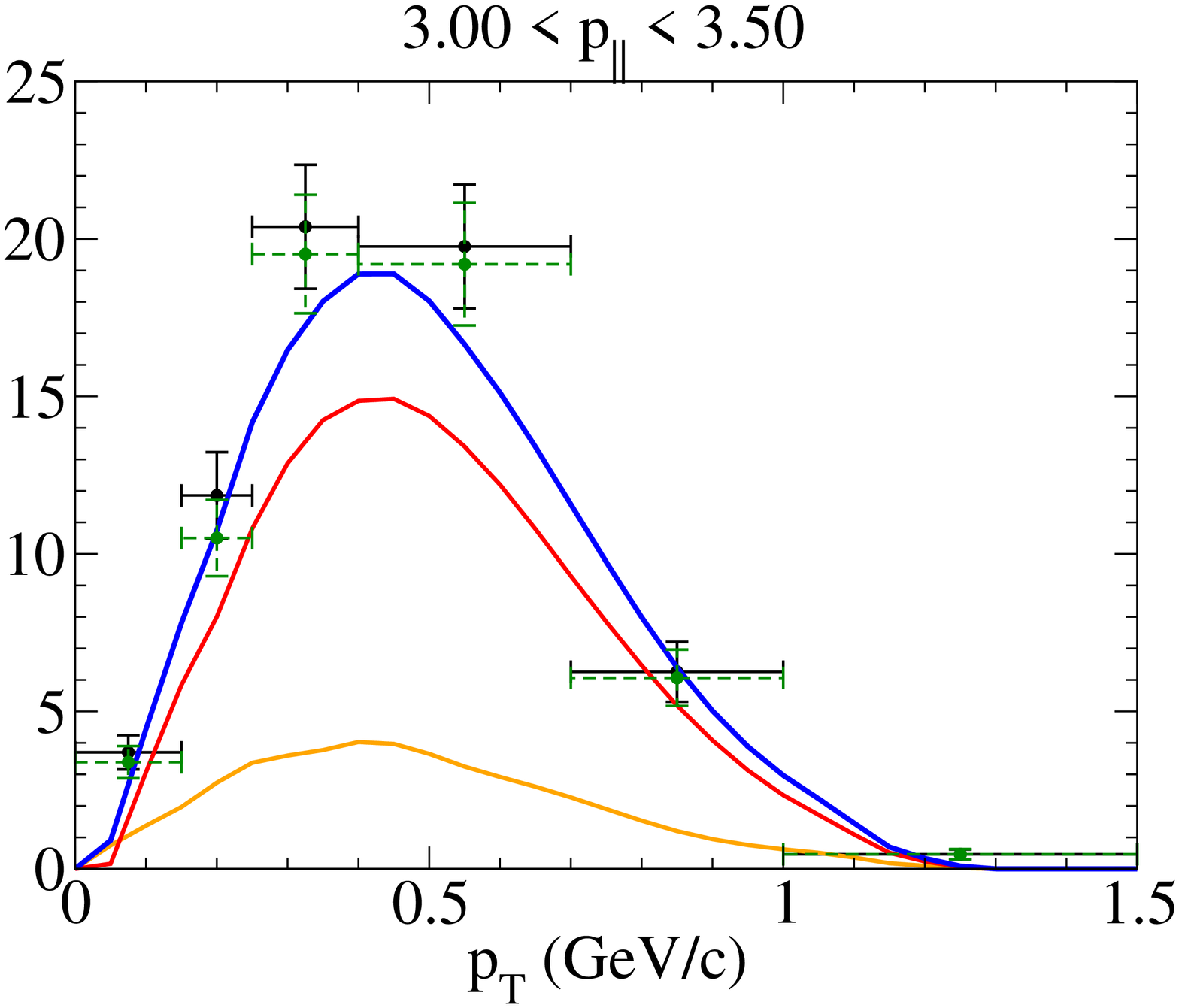}\hspace*{-0.584cm}%
		\hspace*{-0.295cm}\includegraphics[scale=0.192, angle=270]{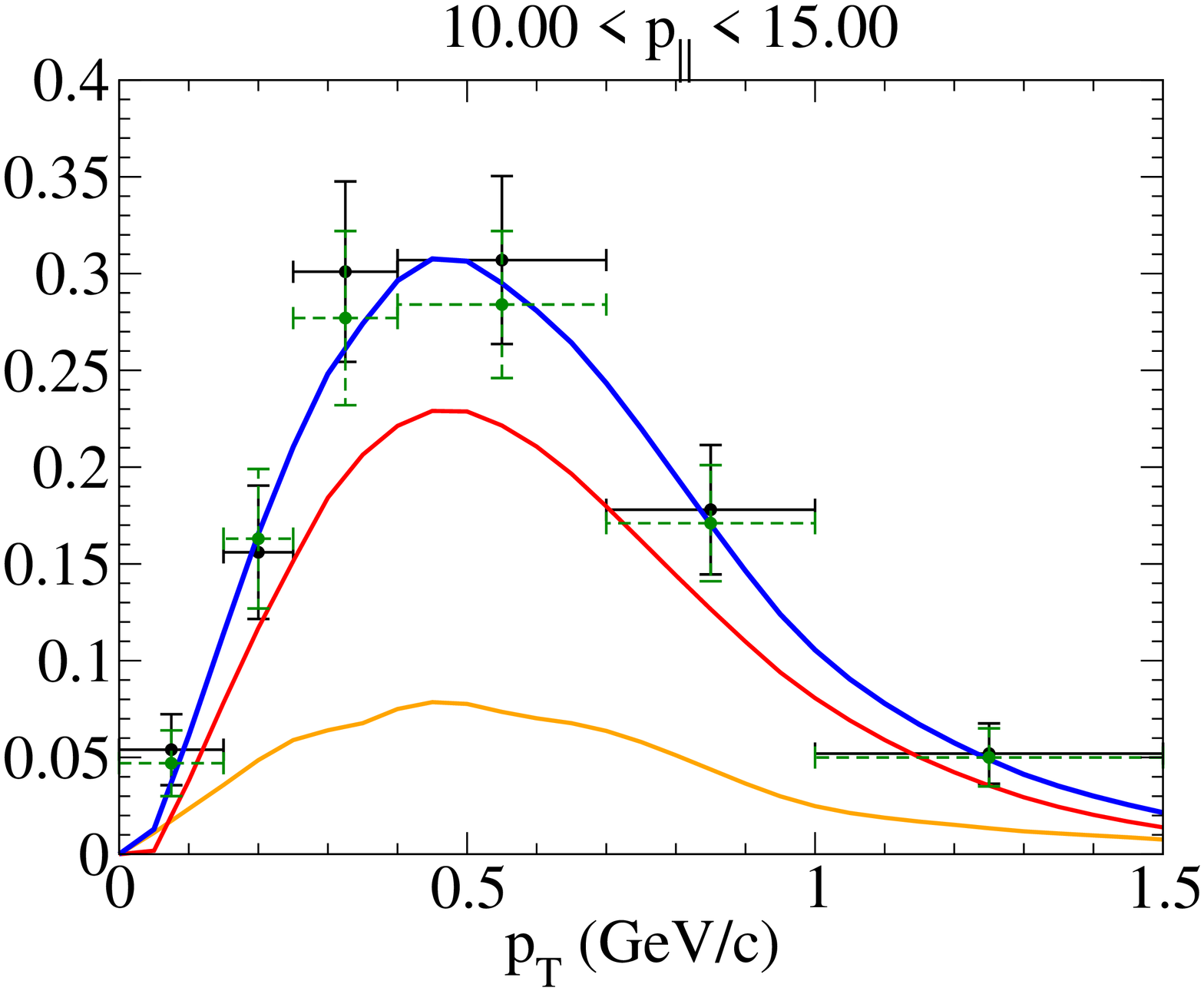}\hspace*{-0.584cm}\\
			\hspace*{-0.295cm}\includegraphics[scale=0.192, angle=270]{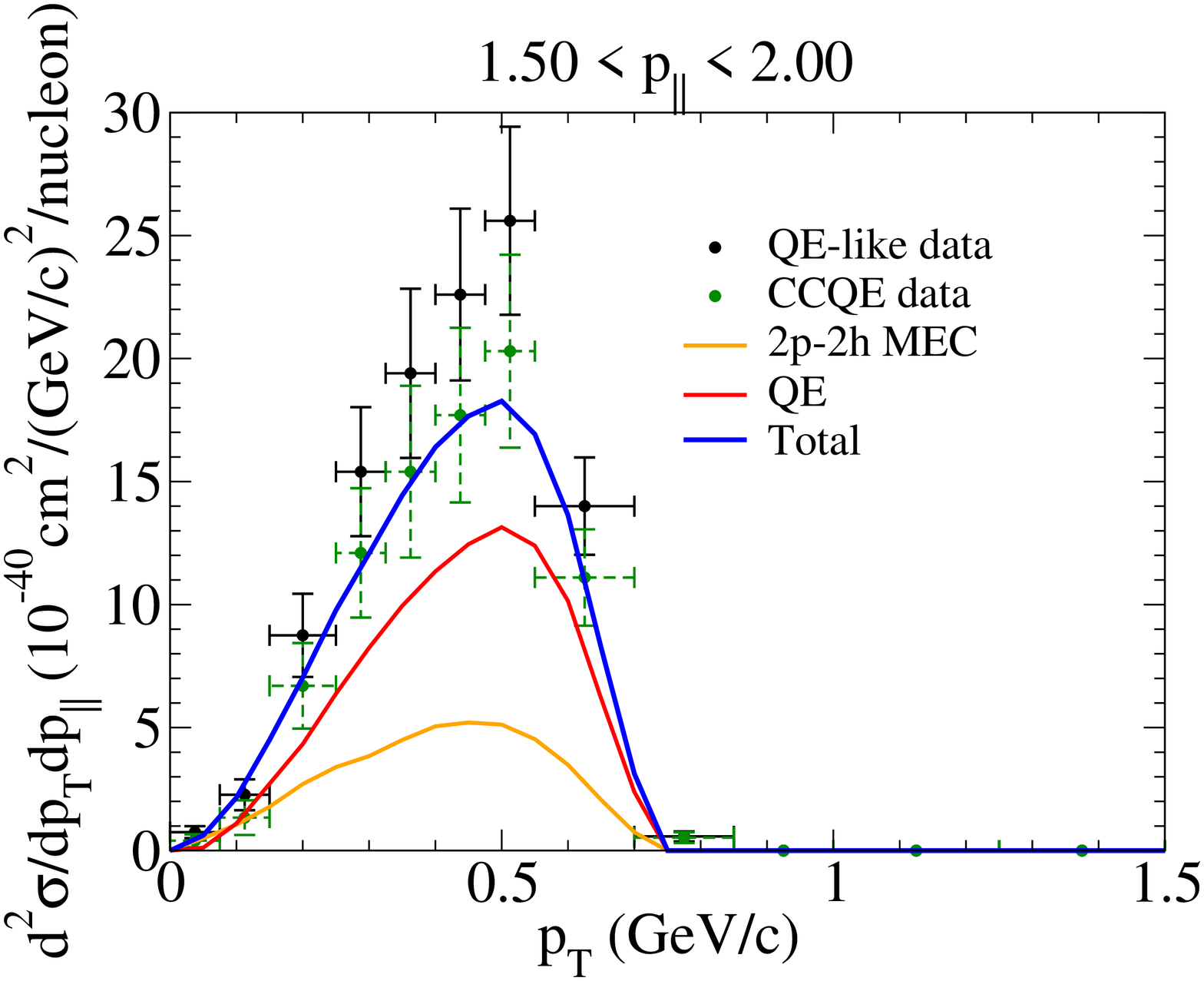}\hspace*{-0.584cm}%
		\hspace*{-0.295cm}\includegraphics[scale=0.192, angle=270]{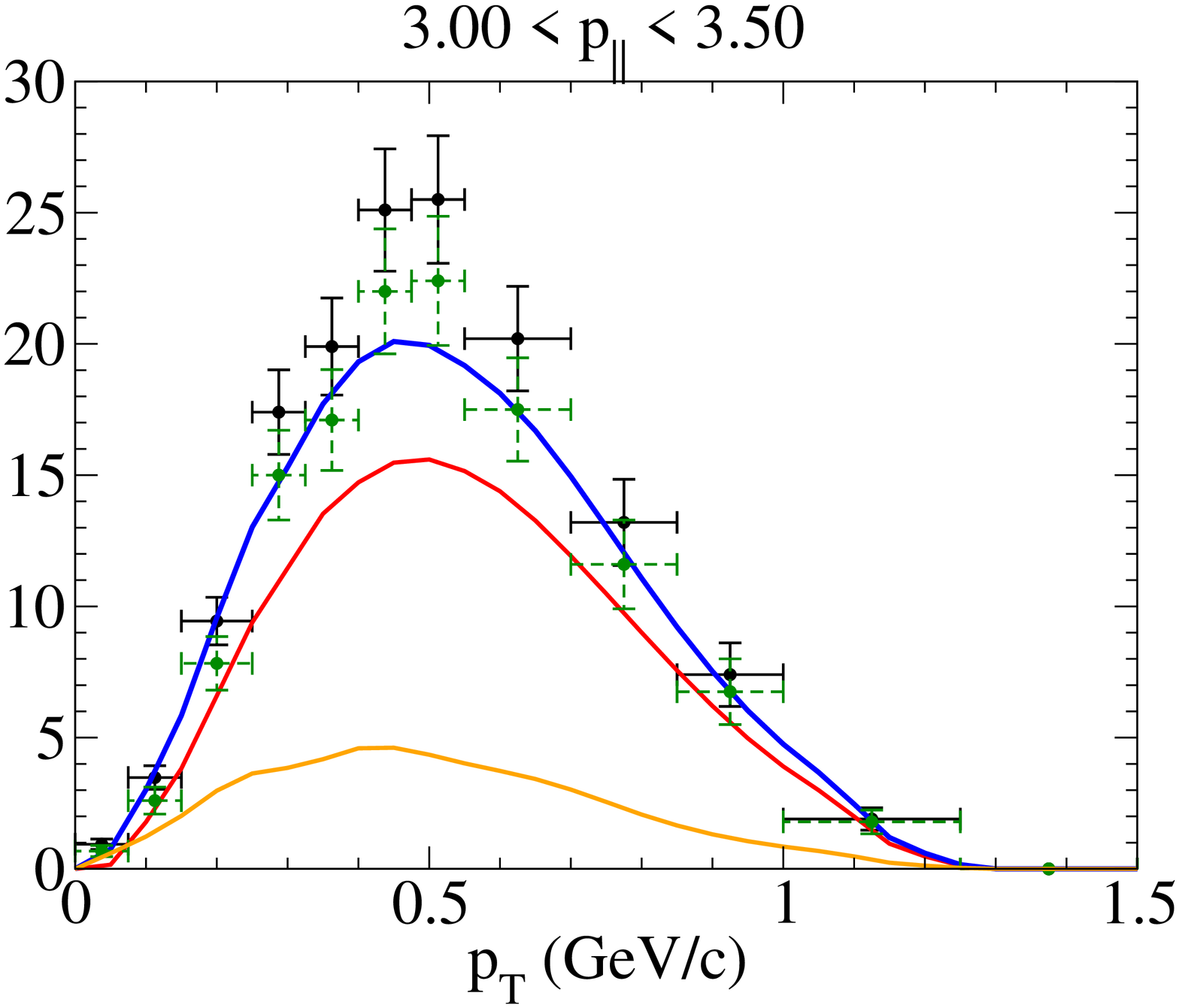}\hspace*{-0.584cm}%
		\hspace*{-0.295cm}\includegraphics[scale=0.192, angle=270]{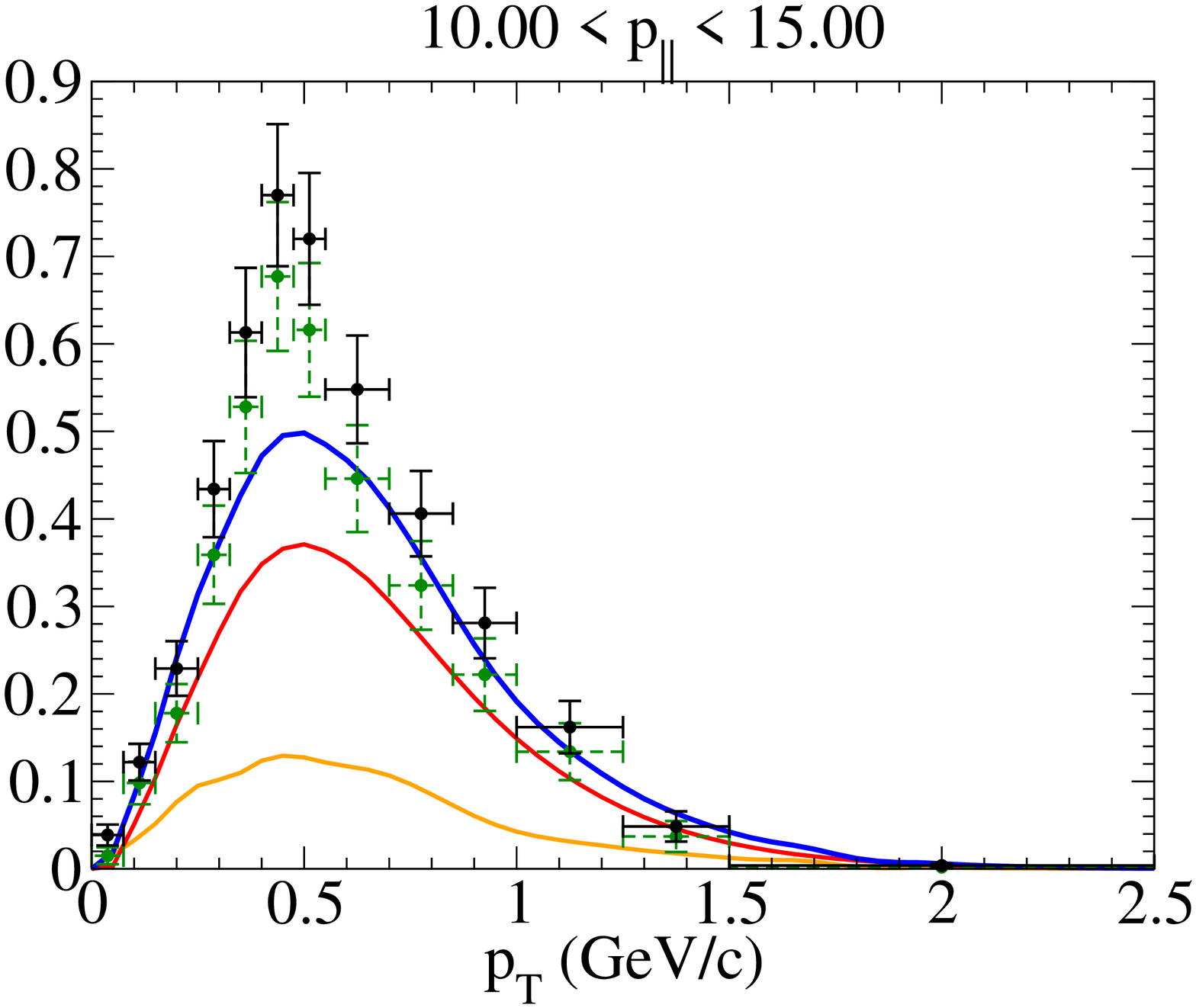}	
	\caption{(Color online) The MINERvA ``QE-like" and ``CCQE"
          double differential cross sections for $\bar\nu_\mu$ (top
          panels) and $\nu_\mu$ (bottom panels) scattering on
          hydrocarbon versus the muon transverse momentum, in bins of
          the muon longitudinal momentum (in GeV/c). The curves
          represent the prediction of the SuSAv2+2p2h-MEC (blue) as
          well as the separate quasielastic (red) and 2p2h-MEC
          (orange) contributions.  The data and the experimental
          fluxes are from Refs.~\cite{Patrick:2018gvi}
          and~\cite{PhysRevD.99.012004} .}
	\label{fig:fig1mnv}
\end{figure*}

Next, in Fig.~\ref{fig:fig1mnv} we compare the SuSAv2-MEC predictions
with the recent MINERvA QE-like double differential (anti)neutrino
cross sections on hydrocarbon (CH), as a function of the transverse (with respect to the antineutrino beam)
momentum of the outgoing muon, in bins of the muon longitudinal momentum. 
The relativistic nature of the model makes it suitable
to describe these data~\cite{PhysRevD.99.113002,Patrick:2018gvi},
which correspond to a mean beam energy of 3.5 GeV. The standard SuSAv2
model predictions agree well with the data without needing any
additional or tuned parameter. For the data comparison we use the same nomenclature employed in the
experimental paper~\cite{Patrick:2018gvi,PhysRevD.99.012004}. The ``QE-like" experimental
points include, besides pure quasielastic contributions, events that
have post-FSI final states without mesons, prompt photons above
nuclear de-excitation energies, heavy baryons, or protons above the
proton tracking kinetic energy threshold of 120 MeV, thus including
zero-meson final states arising from resonant pion production followed
by pion absorption in the nucleus and from interactions on
multinucleon states. This is similar to the so-called CC0$\pi$
definitions used in T2K. On the contrary, the
``CCQE" signal (also defined in other experiments as ``CCQE-like'')
corresponds to events initially generated in the GENIE neutrino
interaction event generator~\cite{Andreopoulos:2009rq} as
quasi-elastic (that is, no resonant or deep inelastic scatters, but
including scatters from nucleons in correlated pairs with zero-meson
final states), regardless of the final-state particles produced, thus
including CCQE and 2p2h interactions.  The difference between the two
data sets, mainly due to pion production plus re-absorption, varies
between $\sim15\%$ and $\sim5\%$ depending on the
kinematics. According to MINERvA's acceptance, the muon scattering
angle is limited to $\theta_\mu<$ 20$^{\circ}$ as well as the muon
kinematics (1.5 GeV $< p_{||} < $ 15 GeV, $p_T<$ 1.5 GeV) in both
experimental and theoretical results, leading to a significant
phase-space restriction for large energy and momentum transfer to the
nuclear target.

The antineutrino-H contribution in the
cross sections only enters through the 1p1h channel and has been
evaluated by computing the elastic antineutrino-proton cross
section. The present calculation does not include processes
corresponding to pion emission followed by re-absorption inside the
nucleus. Therefore the curves are meant to be compared with the
``CCQE" data rather than with the ``QE-like" ones. However, we also
display the QE-like cross sections, to illustrate MINERvA's estimation
of the magnitude of the QE-like resonance component among other minor
effects. A more detailed analysis of these results together with a $\chi^2$ test can be found in~\cite{PhysRevD.99.113002}, where the SuSAv2 $\chi^2$ seems to be very similar to  the MINERvA/GENIE ones and compatible with data.

For completeness, it is worth mentioning that in the specific kinematic
conditions of MINER$\nu$A, it clearly appears that, even if the
neutrino energy is as large as $3$~GeV, the process is largely
dominated by relatively small energy and momentum transfer, namely,
$\omega<500$ MeV, $q<1000$ MeV, whereas contributions below
$\omega<50$ MeV, $q<200$ MeV govern the lowest $Q^2_{QE}$ region. More
specific details can be found in
Refs.~\cite{Megias:2014kia,Megias:2016nu}.

    \subsection{Form Factors analysis}\label{section-ff}
    
\begin{figure}[htbp]
\begin{center}\vspace*{0.02cm}
 \includegraphics[scale=0.327, angle=0, clip]{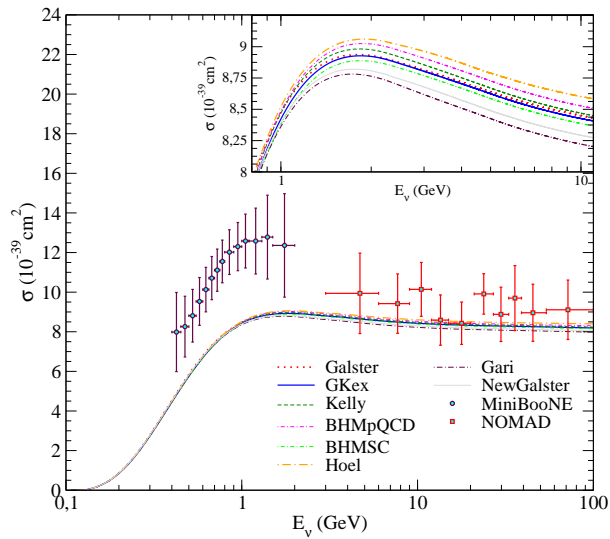} 
\end{center}\vspace*{-0.52cm}
\caption{ CCQE $\nu_\mu-^{12}$C cross section per nucleon evaluated in
  the SuSA model for various parametrizations of the nucleon
  electromagnetic form factors~\cite{Megias:2013aa}. A sub-panel
  zooming in the region near the maximum is inserted on the top. The
  MiniBooNE~\cite{AguilarArevalo:2010zc} and
  NOMAD~\cite{Lyubushkin:2008pe} data are also shown for reference.
\label{fig:emffnu}}
\end{figure}

    As already considered in the analysis of ($e,e'$) reactions, here
    we also adopt the electromagnetic nucleon form factors of the
    extended Gari-Krumpelmann (GKeX)
    model~\cite{Lomon1,Lomon2,Crawford1} for the vector CC current
    entering into neutrino cross sections. As described
    in~\cite{Megias:2017PhD}, this prescription improves the commonly
    used Galster parametrization at $Q^2>1$ GeV$^2$. For
    completeness, we show in Figure~\ref{fig:emffnu} the sensitivity
    of the total CCQE neutrino cross section within the SuSA approach
    for the different up-to-date parametrizations of the nucleon form
    factors (see Refs.~\cite{Megias:2013aa,PRRaul} for details): 
    all of them are essentially equivalent for the MiniBooNE
    kinematics, while some difference emerges at the energies of the
    NOMAD experiment, which implies larger $Q^2$ values. Similar
    comments also apply to the SuSAv2 model.
\vspace*{-0.104cm}

\begin{figure}\vspace{0.08cm}
	\begin{center}\vspace{0.80cm}
			\hspace*{-0.95cm}\includegraphics[scale=0.192, angle=270]{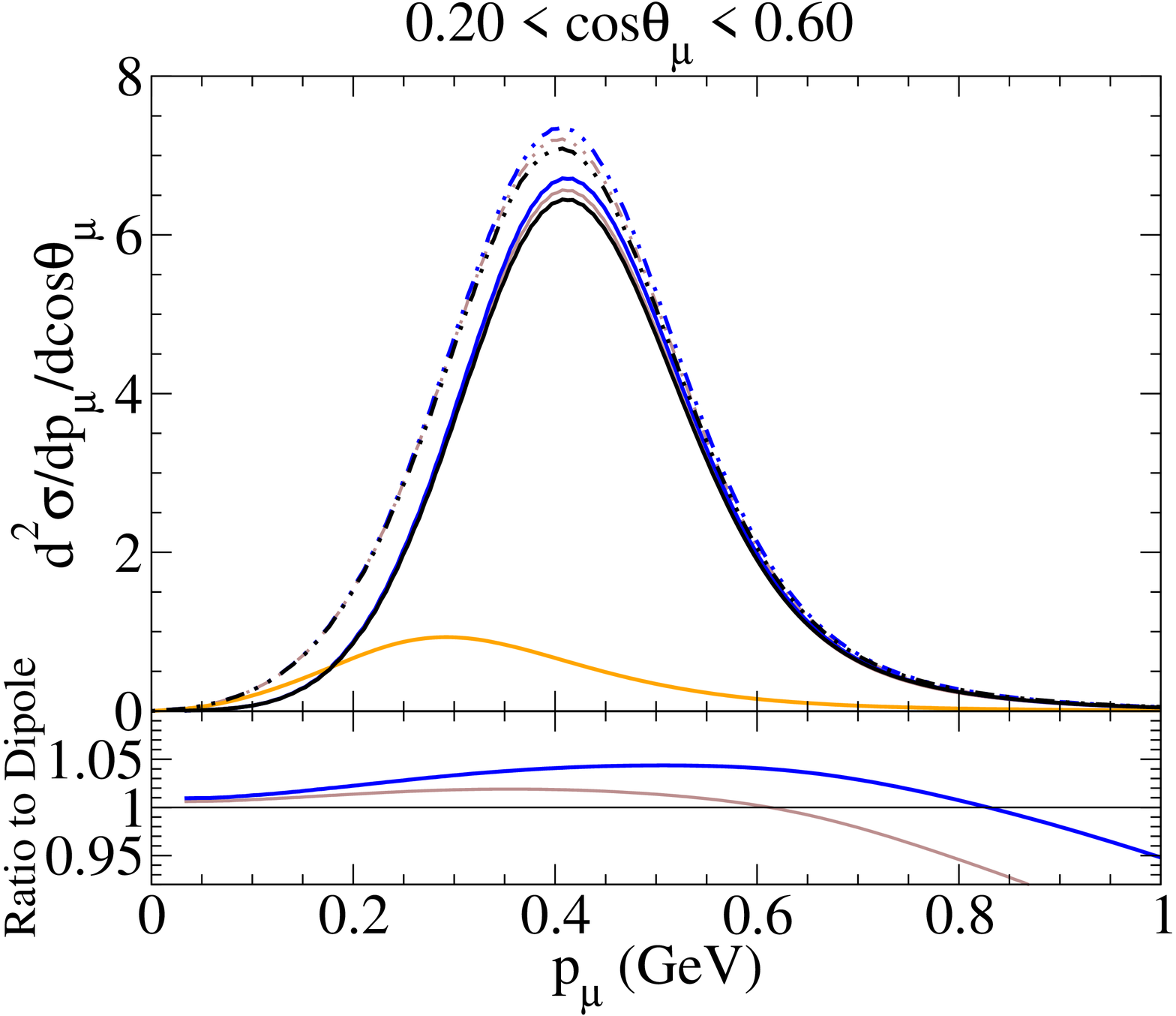}\hspace*{-0.295cm}%
		\includegraphics[scale=0.192, angle=270]{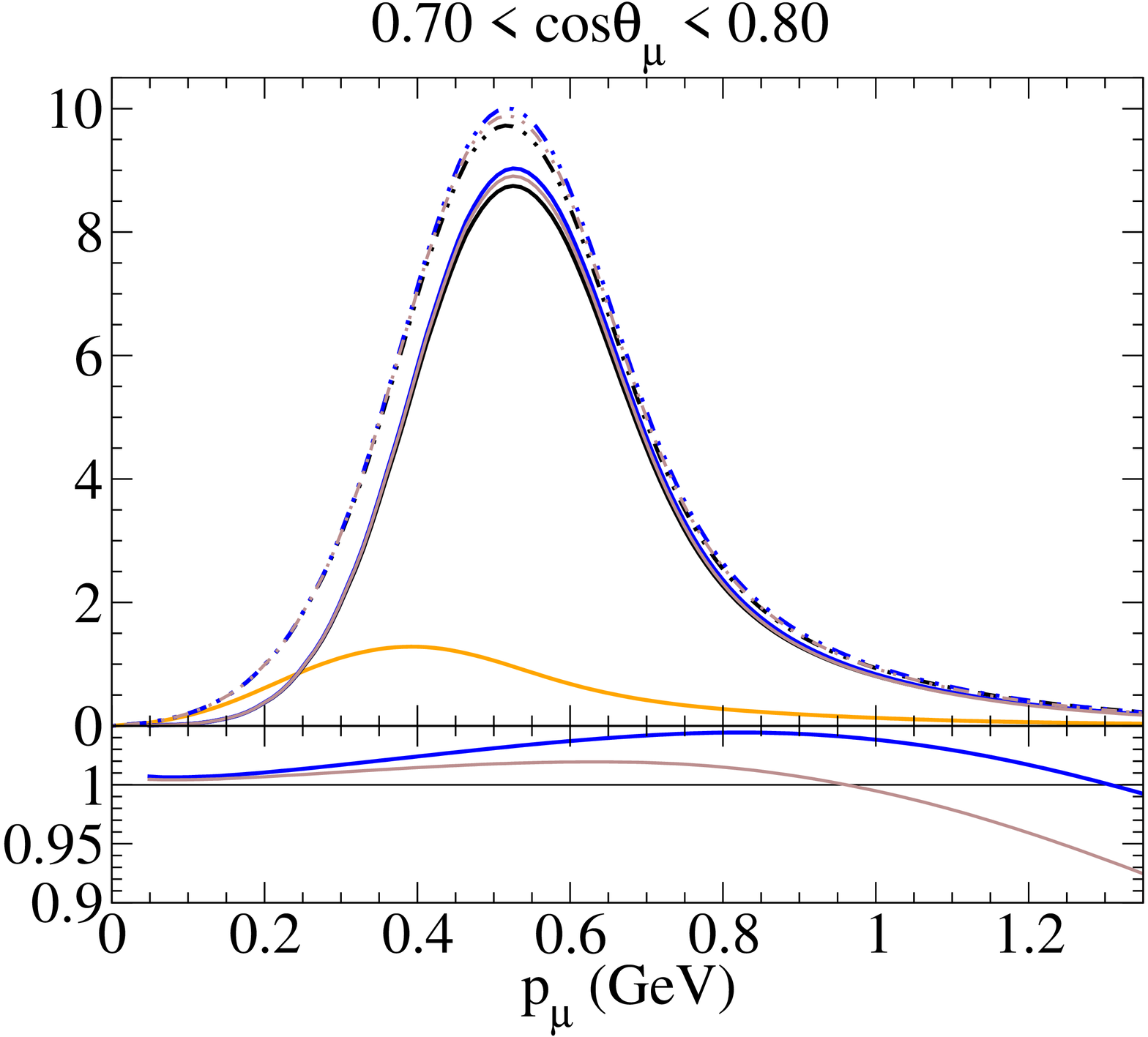}\hspace*{-0.495cm}%
		\includegraphics[scale=0.192, angle=270]{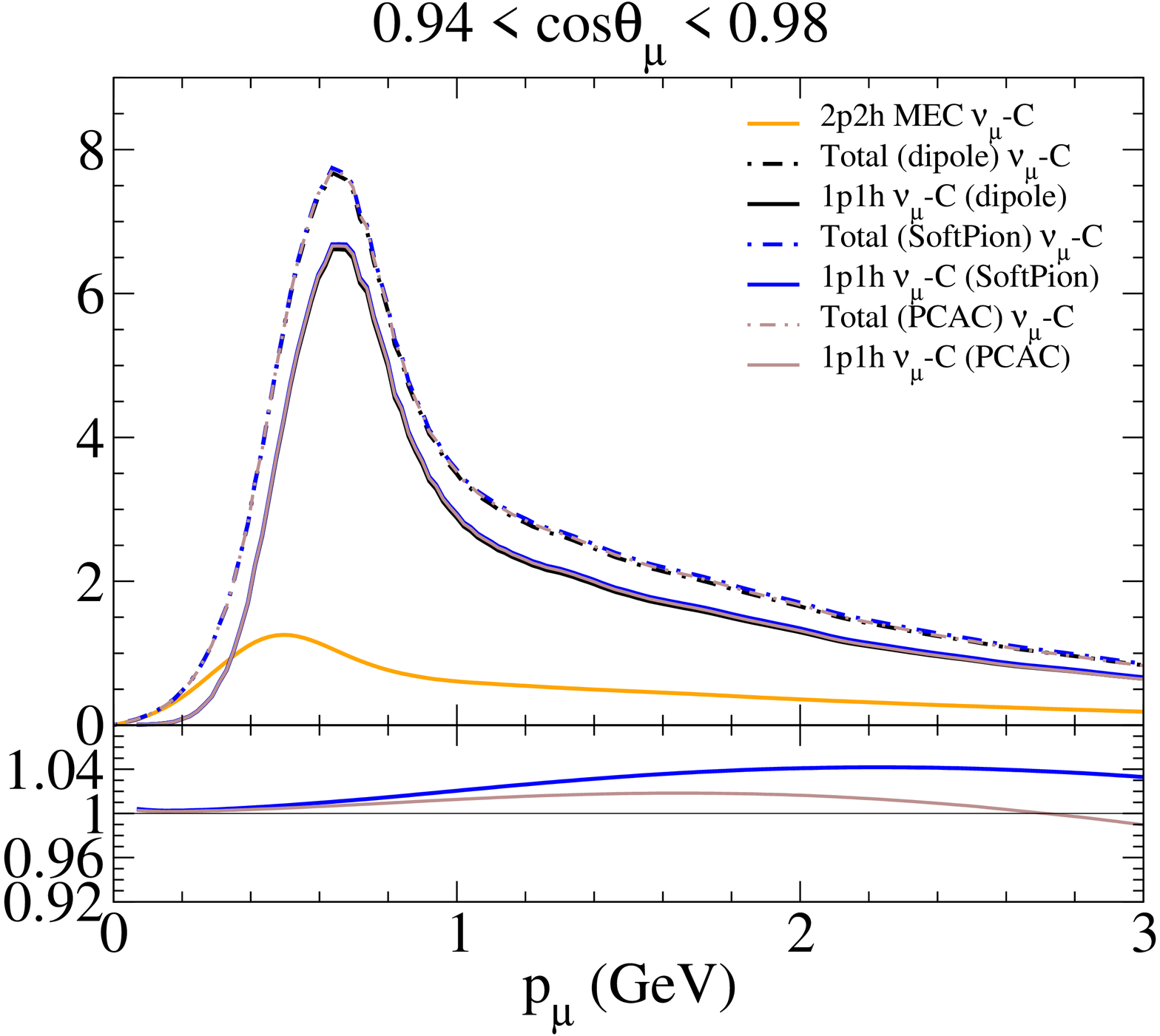}
		\\
		\hspace*{-0.95cm}\includegraphics[scale=0.192, angle=270]{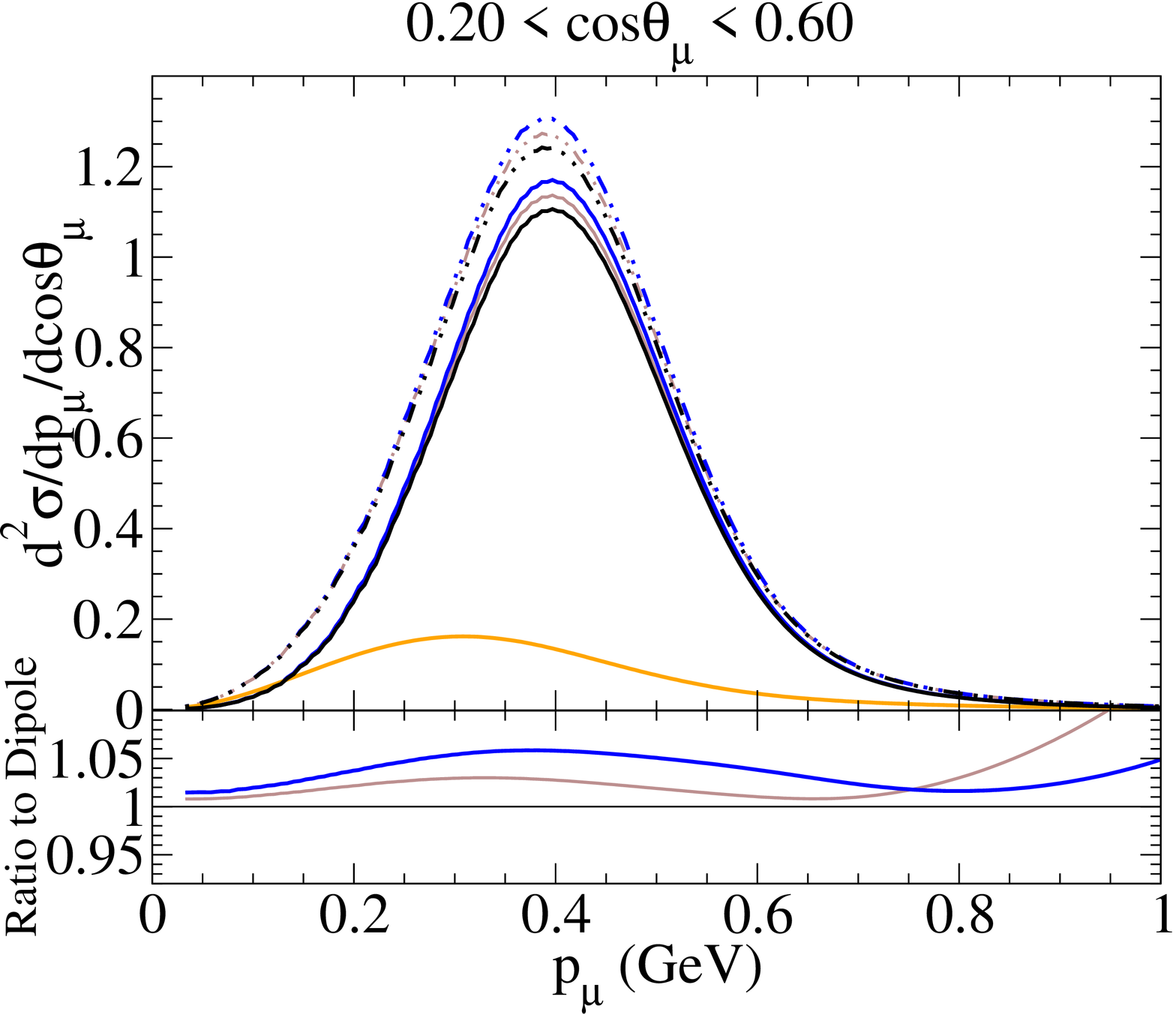}\hspace*{-0.295cm}%
		\includegraphics[scale=0.192, angle=270]{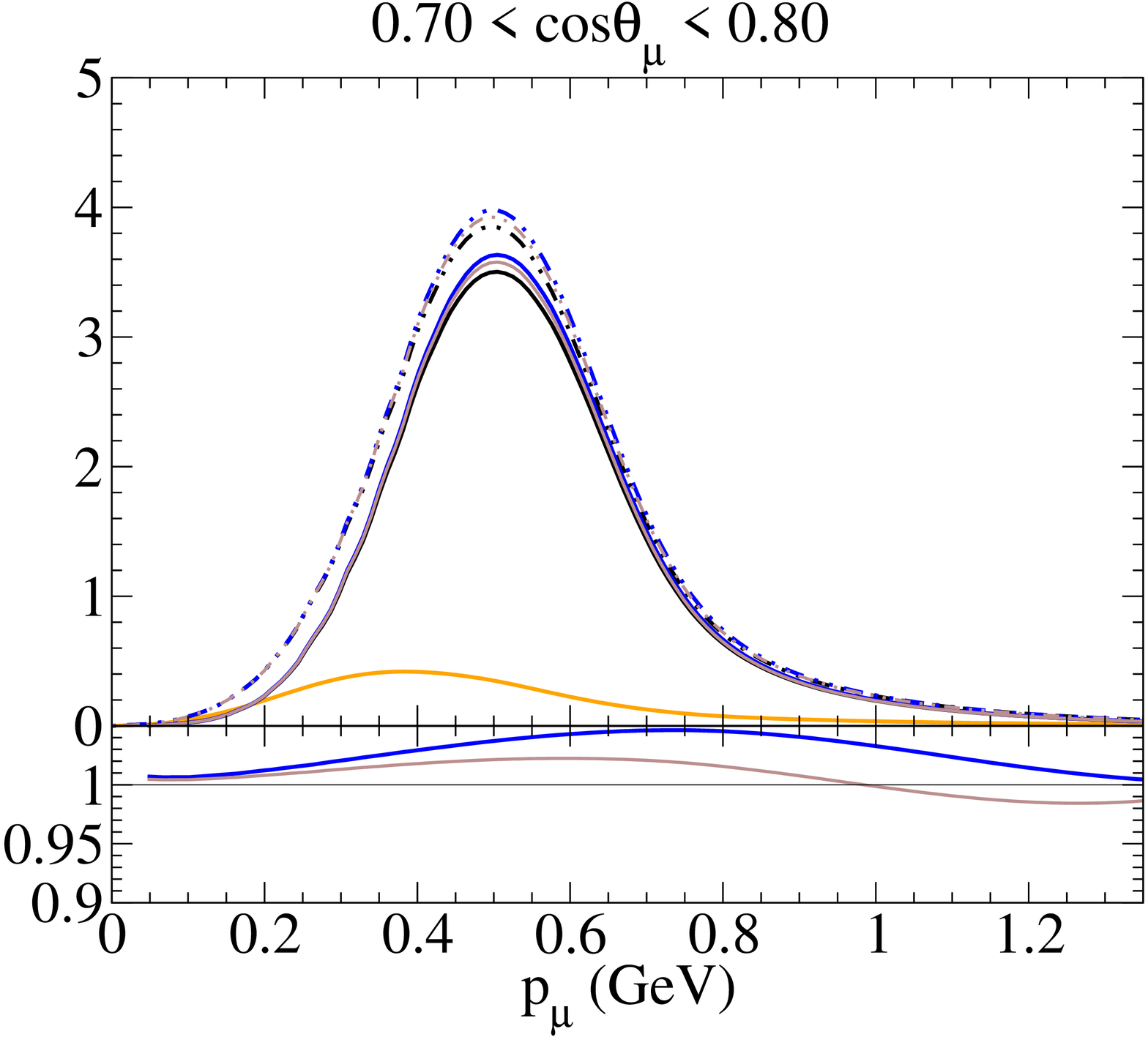}\hspace*{-0.495cm}%
		\includegraphics[scale=0.192, angle=270]{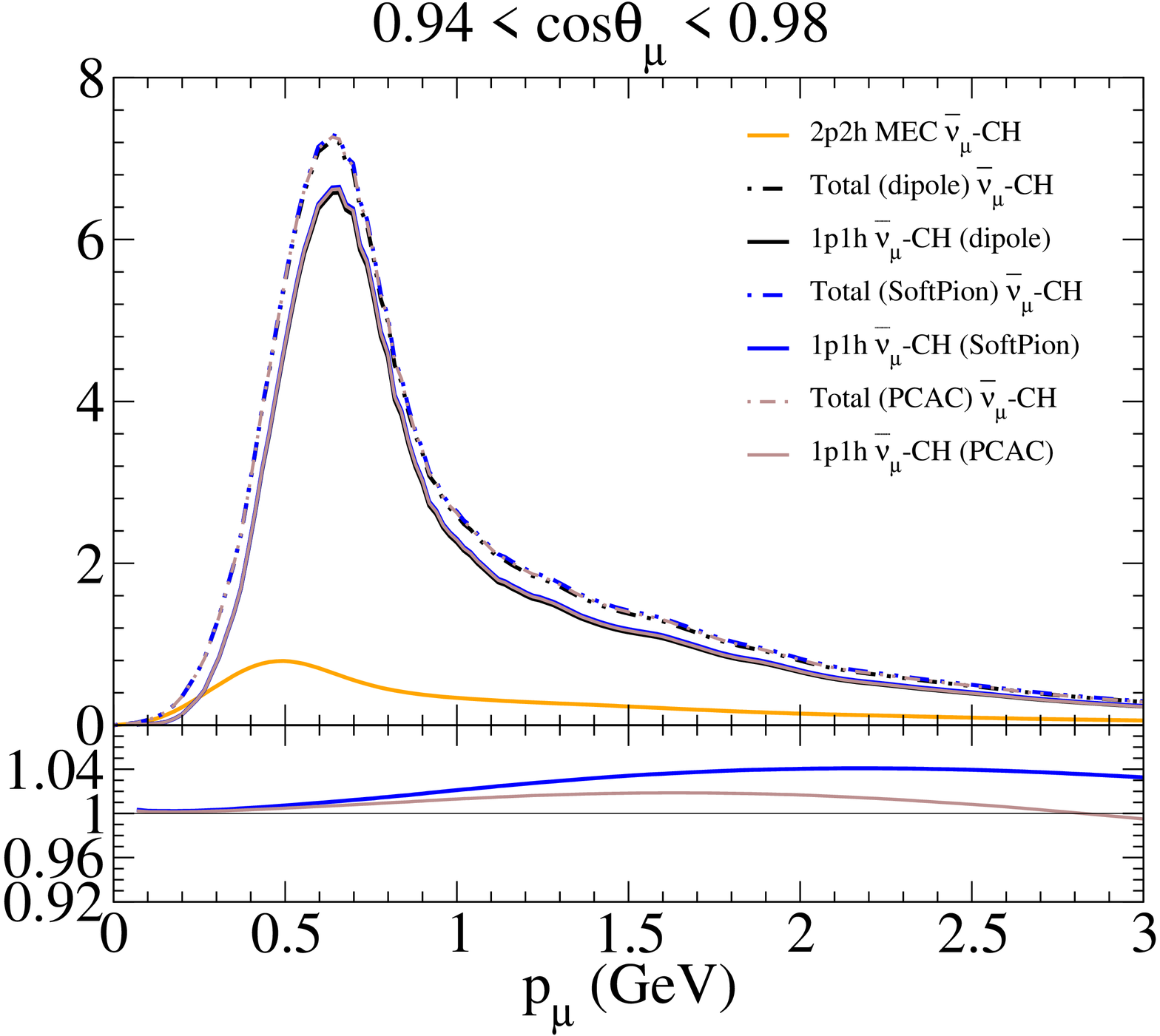}		
	\end{center}
	\caption{The T2K flux-integrated CCQE and 2p2h
          double-differential cross-section for neutrino (first row) and
          antineutrino (second row) 
          within the SuSAv2-MEC model, using different form
          factors. Results are displayed for different bins of the
          muon scattering angle as functions of the muon
          momentum. Double differential cross sections are shown in
          units of 10$^{-39}$ cm$^2$/GeV per nucleon.
        }\label{fig:T2K_d2s}
\end{figure}

\begin{figure}\vspace{0.08cm}
	\begin{center}\vspace{0.80cm}
		\hspace*{-0.95cm}\includegraphics[scale=0.192, angle=270]{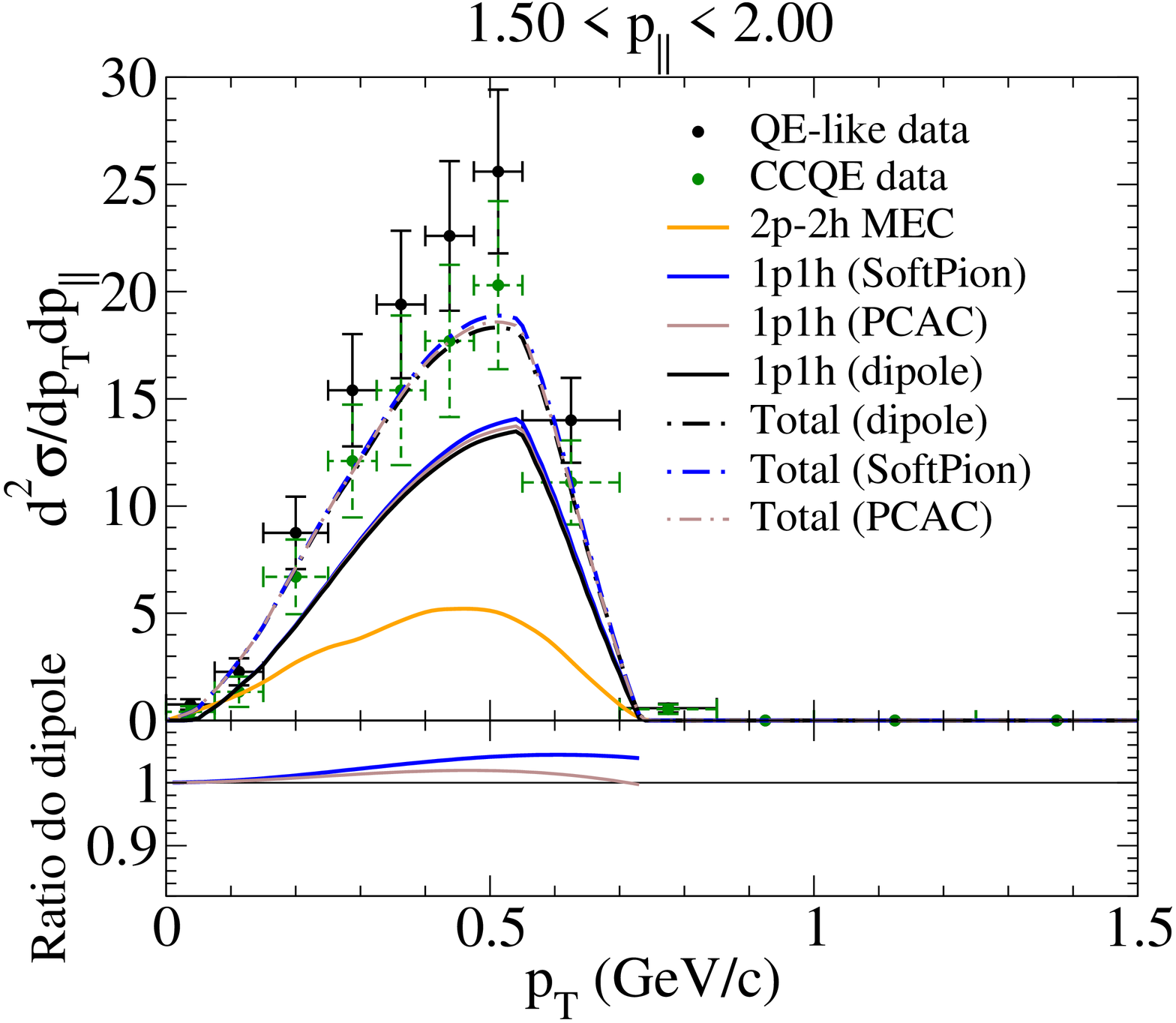}\hspace*{-0.495cm}%
		\includegraphics[scale=0.192, angle=270]{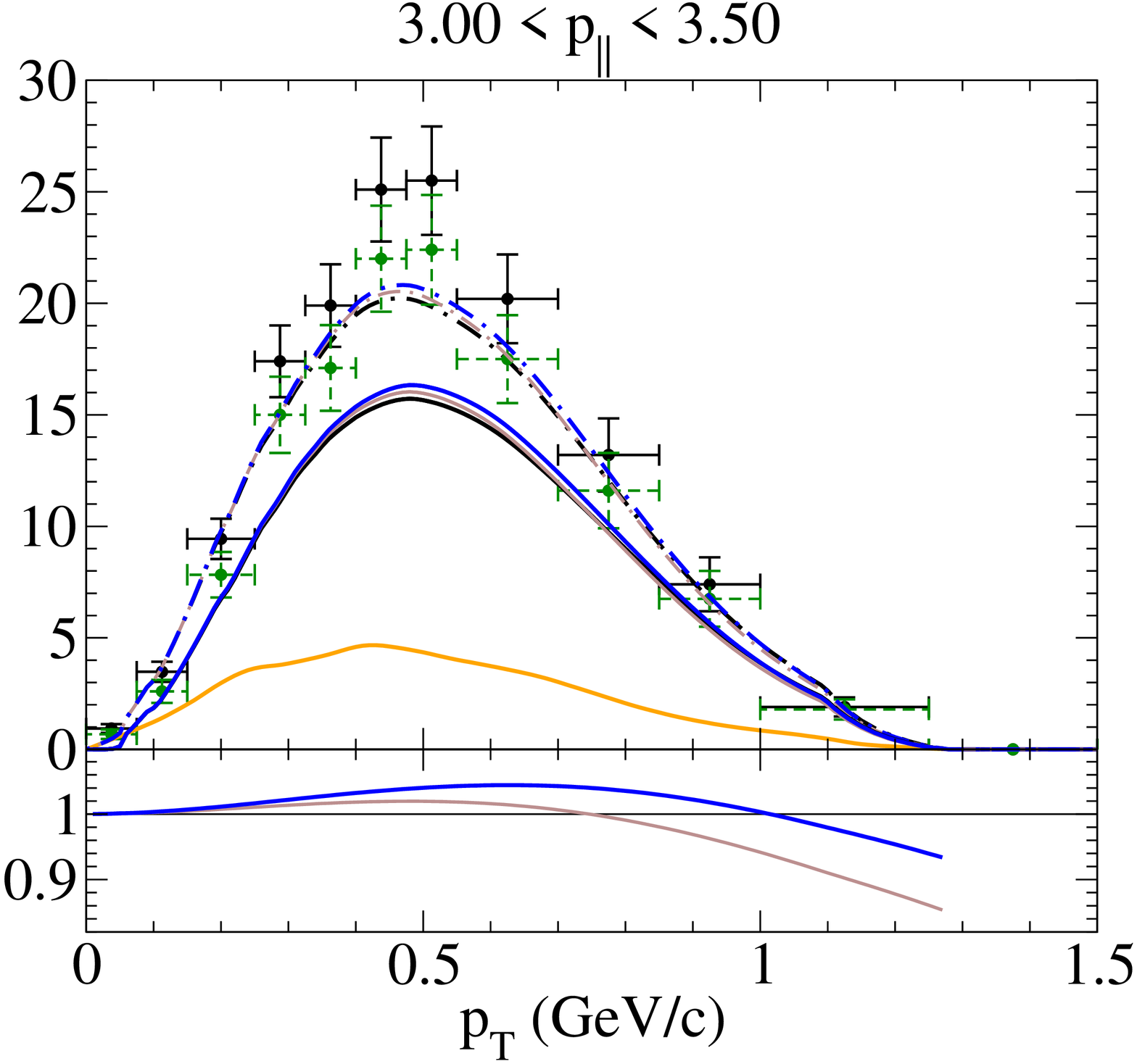}\hspace*{-0.495cm}%
		\includegraphics[scale=0.192, angle=270]{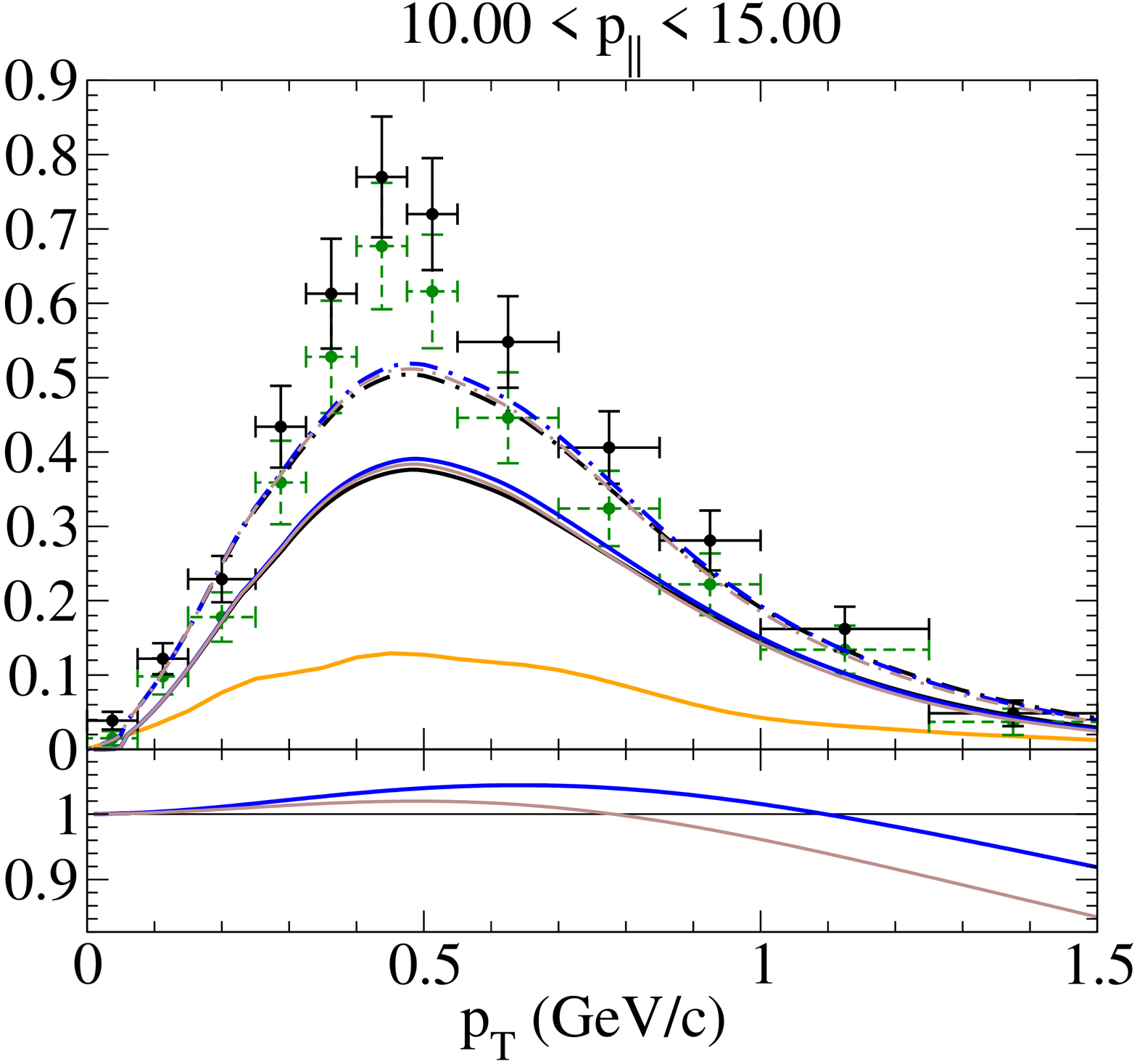}
		\\
		\hspace*{-0.95cm}\includegraphics[scale=0.192, angle=270]{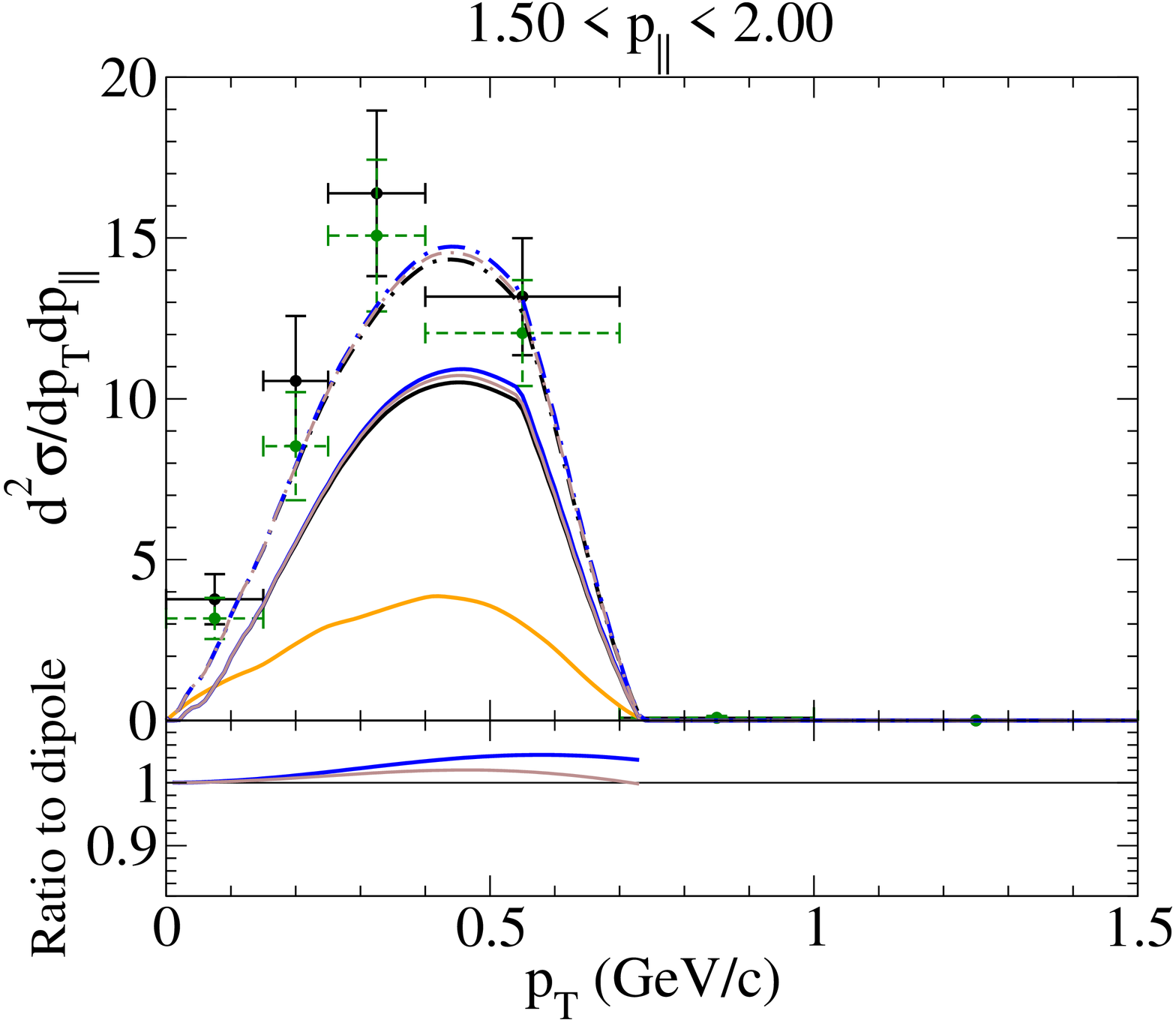}\hspace*{-0.495cm}%
		\includegraphics[scale=0.192, angle=270]{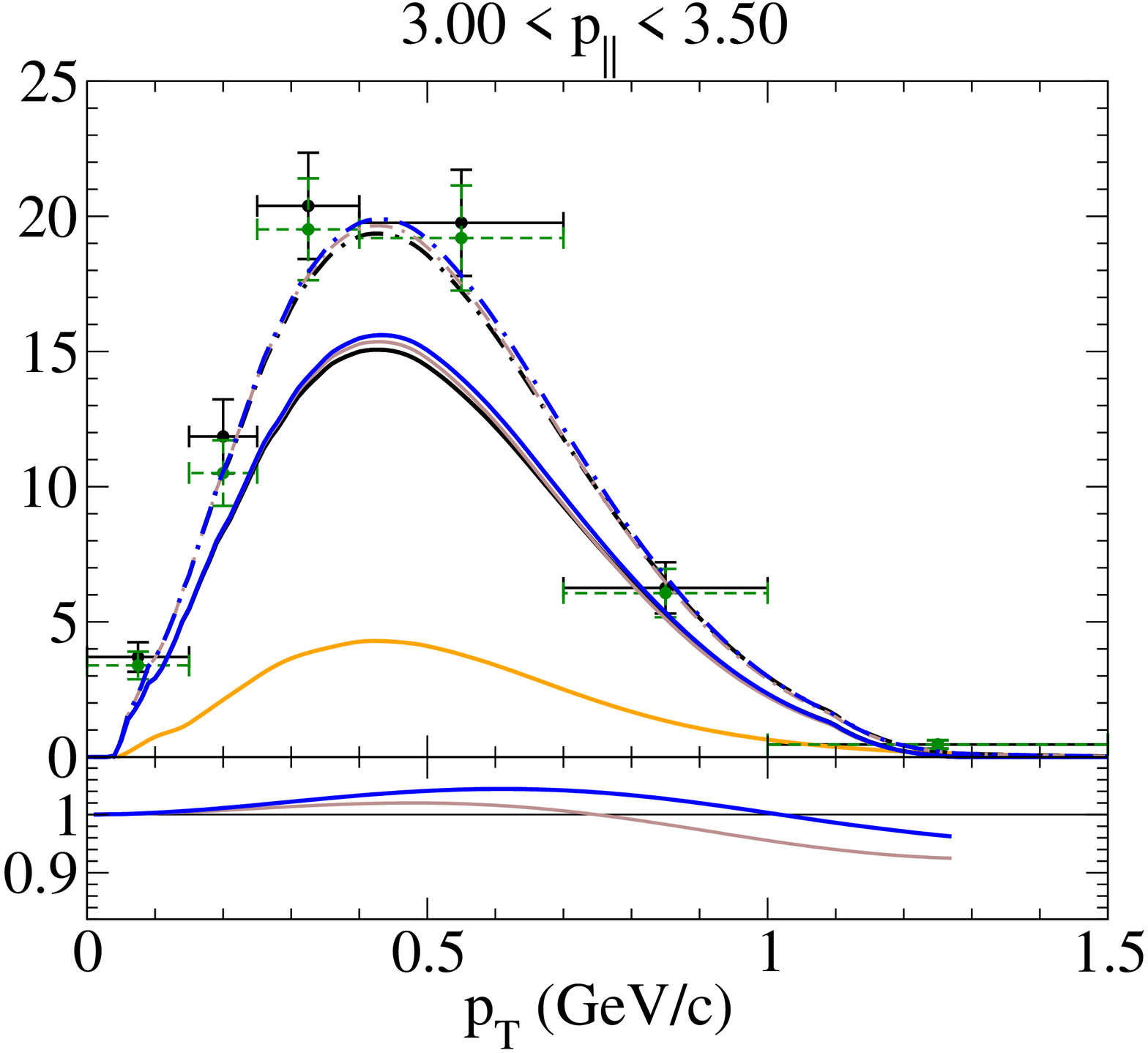}\hspace*{-0.495cm}%
		\includegraphics[scale=0.192, angle=270]{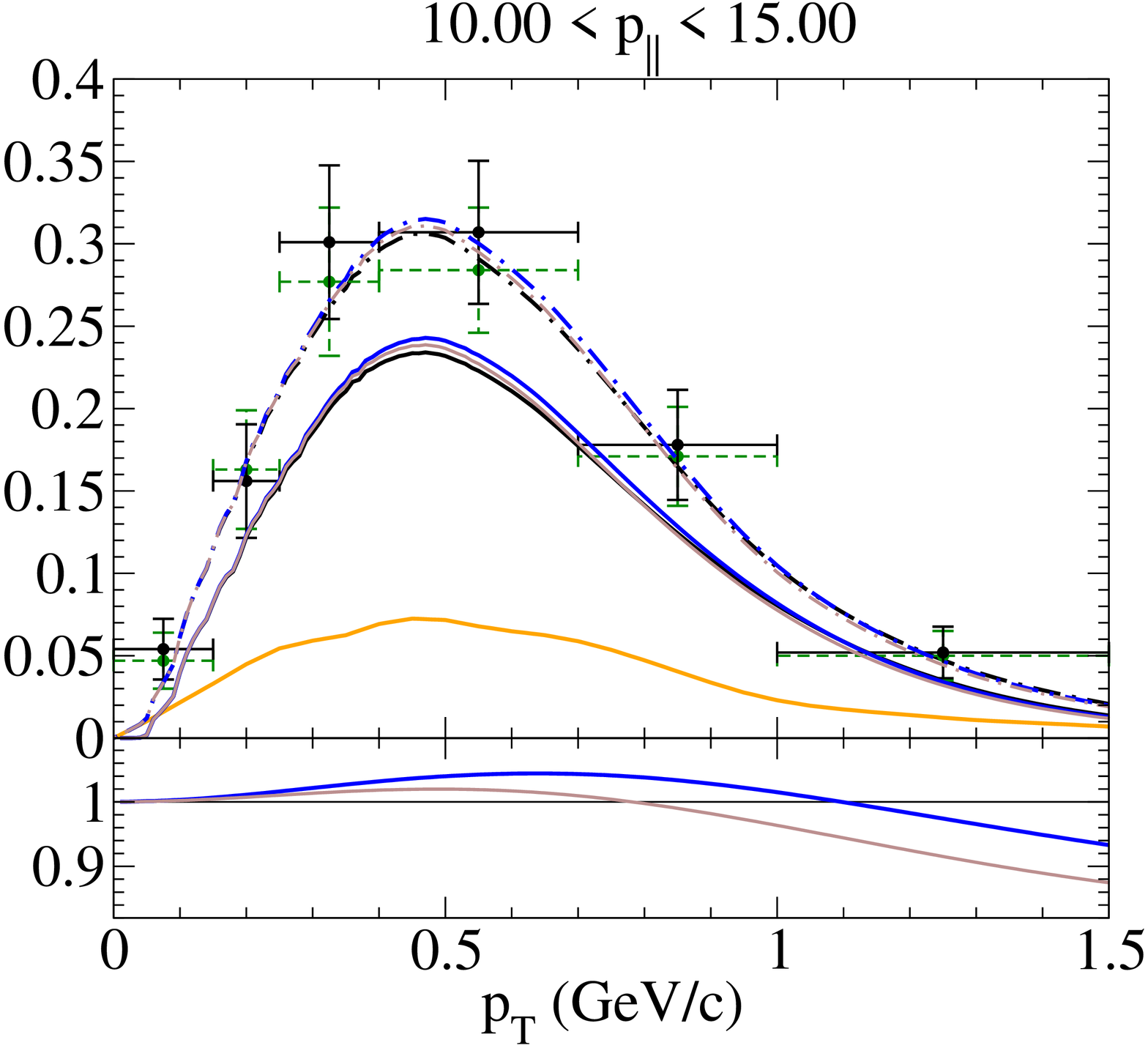}
	\end{center}
	\caption{MINERvA flux-integrated double-differential
          cross-section for different axial form factors, as a
          function of the muon transverse and longitudinal momentum,
          for neutrino (first row) and antineutrino (second row) double
          differential cross sections (10$^{-39}$
          cm$^2$/GeV$^2$ per nucleon).}
    \label{fig:Mnv_d2s}
\end{figure}

Regarding the axial contributions, we employ the commonly used dipole
axial nucleon form factor described in~\cite{Megias:2017PhD} where a
comparison between dipole and monopole axial form factor is shown
together with a discussion on the axial coupling $g_A$ parameter. More
recently, we have also compared the dipole axial form
factor with other choices based on the so-called  two-component
model~\cite{Iachello:1972nu,Bijker:2004yu}, consisting of a three-quark intrinsic structure
surrounded by a meson cloud. In \cite{Megias:2017PhD} a joint fit
to neutrino-nucleon scattering and pion electroproduction data has
been performed to evaluate the nucleon axial form factor in the
two-component model. Further constrains on the model are
obtained by re-evaluating the electromagnetic form factor using
electron scattering data. The results for this new fit of the axial form factor show
sizable differences at some kinematics with respect to the dipole model. The impact of such changes on the CCQE neutrino-nucleus
cross-section is evaluated in the SuSAv2 nuclear model in comparison
with recent T2K and MINERvA measurements in Figs.~\ref{fig:T2K_d2s}
and~\ref{fig:Mnv_d2s}, respectively.  The two-nucleon component approach was
successful in describing the EM form factors~\cite{Bijker:2004yu,Wan:2005ds,Iachello:2004aq}, the strange
form factors of the proton~\cite{Bijker:2005pe} and was applied to the
deuteron as well~\cite{TomasiGustafsson:2005ni}.  Advantages of this
model are that it contains a limited number of parameters and can be
applied both in the space- and time-like regions. The extension to
axial form factors has been done in \cite{Adamuscin:2007fk}.

Following Ref.~\cite{Iachello:1972nu}, the axial nucleon form factor
is parametrized in the two-component approach as:
\begin{eqnarray}
G_A(Q^2) &=& G_A(0) \, g(Q^2) \left[ 1-\alpha +\alpha \frac{m_A^2}{m_A^2+Q^2} \right] ~, 
\nonumber\\
g(Q^2) &=& \left (1+\gamma Q^2\right )^{-2} ~,
\label{eq:eq1}
\end{eqnarray}
where $Q^2>0$ in the space-like region and $\alpha$ is a fitting
parameter which corresponds to the coupling of the photon with an
axial meson.  One can fix $m_A= 1.230$ GeV, corresponding to the mass
of the axial meson $a_1(1260)$ with $I^G(J^{PC})=1^-(1^{++})$. The
form factor $g(Q^2)$ describes the coupling to the intrinsic structure
(three valence quarks) of the nucleon. Note that setting $\alpha=1$
and $\gamma=0$, the usual dipole axial functional form is recovered.

The results of fit to pion electroproduction data and neutrino
scattering data for different extractions of the pion data can be seen
in~\cite{megias2019new}. Here we will focus on the two most extreme
cases, the Soft Pion dataset and the PCAC (Partially Conserved Axial
Current) one, also detailed in~\cite{megias2019new}.  The impact
of the axial form factor choice on the CCQE-like cross section is
illustrated in Figs.~\ref{fig:T2K_d2s} and~\ref{fig:Mnv_d2s} for T2K
and MINERvA kinematics, respectively.

The largest differences
between neutrino cross-sections evaluated with different form factors
is about 5\%. By mapping the muon kinematics ($p_\mu, \theta_\mu$)
into $Q^2$ on the basis of~\cite{megias2019new}, the largest
difference with respect to the cross-section with dipole form factor
appears always in correspondence of $Q^2\sim$ 0.5~GeV$^2$ for
Soft Pion.  For forward angles, the relevant kinematic region is
$Q^2\sim 0.5$~GeV$^2$ where the cross-section is mostly unaffected by
form factor differences. For backward angles, the $Q^2\sim$
0.5~GeV$^2$ region corresponds instead exactly to the region of larger
cross-section with intermediate muon momentum, thus the impact of form
factors difference is larger, as detailed on~\cite{megias2019new}. In
the backward region, the form factor differences can reach 5\%. In
such region the effect in the antineutrino case is even larger, up to
10\%. Still, in the neutrino-antineutrino asymmetry the effect is at
\% level (see~\cite{megias2019new} for a detailed analysis of the $\nu-\bar\nu$ asymmetry).  The region $Q^2>1$~GeV, where the different axial form
factors depart from each other sizeably, is negligible in T2K data.

The effects of axial form factors at MINERvA kinematics are shown via
the analysis of the double differential cross-section as a function of
$p_T, p_L$ in Fig~\ref{fig:Mnv_d2s}. The region of $Q^2 \simeq
0.5$~GeV$^2$ shows differences of the order of 5\%, similarly to T2K,
while in the region of high $p_T$ and lower cross-section effects up
to 10\% and above can be observed.

Overall the agreement of the SuSAv2-MEC model with data, using the
different axial form factors, is positive but the experimental
uncertainties on T2K and MINERvA measurements do not allow yet to
clearly discriminate between the various evaluations. It is
interesting to note that, in the model considered here, the form
factor effects have a different $Q^2$ dependence than the one of 2p2h,
as well as a different neutrino/antineutrino dependence, making the
disentagling of nucleon and nuclear effects feasible in future with
higher statistics measurements. The feasibility of this approach
relies on the capability of exploiting external data to drive the
$Q^2$ dependence of the form factor. For this reason, the
investigation of the earlier data of pion electro-production, as shown
in~\cite{megias2019new}, is of primary importance.

\subsection{Relevance of $L/T$ channels for neutrino reactions}
In this section we study in detail the relevance of the different
longitudinal and transverse channels that contribute to the QE and
2p-2h MEC at MiniBooNE kinematics, also accounting for the
corresponding axial and vector contributions which arise from the
hadronic currents. The conclusions extracted from this analysis are
roughly extensible to the ones from the T2K and MINERvA experiments,
whose relevant kinematic regions are not significantly different.

\begin{figure}
\begin{center}\vspace*{-0.204cm}
\includegraphics[scale=0.225, angle=270]{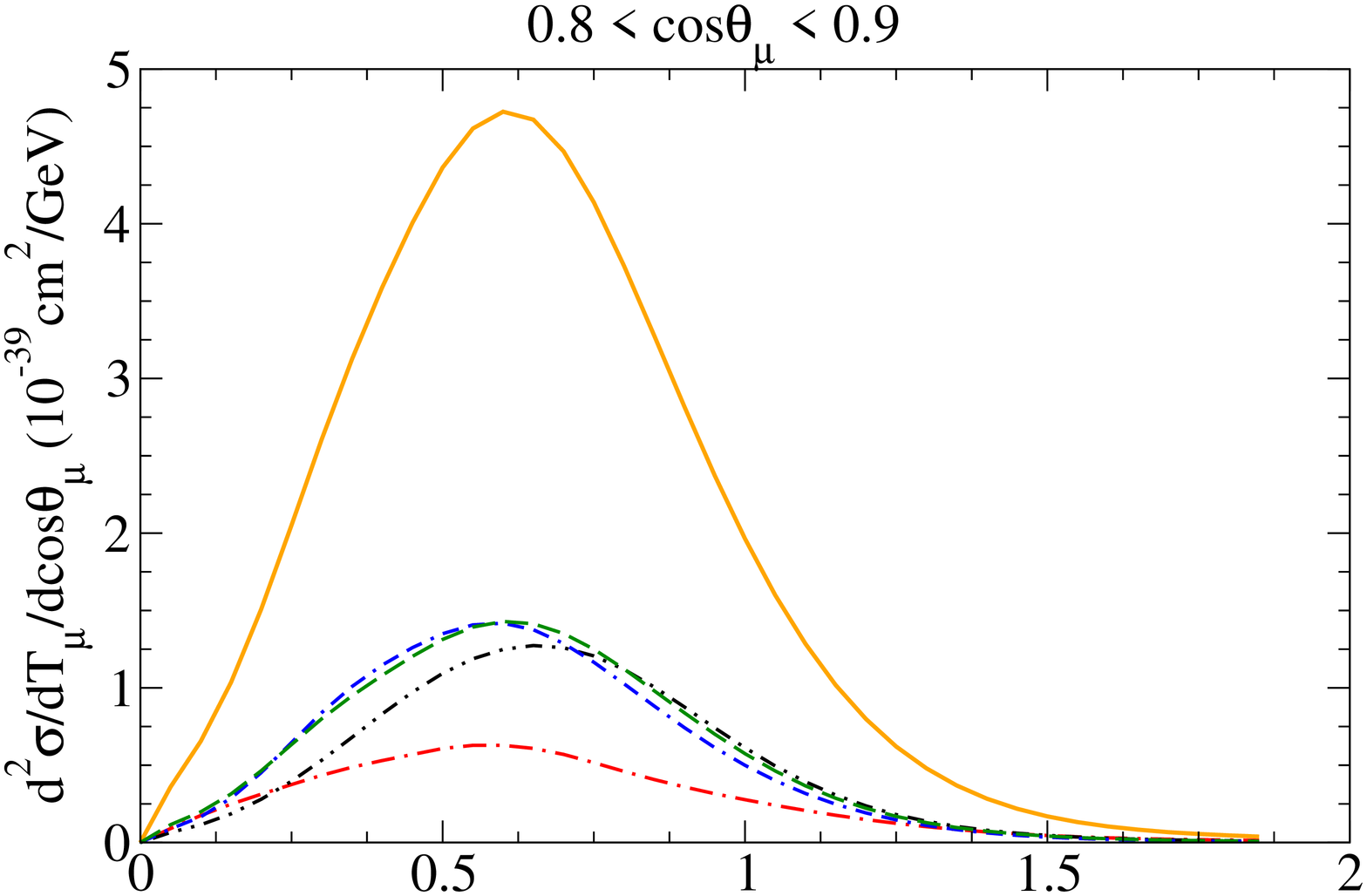}\hspace*{0.0415cm}\includegraphics[scale=0.225, angle=270]{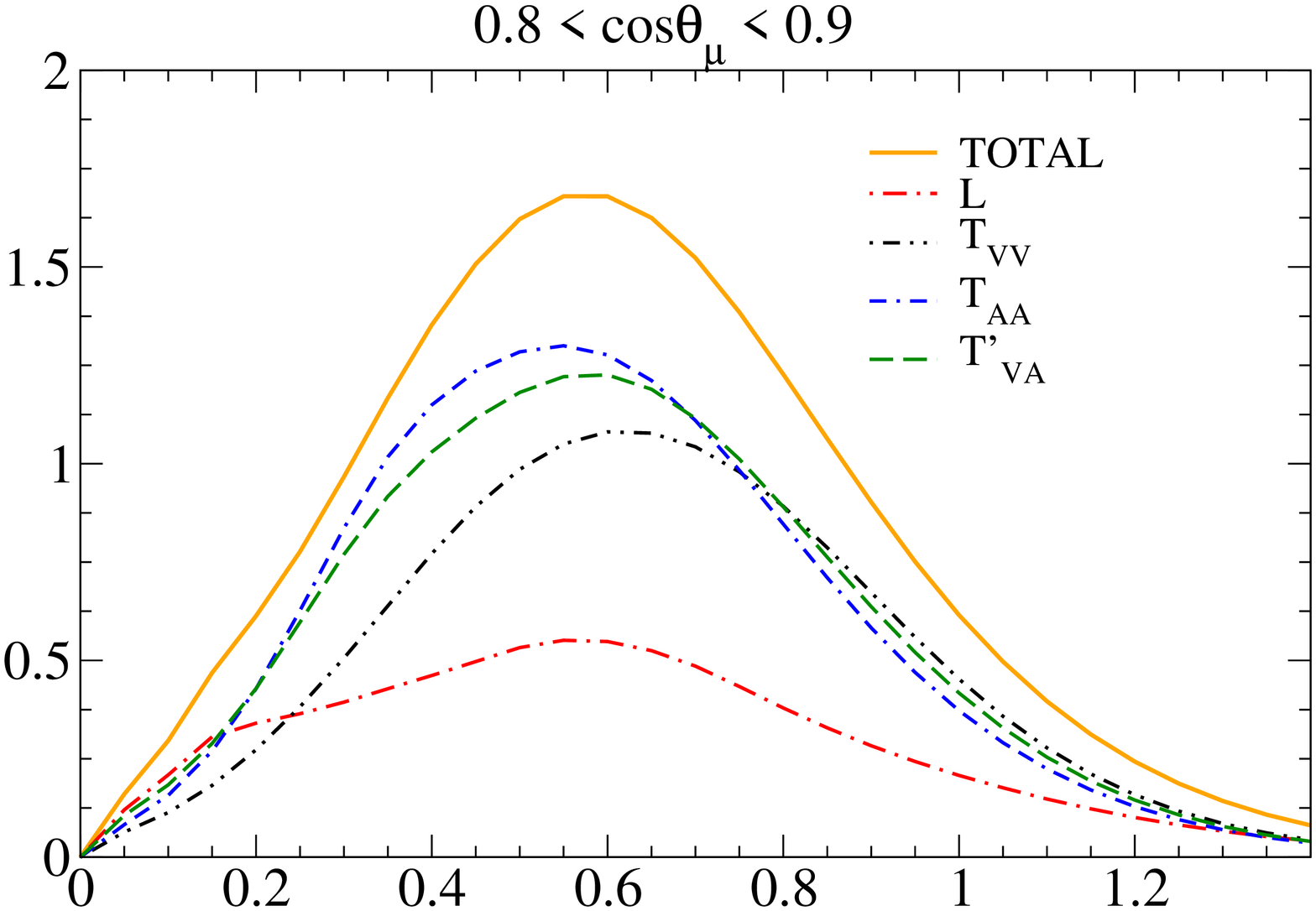}\hspace*{0.0415cm}\\\vspace*{-0.498cm}
\includegraphics[scale=0.225, angle=270]{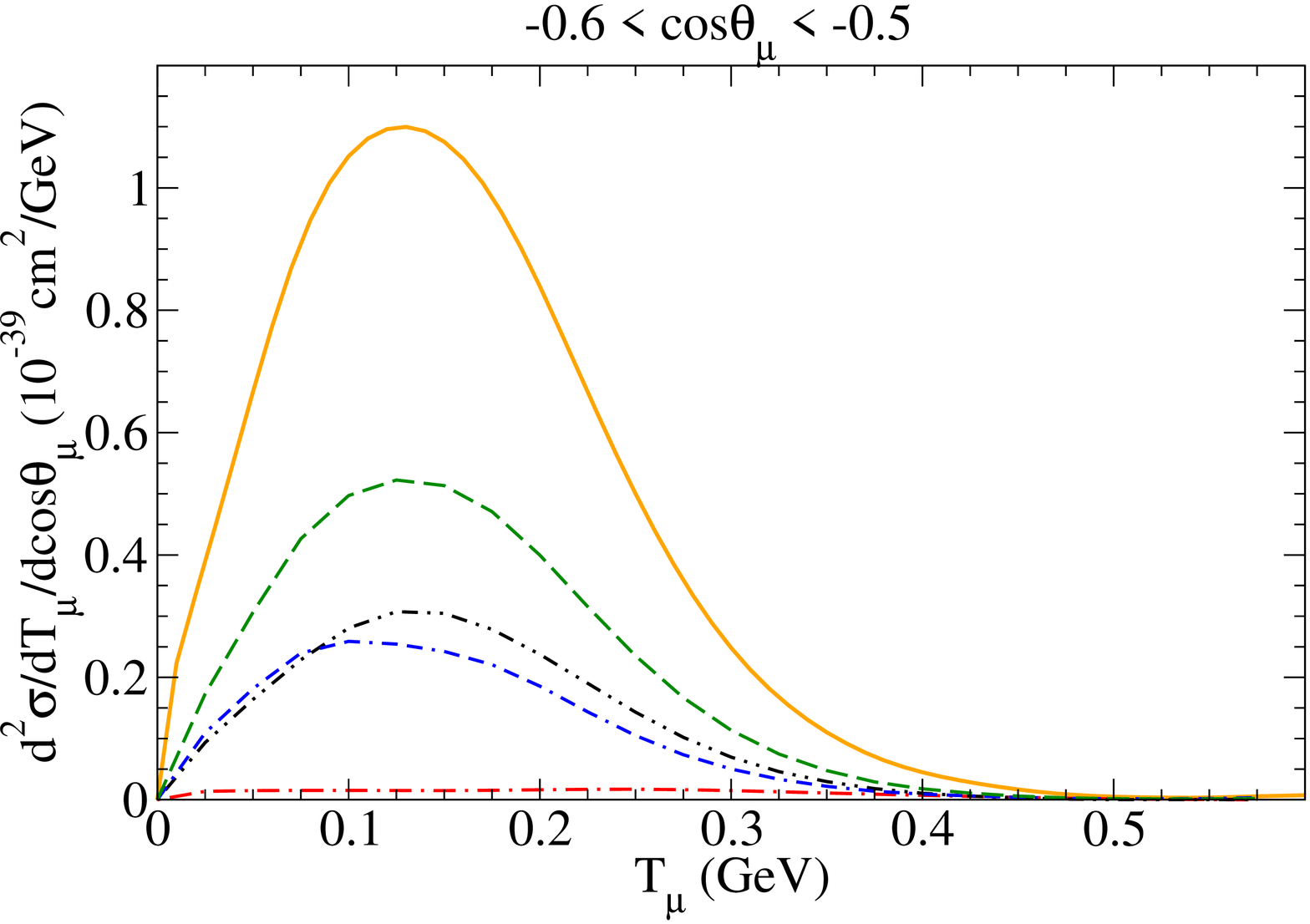}\hspace*{0.0415cm}\includegraphics[scale=0.225, angle=270]{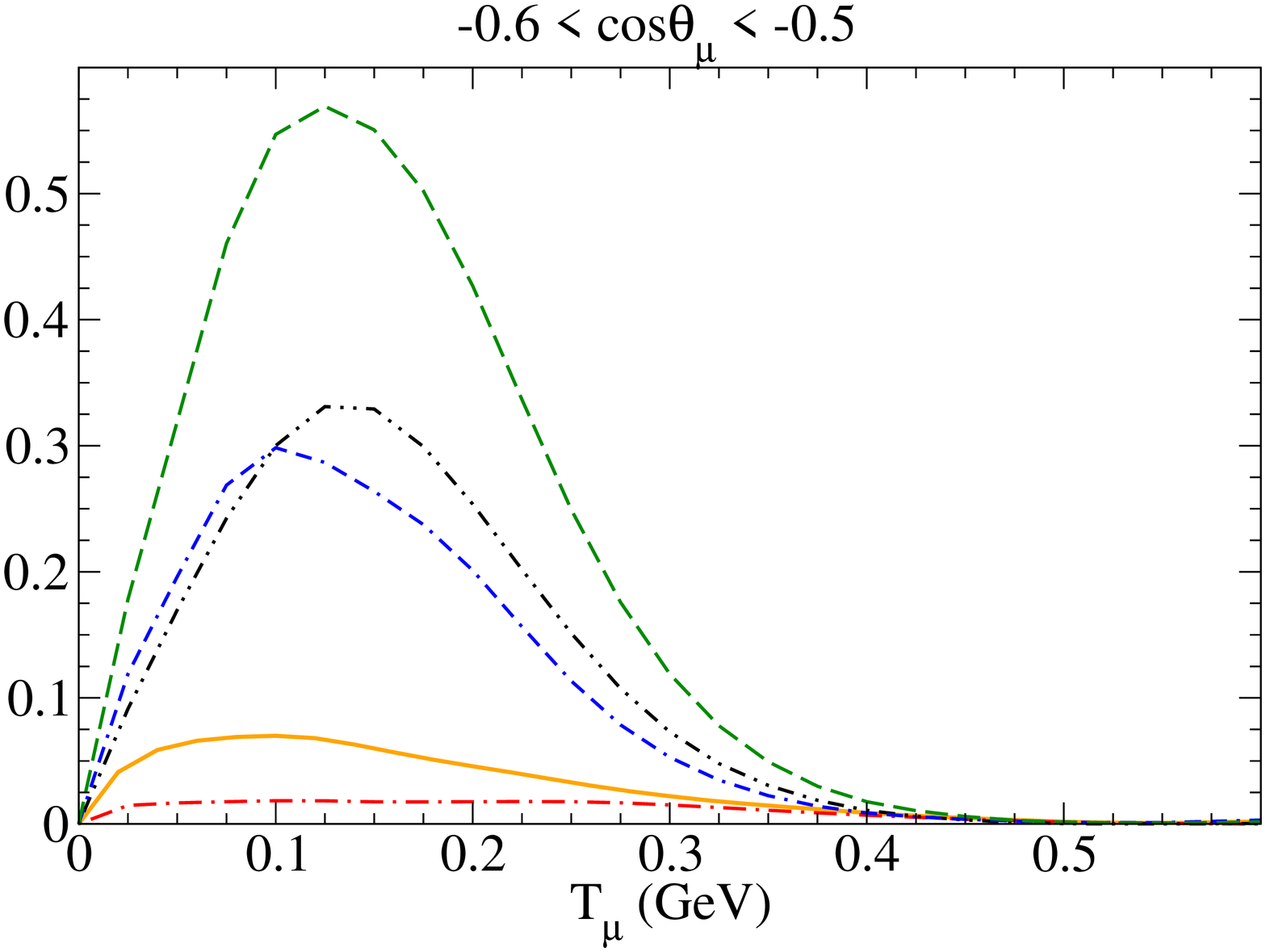}\hspace*{0.0415cm}\\
\end{center}\vspace*{-0.475cm}
\caption{ Comparison of the different 2p-2h MEC channels for the
  $\nu_\mu$ (left panels) and $\overline\nu_\mu$ (right panels)
  MiniBooNE double differential cross
  section.}\label{Miniboone_components}
\end{figure}

An analysis on the different channels for the 2p-2h MEC nuclear
responses and the total cross section was addressed
in~\cite{Megias:2016nu,Megias:2017PhD}, showing a predominance of the
transverse responses over the longitudinal ones. In the latter the
pure vector contributions were negligible in comparison with the axial
ones. Moreover, the separate transverse channels, $T_{VV}, T_{AA}$ and
$T'_{VA}$, while showing some remarkable differences for different $q$
values, contribute in a similar way to the total cross section. This
is due to the relevant kinematic regions explored which goes from 0.3
GeV/c up to 1 GeV/c in $q$ and from 0.3 GeV to 0.8 GeV in
$\omega$~\cite{Megias:2016nu,Megias:2017PhD}. Here we investigate the
relevance of the different channels at kinematics relevant for
MiniBooNE, T2K and MINERvA experiments. In
Fig.~\ref{Miniboone_components} the separate 2p-2h MEC contributions
to the different channels ($L, T_{VV}$, $T_{AA}$ and $T'_{VA}$)
corresponding to the MiniBooNE double differential cross section at
different bins of the muon scattering angle are presented.

\begin{figure*}\vspace*{0.24cm}
	\begin{center}
		\hspace*{-0.295cm}\includegraphics[scale=0.192, angle=270]{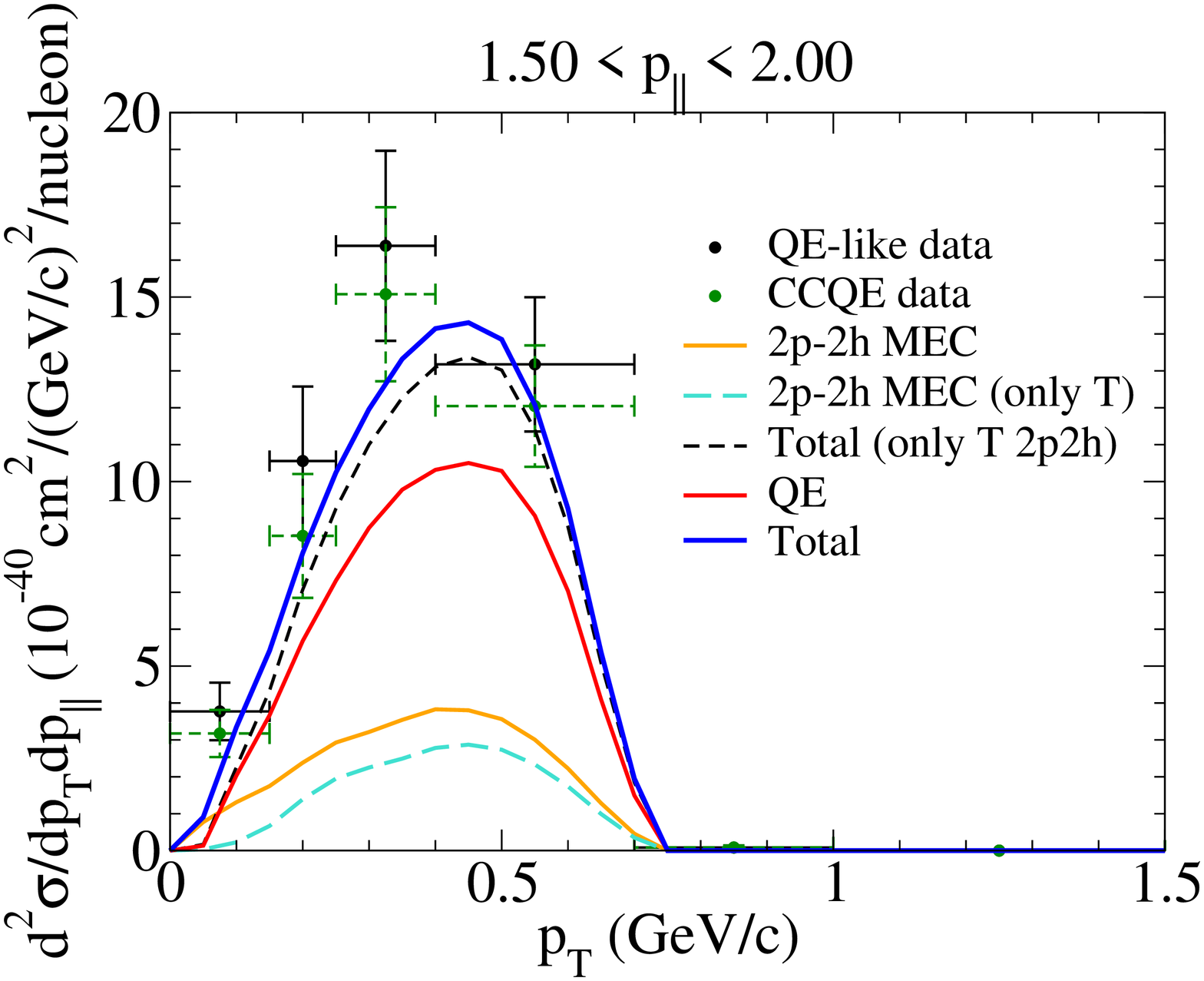}
		\hspace*{-0.65cm}\includegraphics[scale=0.192, angle=270]{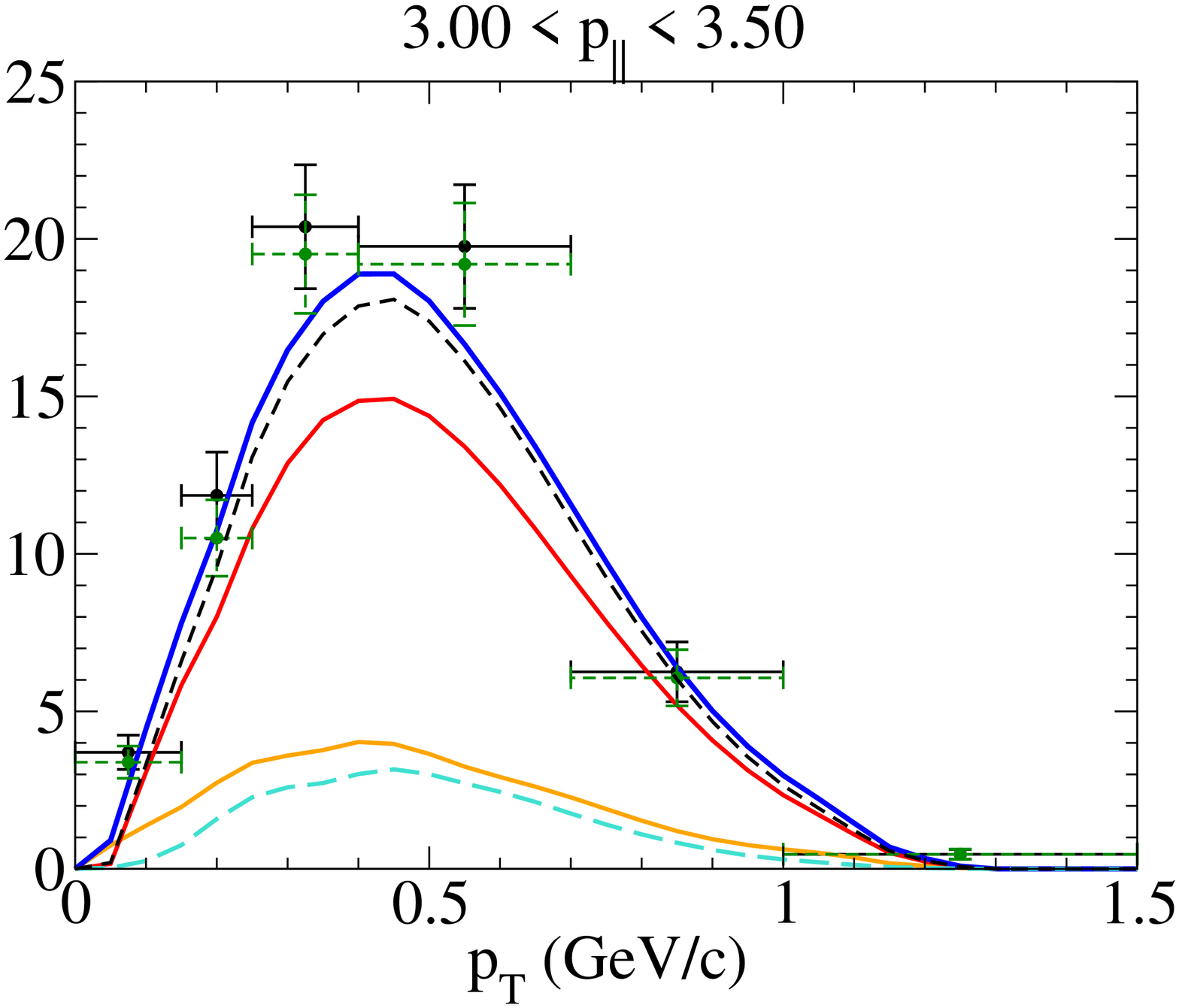}
		\hspace*{-0.65cm}\includegraphics[scale=0.192, angle=270]{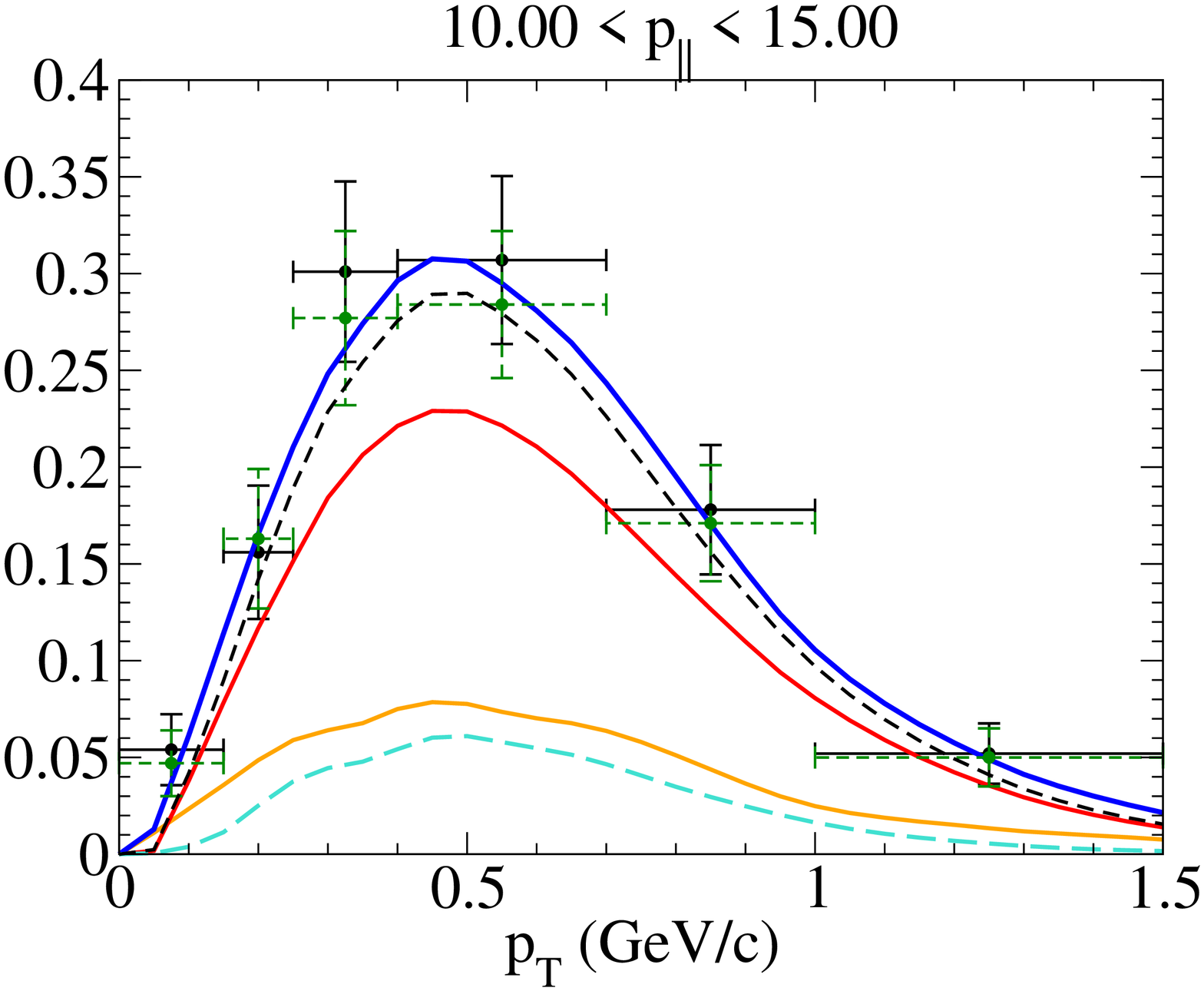}
	\end{center}\vspace*{-0.45cm}
	\caption{(Color online) As Fig.~\ref{fig:fig1mnv}, but showing the separate
          contribution of the pure transverse MEC (dashed curves) to
          also stress the relevance of the longitudinal MEC channel. }
\label{fig:fig2mnv}
\end{figure*}

Results in Fig.~\ref{Miniboone_components} show the differences
between the $T_{AA}$ and $T_{VV}$ contributions, the latter being
shifted to higher $T_\mu$ values by about 50 MeV for all angular
bins. At very forward angles, {\it i.e.}, lower $q$-values, the global
magnitude of the $AA$ channel is greater than the $VV$ one, in
accordance with the results observed in~\cite{Megias:2017PhD}, where
the $T_{VV}$ and $T_{AA}$ responses differ roughly by a factor 2 at
the maximum at $q$ of the order of 400 MeV/c. Concerning the
interference $T'_{VA}$ component, its magnitude is not particularly
different from the $VV$ and $AA$ ones at very forward angles, being on
the contrary the most relevant term at larger angles. Finally,
although the longitudinal channel gives the smallest global
contribution, its role is essential in order to interpret antineutrino
scattering at backward angles. This is a consequence of the negative
$T'_{VA}$ term for antineutrino reactions that almost cancels out the
$T_{VV}+T_{AA}$ contribution. The conclusions that can be extracted
from these results launch a warning to those 2p2h models which neglect
the longitudinal contributions in their analysis as well as to those
ones who extrapolate the results from electrons (purely vector
responses) to the weak ones (vector+axial), making those approaches
rather questionable in some cases. The same conclusions apply to the
recent MINERvA results shown in Fig.~\ref{fig:fig2mnv}.

\begin{figure}
\begin{center}\vspace*{-0.3498cm}
\includegraphics[width=15cm,bb=147 433 500 658]{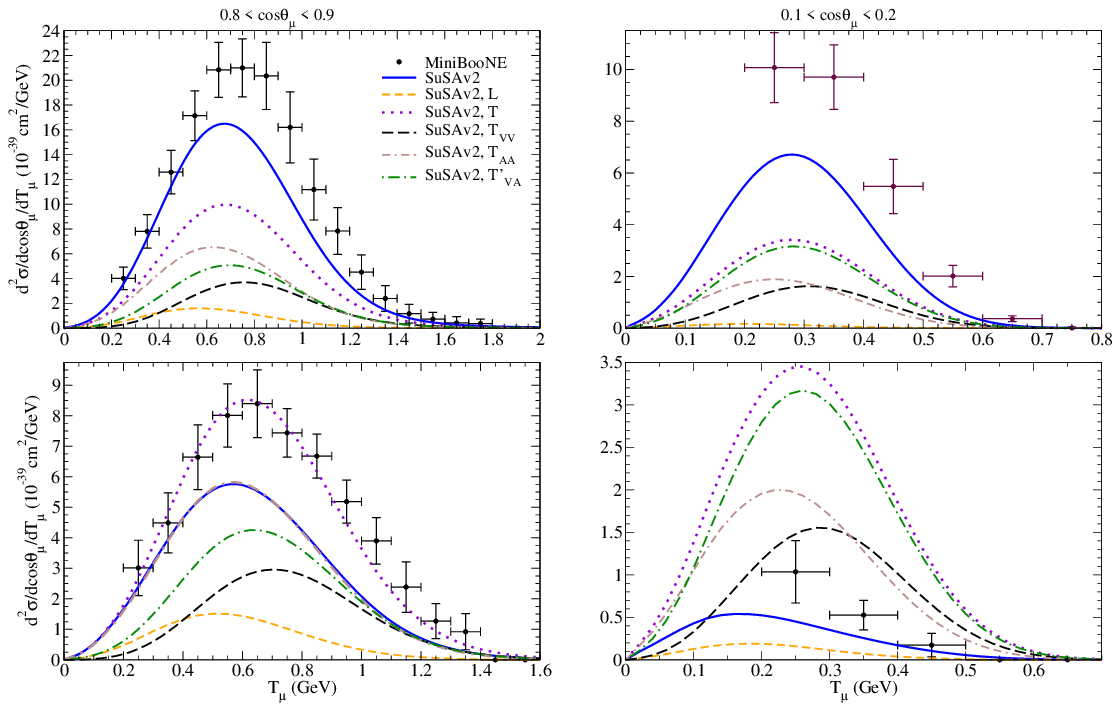}
\caption{ Separation into components of the MiniBooNE CCQE $\nu_\mu$
  (top panel) and $\overline\nu_\mu$ (bottom panel)
  double-differential cross section per nucleon displayed versus
  $T_\mu$ for various bins of $\cos\theta_\mu$ within the SuSAv2
  approach. The MiniBooNE data
  \cite{AguilarArevalo:2010zc,AguilarArevalo:2013hm} are also shown
  for reference.  \label{fig14ch}}
\end{center}\vspace*{-0.298cm}
\end{figure}

The conclusions extracted from the previous analysis on the 2p-2h MEC
cross section also apply for the separate QE contributions to neutrino
and antineutrino cross sections. The different QE channels are
analyzed for the MiniBooNE double differential cross sections in
Fig.~\ref{fig14ch}, where the transverse contribution predominates at
all kinematics whilst the net longitudinal channel, even not being a
relevant contribution, is essential to describe antineutrino data at
backward kinematics together with the 2p-2h longitudinal one.

\begin{figure}[htbp]\begin{center}\vspace*{-0.2598cm}
\includegraphics[width=.375\textwidth,angle=270]{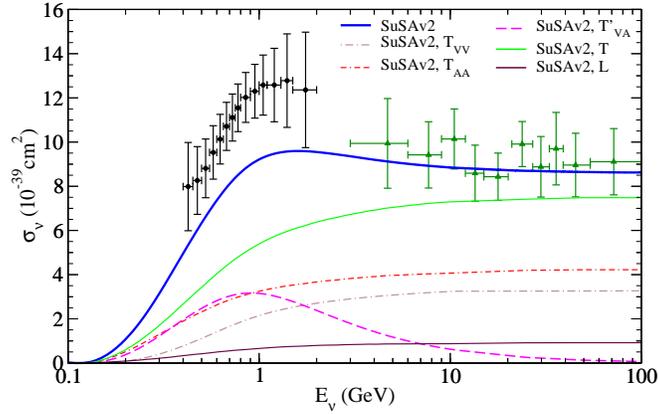}                     \end{center}\vspace*{-0.798cm}
\caption{Separation into components of the CCQE $\nu_\mu$ cross
  section per nucleon on $^{12}$C displayed versus neutrino energy
  $E_\nu$ within the SuSAv2 approach.  The
  MiniBooNE~\cite{AguilarArevalo:2010zc} and
  NOMAD~\cite{Lyubushkin:2008pe} data are also shown for reference.}
     \label{fig:componentsqe}
  \end{figure}

In Fig.~\ref{fig:componentsqe} we show the breakdown of the total
integrated neutrino cross sections into $L(=L_{VV}+L_{AA})$,
$T(=T_{VV}+T_{AA})$, $T_{VV}$, $T_{AA}$ and $T'_{VA}$ contributions,
with the last occurring as a positive (constructive) term in the
neutrino cross section and a negative (destructive) term in the
antineutrino one. The sign of the $T'_{VA}$ channel represents the
main difference between the total neutrino and antineutrino cross
sections. In addition to the opposite sign in the $VA$ response, some
minor differences between neutrino and antineutrino cross sections
arise from the Coulomb distortion of the emitted lepton and the
different nuclei involved in the CC neutrino (nitrogen) and
antineutrino (boron) scattering processes on carbon.  We also notice
that below 1 GeV the $T'_{VA}$ response is higher than the $T_{VV}$
one but of the same order as the $T_{AA}$ one. Note that the maximum
of the $VA$ channel is around the peak of the MiniBooNE neutrino
flux. On the contrary, the VA contribution to the cross section is negligible
at energies above $10$~GeV as a consequence of the small values of the
axial form factor $G_A$ and the lepton factor $V_{T'}$ at high $E_\nu$
and $|Q^2|$ (see~\cite{Megias:2017PhD} for details). This is also in
agreement with some previous QE results \cite{Megias:2013aa}. As a
consequence, for very high $\nu_\mu$ ($\overline{\nu}_\mu$) energies
(above $\sim 10$~GeV) the total cross section for neutrinos and
antineutrinos is very similar. Only the $L$ and $T$ channels
contribute for the higher values explored by NOMAD experiment. On the
contrary, in the region explored by the MiniBooNE collaboration, the
main contributions come from the two transverse $T$ and $T'$
channels.\newpage
\vspace*{-1.198cm}
  \vspace*{0.3798cm}

       \subsection{Implementation of the SuSAv2-MEC and RMF models in MC event generators and extension to semi-inclusive processes }\label{impl-section}

       In Section~\ref{sec:susav2-results}, the SuSAv2-MEC model has
       been shown to be capable of reproducing the nuclear dynamics
       and superscaling properties observed in ($e,e'$) reactions
       which serves as a stringent test for nuclear models, whilst
       also providing an accurate description of existing neutrino
       data. Up to now, SuSAv2-MEC is the only fully relativistic
       model that can be extended without kinematical restrictions or
       further approximations to the full-energy range of interest for
       present and future neutrino experiments. This has motivated the
       implementation of SuSAv2-MEC 1p1h and 2p2h contributions in the
       GENIEv3 MonteCarlo neutrino interaction
       simulation~\cite{Dolan:2019bxf} in order to use it to better
       characterise nuclear effects in T2K and MINERvA neutrino
       scattering cross-section measurements. Work is also in progress to implement it in NEUT~\cite{Hayato:2009zz}.
       
       Accordingly, we present in this section the implementation and
       validation of the SuSAv2-MEC 1p1h and 2p2h models in the GENIE
       neutrino-nucleus interaction event generator and a comparison
       of the subsequent predictions to measurements of lepton and
       hadron kinematics from the T2K experiment. These predictions
       are also compared to those of other available models in
       GENIE. We additionally compare the semi-inclusive predictions
       of the implemented 1p1h model to those of the microscopic model
       on which SuSAv2 is based - Relativistic Mean Field (RMF) - to
       begin to test the validity of widely-used ‘factorisation’
       assumptions employed by generators to predict hadron kinematics
       from inclusive input models. The results highlight that a more
       precise treatment of hadron kinematics in generators is
       essential in order to attain the few-$\%$ level uncertainty on
       neutrino interactions necessary for the next generation of
       accelerator-based long-baseline neutrino oscillation
       experiments. More details about this analysis and the
       implementation can be found in~\cite{Dolan:2019bxf}.

      The recent experimental interest on more exclusive measurements
      is related to the information about the final state nucleons
      that they provide, such as those which have recently been
      performed by T2K~\cite{Abe:2018pwo} and
      Minerva~\cite{Lu:2018stk}, which have been demonstrated to have
      a much more acute sensitivity to the different nuclear effects
      involved in neutrino-nucleus interactions. Unfortunately a
      comparison of these measurements directly to microscopic models
      requires semi-inclusive or exclusive predictions which the
      majority of models are not able to make, as they simplify their
      calculations by integrating over outgoing nucleon kinematics. An
      exception to this is the RMF model, used to construct the SuSAv2
      predictions, which is capable of semi-inclusive predictions for
      neutrino reactions\footnote{The RMF model has proven its
        validity to address exclusive predictions for electron
        scattering~\cite{Udias01} and work is underway to fully extend
        it to neutrino reactions. }. As described in
      \cite{Dolan:2019bxf}, the simulations used by experiments
      circumvent this limitation by factorising the leptonic and
      hadronic components of the interaction. Among other
      approximations, this approach relies strongly on a
      semi-classical description of FSI and the distribution of
      initial state nucleon kinematics seen by the probe being
      independent of its energy and momentum transfer.
       
\begin{figure*}
\begin{center}
\includegraphics[width=15cm,bb=55 420 588 671]{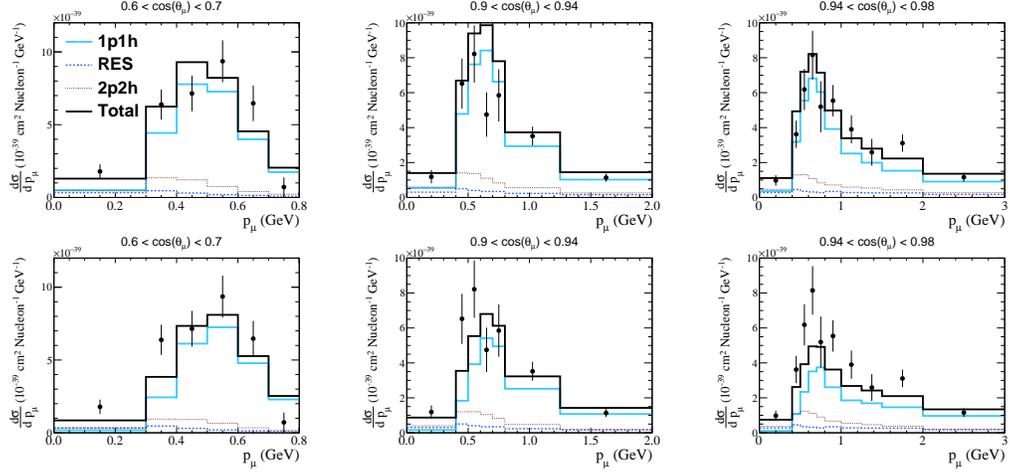}
\end{center}
\caption{Comparison of the T2K CC0$\pi$ measurement of muon-neutrino
  interactions on Carbon with the SuSAv2 and Valencia models
  (1p1h+2p2h) implemented in GENIE with additional pion-absorption
  effects (from GENIE's Berger-Sehgal model). The top plots are the
  SuSAv2 predictions whilst the Valencia ones are below. The data
  points are taken from~\cite{Abe:2016tmq}.}
\label{fig:incT2KModelComp}

\end{figure*}

\begin{figure*}
\begin{center}
\includegraphics[width=15cm,bb=55 420 588 671]{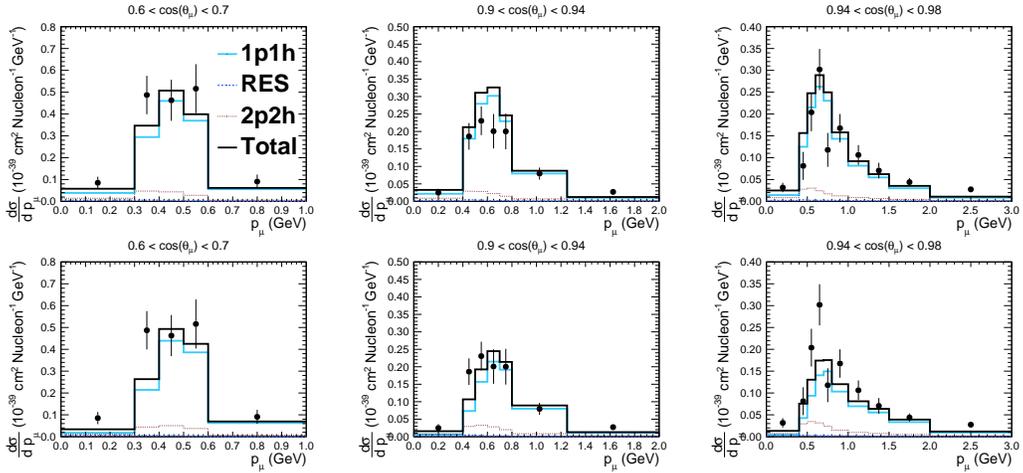}
\end{center}
\caption{Comparison of the T2K CC0$\pi$ measurement of muon-neutrino
  interactions on Carbon where there are no protons above 500 MeV with
  the SuSAv2 and Valencia models (1p1h+2p2h) implemented in GENIE with
  additional pion-absorption effects (from GENIE's Berger-Sehgal
  model). The top plots are the SuSAv2 predictions whilst the Valencia
  ones are below. The data points are taken from~\cite{Abe:2018pwo}.}
\label{fig:ssincT2KModelComp}
\end{figure*}

       The implementation of the SuSAv2 1p1h model in GENIE provides a
       first opportunity to test the factorisation approach in event
       generators as well as to compare with other models available in
       event generators. In Figs.~\ref{fig:incT2KModelComp}
       and~\ref{fig:ssincT2KModelComp} we show a comparison of the
       SuSAv2 and Valencia model~\cite{Nieves:2011pp,Nieves:2011yp} predictions (1p1h and 2p2h) as
       implemented in GENIE on top of GENIE's Berger-Sehgal pion
       production prediction for T2K inclusive CC0$\pi$ and
       `semi-semi-inclusive' CC0$\pi$Np results, being the latter a
       semi-inclusive cross section where the final state proton is
       below 500 MeV/c. This clearly shows that the implemented
       Valencia and SuSA models differ substantially, with only the
       SuSA model able to describe the very forward data and the
       Valencia model describing the mid-angle data a little
       better. The discrepancies between the model and data for
       the inclusive and semi-inclusive results are consistent,
       suggesting that they at least partially stem from the
       underlying inclusive cross section model.

       Beyond the inclusive comparison, the `semi-semi-inclusive'
       predictions within the kinematic region where SuSAv2 is a good
       description of RMF allows us to study the validity of the
       factorisation approach used in event generators. Here it can be
       seen that the implementation with both the kinematic-dependent
       binding energy and with FSI is closest to reproducing the RMF
       microscopic model prediction, but still appears to peak at too
       low muon momentum and also fails to describe the higher
       momentum region. It can also be seen that variations to the
       hadronic component of the interaction cause substantial
       alterations to the predictions, highlighting the role of these
       nonphysical freedoms available within the factorisation
       approach. Further work will focus on more stringent tests
       through the implementation of the RMF spectral function into
       event generators and by exploring the predictions in a wider
       region of hadronic kinematic phase-space.
       
\begin{figure*}
	\begin{center}
\includegraphics[width=15cm,bb=55 420 588 671]{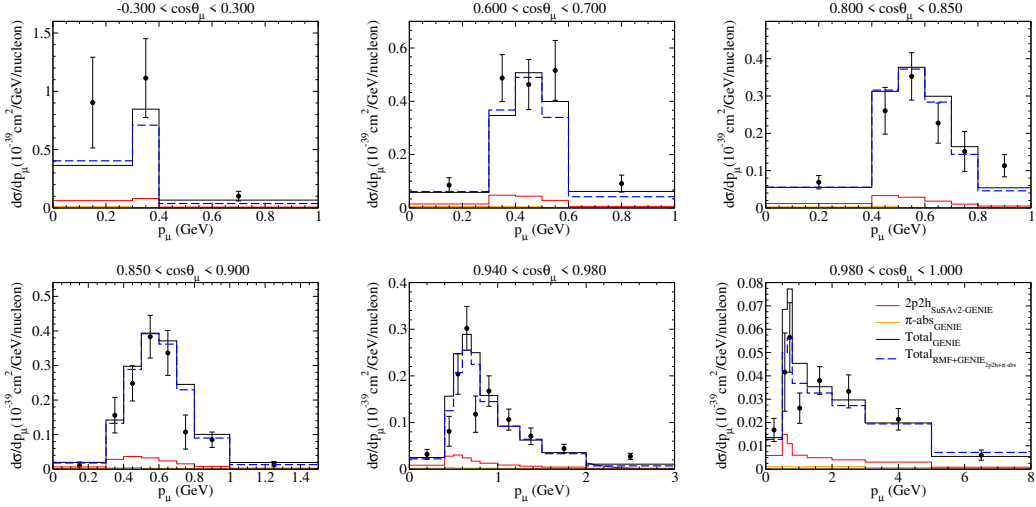}
	\end{center}
	\caption{Comparisons of single differential CC0$\pi$
          muon-neutrino cross sections on Carbon at T2K kinematics as
          a function of the muon kinematics when there are no protons
          (with momenta above 500 MeV). Two 1p1h predictions are used
          (one from RMF, the other from SuSAv2 implemented in GENIE),
          in addition to the SuSAv2 2p2h and Berger-Sehgal pion
          absorption contributions from GENIE. Goodness of fit are
          calculated to be $\chi^2_{RMF}=171.87$ (59 bins) and
          $\chi^2_{SuSA}=168.92$ (60 bins), where the latter includes
          a single extra bin from -1.0 to -0.3 $\cos{\theta}$ (not
          shown). The data points are taken from~\cite{Abe:2018pwo}.}
	\label{fig:ssincT2KComp}
\end{figure*}

       The T2K semi-inclusive CC0$\pi$ measurement of interactions
       with protons less than 500 MeV~\cite{Abe:2018pwo} also provides an
       opportunity to compare the RMF semi-inclusive model
       predictions to data, which is shown in
       Fig.~\ref{fig:ssincT2KComp} alongside the SuSAv2-GENIE
       predictions using the factorisation approach. In order to make
       this comparison the RMF predictions are added to the SuSAv2-MEC
       (2p2h) and pion-absorption predictions from GENIE (for pion
       production the Berger-Sehgal model was
       used~\cite{Berger:2007rq}).  In general, we observe a fair
       agreement of both RMF+GENIE (SuSAv2-2p2h+$\pi$-abs) and GENIE
       (SuSAv2-1p1h+SuSAv2-2p2h+$\pi$-abs). The overestimation of data
       at very forward angles by the SuSAv2-GENIE is ascribed to the
       aforementioned low energy transfer scaling violations absent in
       the SuSAv2-model but present in RMF, thereby explaining the
       better agreement achieved with the latter. On the contrary, the
       larger results from SuSAv2-1p1h at very backward angles
       compared to RMF are related to the previously discussed FSI
       treatment alterations. In general, it is clear that RMF
       performs better within the most forward angular bins (where
       additional RMF effects are most important). The recently
       developed Energy-Dependent RMF (ED-RMF)
       model~\cite{Gonzalez-Jimenez19a,Gonzalez-Jimenez20},
       which keeps the original RMF potentials at low kinematics but
       make them softer for increasing nucleon momenta, following the
       SuSAv2 approach, will solve the limitations of the SuSAv2 model
       at forward angles while solving the drawbacks of the RMF at
       very high kinematics, constituting a promising candidate to be
       implemented in neutrino event generators. This will help to
       reduce nuclear-medium uncertainties and to improve systematics
       in neutrino oscillation experiments.

   \section{Conclusions} \label{sec:concl}

The advent of next generation neutrino oscillation experiments demands
more and more sophisticated nuclear modelling, necessary to extract
significant information on the properties of neutrinos and physics
beyond the Standard Model of electroweak interactions.  Nuclear theory
plays a crucial role in these analyses, which strongly rely on the
accurate description of neutrino interactions with the detector, made
of medium/heavy nuclei, implemented in Monte Carlo generators.

Contrary to the case of electron scattering, where many precise data
from previous experiments exist, neutrino scattering data are rare and
have large error bars. Beyond that, the identification of the primary
reaction suffers from the absence of monochromatic neutrino
beams. Therefore, electron scattering represents not only a necessary
test but also a useful tool to understand and control nuclear effects
in oscillation experiments.

In this paper we have reviewed the current status of the description
of lepton-nucleus scattering at different kinematics, going from the
quasi-elastic up to the deep-inelastic regime. We have focused in
particular on the SuSAv2+MEC model, developed and improved by our
group, collecting the main results obtained in the last few years. The
model is based on the Relativistic Mean Field description of the
nucleus, complemented with the contribution of two-body currents.  We
have shown that the model is capable of describing in a very
satisfactory way both electron and neutrino data in the case of
inclusive reactions, in which only the outgoing lepton is detected.
This outcome, although reassuring, is not sufficient to guarantee the
required precision - of a few \% - on the description of nuclear
effects in neutrino oscillation experiments.  Work is in progress on the study of exclusive reactions,
where one or more hadrons are detected in coincidence with the lepton,
in the same relativistic framework. The extension of nuclear models to
semi-inclusive reactions is a challenge to be faced by theorists
working in the field.  A proper description of the hadrons and mesons
in the final state will be essential for the next-generation of
neutrino experiments. This requires having a reasonable control on the
reconstruction of the energy neutrino which can be only achieved by
analyzing the kinematics of the final particles.

 \begin{acknowledgments}
The authors wish to thank
Igor Kakorin, from JINR (Dubna),
for his careful reading of this
manuscript and for pointing
out several misprints in its
first version.

This work was partially supported by the Spanish Ministerio de
Economia y Competitividad and ERDF (European Regional Development
Fund) under contracts FIS2017-88410-P and FIS2017-85053-C2-1-P, and by
the Junta de Andalucia (grants No. FQM160, FQM225 and
SOMM17/6105/UGR). M.B.B. acknowledges support by the INFN under
project Iniziativa Specifica MANYBODY and the University of Turin
under Project BARM-RILO-18.  R.G.J. was supported by Comunidad de
Madrid and U.C.M. under the contract No. 2017-T2/TIC-5252.
G.D.M. acknowledges support from a P2IO-CNRS grant and from CEA,
CNRS/IN2P3, France; and by the European Unions Horizon 2020 research
and innovation programme under the Marie Sklodowska-Curie grant
agreement No. 839481.
\end{acknowledgments}

\appendix

\section{Scaling and superscaling: definitions}
\label{scaling-appendix}
In this appendix we recall the definitions of the scaling and
superscaling functions, along with the $\psi$-scaling variable
\cite{Alberico:1988bv,Day:1990mf,Barbaro:1998gu,Donnelly99a,Donnelly99b}
for  inclusive electron scattering off a nucleus. We work in
the target rest frame and we use fully relativistic kinematics.

The scaling variable $\psi$ naturally emerges in the Relativistic
Fermi Gas calculation of the quasielastic responses (see
Sect. \ref{section-rfg} and Appendix \ref{RFG-appendix}).  It is
defined in terms of the lowest kinetic energy $T_{min}\equiv
T_{min}(q,\omega)$ of the struck nucleon inside the Fermi sphere for
given $q$ and $\omega$, according to Eqs (\ref{epsilon0},\ref{psi}))
\begin{equation}
      \psi =
      \pm \sqrt{\frac{T_{min}}{T_F}} =
       {\rm sgn} (\lambda-\tau) \sqrt{\frac{\epsilon_0-1}{\epsilon_F-1}} \,,
       \label{eq:psi1}
\end{equation}
where  $     \epsilon_F=(k_F^2+m_N^2)^{\frac12}/m_N$,
$k_F$ being the Fermi momentum,
and 
\begin{equation} \label{eq:eps0}
	  \epsilon_0 =
                  {\rm Max}
                  \left\{
                  \kappa\sqrt{1+\frac1{\tau}}-\lambda , \epsilon_F -2\lambda 
                  \right\} \,.
\end{equation}
We recall the definition of the
dimensionless energy, momentum and four-momentum transfers,
   \begin{equation}
     \lambda=\omega/2m_N, \kern 5mm
     \kappa=q/2m_N, \kern 5mm
     \tau=\kappa^2-\lambda^2. \kern 5mm
\label{eq:adimvar}
   \end{equation}
   The second value inside the Max in \eqref{eq:eps0}
   takes into account Pauli blocking, which prevents the ejected
   nucleon from occupying a state with $k<k_F$. For $q>2 k_F$ no Pauli
   blocking effects are present in the RFG model and
   $\epsilon_0=\kappa\sqrt{1+1/\tau}$.
   
   The variable $\psi$ vanishes at the QEP and is negative
   (positive) for transferred energies lower (higher) than the energy of the
   maximum of the QEP
   \begin{equation}
     \omega_{QEP} = \sqrt{q^2+m_N^2} -m_N .
     \end{equation}
   When comparing to experimental data one finds that the maximum of
   the QEP is shifted by the nucleon separation energy. 
This  energy shift $E_{shift}$ is taken into account
    by modifying the scaling variable  $\psi \rightarrow \psi'$,
     with $\lambda$ replaced by
    $\lambda'=(\omega-E_{shift})/(2m_N)$
    and likewise $\tau \rightarrow \tau'=\kappa^2-\lambda'^2$.

   An alternative expression for $\psi$,
   valid only in the non Pauli-blocked regime (the interesting regime
    for scaling arguments) is
    \begin{equation}
	\psi = \frac{1}{\sqrt{\xi_F}} \frac{\lambda-\tau}{\sqrt{(1+\lambda)\tau+\kappa\sqrt{\tau(1+\tau)}}} \,,
    \label{eq:psi2}
    \end{equation}    
    where $\xi_F= \epsilon_F-1$.
        
    The variable $\psi$ can be extended to the inelastic region.
    For the excitation of a resonance $N^*$ of mass $m^*$ one can introduce the inelasticity
    function~\cite{Barbaro:2003ie}
    \begin{equation}
        \label{eq:inelpar}
        \rho = 1 + \frac{\mu^2-1}{4\tau}\,, \ \ \ \ \ \mu=\frac{m^*}{m_N}
    \end{equation}
    through the replacements
    \begin{equation}
        \lambda\to\lambda\rho\,,\ \ \ \kappa\to\kappa\rho\,,\ \ \ \tau\to\tau\rho^2
    \end{equation}
    in \eqref{eq:psi2}. This yields the resonance scaling variable
    \begin{equation}
	\psi^* = \frac{1}{\sqrt{\xi_F}} \frac{\lambda-\tau\rho}{\sqrt{(1+\lambda\rho)\tau+\kappa\sqrt{\tau(1+\tau\rho^2)}}} \,.
    \end{equation}
Likewise, the inelastic scaling variable $\psi_X$ can be defined for any inelastic channel by setting $\mu=\mu_X=\frac{W_X}{m_N}$,
being $W_X$ the invariant mass of the final state.

The {\it superscaling function}
 becomes a function the variable
      $\psi$ only if superscaling occurs.
It is defined as
\begin{equation}
  f(q,\omega) =  k_F
  \frac{d^2\sigma/d\Omega_e d\omega}{\overline{\sigma_{eN}}} \,,
    \label{eq:F}
\end{equation}
where $d^2\sigma/d\Omega_e d\omega$ is the double differential cross
section with respect to the outgoing electron solid angle $\Omega_e$
and energy transfer $\omega$.  The dividing factor is some average
value of the single-nucleon cross section compatible with the
kinematics.  Note that $\overline{\sigma_{eN}}$ is not the free
stationary elastic $eN$ cross section, because the hit nucleon is
moving inside the nucleus and it is off-shell. The treatment of
off-shell effects introduces ambiguities which are more or less
important depending on the kinematics. In this work we use the
averaged single-nucleon cross section arising from the RFG model,
given by
\begin{eqnarray} \label{averagesn}
    \overline{\sigma_{eN}} &=& \sigma_{Mott}(v_L G_L + v_T G_T)
\\
G_{L,T} &=& \frac{\xi_F}{\eta_F^2 \kappa} ( Z U^p_{L,T}+ N U^n_{L,T}) \, ,
\end{eqnarray}
where the proton and neutron structure functions $U^{p,n}_{L,T}$
are  defined in Eqs. (\ref{ul},\ref{ut}).

\section{Single nucleon tensor}
\label{sn-appendix}

In this appendix we derive the single nucleon tensor and responses for
a nucleon transition $\np \rightarrow \np'$ in the Fermi gas before
integration. The starting point is the single nucleon tensor given in
Eq. (\ref{single-nucleon-tensor}). We consider for definiteness the
tensor induced by the weak CC interaction, which is the sum of vector
and axial currents, given in
Eqs. (\ref{vector-current},\ref{axial-current}). The electromagnetic
tensor can be easily obtained from the below expressions by
considering only the vector part of the current.

We start, using the properties of the Dirac spinors, writing the sum
over spin indices in Eq. (\ref{single-nucleon-tensor} as a trace of
Dirac matrices
\begin{equation}
w^{\mu\nu}_{s.n.}(\np',\np) = \frac12 {\rm Tr} 
\left[ 
\frac{\pbar+m_N}{2m_N}
(\Gamma_V-\widetilde{\Gamma}_A)^\mu
\frac{\pbar'+m_N}{2m_N}
(\Gamma_V-{\Gamma}_A)^\nu
\right] \, ,
\end{equation}
where $\Gamma^\mu_V$ is the spin matrix of the vector current after
using Gordon identity
\begin{equation}
\label{Amaro:vector_current}
\Gamma^{\mu}_V = 2G_M^V \gamma^\mu - 2F_2^V \frac{(p+p')^\mu}{2m_N},
\end{equation}
while  $\Gamma^\mu_A$ is the spin matrix of the axial current 
\begin{equation}
\label{Amaro:axial_current}
\Gamma^{\mu}_A = G_A \gamma^\mu\gamma^5 + G_P \frac{Q^\mu}{2m_N}\gamma^5 .
\end{equation}
Finally   $\widetilde{\Gamma}^{\mu}_A = \gamma^0{\Gamma}^{\mu\dagger}_A\gamma^0$ or  
\begin{equation}
\widetilde{\Gamma}^{\mu}_A 
= G_A \gamma^\mu\gamma^5 - G_P \frac{Q^\mu}{2m_N}\gamma^5 .
\end{equation}
We see that the $V-A$ dependence of the weak current allows to single
out four contributions labeled VV, AA, VA and AV
 \begin{equation}
w^{\mu\nu}_{s.n.}(\np',\np) =
w^{\mu\nu}_{VV}+
w^{\mu\nu}_{AA}+
w^{\mu\nu}_{VA}+
w^{\mu\nu}_{AV} .
\end{equation}

Performing the corresponding traces, it is straightforward to obtain the
result:
\begin{eqnarray}
w_{VV}^{\mu\nu} &=&
\frac{1}{m_N^2} 
\left[ (G_M^V)^2(Q^2 g^{\mu\nu} - Q^\mu Q^\nu) 
     + \left((F_1^V)^2 + \tau (F_2^V)^2\right)
       (p+p')^\mu (p+p')^\nu 
\right]
\label{wvv}
\\
w_{AA}^{\mu\nu} &=&
\frac{1}{4m_N^2} 
\left[ G_A^2 \left( (p+p')^\mu (p+p')^\nu 
                    -  Q^\mu Q^\nu
                   - \left(4m_N^2 - Q^2\right) g^{\mu\nu} 
             \right)
-\left(2G_AG_P + G_P^2\frac{Q^2}{4m_N^2}\right) Q^\mu Q^\nu
\right]
\label{waa}
\\
w_{VA}^{\mu\nu} +w_{AV}^{\mu\nu} &=&
i \frac{2G_M^V G_A }{m_N^2} 
\epsilon^{\alpha\beta\mu\nu}p_{\alpha}p'_{\beta} .
\label{wva}
\end{eqnarray}
Note that the VA interference tensor is purely imaginary and
antisymmetric and therefore it only contributes to the $T'$ response
function, see Eqs.  (\ref{rcc}-\ref{rtprima}).

Using $p'_\mu= p_\mu+ Q_\mu$ we can write the hadronic tensor in the
more standard form
\begin{equation}  \label{wsn}
w_{s.n.}^{\mu\nu}
= - w_1 \left(  g^{\mu\nu} - \frac{Q^\mu Q^\nu}{Q^2} \right)+
    w_2 K^\mu K^\nu             
  -\frac{i}{m_N}w_3\epsilon^{\alpha\beta\mu\nu}Q_{\alpha}K_{\beta}
  + w_4 \frac{Q^\mu Q^\nu}{m_N^2} \, ,
\end{equation}
where the invariant structure functions $w_i = w_i(Q^2)$ only depend
on $Q^2$, and we have defined the four vector
\begin{equation} \label{kmu}
K^\mu = \frac{1}{m_N}\left( p^\mu - \frac{p\cdot Q}{Q^2}Q^\mu \right) .
\end{equation}
After a straightforward calculation, using
Eqs. (\ref{wvv},\ref{waa},\ref{wva}), we obtain
\begin{eqnarray}
w_1 &=& \tau (2G_M^V)^2 + (1+\tau)G_A^2  \label{w1}  \\
w_2 &=& \frac{ (2G_E^V)^2 + \tau (2G_M^V)^2}{1+\tau} + G_A^2 \\
w_3 &=& 2G_M^V G_A \\
w_4 &=& \frac{1}{4\tau}(G_A - \tau G_P )^2 \label{w4} .
\end{eqnarray}

Finally we extract the suitable components of the hadron tensor to
obtain the corresponding single nucleon responses to be used later for
computing the nuclear responses in the RFG.

The single-nucleon $CC$ response function, using Eq. (\ref{wsn}), can
be written as
\begin{equation}
r_{CC}=
 w_{s.n.}^{00} =
 -w_1\left(1-\frac{\omega^2}{Q^2}\right)
+w_2 (K^0)^2+w_4\frac{\omega^2 }{m_N^2} .
\end{equation}
Writing it in terms of the dimensionless variables
$\lambda,\kappa,\tau$, introduced in Eq. (\ref{eq:adimvar}), and the
reduced initial nucleon energy $\varepsilon \equiv E/m_N$, we find
\begin{equation} \label{rcc-sn}
r_{CC} = - w_1 \frac{\kappa^2}{\tau}+ w_2(\varepsilon+\lambda)^2+4w_4\lambda^2.
\end{equation}
Analogously, for the case of the $CL$ and $LL$ single-nucleon
responses
\begin{eqnarray}
r_{CL} &=&
-\frac12(w_{s.n.}^{03} +w_{s.n.}^{30}) =
-\frac{\lambda}{\kappa}
\left[ - w_1 \frac{\kappa^2}{\tau}+ w_2(\varepsilon+\lambda)^2
\right]
-4w_4\lambda\kappa
\label{rcl-sn}\\
r_{LL} &=&
 w_{s.n.}^{33} =
\frac{\lambda^2}{\kappa^2}
\left[ - w_1 \frac{\kappa^2}{\tau}+ w_2(\varepsilon+\lambda)^2
\right]
+4w_4\kappa^2 .
\label{rll-sn}
\end{eqnarray}
Note that the first term of $r_{CL}$, between squared parentheses, and
multiplied by $-\lambda/\kappa$, comes from current conservation of
the terms with $w_1$ and $w_2$ of the hadronic tensor. The same
comment applies to the first term of the $r_{LL}$ response, with the
factor $\lambda^2/\kappa^2$.

The transverse response is the component 
\begin{equation}
r_T = w_{s.n.}^{11} +w_{s.n.}^{22}
= 2w_1 + w_2[ (K^1)^2+ (K^2)^2 ] .
\end{equation}
Using the definition of the vector $K^\mu$, Eq. (\ref{kmu}), we have
\begin{equation}
 (K^1)^2+ (K^2)^2 = \eta^2-\eta_3^2 \, ,
\end{equation}
where $\eta_i = p_i/m_N$ is the reduced initial nucleon momentum
vector with length $\eta$. Using the condition for on-shell nucleons
$\np\cdot\nq= E\omega + Q^2/2$ written in terms of dimensionless
variables
\begin{equation}
\eta_3 = \frac{\varepsilon\lambda-\tau}{\kappa}
\end{equation}
and $\eta^2= \varepsilon^2 -1$, we can finally write
\begin{equation} \label{rt-sn}
r_T = 
 2w_1 + w_2
\frac{\varepsilon^2\tau-\kappa^2-\tau^2+2\varepsilon\lambda\tau}{\kappa^2} .
\end{equation}
Finally, the $r_{T'}$ response comes from the antisymmetric part of the
hadronic tensor\footnote{Our convention is $\epsilon_{0123}=1$ for the
  Levi-Civita four tensor.}
\begin{equation}
r_{T'}
= -\frac{i}{2}(w_{s.n}^{12}-w_{s.n}^{21})
= -\frac{1}{m_N} w_3 (-q K^0 + \omega K^3) .
\end{equation}
Using $K^0= \varepsilon+\lambda$ 
and $K^3= \omega K^0 / q$, we obtain
\begin{equation} \label{rtp-sn}
r_{T'} = 2 w_3 \frac{\tau}{\kappa}(\varepsilon+\lambda) .
\end{equation}

\section{Derivation of response functions in the RFG model}
\label{RFG-appendix}

We outline here the main steps used to obtain the analytical
expression of the response functions in the RFG, as this material 
can only be found for electron scattering and not for neutrino scattering
in such detailed form.

We start with the definition of the response function $R_K$ for
neutrino (antineutrino) CC quasielastic scattering.  Taking the
corresponding components of the hadronic tensor from
Eq. (\ref{hadronicorfg}), we have
\begin{equation}
R_K(q,\omega) 
= \frac{3{\cal N} m_N^2 }{4\pi k_F^3}
 \int d^3p \delta(E'-E-\omega)
\frac{1}{EE'} 
r_{K}(\np,q,\omega)
\theta(k_F-p)\theta(p'-k_F) \, ,
\end{equation}
where ${\cal N}= N$ for neutrinos and $Z$ for antineutrinos.  The
single nucleon responses $r_K$ have been obtained in the previous
appendix \ref{sn-appendix} for on shell nucleons with momenta
$\np'=\np+\nq$. Note that $r_K$ depends on $q,\omega$ via the
variables $\kappa,\lambda$, and $\tau$, and that it depends on $\np$
only through the reduced initial energy $\varepsilon$, see
Eqs. (\ref{rcc-sn}, \ref{rcl-sn}, \ref{rll-sn}, \ref{rt-sn},
\ref{rtp-sn}).

To perform the integral over the initial nucleon momentum, we change
variables $(p,\theta,\phi) \longrightarrow (E,E',\phi)$, where
$E^2=m_N^2+\np^2$ and $E'{}^2= m_N^2+ (\np+\nq)^2$ are the initial and
final nucleon energies.  The volume element transforms as
\begin{equation}
d^3p = \frac{EE'}{q}dE dE' d\phi
\end{equation}
and the integral with the appropriate integration limits becomes
\begin{equation}
R_K(q,\omega) 
= \frac{3{\cal N} m_N^2 }{4\pi k_F^3}
\frac{1}{q}
 \int_{m_N}^{E_F} d E \int_{E_{p-q}}^{E_{p+q}} dE' 
\int_0^{2\pi} d\phi
\delta(E'-E-\omega)
r_{K}(\np,q,\omega)
\theta(E'-E_F) \, ,
\end{equation}
where $E_F$ is the relativistic Fermi energy,
$E_{p+q}^2=m_N^2+(p+q)^2$, and $E_{p-q}^2=m_N^2+(p-q)^2$.  The
integral over $\phi$ gives a factor $2\pi$ because the single nucleon
responses do not depend on the azimuthal angle. Using the delta
function to integrate over $E'=E+\omega$, the result can be written as an integral
over the initial energy
\begin{equation}
R_K(q,\omega) 
= \frac{3{\cal N}}{4\eta_F^3\kappa m_N^2}
 \int_{m_N}^{E_F} d E 
\theta(E_{p+q}-E-\omega)
\theta(E+\omega-E_{p-q})
\theta(E'-E_F)
r_{K}(E,q,\omega) \, ,
\end{equation}
where we have used the reduced variable $\eta_F=k_F/m_N$ and
$\kappa=q/2m_N$.  Note that the single nucleon responses only depend
on the initial energy.

The step functions and the integration limits are equivalent to the
following constraints to the energy:
\begin{eqnarray}
E_{p-q} & < & E + \omega < E_{p+q} \label{desigualdad1} \\
E_{F} & < & E + \omega < E_{F}+ \omega . \label{desigualdad2}
\end{eqnarray}
Taking the square of (\ref{desigualdad1}) and introducing the reduced
variables: momentum, $\eta=p/m_N$, energy, $\varepsilon$ and
$\lambda$, $\kappa$ and $\tau$, it is straightforward to obtain
\begin{equation}
|\varepsilon \lambda -\tau | < \kappa \eta
\end{equation}
which in turn can be squared again, using $\eta^2= \varepsilon^2 -1$,
in terms of $\varepsilon$, yielding
\begin{equation}
\kappa^2\left(1+\frac1{\tau}\right) < (\varepsilon+\lambda)^2 .
\end{equation}
Solving for $\varepsilon$ we get the following lower bound for given $q,\omega$
\begin{equation}
\kappa\sqrt{1+\frac1{\tau}}-\lambda < \varepsilon .
\end{equation}
On the other hand, from (\ref{desigualdad2}), $\varepsilon$ must
verify the Pauli blocking condition as a second lower bound
\begin{equation}
\varepsilon_F -2\lambda < \varepsilon \, ,
\end{equation}
where $\varepsilon_F= E_F/m_N$.  
Both lower bounds for $\varepsilon$ imply
\begin{equation}
{\rm Max} \left\{
\kappa\sqrt{1+\frac1{\tau}}-\lambda , 
\varepsilon_F -2\lambda 
\right\} \equiv \varepsilon_0 <  \varepsilon .
\end{equation}
Note now that the following inequality always holds
\begin{equation}    
\kappa\sqrt{1+\frac1{\tau}} -\lambda  \geq 1 .
\end{equation}
This can be demonstrated by moving $\lambda$ to the right-hand side
and taking the square on both sides of the inequality.  Hence $1 <
\varepsilon_0 < \varepsilon$ and therefore $\varepsilon$ correspond to
an allowed nucleon energy. The response function can thus be written
as the following integral over $\varepsilon$
\begin{equation}
R_K(q,\omega) 
= \frac{3{\cal N}}{4\eta_F^3\kappa m_N}
\theta(\varepsilon_F-\varepsilon_0)
 \int_{\varepsilon_0}^{\varepsilon_F} 
r_{K}(\varepsilon,\kappa,\lambda)  d \varepsilon .
\end{equation}

Note that there is a dependence of the response functions on the
variable $\varepsilon_0$ through the lower limit of the 
integral. In the scaling approach $\varepsilon_0$ is written in terms of
the scaling variable $\psi$, defined by
\begin{equation}
\psi = \sqrt{\frac{\varepsilon_0-1}{\varepsilon_F-1}} 
{\rm sign}(\lambda-\tau). 
\end{equation}
The scaling variable is defined such as it is positive to the right
side of the QE peak ($\lambda=\tau$) and negative otherwise.  The
inverse relation is
\begin{equation}
\varepsilon_0 = 1 + \psi^2\xi_F \, ,
\end{equation}
where $\xi_F=\varepsilon_F -1$ is the kinetic Fermi energy in units of
the nucleon mass.  The single nucleon responses, obtained in
App. \ref{sn-appendix}, depend on $\varepsilon$ as a second-degree
polynomial at most. Therefore, in terms of this scaling variable, the
only relevant integrals are the following:
\begin{eqnarray}
 \int_{\varepsilon_0}^{\varepsilon_F} 
d \varepsilon 
&=& 
\xi_F(1-\psi^2)
\\
 \int_{\varepsilon_0}^{\varepsilon_F} \varepsilon \,
d \varepsilon 
&=& 
\xi_F(1-\psi^2)(1+\frac12 \xi_F(1+\psi^2))
\\
 \int_{\varepsilon_0}^{\varepsilon_F} \varepsilon^2 \,
d \varepsilon 
&=& 
\xi_F(1-\psi^2)(1+\xi_F(1+\psi^2)+\frac13 \xi_F^2(1+\psi^2+\psi^4)) .
\end{eqnarray}
Applying these results to the response $R_{CC}$ we can directly write
\begin{eqnarray}
R_{CC} &=& 
\frac{3 {\cal N}}{4m_N\eta_F^3 \kappa}
\theta(1-\psi^2)\xi_F(1-\psi^2)
\left\{ 
-w_1\frac{\kappa^2}{\tau} + w_2\lambda^2+4w_4\lambda^2
\right. \nonumber\\
&& 
+ 2w_2 \lambda (1+\frac12 \xi_F(1+\psi^2))
\left. +
w_2 
(1+\xi_F(1+\psi^2)+\frac13 \xi_F^2(1+\psi^2+\psi^4))
\right\}
\\
&=&  \frac{1}{k_F}
\frac{{\cal N}\xi_F}{\eta_F^2 \kappa}
f(\psi) U_{CC} \, ,
\label{amaro-factorize}
\end{eqnarray}
where $f(\psi)$ is the RFG scaling function  defined in
Eq. (\ref{scalingfun}).

To write this expression in terms of the quantity $\Delta$ introduced
in Eq.  (\ref{amaro-delta}) we add and subtract the elastic limit of
the function $U_{CC}$, corresponding to $\eta_F= \xi_F=0$ and
$\lambda=\tau$:
\begin{equation}
U_{CC}^0 = U_{CC}(\xi_F=0,\lambda=\tau) = \frac{\kappa^2}{\tau}
\left[-w_1+(1+\tau)w_2+4\tau w_4 \frac{\lambda^2}{\kappa^2}
\right] .
\end{equation}
The small quantity $\Delta$, reflecting the deviation from the elastic
value due to the Fermi motion of the nucleons, is $U_{CC}-U_{CC}^0=
(\kappa^2/\tau) w_2 \Delta$. Therefore
\begin{equation}
U_{CC} = U_{CC}^0 + U_{CC} - U_{CC}^0 = 
\frac{\kappa^2}{\tau}
\left[-w_1+(1+\tau)w_2+4\tau w_4 \frac{\lambda^2}{\kappa^2} + w_2\Delta
\right] .
\end{equation}
Finally, we replace the values of the $w_i$ structure functions, given
in Eqs.  (\ref{w1}--\ref{w4}), obtaining
Eqs. (\ref{ucc}--\ref{amaro-uccnc}).

Proceeding similarly to the $R_{CC}$ responses, we find that all the responses
$R_{CL}$,  $R_{LL}$, $R_T$ and $R_{T'}$ can be factorized as in
Eq. (\ref{amaro-factorize}),
\begin{equation}
R_K =  \frac{1}{k_F}\frac{{\cal N}\xi_F}{\eta_F^2 \kappa} f(\psi) U_K  .
\end{equation}
The  functions $U_{CL}$ and
$U_{LL}$ can be obtained more easily by writing them
as sums of two pieces, coming from the conserved and non-conserved
parts of the hadron tensor. First note that
\begin{equation}
U_{CC}= 
(U_{CC})_{c.}
+ (U_{CC})_{n.c.} \, ,
\end{equation}
where the conserved and non conserved parts are
\begin{eqnarray}
(U_{CC})_{c.} &=&
\frac{\kappa^2}{\tau}
\left[-w_1+(1+\tau)w_2
+ w_2\Delta
\right]
\\
(U_{CC})_{n.c.} &=&
4\lambda^2  w_4 .
\end{eqnarray}
Note that the non conserved part $(U_{CC})_{n.c.}$ is precisely the
term of $r_{CC}$, in Eq. (\ref{rcc-sn}), containing the $w_4$
structure function. This is so because this single-nucleon
contribution is independent of $\varepsilon$ and it factorizes out of
the energy integration.  The same can be said for the non conserved
parts of $r_{CL}$ and $r_{LL}$ (Eqs. (\ref{rcl-sn},\ref{rll-sn}).

Consequently, applying current conservation to the conserved part, we
can write
\begin{eqnarray}
(U_{CL})_{c.} 
&=& -\frac{\lambda}{\kappa}
(U_{CC})_{c.} 
\\
  (U_{LL})_{c.} 
&=& \frac{\lambda^2}{\kappa^2}
(U_{CC})_{c.} 
\end{eqnarray}
On the other hand for the non conserved part due to the factorization
property
\begin{eqnarray}
 (U_{CL})_{n.c.} &=&  -4\lambda\kappa w_4 \\
 (U_{LL})_{n.c.} &=&  4\kappa^2 w_4. \\
 \end{eqnarray}

Proceeding in the same lines for the $T$ response
we obtain
\begin{eqnarray}
U_T  &=& 
2w_1 - w_2 \frac{\kappa^2+\tau^2}{\kappa^2} 
+ w_2 \frac{2\lambda\tau}{\kappa^2}
\left[1+\frac12 \xi_F(1+\psi^2)\right]
+ w_2 \frac{\tau}{\kappa^2}
\left[1+\xi_F(1+\psi^2)+ \frac13 \xi_F^2(1+\psi^2+\psi^4)\right]
\\
&=& 2w_1 +w_2\Delta \, ,
\end{eqnarray}
where the same term $\Delta$ as before is obtained as the deviation
from the elastic limit for $k_F=0$.

Finally, the $T'$ response gives
\begin{eqnarray}
U_{T'} &=& 
2 w_3 \frac{\tau}{\kappa} \left[ 1+\lambda + \frac12\xi_F(1+\psi^2) \right]
\nonumber
\\
&=&
2w_3  \sqrt{\tau(\tau+1)} \left[1+\widetilde{\Delta}\right] \, ,
\end{eqnarray}
where $\widetilde\Delta$ was defined in Eq. (\ref{amaro-deltatilde})
and results from adding and subtracting the elastic limit for $k_F=0$
and $\lambda=\tau$.

\bibliography{biblio}


\end{document}